\documentclass[a4paper,11pt]{scrbook}

\usepackage[dvips]{graphics}
\usepackage{epsfig}
\usepackage{axodraw}

\usepackage{amsmath, amsthm, amsfonts, amssymb, bbm}
\usepackage[mathscr]{eucal}

\theoremstyle{plain}
\newtheorem{proposition}{Proposition}[section]

\newtheorem{lemma}[proposition]{Lemma}

\theoremstyle{definition}

\newtheorem{example}[proposition]{Example}

\newtheorem{remark}[proposition]{Remark}

\newcommand{\ud}{\mathrm{d}}
\newcommand{\del}{\partial}
\newcommand{\bra}[1]{\langle #1 |}
\newcommand{\ket}[1]{| #1 \rangle}

\newcommand{\skal}[2]{\langle #1 , #2 \rangle}

\DeclareMathOperator{\tr}{tr}
\DeclareMathOperator{\Tr}{Tr}

\DeclareMathOperator{\si}{si}
\DeclareMathOperator{\ci}{ci}
\DeclareMathOperator{\supp}{supp}
\DeclareMathOperator{\id}{id}
\DeclareMathOperator{\ad}{ad}

\newcommand{\V}[1]{\mathbf{#1}}

\newcommand{\order}{\mathcal{O}}

\newcommand{\M}{\mathcal{M}}
\newcommand{\schw}{\mathcal{S}}
\newcommand{\stet}{\mathcal{L}}
\newcommand{\R}{\mathbb{R}}
\newcommand{\1}{\mathbbm{1}}

\newcommand{\A}{\mathscr{A}}
\newcommand{\E}{\mathscr{E}}

\newcommand{\Hom}{\text{Hom}}
\newcommand{\End}{\text{End}}
\newcommand{\Der}{\text{Der}}

\newcommand{\F}{\mathcal{F}}

\newcommand{\tensor}{\otimes}
\newcommand{\betrag}[1]{\left| #1 \right|}
\newcommand{\norm}[1]{\left\| #1 \right\|}

\newcommand{\pv}[1]{\mathrm{P} \frac{1}{#1}}
\newcommand{\iep}[1]{\frac{1}{#1+i\epsilon}}

\newcommand{\WDp}[1]{\colon \negthickspace #1 \! \colon \negthickspace }

\renewcommand{\slash}[2][4]{\ensuremath{\rlap{\raisebox{1pt}{$\mspace{#1mu}/$}}#2}}

\newcommand{\dslash}{\slash[2]{\partial}}
\newcommand{\kslash}{\slash[2]{k}}
\newcommand{\pslash}{\slash[2]{p}}
\newcommand{\lslash}{\slash[0]{l}}

\newcommand{\dotalpha}{{\dot \alpha}}
\newcommand{\dotbeta}{{\dot \beta}}

\newcommand{\Vmat}[1]{\underline{#1}}

\begin{document}

\begin{titlepage}
\begin{center}

{ \sf \huge 
Dispersion relations in quantum electrodynamics on the noncommutative Minkowski space \\}

\vspace{5cm}

{ \Large  Dissertation 
\\
 zur Erlangung des Doktorgrades 
\\
 des Departments Physik 
\\
 der Universit\"at Hamburg \\ }

\vspace{4cm}

{ \Large vorgelegt von 
\\
 Jochen Wolfgang Zahn 
\\
 aus Karlsruhe \\ }

\vspace{3cm}

{ \large \bf Hamburg \\ 2006}

\end{center}
\end{titlepage}

\thispagestyle{empty}


\begin{tabular}{ll}
Gutachter der Dissertation: & Prof.~Dr.~K.~Fredenhagen \\
                                             & Prof.~Dr.~V.~Schomerus \\
                                             & Prof.~Dr.~H.~Grosse \\
                                             & \\
Gutachter der Disputation: & Prof.~Dr.~K.~Fredenhagen \\
                                             & Prof.~Dr.~J.~Louis \\
                                             & \\                
Datum der Disputation: & 12.~12.~2006 \\
                                             & \\            
Vorsitzende des Pr\"ufungsausschusses: & Prof.~Dr.~C.~Hagner \\
                                             & \\
Vorsitzender des Promotionsausschusses: & Prof.~Dr.~G.~Huber \\
                                             & \\
Leiter des Departments Physik: & Prof.~Dr.~R.~Klanner \\
				            & \\
Dekan der Fakult\"at f\"ur Mathematik, & \\ Informatik und Naturwissenschaften: & Prof.~Dr.~A.~Fr\"uhwald
\end{tabular}

\newpage

\thispagestyle{empty}

\begin{center}
{\sf \large Zusammenfassung}
\end{center}

\noindent
Wir untersuchen Feldtheorien auf dem nichtkommutativen Minkowskiraum mit nichtkommutierender Zeit. Das Hauptaugenmerk liegt dabei auf Dispersionsrelationen in quantisierten wechselwirkenden Modellen im Yang--Feldman--Formalismus. Insbesondere berechnen wir die Zwei--Punkt--Korrelationsfunktion der Feldst\"arke in der nichtkommutativen Quantenelektrodynamik in zweiter Ordnung. Hierbei ber\"ucksichtigen wir die kovarianten Koordinaten zur Konstruktion lokaler eichinvarianter Gr\"o{\ss}en (Observablen). Es stellt sich heraus, dass dies die bekannten schwerwiegenden Infrarot-Probleme nicht behebt, wie man h\"atte hoffen k\"onnen, sondern im Gegenteil noch verschlimmert, da nichtlokale Divergenzen auftreten. Wir zeigen auch, dass diese Divergenzen in einer supersymmetrischen Version der Theorie wegfallen, wenn die kovarianten Koordinaten entsprechend angepasst werden.

Dar\"uber hinaus untersuchen wir das $\phi^3$-- und das Wess--Zumino--Modell und zeigen, dass die Verzerrung der Dispersionrelation moderat ist f\"ur Parameter die typisch f\"ur das Higgs--Feld sind. Wir diskutieren auch die Formulierung von Eichtheorien auf nichtkommutativen R\"aumen und betrachten klassische Elektrodynamik auf dem nichtkommutativen Minkowskiraum unter Verwendung kovarianter Koordinaten. Insbesondere berechnen wir die \"Anderung der Lichtgeschwindigkeit durch nichtlineare Effekte bei Anwesenheit eines Hintergrundfeldes. Schliesslich untersuchen wir den sogenannten Twist--Ansatz f\"ur Quantenfeldtheorien auf dem nichtkommutativen Minkowskiraum und weisen auf einige konzeptionelle Probleme dieses Ansatzes hin.

\begin{center}
{\sf \large Abstract}
\end{center}

\noindent
We study field theories on the noncommutative Minkowski space with noncommuting time. The focus lies on dispersion relations in quantized interacting models in the Yang--Feldman formalism. In particular, we compute the two--point correlation function of the field strength in noncommutative quantum electrodynamics to second order. At this, we take into account the covariant coordinates that allow the construction of local gauge invariant quantities (observables). It turns out that this does not remove the well--known severe infrared problem, as one might have hoped. Instead, things become worse, since nonlocal divergences appear. We also show that these cancel in a supersymmetric version of the theory if the covariant coordinates are adjusted accordingly.

Furthermore, we study the $\phi^3$ and the Wess--Zumino model and show that the distortion of the dispersion relations is moderate for parameters typical for the Higgs field. We also disuss the formulation of gauge theories on noncommutative spaces and study classical electrodynamics on the noncommutative Minkowski space using covariant coordinates. In particular, we compute the change of the speed of light due to nonlinear effects in the presence of a background field. Finally, we examine the so--called twist approach to quantum field theory on the noncommutative Minkowski space and point out some conceptual problems of this approach.

\newpage

\

\newpage

\pagenumbering{roman}

\tableofcontents

\newpage

\

\newpage

\pagenumbering{arabic}

\chapter{Introduction}
\label{chapter:Introduction}

In recent years, the study of noncommutative spacetimes has attracted a lot of attention among theoretical physicists. The earliest such model was introduced already 60 years ago by Snyder~\cite{Snyder}. The main motivation was to cure the bad ultraviolet behavior of quantum field theories. However, because of the success of renormalization techniques, it was soon forgotten.


In the 1990s, Connes and Lott showed that the standard model of elementary particle physics, in particular the Higgs sector, can be naturally described in the setting of noncommutative geometry \cite{Connes}. However, these models were almost commutative in the sense that one replaced functions on spacetime by matrix-valued functions.

Another approach, due to Doplicher, Fredenhagen and Roberts \cite{DFR}, is to discuss the operational meaning of events in spacetime. Taking quantum field theory and general relativity as a starting point, one can do the following Gedanken experiment:
Suppose we want to measure  the coordinates of an event with very high precision. Because of Heisenbergs uncertainty relation, we have to concentrate a high amount of energy-momentum in the corresponding spacetime region.
Assuming general relativity to be correct on such small scales, this will create a black hole if the typical spatial extension of the region is the Planck length $\lambda_P$. But this will prevent us from capturing any signal from that region. In this sense, the usual description of spacetime loses its operational meaning at the Planck scale. However, a more refined analysis shows that the measurement of a single coordinate is not restricted. Instead, the following space-time uncertainty relations were found~\cite{DFR}:
\begin{subequations}
\begin{align}
\label{eq:STUR1}
  \Delta x^0 \sum_i \Delta x^i & \geq \lambda_P^2, \\
\label{eq:STUR2}
  \sum_{i<j} \Delta x^i \Delta x^j & \geq \lambda_P^2.
\end{align}
\end{subequations}
A natural way to implement such uncertainty relations is to introduce noncommuting coordinates $q^{\mu}$:
\begin{equation}
\label{eq:qqQ}
  [q^\mu, q^\nu] = i Q^{\mu \nu}.
\end{equation}
As shown in~\cite{DFR}, the space-time uncertainty relations~(\ref{eq:STUR1},b) are fulfilled if the commutator $Q^{\mu \nu}$ fulfills the so-called quantum conditions
\begin{subequations}
\begin{align}
\label{eq:QC1}
  Q_{\mu \nu} Q^{\mu \nu} & = 0, \\
\label{eq:QC2}
 \left( \epsilon_{\mu \nu \lambda \rho} Q^{\mu \nu} Q^{\lambda \rho} \right)^2 & = 16 \lambda_P^8, \\
\label{eq:QC3}
  [q^\mu, Q^{\lambda \rho}] & = 0.
\end{align}
\end{subequations}
Since the commutator transforms as a tensor, these conditions are invariant under the full Lorentz group.

Another motivation for the study of noncommutative spaces comes from string theory.
As was shown in~\cite{Tori}, noncommutative tori appear naturally in compactifications of matrix theories.
In~\cite{Chu,Schomerus} it was proven that the coordinates of the endpoints of an open string, confined to lie on a worldvolume D-brane, become noncommutative upon switching on a background $B$-field perpendicular to that brane. This was further elaborated in \cite{SW}.
In particular, it was shown that in a certain limit the string theory reduces to field (gauge) theory on a noncommutative spacetime.
The noncommutativity found in these models is of the form~(\ref{eq:qqQ}), with $Q$ replaced by an antisymmetric matrix $\theta$. Obviously, the introduction of a background field breaks Lorentz invariance.

Having a noncommutative spacetime, it is natural to study dynamics on it, i.e., to do (quantum) field theory on it. One hope was that the fuzzyness of noncommutative spaces regularizes the ultraviolet divergences inherent in quantum field theory. While this was shown to be the case for some compact noncommutative spacetimes (see, e.g., \cite{ChaichianDivergence}), this hope was not fulfilled in general, in particular not on the noncommutative Minkowski space with central commutator $[q^\mu, q^\nu] = i \sigma^{\mu \nu}$. It was proven by Filk~\cite{Filk}, that only a subclass of the graphs, the so--called nonplanar graphs, are regularized. It was later shown in~\cite{Minwalla} that these graphs lead to a strange phenomenon called UV/IR--mixing: In the nonplanar graphs the external momentum serves as a regulator. Thus, in the limit of vanishing external momentum, one recovers a divergence. In this sense a UV--divergence has been converted into an IR--divergence. This effect is not only a threat to renormalizability, but can also lead to severe distortions of the dispersion relations~\cite{Matusis}.  

These studies, however, used the set of modified Feynman rules proposed in \cite{Filk}. While these can be derived formally in a Euclidean path integral formalism, the connection to the Lorentzian signature is unclear in the case of space/time noncommutativity $\sigma^{0i} \neq 0$. In fact it was shown in \cite{GomisMehen} that in this case the modified Feynman rules lead to a violation of unitarity. This was often taken as an argument to study only space/space noncommutativity $\sigma^{0i}=0$.\footnote{This argument was supported by the fact that a field theory on a spacetime with space/time noncommutativity can not be obtained as a limit of string theory in the presence of a background electromagnetic field \cite{EField1, EField2}, contrary to the case of space/space noncommutativity.} However, for the space-time uncertainty relation (\ref{eq:STUR1}) it is of course crucial to have $\sigma^{0i} \neq 0$. Furthermore, it has been shown in \cite{BDFP02} that the violation of unitarity is due to an inappropriate time--ordering and does not occur if one uses the Hamiltonian or Yang--Feldman formalism.

It turns out that also in these formalisms a crucial effect of the noncommutativity is the distortion of dispersion relations in interacting models. This thesis is mainly concerned with the investigation of this effect for various models in the Yang--Feldman formalism. It was shown in \cite{DorosDiss} that in the $\phi^4$ model the distortion affects mainly the infrared and is so strong that realistic dispersion relations require a mass close to the noncommutativity scale. Here we will show that in the $\phi^3$ and the Wess--Zumino model the deviations are moderate (at the \%-level) for parameters typical for the Higgs field. The reason is that these models are only logarithmically divergent. But since the dispersion relation of the photon is known very accurately and the formulation of electrodynamics is highly constraint by the gauge principle, quantum electrodynamics is the ultimate testbed for quantum field theory on the noncommutative Minkowski space (NCQFT).


As a consequence of the noncommutativity, pure electrodynamics on the noncommutative Minkowski space is self--interacting. There are two main approaches to treat this model (and other gauge theories). In the approach via the Seiberg-Witten map~\cite{SW}, fields and interactions are expanded in a formal power series in the commutator $\sigma$. At each finite order in $\sigma$, the theory is local, and can be treated as field theory on the ordinary Minkowski space.

In the unexpanded approach, one treats the model as other noncommutative field theories and has to cope with the problems mentioned above. However, there is an additional difficulty to define local gauge invariant quantities, i.e., observables. This can be done using covariant coordinates~\cite{Madore}.

The self--interaction of pure electrodynamics on the noncommutative Minkowski space leads to a change of the dispersion relation in the presence of a background electromagnetic field, already at the classical level. We will investigate this effect in the formalism of covariant coordinates. It turns out that one obtains the same result as in the approach via the Seiberg-Witten map. However, the effect seems to be far too small to be measurable.

For the quantum theory (NCQED) it was shown in \cite{Wulkenhaar} that in the Seiberg-Witten approach the model is not renormalizable, already at first order in~$\sigma$.
In the unexpanded approach, on the other hand, the self--interaction leads to a quadratically IR--divergent photon self--energy \cite{Hayakawa}.
The resulting distortion of the dispersion relation is so strong that the model is ruled out. However, this calculation has two shortcomings. It was done with the modified Feynman rules and is thus not valid for the case of space/time noncommutativity.
Furthermore, it was not dealing with gauge invariant quantities, since the covariant coordinates were not used.
It is conceivable that these cure the bad infrared behaviour indicated above. One of the main goals of this thesis is to check this. Unfortunately, it turns out that the opposite is true: The covariant coordinates bring in even more dangerous terms, in particular nonlocal divergences.

It is well known that the quadratic infrared divergences mentioned above disappear in a supersymmetrized version of the model, at least in the setting of the modified Feynman rules. We will show that this is also true in the Yang--Feldman formalism. Furthermore, we show that if one uses covariant coordinates for observables that are invariant under supersymmetry transformations, (most of) the nonlocal divergences from the covariant coordinates are cancelled, too.

A distortion of the dispersion relation is also a threat to Lorentz invariance. Following \cite{Colladay,LorentzSymmetry} we distinguish observer and particle Lorentz transformations. The latter correspond to the observation of particles with different momenta in the same reference frame. The distortion of the dispersion relations due to quantum effects shows that particle Lorentz invariance is broken. Observer Lorentz transformations, on the other hand, correspond to a change of the reference frame. There, one also has to transform any background fields, in the present case the spacetime commutator $Q^{\mu \nu}$ (or the matrix $\theta^{\mu \nu}$ in the string inspired models). Invariance under these transformations is not broken. However, in the case of a matrix $\theta^{\mu \nu}$ as commutator, one would then have to live with the fact that it takes different values for different observers. As a way to avoid this, it has been proposed to twist the coproduct of the Poincar\'e group such that $\theta$ stays constant. This so-called twist approach has become quite popular recently, see, e.g., \cite{WessEtAl, Alvarez-GaumeMeyer, Chaichian2}.
In particular, there have been claims about the absence of UV/IR--mixing and a violation of Pauli's principle in this setting~\cite{Balachandran, BalachandranUVIR}. We will also critically review this approach.

This thesis is organized as follows: In Chapter~\ref{chapter:NCMink}, we introduce the mathematical description of the noncommutative Minkowski space. We also review some aspects of classical field theory on it and discuss the different approaches towards quantization. In Chapter~\ref{chapter:TwistedNCQFT}, we discuss how to embed twisted NCQFT into a general framework for the quantization of systems with twisted symmetries. We point out some conceptual difficulties of this approach and in particular comment on claims about the violation of the Pauli principle and the absence of UV/IR--mixing in this setting. In Chapter~\ref{chapter:NCGaugeTheory}, we review the different approaches to formulate gauge theories on noncommutative spaces. We also study classical electrodynamics on the noncommutative Minkowski space in the unexpanded approach. Chapter~\ref{chapter:NCQFT} deals with NCQFT in the Yang--Feldman formalism. We discuss the adiabatic limit and how to compute the distortion of the dispersion relations in interacting models in this approach. With these tools, we treat the $\phi^3$ and Wess--Zumino model at the one-loop level and show that the distortion of the dispersion relation is moderate. In Chapter~\ref{chapter:NCQED}, we study NCQED in the Yang--Feldman approach at second order. We show that the quadratic infrared divergences known from the treatment with the modified Feynman rules also show up in this approach. We interpret them in a new way by proving that they require nonlocal counterterms. We also show that the covariant coordinates bring in new nonlocal divergences. In Chapter~\ref{chapter:SNCQED}, we investigate whether in a supersymmetrized version of NCQED the nonlocal divergences are absent. We show that this is indeed the case if one uses covariant coordinates for observables that are invariant under supersymmetry transformations. The only effect of the noncommutativity is then a momentum--dependent field strength renormalization. It can be interpreted as giving rise to acausal effects. We conclude with a summary and an outlook.

\chapter{The noncommutative Minkowski space and field theory}
\label{chapter:NCMink}

This chapter provides an introduction to the mathematical description of the noncommutative Minkowski space. Furthermore, we review aspects of classical field theory on the noncommutative Minkowski space and discuss the different approaches towards quantization.
We start with a very brief review of the framework of noncommutative geometry.

\section{Noncommutative Geometry}

One of the cornerstones of noncommutative geometry is the Gelfand--Naimark theorem. For a (locally) compact Hausdorff space $X$, the algebra $\mathcal C_0(X)$ of continuous functions (vanishing at infinity), equipped with pointwise multiplication as product, the usual involution and the supremum norm, is a (non-) unital commutative $C^*$-algebra. On the other hand, the theorem of Gelfand and Naimark (cf., e.g.,~\cite[Thm. 1.4]{ElementsOfNCG}) states that for each (non-) unital, commutative $C^*$-algebra $\A$ there is a  (locally) compact Hausdorff space $X_\A$ such that $\A$ is isomorphic to $\mathcal C_0(X_\A)$.

It is thus natural to interpret the elements of a (non-) unital, noncommutative $C^*$-algebra $\A$ as ``functions'' (vanishing at infinity) on a noncommutative space. This idea proved to be very fruitful and was the starting point for the field of noncommutative geometry. We will see in the next chapter how the Serre--Swan theorem motivates the definition of vector bundles over noncommutative spaces. The spectral triples introduced by Connes~\cite{Connes} also allow the definition of a metric on noncommutative spaces. However, this only works for compact and riemannian spaces. Obviously, both restrictions are problematic if one wants to construct a realistic field theory. There are promising attempts to overcome these restrictions~\cite{Moretti, PaschkeVerch}. We will not pursue them here and introduce the metric on the noncommutative Minkowski space in a rather ad hoc way.

\section{The noncommutative Minkowski space $\E$}
\label{sec:E}

Now we want to describe the noncommutative Minkowski space introduced in~\cite{DFR} in more detail. In particular, in view of the discussion in the previous section, it is desirable to introduce a $C^*$-algebra $\E$ to which the coordinates $q^\mu$ are affiliated\footnote{From the commutation relations it follows that the coordinates $q^\mu$ are unbounded, thus, they can not be elements of $\E$. However, they will be affiliated to $\E$ in the sense explained in~\cite[App. A]{DFR}.}.

We start by looking at the joint spectrum $\Sigma$ of the commutators $Q^{\mu \nu}$. 
Since the commutators are central, cf. (\ref{eq:QC3}), elements $\sigma \in \Sigma$ are real antisymmetric matrices. Expressing these in the usual way by their ``electric'' and ``magnetic'' parts $\V e$ and $\V m$, the quantum conditions~(\ref{eq:QC1}) and~(\ref{eq:QC2}) can be written as
\begin{align*}
  \betrag{\V e}^2 - \betrag{\V m}^2 & = 0, \\
  \V e \cdot \V m & = \pm \lambda_{nc}^4.
\end{align*}
Here we replaced the Planck length by a length scale $\lambda_{nc}$. 
It is easy to see that all such matrices can be reached from
\begin{equation}
\label{eq:sigma_0}
  \sigma_0 = \lambda_{nc}^2 \begin{pmatrix} 0 & - \1 \\ \1 & 0 \end{pmatrix}
\end{equation}
by proper orthochronous Lorentz transformation and the parity operator. The matrix $\sigma_0$ is, up to the scale $\lambda_{nc}$, just the symplectic matrix for an ordinary point particle in two spatial dimensions. It is thus natural to proceed as in quantum mechanics and require, instead of~(\ref{eq:qqQ}), the Weyl relation
\begin{equation}
\label{eq:Weyl}
  e^{ikq} e^{ipq} = e^{i(k+p)q} e^{-\frac{i}{2} k Q p},
\end{equation}
where we used the notation $k Q p = k_\mu Q^{\mu \nu} p_\nu$.
One can now proceed to define functions of the coordinates $q^\mu$, $Q^{\mu \nu}$. Let $f(\sigma, x)$ be such that $\sigma \mapsto f(\sigma, \cdot)$ is a continuous map, vanishing at infinity, from $\Sigma$ to ${\cal F}(L^1(\R^4))$. Here ${\cal F}$ denotes Fourier transformation. Then we define
\begin{equation*}
  f(Q,q) = (2\pi)^{-2} \int \ud^4k \ \check f(Q,k) e^{ikq}. 
\end{equation*}
Here $\check f(\sigma, \cdot)$ is the inverse Fourier transform of $f(\sigma, \cdot)$. 
The function $f(\sigma, x)$ is also called the \emph{Weyl symbol} of $f(Q,q)$.
Using~(\ref{eq:Weyl}), the multiplication of two such ``functions'' is given by
\begin{equation*}
  f(Q, q) h(Q,q) = \widehat{(\check f \times_Q \check h)}(Q,q)
\end{equation*}
where the hat denotes Fourier transformation and
\begin{equation}
\label{eq:TwistedConvolution}
  (\check f \times_\sigma \check h)(\sigma, k) = (2\pi)^{-2} \int \ud^4l \ \check f(\sigma, l) \check h(\sigma, k-l) e^{\frac{i}{2} k \sigma l}.
\end{equation}
The product $\times_\sigma$ is called the \emph{twisted convolution}. Defining an involution and a norm by
\begin{align*}
  f(\sigma, q)^* & = \bar{f}(\sigma, q), \\
  \norm{f} & = \sup_{\sigma \in \Sigma} \norm{\check f(\sigma, \cdot)}_{L^1},
\end{align*}
one obtains a Banach $*$-algebra $\E_0$. One can now show~\cite[Thm. 4.1.]{DFR} that there is a unique $C^*$-norm on $\E_0$. The completion of $\E_0$ in this norm is called $\E$ and is isomorphic to $\mathcal{C}_0(\Sigma, \mathcal K)$, where $\mathcal K$ is the algebra of compact operators.

The action $\tau$ of the Poincar\'e group on functions of the noncommutative coordinates is defined by
\begin{equation}
\label{eq:Poincare}
  (\tau_{(a, \Lambda)} f)(\sigma, q) = f( \Lambda^{-1} \sigma \Lambda^{-1 T}, \Lambda^{-1} (q-a)).
\end{equation}

The set $I_\sigma = \{ f \in \E | f(\sigma) = 0 \}$ is a closed two-sided ideal in~$\E$. Thus, the quotient algebra $\E_\sigma = \E / I_\sigma$ is also a $C^*$-algebra~\cite[Prop. 2.2.19]{BratteliRobinson}.
It is isomorphic to $\mathcal K$ and can also be obtained by a $C^*$-completion of the algebra ${\cal F}(L^1(\R^4))$ using the Fourier transform of the twisted convolution $\times_\sigma$ as the product. This algebra, upon replacing $\sigma$ by $\theta$, describes the noncommutative Minkowski space arising in string theory.

Since the product is point-wise in $\Sigma$ and we do not introduce any kinetic term involving derivatives in $\Sigma$, the dynamics of fields does not take place in $\Sigma$. For the purpose of field theory, it is thus sufficient to work at a fixed $\sigma$, since results obtained there can be easily carried over to the whole of $\Sigma$ using (\ref{eq:Poincare}).\footnote{Note, however, that in the Hamiltonian approach to NCQFT, which is discussed in the next section, the situation is different.} As already noted, $\sigma_0$ is the symplectic matrix of a point particle in two dimensions. Thus, because of von Neumann's uniqueness theorem, all finite dimensional irreducible representations of $\E_{\sigma_0}$ are isomorphic to the Schr\"odinger representation
\begin{equation}
\label{eq:Schroedinger}
  \pi(q^0) = P_1, \quad \pi(q^1) = P_2, \quad \pi(q^2) = Q_1, \quad \pi(q^3) = Q_2.
\end{equation}
Here $P_i$ and $Q_i$ are the usual momentum and position operators in two dimensions.

Although it is nice, in particular to make contact with the framework of noncommutative geometry, to have a $C^*$-algebra at our disposal, it is sometimes advantageous to use two other algebras. We introduce them here for $\sigma = \sigma_0$, but it is possible to do this for other $\sigma$'s and glue these together to get an analog of $\E$.
In the representation~(\ref{eq:Schroedinger}), the coordinates are represented as continuous operators on the Schwartz space $\schw(\R^2)$. It is thus natural to consider the algebra $\stet(\schw(\R^2))$ of continuous operators on $\schw(\R^2)$. Using the standard $L^2(\R^2)$ scalar product, one can define an adjoint. Then $\M = \stet(\schw(\R^2)) \cap \stet(\schw(\R^2))^*$ is a $*$-algebra. It can be equipped with a locally convex topology, in which it is complete. Details can be found in~\cite{Varilly,Diplom}. This algebra has the advantage of containing the coordinates and plane waves. It has been used in~\cite{Diplom} as the algebra of classical fields.

Furthermore, one can consider the algebra of continuous operators on $\schw(\R^2)$ whose integral kernel is in $\schw(\R^4)$. In fact these are in one--to--one correspondence with the elements of $\E_{\sigma_0}$ whose Weyl symbols are Schwartz functions~\cite{Varilly}. We call this algebra $\schw_2$. It is a dense $*$-ideal of $\M$ and can be equipped with a locally convex topology. Again, details can be found in~\cite{Varilly,Diplom}.

Finally, we note that for analytic functions $f$ and $h$, the Fourier transform of the twisted convolution is given by the Moyal $\star$-product:
\begin{equation}
\label{eq:Star_Product}
  f \star_\sigma h = \widehat{\check f \times_\sigma \check h} = (\mu \circ e^{\frac{i}{2} \sigma^{\lambda \nu} \del_\lambda \otimes \del_\nu} ) (f \otimes h).
\end{equation}
Here $\mu$ stands for the pointwise product of the two tensor factors. Sometimes the restriction to analytic functions is dropped and the formula is interpreted as a formal power series in~$\sigma$. This is of course in the spirit of deformation quantization~\cite{BFFLS}.

\subsection{Calculus on $\E$}
\label{sec:Calculus}

Given the action~(\ref{eq:Poincare}) of the Poincar\'e group, it is natural to define partial derivatives of functions of the noncommutative coordinates by
\begin{equation*}
  \del_\mu = - \frac{\ud}{\ud t} \tau_{(t e_\mu,\1)}.
\end{equation*}
This is not well-defined on all elements of $\E$. However, it is well-defined on $\schw_2$ and $\M$, and it is easy to see that
\begin{equation*}
  \del_\mu f(Q,q) = ( \del_\mu f )(Q,q)
\end{equation*}
holds. Furthermore, because of the commutation relation~(\ref{eq:qqQ}), it can be written as
\begin{equation}
\label{eq:del_mu_commutator}
  \del_{\mu} f(\sigma,q) = - i \sigma^{-1}_{\mu \nu} [q^{\nu}, f(\sigma,q)].
\end{equation}
One can also define an ``integral'' for functions of the noncommutative coordinates:
\begin{equation}
\label{eq:Integral}
  \int \ud^4q \ f(q) = (2\pi)^2 \check f (0).
\end{equation}
This is well-defined on $\schw_2$, but not on all elements of $\E_\sigma$ and $\M$. It is easy to check that this map is cyclic, i.e., a trace, and positive. Furthermore, using (\ref{eq:TwistedConvolution}), it is straightforward to show that
\begin{equation}
\label{eq:IntegralIdentity}
  \int \ud^4q \ f(q) h(q) = \int \ud^4x \ f(x) h(x)
\end{equation}
holds for $f, h \in \schw(\R^4)$. If we consider elements of the full algebra, i.e., functions that depend also on $\sigma \in \Sigma$, the integral is a map to the continuous functions on $\Sigma$. Finally, it possible to define a ``spatial integral'' at fixed time~$t$:
\begin{equation}
\label{eq:Spatial_Integral}
  \int_{q^0=t} \ud^3q \ f(q) = 2 \pi \int \ud k_0 \ \check{f}((k_0, \V 0)) e^{i k_0 t}.
\end{equation}
This map is well-defined on $\schw_2$ and positive, but not cyclic. On the full algebra, also this integral is a map to the continuous functions on~$\Sigma$.

\subsection{Noncommutative superspace}
\label{sec:SUSYMink}
Since we will investigate supersymmetric models in this thesis, we introduce the supersymmetric noncommutative Minkowski space. The easiest way to introduce supersymmetry is to add the usual anticommuting coordinates $\theta^\alpha$ and $\bar \theta^\dotalpha$ and postulate the (anti-) commutation relations
\begin{subequations}
\begin{gather}
\label{eq:NCSUSYMink1}
  \{ \theta^\alpha, \theta^\beta \} = \{ \bar \theta^\dotalpha, \theta^\alpha \} = \{ \bar \theta^\dotalpha, \bar \theta^\dotbeta \} = 0, \\
\label{eq:NCSUSYMink2}
  [q^\mu, \theta^\alpha] = [q^\mu, \bar \theta^\dotalpha] = 0.
\end{gather}
\end{subequations}
These are the relations that are mostly used in the literature. In~\cite{ChuZamora} it has been shown that these (anti-) commutation relations arise in superstring theory on D-branes in the presence of a background $B$-field. Noncommutative superspaces where the $\theta$'s no longer anticommute have been considered for example in~\cite{FerraraLledo, Seiberg}.

Calculus on the noncommutative superspace is very similar to the usual calculus on superspace. Dropping the dependence on $\sigma$, we can write ``superfunctions'' on it as $f(q, \theta, \bar \theta)$. Because of~(\ref{eq:NCSUSYMink1}), an expansion in $\theta$ stops at $\theta^2 \bar \theta^2$ (our conventions for the handling of $\theta$s are summarized in Appendix~\ref{app:SUSY}). The ``coefficients'' at each order of $\theta$ are then simply functions of the noncommutative coordinates $q^\mu$. Because of~(\ref{eq:NCSUSYMink2}), the product of two such superfunctions can be calculated easily.

In order to write down actions or observables, one also needs an integral on noncommutative superspace. Integrals over the $\theta$'s are defined by
\begin{equation*}
  \int \ud^2 \theta \ (a + b^\alpha \theta_\alpha + c \theta^2 ) = c,
\end{equation*}
and analogously for $\bar \theta$. We also introduce the notation
\begin{equation*}
  \int \ud^6q  = \int \ud^4q \ud^2\theta, \quad \int \ud^8q  = \int \ud^4q \ud^2\theta \ud^2 \bar \theta.
\end{equation*}

\section{Field theory}

The aim of this section is to give a brief introduction to field theory on the noncommutative Minkowski space. We start by considering classical fields, discuss the question of current conservation, and finally introduce free quantum fields.

\subsection{Classical fields}
\label{sec:ClassicalFields}

Having partial derivatives and an integral at our disposal, we can write down an action, e.g.,
\begin{equation}
\label{eq:Phi4}
  \int \ud^4q \ \left\{ \frac{1}{2} \del_\mu \phi^* \del^\mu \phi - \frac{m^2}{2} \phi^* \phi - \frac{\lambda}{4} \phi^* \phi \phi^* \phi + f^* \phi + f \phi^* \right\}.
\end{equation}
It has been shown in~\cite{Diplom} how to rigorously formulate an action principle. The upshot is that the resulting equations of motion are just what one naively expects (of course one has to take care about the order of the fields and use the cyclicity of the integral). Thus, from the above action, one obtains the equation of motion
\begin{equation*}
  ( \Box + m^2 ) \phi + \lambda \phi^* \phi \phi^* = f.
\end{equation*}
Solutions can be constructed quite analogously to the commutative case: Assuming $\lambda = 0$, the retarded solution to the above equation is
\begin{equation}
\label{eq:Convolution}
  \phi = \Delta_R \times f := \int \ud^4x \ \Delta_R(x) f_x.
\end{equation}
Here $\times$ stands for convolution and the subscript $x$ for translation, i.e., $f(q)_x = f(q-x)$. It has been shown in~\cite{Diplom} that for $f \in \schw_2$, the above is well-defined and $\phi \in \M$.

A strange phenomenon occurring in interacting field theories on the noncommutative Minkowski space is the nonconservation of local currents. As an example, consider the above action for $f = 0$. The usual energy-momentum tensor in this case would be
\begin{equation*}
  T_\nu^\mu = \frac{1}{2} ( \del_\nu \phi^* \del^\mu \phi + \del^\mu \phi^* \del_\nu \phi ) - \delta_\nu^\mu L.
\end{equation*}
Here $L$ is the Lagrangean, i.e., the expression in curly brackets in (\ref{eq:Phi4}).
But using the equation of motion, we find
\begin{equation*}
  \del_\mu T_\nu^\mu = \frac{\lambda}{2} [\phi^* \del_\nu \phi - \del_\nu \phi^* \phi, \phi^* \phi].
\end{equation*}
The right hand side does not vanish in general, so $T^\mu_\nu$ is not conserved.
This phenomenon was first observed in~\cite{Micu}, see also~\cite{EIT,Diplom}.
Note, however, that not all currents are affected. In the model above, e.g., the current $j_\mu = - i \phi^* \del_\mu \phi + i \del^\mu \phi^* \phi$ is still conserved. The same is true for the current in electrodynamics on the noncommutative Minkowski space \cite{NCED_EMT}.

We remark that the source term of the currents is always a commutator. In the $\star$-product formulation~(\ref{eq:Star_Product}), such terms can then be written as a divergence:
\begin{equation*}
  f \star h - h \star f = i \sigma^{\mu \nu} \del_{\mu} \left( f \star' \del_\nu h \right).
\end{equation*}
Here we used the antisymmetry of $\sigma$ and
\begin{equation*}
  f \star' h = \left( \mu \circ \frac{\sin \frac{1}{2} \sigma^{\mu \nu} \del_\mu \otimes \del_\nu}{\frac{1}{2} \sigma^{\mu \nu} \del_\mu \otimes \del_\nu} \right) (f \otimes h).
\end{equation*}
This was used in~\cite{Pengpan, AbouZeid} to include the source term of the current into the divergence of the energy-momentum tensor. Thus, one obtains conserved energy momentum tensors. Here, however, we stick to the point of view that the energy-momentum tensor (and other currents) should be local quantities. In the present context, we mean by this that they should be given by ($\star$-) products of (derivatives of) fields. Then one can not use the $\star'$-product and has to live with the fact that some local currents are not conserved. We also remark that, since the source term is always given by a commutator, the nonconservation of the corresponding charge is relevant only at the noncommutativity scale. Furthermore, such an effect is to be expected by heuristic considerations~\cite{Doplicher}:
Charge conservation requires that the production of a particle with positive charge is always accompanied by the production of a particle with opposite charge at the same place. But because of the noncommutativity, it is not possible to localize two particles at the same place, see, e.g., the discussion in~\cite{UVfinite}.
We also remark that in perturbative treatments of quantized theories, one still has energy-momentum conservation at each vertex.
Finally, we mention that similar problems appeared in the context of nonlocal field theories, see, \mbox{e.g., \cite{Pauli, Marnelius}.}

\subsection{Quantum fields}
\label{sec:QuantumFields}

We now come to the formulation of quantum field theory on the noncommutative Minkowski space (NCQFT). Already in~\cite{DFR}, the free scalar field was defined as
\begin{equation*}
  \phi(q) = (2\pi)^{-2} \int \ud^4k \ \hat \phi(k) \otimes e^{-ikq},
\end{equation*}
with $\hat{\phi}(k)$ as usual, i.e., given by
\begin{equation*}
  \hat \phi(k) = (2\pi)^{\frac{1}{2}} \delta(k^2-m^2) \left( \theta(k_0) a(\V{k}) + \theta(-k_0) a^*(-\V k) \right),
\end{equation*}
where $a^*$ and $a$ are the usual creation and annihilation operators. Thus, quantum fields are elements of (or rather affiliated to) $\mathfrak{F} \otimes \E_\sigma$, where $\mathfrak{F}$ is the algebra of operators on the Fock space $\mathcal{H}$. The product of such quantum fields is then given by
\begin{equation}
\label{eq:OldProduct}
  (\phi \otimes f) \cdot (\psi \otimes g) = \phi \psi \otimes f g. 
\end{equation}

Observables can now be constructed using the integral (\ref{eq:Integral}). For example we define, for $f \in \schw_2$,
\begin{equation*}
  \phi(f) = \int \ud^4q \ f(q) \phi(q) = \int \ud^4k \ \hat{f}(-k) \hat \phi(k).
\end{equation*}
This is an element of $\mathfrak F$. More precisely, it is an unbounded operator with invariant dense domain $D \subset \mathcal{H}$, see, e.g., \cite[Section~IX.8]{ReedSimon2}.

We mention that there are several proposals to use different classes of test functions, in particular analytic ones, see, e.g., \cite{Franco, Soloviev}.

\begin{remark}
\label{rem:OldProduct}
In~\cite{Diplom} it was shown that it is natural to use localized interaction terms (or observables) of the form
\begin{equation*}
  \int \ud^4q \ f_1 \phi \dots f_n \phi.
\end{equation*}
Because of the cyclicity of the integral, the classical expression is symmetric under cyclic permutations of the $f_i$. However, this is not the case for the quantized expression if the product~(\ref{eq:OldProduct}) is used. This may be taken as an indication that this product is not always appropriate. We will come back to this point in Chapter~\ref{chapter:NCQED}.
\end{remark}

\begin{remark}
\label{rem:WickProducts}
Regarding the issue of products of quantum fields on the noncommutative Minkowski space, we also note that there are several definitions for Wick products. The simplest choice is to set
\begin{equation*}
  \WDp{\phi^n}(q) = \int \prod_{i=1}^n \ud^4k_i \ \WDp{ \hat \phi(k_1) \dots \hat \phi(k_n)} \otimes e^{-ik_1q} \dots e^{-ik_nq}.
\end{equation*}
While this is well-defined, it has the disadvantage that one also subtracts terms that are finite. Furthermore, these terms are also not local in the sense of the $q$--locality concept introduced in~\cite{Quasiplanar}. It has thus been proposed in~\cite{Quasiplanar} to use so--called quasiplanar Wick products. These arise by the subtraction of only local and infinite terms. In this thesis, we follow the philosophy to only subtract local quantities. However, in the interactions we consider in the following, the distinction between the ordinary and the quasiplanar Wick product is irrelevant. Finally, we note that in~\cite{UVfinite} yet another Wick product, the so--called quantum Wick product, has been introduced. There, the products of fields are defined by evaluating the differences of coordinates in best--localized states. It is particularly suited for the Hamiltonian approach (see below) and has up to now only been used in this context.
\end{remark}

There are several ans\"atze for a perturbative treatment of interacting NCQFT. Unfortunately, they all give different results in the case of space/time noncommutativity $\sigma^{0i} \neq 0$. We will discuss them in the following. The more recent twisted NCQFT is treated in the next chapter.

\subsubsection{The modified Feynman rules}
were first proposed in~\cite{Filk} and can be derived formally from a Euclidean path integral. They amount to using the usual Feynman rules, but to attach to each vertex a phase factor $e^{-\frac{i}{2} \sum_{l<m} k_l \sigma k_m}$, where $k_l$ are the incoming momenta. Thus, this approach is quite easy to handle computationally.

Some of the main features of this approach can be discussed by looking at the fish graph in the $\phi^3$-model. Since the order of the momenta at the vertices is now important, there are two such graphs:
\begin{center}
\begin{picture}(250,50)
\Line(0,25)(20,25)
\Curve{(20,25)(50,5)(80,25)}
\Curve{(20,25)(50,45)(80,25)}
\Line(80,25)(100,25)
\Line(150,25)(170,25)
\Curve{(170,25)(185,10)(200,25)(215,40)(230,25)}
\Curve{(170,25)(185,40)(198,27)}
\Curve{(202,23)(215,10)(230,25)}
\Line(230,25)(250,25)
\end{picture}
\end{center}
Here the momenta at the vertices are assigned clockwise. It is easy to check that in the first graph the phase factors cancel. Thus, it is as in the commutative case. In particular, a mass renormalization is necessary. One speaks of a \emph{planar} graph. In the second graph, on the other hand, the phases add up. One speaks of a \emph{nonplanar} graph. The self-energy for this graph is given by
\begin{equation}
\label{eq:Feynman_Fish}
  ( \hat \Delta_F \times_{2 \sigma} \hat \Delta_F ) (k) = \int \ud^4l \ \frac{1}{l^2+m^2} \frac{1}{(k-l)^2+m^2} e^{i k \sigma l}.
\end{equation}
The phase factor regularizes the integral\footnote{We remark that a rigorous definition of the above integral in the sense of oscillatory integrals has been given only recently~\cite{NCDispRel}.}. The integral is effectively cut off at a momentum scale $((k \sigma)^2)^{-\frac{1}{2}}$. In this way, an original ultraviolet divergence has been converted into an infrared divergence. This is the so-called UV/IR--mixing, which has first been observed in~\cite{Minwalla}. It may spoil renormalizability if such a graph is embedded into another graph, so that one has to integrate over $k$. Furthermore, it leads to a distortion of the dispersion relations~\cite{Matusis}.

While such phenomena also occur in other approaches to NCQFT, the modified Feynman rules have particular drawbacks in the case of space/time noncommutativity: Because of the absence of an Osterwalder--Schrader theorem for the noncommutative spacetime, the relation to the Lorentzian signature is not clear. Obviously, the change $k_0 \to i k_0$, $l_0 \to i l_0$ in~(\ref{eq:Feynman_Fish}) makes the oscillating phase highly divergent if $\sigma^{0 i} \neq 0$. Further arguments against a simple relation between the Euclidean and Lorentzian signature can be found in~\cite{DorosDiss}. Finally, it has been shown in~\cite{GomisMehen} that the modified Feynman rules lead to a nonunitarity $S$-matrix in the case of space/time noncommutativity.

A more recent proposal is to consider self--dual models in the sense of \cite{LangmannSzabo}. This amounts to adding a confining potential to the Lagrangean. In this framework it has been shown that the $\phi^4$ model is renormalizable to all orders \cite{GrosseWulkenhaar}.
But also for these models there is no indication for a connection to the Lorentzian metric.

\subsubsection{The Hamiltonian approach}
has already been proposed in~\cite{DFR}. The idea is to use the spatial integral~(\ref{eq:Spatial_Integral}) to define an interaction Hamiltonian $H_I(t)$. One can then formally define the $S$-matrix as usual, i.e., via the Dyson series:
\begin{equation*}
  S = \text{T} e^{-\frac{i}{\hbar} \int \ud t \ H_I(t)}.
\end{equation*}
This $S$-matrix is formally unitary if the Hamiltonian is symmetric. Thus, there is no problem with violation of unitarity in this approach.

There are, however, many possibilities to define the Hamiltonian. Since the spatial integral maps into the continuous functions on $\Sigma$, one still has to choose some measure on $\Sigma$. Because the Lorentz group is not amenable, one has to break Lorentz invariance at this point. A simple choice would be to pick a single $\sigma$. This was discussed in~\cite{TOPT, Christoph}. Another possibility, proposed in \cite{DFR}, is to smear over $\Sigma_1$, which is the orbit of $\sigma_0$ under rotations and parity. This retains rotational invariance. In \cite{DoroUVFinite} it was shown that this model is ultraviolet finite. In~\cite{UVfinite}, a similar model was discussed, where the interaction term is defined with the quantum Wick product, cf.~Remark~\ref{rem:WickProducts}. This model is also ultraviolet finite, but has some unresolved infrared problems. The distortion of the dispersion relation in this model was investigated in~\cite{Marcel}.

A particular drawback of the Hamiltonian approach is that the interacting field given by the time evolution generated by the Hamiltonian does not fulfill the classical equations of motion at tree level~\cite[Remark 2.2]{DorosDiss}, see also~\cite{Heslop}. This is certainly in conflict with fundamental principles of quantum theory. Of course one can argue that these may have to be altered in a quantum spacetime. However, the classical equations of motion are important for current conservation. Thus, it is not surprising that in NCQED in the Hamiltonian framework the Ward identities are violated at tree level in Compton scattering~\cite{Ohl}.\footnote{Note that this problem is not connected with the phenomenon of current nonconservation discussed in Section~\ref{sec:ClassicalFields}. In the Yang--Feldman approach this problem is absent.}

In~\cite{Heslop}, a variation of this approach was proposed, where time--ordering is defined with respect to light cone coordinates. While Feynman rules can be formulated quite elegantly in this setting, actual computations seem to be rather involved.

\subsubsection{The Yang--Feldman approach}
dates back to the 1950s, and is an attempt to directly use the equations of motion for quantization~\cite{YF,Kaellen}. It can also be employed in situations where a Hamiltonian quantization is problematic. In particular, it was used in the context of nonlocal field theories, see, e.g., \cite{Moeller, Marnelius}. On the ordinary Minkowski space with local interactions, it has not been very successful, since it is combinatorically more difficult than the Feynman rules. Thus, even basic questions about the adiabatic limit or renormalization in this framework have not been discussed\footnote{However, we remark that the related approach via retarded products~\cite{Steinmann} has matured considerable in recent years~\cite{DuetschFredenhagen}.}. Some of these issued will treated in Chapter~\ref{chapter:NCQFT}.

The use of the Yang--Feldman formalism for NCQFT has been proposed in~\cite{BDFP02}. In this approach, the interacting field is hermitean and in this sense no problems with unitarity arise\footnote{The existence of a unitary $S$-matrix is a delicate question however, because the asymptotic outgoing field fulfills different dispersion relations, as we will see in Chapter~\ref{chapter:NCQFT}.}. In~\cite{Quasiplanar} it was shown that the distortion of the dispersion relation in the $\phi^4$-model in this approach is very strong in the infrared. In a sense this had to be expected, since it is a manifestation of the UV/IR--mixing and the $\phi^4$-model is quadratically divergent. One of the main goals of this thesis is to investigate the strength of this effect in other models, namely $\phi^3$, Wess--Zumino and (supersymmetric) NCQED.

\chapter{Twisted NCQFT}
\label{chapter:TwistedNCQFT}


In this chapter, which is based on \cite{TwistedNCQFT}, we study in detail the twist approach to NCQFT. It was proposed as a way to circumvent the breaking of Lorentz invariance that follows from the choice of a particular noncommutativity matrix $\sigma$, and has recently gained a lot of popularity.
It was triggered by the realization that it is possible to twist the coproduct of the universal envelope $U \mathcal{P}$ of the Poincar\'e algebra such that it is compatible with the $\star$-product.  Already in \cite{Watts} it was shown that the $\star$-product naturally arises from a quasitriangular structure in the Hopf algebra corresponding to the translation group. Soon afterwards, it was shown that this quasitriangular structure is generated by a twist \cite{Oeckl}. 
Also the embedding into the (Euclidean) Poincar\'e group was discussed there.
In~\cite{Chaichian, Wess} this was reformulated in dual language for the proper Poincar\'e algebra.
Subsequently, there have been claims about the violation of the Pauli principle \cite{Balachandran} and the absence of UV/IR--mixing \cite{BalachandranUVIR} in this twisted setting. Also an axiomatic characterization of twisted NCQFT was attempted \cite{Chaichian2}.

One aim of this chapter is to reach a more general understanding of QFT in the presence of a twisted symmetry. In this sense, it is related to the field of $q$-deformed quantum mechanics (see, e.g., \cite[Chapter 2]{ChaichianDemichev} for an overview) and the study of quantum systems with quantum symmetry (see, e.g., \cite{PW, Fiore, BraidedQFT,FuzzySphereII}). We follow the philosophy outlined in \cite{WessEtAl}: Each time we encounter a bilinear map involving two spaces carrying a representation of the symmetry group, we deform this map with the twist. 
In the following section, we present the general setup and apply it to NCQFT in Section~\ref{sec:NCQFT}.
In Section~\ref{sec:commutator}, we consider two different commutators that appear naturally in the twisted setting. Unfortunately, both are lacking some crucial properties of the usual commutator.
Thus, we are leaving the safe grounds of established quantum mechanics.
This becomes even more evident in Section~\ref{sec:time-evolution}. There, it is shown that it is in general not possible to add a localized interaction (a source, for example) to the Hamiltonian without getting into serious trouble with the correspondence principle. We argue that this makes it extremely difficult to derive any predictions in the twisted setting (at least none that are not already present in conventional NCQFT). 
In Section~\ref{sec:interactions}, we make some remarks on the effect of the twist in the interacting case and in particular on the claim that the UV/IR--mixing is absent in the twisted setting~\cite{BalachandranUVIR}.


\section{Setup}
\label{sec:setup}

Let $\mu: \schw(\R^4) \otimes \schw(\R^4) \to \schw(\R^4)$ be the point-wise product of Schwartz functions. As is obvious from~(\ref{eq:Star_Product}), the $\star$-product can be defined as\footnote{Note that we use a different notation than in~\cite{WessEtAl}. Our $\F$ corresponds to $\F^{-1}$ there.} $\mu_{\star} = \mu \circ \F$ with
\begin{equation*}
  \F= e^{-\frac{i}{2} \sigma^{\mu \nu} P_{\mu} \otimes P_{\nu}}.
\end{equation*}
In~\cite{WessEtAl}, the twist was interpreted as a formal power series in some deformation parameter.
Here, we define it rigorously by going to momentum space, cf. the twisted convolution~(\ref{eq:TwistedConvolution}).

The Poincar\'e algebra $\mathcal{P}$ can be embedded into the Lie algebra $\Xi$ of vector fields and this in turn, into the algebra $U \Xi$ that is obtained from the universal enveloping algebra by dividing out the ideal generated by the commutation relations in $\Xi$. Following \cite{WessEtAl}, one can equip $U \Xi$ with the structure of a Hopf algebra by defining the coproduct, counit and antipode through
\begin{align*}
  \Delta(u) =& u \otimes 1 + 1 \otimes u & \Delta(1) =& 1 \otimes 1 \\
  \varepsilon(u)  = &0 & \varepsilon(1) = & 1 \\
  S(u)  = & -u & S(1)  = & 1
\end{align*} 
where $u \in \Xi$. This definition can be extended to $U \Xi$ by requiring $\Delta$ and $\varepsilon$ to be algebra homomorphisms and $S$ to be an antialgebra homomorphism. Furthermore, one can give $U \Xi$ a $*$-structure by defining
\begin{equation}
\label{eq:star-structure}
  u^*(f) = (S(u)(f^*))^*
\end{equation}
for $f \in \schw(\R^4)$, $u \in \Xi$ and extending this as an antialgebra homomorphism. Moreover, we note that $\F$ fulfills
\begin{align}
\label{eq:2cocycle}
  (\id \otimes \Delta)\F (\1 \otimes \F) = & (\Delta \otimes \id)\F (\F \otimes \1), \\
\label{eq:counital}
  (\varepsilon \otimes \id) \F = & \1, \\
\label{eq:real}
  (S \otimes S) (\F^{* \otimes *}) = & \F_{21},
\end{align}
where $\F_{21}$ is the transposed $\F$ (in our case $\F_{21} = \F^{-1}$). Thus, $\F^{-1}$ is a real (\ref{eq:real}), counital (\ref{eq:counital}) 2-coclycle (\ref{eq:2cocycle}), see, e.g., \cite{Majid}. From (\ref{eq:2cocycle}) it follows that the $\star$-product is associative, and because of (\ref{eq:real}) it respects the $*$-structure\footnote{Note that the notations $\star$ and $\mu_\star$ are used interchangeably here.}: $(f \star g)^* = g^* \star f^*$.

Now we consider the compatibility of the $\star$-product with Poincar\'e transformations.
Let $\xi \in U \mathcal{P}$. The point-wise product fulfills
\begin{equation}
\label{eq:a_mu}
 \xi \circ \mu = \mu \circ \Delta(\xi).
\end{equation} 
Here we identified $\xi$ with its action on $\schw(\R^4)$.
Now we want to find a deformed coproduct $\Delta_{\star}$ that fulfills $\xi \circ \mu_{\star} = \mu_{\star} \circ \Delta_{\star}(\xi)$. Using (\ref{eq:a_mu}), we obtain
\begin{equation*}
  \xi \circ \mu_{\star} = \mu \circ \Delta(\xi) \circ \F = \mu_{\star} \circ \F^{-1} \circ \Delta(\xi) \circ \F.
\end{equation*}
Thus, one defines
\begin{equation*}
 \Delta_{\star}(\xi) =  \F^{-1} \circ \Delta(\xi) \circ \F.
\end{equation*}
Since $\F \in U \mathcal{P} \otimes U \mathcal{P}$, it is clear that $\Delta_{\star}(\xi) \in U \mathcal{P} \otimes U \mathcal{P}$. In fact one explicitly finds \cite{Chaichian, Wess}
\begin{align}
\label{eq:DeltaP}
  \Delta_{\star}(P_{\mu}) = & P_{\mu} \otimes 1 + 1 \otimes P_{\mu}, \\
  \Delta_{\star}(M_{\mu \nu}) = & M_{\mu \nu} \otimes 1 + 1 \otimes M_{\mu \nu} \nonumber \\
   - & \frac{1}{2} \sigma^{\alpha \beta} \left[ g_{\mu \alpha} \left( P_{\nu} \otimes P_{\beta} -P_{\beta} \otimes P_{\nu} \right) - g_{\nu \alpha} \left( P_{\mu} \otimes P_{\beta} - P_{\beta} \otimes P_{\mu} \right) \right]. \nonumber
\end{align}
There is a general theorem (see, e.g., \cite{Majid}, Thm 2.3.4) stating that $\Delta_\star$, together with $\epsilon_\star = \epsilon$ and $S_\star(\xi) = \chi^{-1} S(\xi) \chi$, where $\chi = S (\F_{(1)}) \F_{(2)}$,\footnote{Here we use the notation $\F = \F_{(1)} \otimes \F_{(2)}$ with an implicit summation.} again define a Hopf algebra. Note that in our particular case $\chi = \1$, i.e., $S_\star = S$. From this and (\ref{eq:real}) it follows that we still have a Hopf $*$-algebra with the old $*$-structure (\cite{Majid}, Prop 2.3.7). Furthermore, one can show \cite{WessEtAl} that there is a triangular structure (or $R$-matrix) $R = \F_{21}^{-1} \F$. In our particular case we have
\begin{equation}
\label{eq:R}
  R = e^{- i \sigma^{\mu \nu} P_{\mu} \otimes P_{\nu}}.
\end{equation}

\begin{remark}
Note that one could also describe the action of a Poincar\'e transformation by a change of the twist $\F$ and thus the product $\mu_{\star}$:
\begin{equation*}
  a \circ \mu_{\star} = \mu \circ \Delta(a) \circ \F = \mu  \circ \Delta(a) \circ \F \circ \Delta(a)^{-1} \circ \Delta(a).
\end{equation*}
Defining $\F' =  \Delta(a) \circ \F \circ \Delta(a)^{-1}$ and $\mu_{\star'} = \mu \circ \F'$, one can write this as
\begin{equation*}
  a \circ \mu_{\star} = \mu_{\star'} \circ \Delta(a).
\end{equation*}
In our concrete case, this would mean that one has to transform $\sigma$ as a tensor, 
as proposed in \cite{DFR} (cf. Section~\ref{sec:E}). 
See also~\cite{LizziVitale,Alvarez-GaumeMeyer}.
\end{remark}

Now suppose we are given vector spaces $A, B, C$ that carry a representation of the Poincar\'e algebra and a map $\nu: A \otimes B \to C$ that is compatible with this action. Then it is in the spirit of \cite{WessEtAl} to deform this map to $\nu_{\star} = \nu \circ \F$. As above, one then has $a \circ \nu_{\star} = \nu_{\star} \circ \Delta_{\star}(a)$. Now consider some special cases: 

\begin{itemize}
\item Let $A$ be an algebra carrying a representation of the Poincar\'e algebra and $\cdot$ the product $A \otimes A \to A$. Applying the above principle, one gets a new algebra $A_{\star}$, being identical to $A$ as a vector space, but with product $\star = \cdot \circ \F$. Due to (\ref{eq:2cocycle}) this product is associative. Note that if $A$ is a $*$-algebra, then, because of (\ref{eq:real}), the new $\star$-product is compatible with the old $*$-structure: $(a \star b)^* = b^* \star a^*$.

\item Let $A$ be an algebra with a representation on a vector space $V$ and $\cdot : A \tensor V \to V$ be the corresponding left action. Applying the above principle, one first deforms $A$ to $A_{\star}$. One then defines the action $\star : A_{\star} \otimes V \to V$ by $\star = \cdot \circ \F$. That this action defines a representation, i.e., $(a \star b) \star v = a \star (b \star v)$, follows again from (\ref{eq:2cocycle}).

\item If $V$ is even a Hilbert space, one should also define a new scalar product that is compatible with the adjoint in $A_{\star}$. The scalar product can be viewed as a bilinear map $\bar{V} \otimes V \to \mathbbm{C}$, where $\bar{V}$ is the conjugate vector space. The new scalar product $( \cdot , \cdot )_\star$ can then be defined in the obvious way. It remains to be shown that it is positive definite and compatible with the $*$-structure of $A_{\star}$. Note that in order to be consistent with~(\ref{eq:star-structure}), one defines the action of $U \Xi$ on $\bar{V}$ by $\xi \bar{v} = \overline{S(\xi^*) v}$. Also note that to the above left action of $A$ on $V$ there corresponds the right action $\bar{v} \cdot a = \overline{a^* \cdot v}$ on $\bar{V}$. This action can also be deformed in the obvious way. Due to~(\ref{eq:real}), we have $\bar{v} \star a = \overline{a^* \star v}$. The compatibility with the $*$-structure of $A_{\star}$, i.e., $(v, a \star w)_\star = (a^* \star v, w)_\star$, is now again a consequence of~(\ref{eq:2cocycle}). Unfortunately, there seems to be no general proof that $( \cdot, \cdot )_\star$ is positive definite\footnote{If one interprets $\F$ as a formal power series, one has positive definiteness in the sense of formal power series, since the  zeroth order component is the old one.}. Thus, this has to be checked explicitly in each example.

\end{itemize}

\section{The application to NCQFT}
\label{sec:NCQFT}

It is now fairly obvious how to apply the above to NCQFT. Identifying $A$ with the free field algebra and $V$ with the Fock space, we get a new product of quantum fields and a new action on the Fock space\footnote{Note that the action of the twist on tensor products of $L^2$-functions is well-defined in momentum space. Thus, we do not have to restrict to Schwartz functions, since no point-wise products are involved.}. It only remains to be checked that the new scalar product is positive definite. This is indeed the case, in fact it is the old one: Let $f \in L^2(\R^{3m}), g \in L^2(\R^{3n})$. Then
\begin{equation*}
  (f, g)_{\star} = \delta_{m n} \frac{1}{m!} \sum_{\pi \in S_m}  \int \prod_{i=1}^m \frac{\ud^3 \V{k_i}}{2 \omega_i} \ \bar{f}(\V{k_1}, \dots , \V{k_m}) g(\V{k_{\pi(1)}}, \dots , \V{k_{\pi(m)}} ) e^{-\frac{i}{2} (-\sum_i k^+_i) \sigma (\sum_j k^+_j)}.
\end{equation*}
Here we used the notation $k^+ = (\omega_{\V k}, \V{k})$. Because of the antisymmetry of $\sigma$, the twisting drops out. This is analogous to the property~(\ref{eq:IntegralIdentity}). 

The same construction can be done for a fermionic Fock space.

\begin{remark}
\label{rem:BDFP}
The new product of quantum field follows naturally from the smeared field operators introduced in \cite{Quasiplanar}:
\begin{equation*}
  \phi_f(q) = \int \ud^4x \ \phi(q+x) f(x) = \int \ud^4k \ \hat{f}(k) \check{\phi}(k) \otimes e^{ikq}.
\end{equation*}
One then has
\begin{align*}
  \phi_f(q) \phi_g(q) = & \int \ud^4k_1 \ud^4k_2 \ e^{-\frac{i}{2} k_1 \sigma k_2}  \hat{f}(k_1) \hat{g}(k_2) \check{\phi}(k_1) \check{\phi}(k_2)  \otimes e^{i(k_1+k_2)q} \\
  = & \phi_{\F f \otimes g}^2(q).
\end{align*}
Here we used the notation\footnote{Note that this notation deviates from the one used in \cite{Quasiplanar}.}
\begin{equation*}
  \phi^n_f(q) = \int \prod_{i=1}^n \ud^4k_i \ \hat{f}(k_1, \dots, k_n)  \prod_i \check{\phi}(k_i) \otimes e^{i(k_1 + \dots + k_n)q}.
\end{equation*}
The equation above can then be generalized to
\begin{equation*}
  \phi^m_f(q) \phi^n_g(q) = \phi^{m+n}_{\F f \otimes g}(q)
\end{equation*}
for $f \in L^2(\R^{3m})$, $g \in L^2(\R^{3n})$.
However, these quantum fields are elements of (or rather affiliated to) $\mathfrak{F} \otimes \mathcal{E}_\sigma$, cf. Section~\ref{sec:QuantumFields}.
An element of $\mathfrak{F}$, i.e., an operator on Fock space, is obtained by applying $\text{id} \otimes \omega$, where $\omega$ is a functional on $\mathcal{E}_\sigma$. Thus, the action of the field algebra on the Fock space is different than in the twisted approach considered in this chapter.
\end{remark}

\begin{remark}
\label{rem:Balachandran}
In \cite{Balachandran} the twisted product was realized by a new definition of the creation and annihilation operators:
\begin{equation*}
  \tilde{a}(\V{k}) = a(\V{k}) e^{\frac{i}{2} k^+ \sigma P} , \quad \tilde{a}(\V{k})^* = a(\V{k})^* e^{-\frac{i}{2} k^+ \sigma P}.
\end{equation*}
One then has $\tilde{a}(f) \Psi = a(f) \star \Psi$ for any Fock space vector $\Psi$ (and analogously for $a(f)^*$).
\end{remark}

\section{The twisted commutators}
\label{sec:commutator}

We turn to a question that is very important for finding a consistent interpretation of the new field algebra. Of course one is inclined to keep the interpretation of $\phi(f)$ as a field operator and of $a(f)^*, a(f)$ as creation and annihilation operators. But then they should fulfill some commutation relation that is compatible with the classical Poisson bracket. Since this classical bracket is not affected by the twist (at least if one uses Peierls' definition, see~\cite{Diplom}), we would like the $\star$-commutator to give the usual result. This, however, is not the case for
\begin{align*}
  [\phi(f) \overset{\star}{,} \phi(g) ] = & \phi(f) \star \phi(g) - \phi(g) \star \phi(f) \\
  = & \int \ud^4k_1 \ud^4k_2 \ \hat{f}(k_1) \hat{g}(k_2) \\
  & \times \left( \check{\phi}(k_1) \check{\phi}(k_2) e^{- \frac{i}{2} k_1 \sigma k_2} - \check{\phi}(k_2) \check{\phi}(k_1) e^{\frac{i}{2} k_1 \sigma k_2} \right),
\end{align*}
as has already been noted in \cite[p.73f]{DorosDiss}. It is not even a $c$-number. But it fulfills the usual algebraic requirements antisymmetry, Leibniz rule and Jacobi identity.

\begin{remark}
This is the form of the commutator considered in \cite{Chaichian2} and denoted by $[\phi(f), \phi(g)]_{\star}$. Thus, our twisted NCQFT does not fulfill the locality axiom posed in there, even in the case of space/space noncommutativity.
\end{remark}

One can also consider the commutator as a bilinear map
\begin{align*}
  [ \cdot, \cdot ] : & \ A \otimes A \to A \\
  [ \cdot, \cdot ] : & \ a \otimes b \mapsto ab - ba.
\end{align*}
Then it is natural to deform it to the twisted commutator
\begin{equation*}
  [ \cdot, \cdot ]_{\star} = [ \cdot, \cdot ] \circ \F = \mu_{\star} - \mu_{\star} \circ R \circ \tau.
\end{equation*}
Here $\tau$ is the transposition and $R$ is the triangular structure (\ref{eq:R}). Note that this commutator was also used for a deformed Lie bracket of vector fields in \cite{WessEtAl}. In the context of NCQFT, it has already been proposed in \cite{DorosDiss} in the language of \cite{Quasiplanar} (cf. Remark \ref{rem:BDFP}). There, it simply amounts to postulating the commutator
\begin{equation*}
  [\phi \otimes f, \psi \otimes g] = [\phi, \psi] \otimes f g
\end{equation*}
for elements of $\mathfrak{F} \otimes \mathcal{E}_\sigma$. It has already been remarked in \cite{DorosDiss} that it is neither antisymmetric nor fulfilling the Jacobi identity. However, it is possible to prove a Jacobi identity that involves the $R$-matrix \cite{WessEtAl}:
\begin{equation*}
  [a,[b,c]_\star]_\star = [[a,b]_\star , c]_\star + [R_{(1)} b, [ R_{(2)} a, c]_\star ]_\star.
\end{equation*}
There is a similar formula expressing a twisted antisymmetry. While these formulae are general, there seems to be no analogous general formula for the Leibniz rule. But in the concrete example of NCQFT we have
\begin{equation*}
  [ a, b \star c]_\star = [a,b]_\star \star c + \F^{-2}_{(1)} b \star [ \F^{-2}_{(2)} a, c]_\star .
\end{equation*}
This can elegantly be seen in the notation of~\cite{Quasiplanar}.

We can now compute the twisted commutator of two fields:
\begin{align*}
  [\phi(f), \phi(g)]_{\star} = & \int \ud^4k_1 \ud^4k_2 \ e^{- \frac{i}{2} k_1 \sigma k_2} \hat{f}(k_1) \hat{g}(k_2) [\check{\phi}(k_1), \check{\phi}(k_2)] \\
  = & i (2\pi)^2 \int \ud^4k_1 \ \hat{f}(k_1) \hat{g}(-k_1) \check{\Delta}(k_1).
\end{align*}
We see that the twisting drops out and we obtain the usual result. In particular, we have twisted commutativity if the supports of $f$ and $g$ are spacelike separated. This seems to indicate that in the case of a twisted symmetry one should demand the correspondence principle between the classical Poisson bracket and the twisted commutator of the basic variables. We remark that Pusz and Woronowicz \cite{PW} found completely analogous twisted canonical commutation relations involving the $R$-matrix in a second quantization of a finite system with $SU_q(N)$-symmetry.
This may be seen as a hint that this is a general structure (see also \cite[Chapter 2]{ChaichianDemichev} and references therein). However, it should be noted that the vanishing of the commutator has a physical meaning, the possibility of simultaneous measurement. It is not clear whether the vanishing of the twisted commutator can be given a similar meaning.

It should also be noted that, as has already been remarked in \cite{DorosDiss}, the twisted commutator of products of fields does not coincide with the usual one, and does in particular not vanish for spacelike separated supports. This is illustrated in the following example:
\begin{align*}
  [\phi^2(f_1 \otimes f_2), \phi(f_3)]_{\star} = & \int \left( \prod_{i=1}^3 \ud^4k_i \hat{f}_i(k_i) \right) e^{-\frac{i}{2} ( k_1 + k_2 ) \sigma k_3} [\check{\phi}(k_1) \check{\phi}(k_2), \check{\phi}(k_3) ] \\
  = & i (2\pi)^2 \int \ud^4k \ \hat{f}_1(k) \check{\phi}(k) \int \ud^4p \ e^{\frac{i}{2} k \sigma p} \hat{f}_2(p) \hat{f}_3(-p) \check{\Delta}(p) \\
   + & i (2\pi)^2 \int \ud^4k \ \hat{f}_2(k) \check{\phi}(k) \int \ud^4p \ e^{\frac{i}{2} k \sigma p} \hat{f}_1(p) \hat{f}_3(-p) \check{\Delta}(p).
\end{align*}
We emphasize once more that this is also in conflict with the correspondence principle.
Note that it does not help to use $\phi(f_1) \star \phi(f_2)$ instead of $\phi^2(f_1 \otimes f_2)$.

\begin{remark}
\label{rem:formal}
If one interprets the twisting as a formal power series, then the $[ \cdot , \cdot ]_{\star}$-commutator is local at every order, i.e., it vanishes for spacelike separated supports (this is not the case for $[ \cdot \overset{\star}{,} \cdot]$). If one identifies the scale of noncommutativity with the Planck scale, then $\sigma$ could be interpreted to be of $\order(\hbar)$. In this sense the twisting would yield higher powers of $\hbar$. It may be interesting to investigate this further.
\end{remark}

Thus, the upshot of this section is that we have two natural commutators in the twisted setting. The first one fulfills the usual algebraic requirements but deviates from the classical Poisson bracket. The other one does not have very nice algebraic properties, but at least reproduces the classical Poisson bracket for simple fields (but not for products of fields). In any case the correspondence principle has to be modified considerably, so one is leaving the safe grounds of established quantum mechanics.

\section{Time evolution}
\label{sec:time-evolution}

In ordinary quantum theory, the time evolution of observables is given by the commutator with the Hamiltonian $H$. If we want to keep this in the twisted setting, we have to decide which commutator to use. Because of~(\ref{eq:DeltaP}) one expects that the time evolution fulfills the Leibniz rule, at least if $H$ is time--independent. It follows that one should use the $[ \cdot \overset{\star}{,} \cdot ]$-commutator. The classical equations of motion, however, do not change. Thus, the requirement that the time evolution is, to zeroth order in $\hbar$, identical to the classical evolution, leads to the condition $[H \overset{\star}{,}a] = [H,a] + \order (\hbar^2)$ for all observables $a$. But in the preceding section we have seen that the $[ \cdot \overset{\star}{,} \cdot ]$-commutator in general deviates from $[\cdot , \cdot]$ already at zeroth order\footnote{Even if one interprets the twist as a formal power series and assumes that $\sigma$ is of $\order(\hbar)$ (cf. remark~\ref{rem:formal}), then the two commutators still deviate at first order in $\hbar$.}. So the only general way to make things consistent seems to be to require that $H$ is invariant under the symmetry operation involved in the twist (in our case the translations), since then we have $[H \overset{\star}{,}a] = [H,a]$. But this makes the structure very rigid, because a change of $H$ must be accompanied by a change of the twist. It is not even clear if there exists such a new $\F$ in general.

\begin{remark}
There seems to be some similarity to the observation~\cite[Remark 2.2]{DorosDiss} that in the Hamiltonian approach to NCQFT the interacting Heisenberg field does, at tree level, not fulfill the classical equation of motion (see also~\cite{Heslop}). But there the problem appears only if the interaction is at least quadratic and if time does not commute with space. Here, instead, the problem is connected only to translation invariance and thus already arises for a source term and also in the case of space/space noncommutativity.
\end{remark}

In the case when $H$ is not invariant under the symmetry involved in the twist, one could of course simply postulate the time evolution $\dot{a} = i [H,a]$. But this time-evolution would in general be incompatible with the twisted algebra structure:
\begin{equation*}
 [H,a \star b] \neq [H,a] \star b + a \star [H,b].
\end{equation*}
Thus, one would have to use the old algebra and nothing would have changed.

In order to see how this rigidity makes a meaningful exploration of the new framework impossible, consider the following example: Applying a creation operator $a(g)^*$ on the vacuum twice, one gets the two-particle wave function
\begin{equation*}
  \Psi_{\F g \otimes g}(\V{k_1}, \V{k_2}) = \sqrt{2} g(\V{k_1}) g(\V{k_2}) \cos \frac{k_1^+ \sigma k_2^+}{2}.  
\end{equation*}
Thus, the modulus $\betrag{ \Psi_{ \F g \otimes g }(\V{k_1}, \V{k_2})}$ is reduced for momenta $\V{k_1}, \V{k_2}$ such that $k_1^+ \sigma k_2^+ \sim 1$. This can only happen if $\Delta_i \Delta_j \sim \lambda_{nc}^{-2}$ for $i,j$ in noncommuting directions. Here $\Delta_i$ denotes the typical width of $g$ in the direction~$i$. In this sense the wave function $\Psi_{\F g \otimes g}$ is more narrow in momentum space and thus has a wider spread in position space in the noncommuting direction (of course the effect is tiny for realistic energies if $\lambda_{nc}$ is identified with the Planck length). Thus, one gets the impression that the twisting disfavors the occurrence of several particles with the same wave function if this wave function is simultaneously localized in noncommuting directions. If this was true, this might be an elegant resolution of the uncertainty problem posed in \cite{DFR}.

But the discussion above was at best heuristic. On the shaky ground we are exploring, we do not have any good intuition about what the repeated action of $a(g)^*$ might actually signify. And the two-particle wave function $\Psi_{g \otimes g}$ is still an element of our Fock space. Thus, we would like to make a statement in more operational terms. Now $\Psi_{g \otimes g}$ is, up to normalization, the two-particle component of the coherent state $e^{\lambda a(g)^*} \Omega$. This, in turn, can be characterized by being the ground state corresponding to the Hamiltonian
\begin{equation*}
  H = H_0 + \lambda \left[ a(f)^* + a(f) \right],
\end{equation*}
where $H_0$ is the usual free Hamiltonian and $f = - P_0 g$. Taking this as a motivation, it would be interesting to find the ground state corresponding to this Hamiltonian in our twisted setting, i.e., the eigenvector $\Psi$ with the lowest eigenvalue $H \star \Psi = E \Psi$. This can be done and it turns out that the corresponding two-particle wave function is indeed more narrow than $\Psi_{g \otimes g}$ (it is even more narrow than $\Psi_{\F g \otimes g}$). However, it is not clear if this result has any meaning, because $H$ is, in the twisted setting, not the generator of the time evolution. In the present example, this is easily seen for the time evolution of a field, as we already computed the $[ \cdot \overset{\star}{,} \cdot ]$-commutator of two fields in the preceding section.

\begin{remark}
In \cite{Balachandran} it has been claimed that in the twisted setting Pauli's exclusion principle is no longer valid. The authors conclude this from the fact that in the case of twisted anticommutation relations one has in general $a(g)^* \star a(g)^* \neq 0$. But the fermionic wave functions are still antisymmetric. It is simply not clear what Pauli's principle tells us in the twisted case (as in the example above, we do not know what the repeated action of $a(g)^*$ actually means). One should rather look for a statement in operational terms. First steps in this direction have been taken in~\cite{Chakraborty}, see also~\cite{TwistedNCQED}.
However, in the light of the preceding discussion it is doubtful that this can be done consistently.
\end{remark}

\begin{remark}
\label{rem:YF}
In view of the problems discussed here, one might be tempted to define the interacting directly by the equation of motion, i.e., use the Yang--Feldman formalism.
But one has to bear in mind that if the interaction is not translation invariant, the interacting field will not transform covariantly under translations, i.e., $\phi_{int}(\tau_a f) \neq U_a \phi_{int}(f) U_a^{-1}$, where $\tau$ is the action on test functions and $U$ the Hilbert space representation of the translation group. This will make the $\star$-product of interacting fields more complicated.
\end{remark}

\section{Interactions}
\label{sec:interactions}

The effect of interactions can be studied by formally computing the $n$-point functions of the interacting field, defined, e.g., by the Yang--Feldman formalism. If the interaction is translation invariant, the interacting field transforms covariantly under translations (see remark~\ref{rem:YF}), and we have
\begin{multline*}
  \skal{\Omega}{\phi_{int}(f_1) \star \dots \star \phi_{int}(f_n) \Omega} \\ = \int \left( \prod_{i=1}^n \ud^4k_i \hat{f}_i(k_i) \right) e^{-\frac{i}{2} \sum_{i < j} k_i \sigma k_j } \skal{\Omega}{\check{\phi}_{int}(k_1) \dots \check{\phi}_{int}(k_n) \Omega}.
\end{multline*}
On the right hand side, all the loops are contained in the vacuum expectation value. Obviously, the twisting factor does not interfere at all with these and has no effect on the divergences, and in particular does not influence the UV/IR--mixing. Thus the absence or presence of the UV/IR--mixing does only depend on the choice of the interaction term\footnote{This also seems to be at odds with the results of \cite{Oeckl}.}.

In \cite{BalachandranUVIR} the old (pointwise) product of fields was used\footnote{More precisely, the inverse $\star$-product between the operators $\tilde{a}, \tilde{a}^*$ (see Remark \ref{rem:Balachandran}) was used. But since these operators already realize the $\star$-product, the combined effect amounts to the pointwise product.}. Thus, it is not surprising that the UV/IR--mixing is absent there. The same conclusion was also reached in~\cite{Tureanu}.

If, however, the $\star$-product of fields is used for the interaction term, the UV/IR--mixing will be exactly as usual.

\chapter{Noncommutative gauge theories}
\label{chapter:NCGaugeTheory}

The main goal of this thesis is to study quantum electrodynamics on the noncommutative Minkowski space. In order to set the stage, we give a brief introduction to gauge theories on noncommutative spaces and study classical electrodynamics on the noncommutative Minkowski space in some detail.

There are different approaches to gauge theories on noncommutative spaces. The unexpanded (module) approach is rooted in the framework of noncommutative geometry and is intrinsic in the sense that it only requires a noncommutative algebra. It has the disadvantage that there are severe restrictions on the gauge groups and representations. In the approach via the Seiberg--Witten map, one always needs a classical spacetime that is deformed by a $\star$-product. The Seiberg--Witten map is then a formal map from ``commutative'' gauge fields to ``noncommutative'' ones. It has the advantage that there are no restrictions on the gauge groups and representations. There are also attempts to formulate gauge theories in the twisted setting introduced in the previous chapter. In this chapter, we will present these different approaches (with a focus on the module approach), and discuss classical electrodynamics on the noncommutative Minkowski space in the module approach.

\section{The module approach}

We already mentioned the Gelfand--Naimark theorem as one of the cornerstones of noncommutative geometry. Another cornerstone is the Serre--Swan theorem (see, e.g., \cite[Thm. 2.10]{ElementsOfNCG}), which states that there is a one--to--one correspondence between vector bundles\footnote{It is one of the biggest problems for the construction of noncommutative gauge theories that there is no analogous theorem for principle bundles.} over a compact topological space $M$ and finitely generated projective $C(M)$-modules. More precisely, the space $\Gamma(M,V)$ of continuous sections of the bundle $V \to M$ is a finitely generated projective $C(M)$-module, and to each finitely generated projective $C(M)$-module $E$ corresponds a vector bundle $V \to M$ such that $E \simeq \Gamma(M,V)$. If $M$ is only locally compact, the situation is more involved~\cite{Rennie}. If the bundle can be extended to a bundle $V \to M^c$ on a compactification $M^c$ of $M$,\footnote{On the Stone--\v{C}ech compactification of $M$, this is possible if $M$ is normal and $V$ has a finite open cover of trivializations.} then the space $\Gamma_0(M,V |_M)$ of sections vanishing at infinity is a $C_0(M)$-module of the form $p C_0(M)$ with $p \in M_n(C(M^c))$ a projection. Conversely, to each $C_0(M)$-module of this form, there corresponds a bundle $V  \to M^c$ such that $E \simeq \Gamma_0(M,V|_M)$.

The application to noncommutative geometry is straightforward: The analog of a vector bundle over a compact ``noncommutative space'' described by a noncommutative, unital $C^*$-algebra $\A$ is a finitely generated projective right\footnote{Of course one could also consider left $\A$-modules. But since in the commutative case there is only one multiplication, it is natural not to consider $\A$-bimodules.} $\A$-module $E$. For nonunital $\A$, we choose some unitization (typically the multiplier algebra $M(\A)$). As discussed in Section~\ref{sec:E}, it is sometimes advantageous to work with general algebras (not necessarily $C^*$). Thus, we also allow for such algebras in the following.


\subsection{Finitely generated projective modules}

A crucial ingredient for doing field theory is a \emph{metric} on the right $\A$-module $E$, i.e., a pairing $E \times E \to \A$ fulfilling
\begin{align*}
  (\phi, \psi + \psi') & = (\phi, \psi) + (\phi, \psi'), \\
  (\phi, \psi a) & = (\phi, \psi) a, \\
  (\phi, \psi ) & = \overline{ (\psi, \phi) }.
\end{align*}
It should also be positive, i.e., $(\phi, \phi) > 0$ for $\phi \neq 0$. In the commutative case, this obviously corresponds to a hermitean metric on the bundle.

\begin{remark}
\label{rem:Endomorphisms}
If $\A$ is a $C^*$-algebra and $E$ is complete in the norm $\phi \mapsto (\phi, \phi)^{\frac 1 2}$, one speaks of a right $C^*$ $\A$-module. A map $T:E \to E$ is called \emph{adjointable}, if is has an adjoint with respect to the metric. One can show that such maps are always bounded module endomorphisms~\cite[Prop.~2.16]{ElementsOfNCG}. Furthermore, for finitely generated projective modules $E = p \A^n$, one has\footnote{Strictly speaking, one has $\End_\A(E)=M(pM_n(\A)p)$, where $M$ stands for the multiplier algrebra~\cite[Section~3.1]{ElementsOfNCG}. In the case $\A = M(\A)$, $p = id$, with which we will be concerned mostly, the above is thus correct.} $\End_\A(E) = p M_n(\A) p$.
\end{remark}

\begin{example}
The simplest example for a finitely generated projective right $\A$-module is of course $E = \A^n$. Choosing a set $\{s_i\}$ of basis sections, one can define the metric as
\begin{equation*}
  (\sum_i s_i a_i, \sum_j s_j b_j) = \sum_i a_i^* b_i.
\end{equation*}
A transformation $s_i \to \sum_j s_j \Lambda_{j i}$ with $\Lambda_{j i} \in \A$ that yields another set of orthonormal basis sections can then be interpreted as a gauge transformation. Thus, we have constructed a $U(N)$ gauge symmetry. The module $E$ then describes fields in the fundamental representation. Unfortunately, there is no suitable definition of a determinant for such transformations. Therefore, it is not possible to construct an $SU(N)$ gauge symmetry in this way.
\end{example}

\begin{example}
A noncommutative algebra $\A$ in general contains nontrivial projectors. Given such a projection $p$, one can define the finitely generated projective module $E = p \A$. Such modules have no analog in the commutative case. We will not consider this possibility further in the following.
\end{example}

Matter fields in the adjoint representation can be described as $\A$-module endomorphisms. As discussed in Remark~\ref{rem:Endomorphisms}, these are elements of $p M_n(\A) p$. In order to write down an action for these fields, one then also needs a trace on $\End_\A(E)$. Given a trace $\tr_\A$ on $\A$, one can simply define $\Tr_E = \tr_\A \circ \tr$, where $\tr$ is the matrix trace.

In~\cite{NC_SM} a noncommutative version of the standard model has been proposed where matter fields transform under different $U(N)$'s from left and right. These can be implemented as elements of $\Hom_\A(E,E')$. In principle one can then form two different actions, either $\Tr_E \phi^* \phi$ or $\Tr_{E'} \phi \phi^*$.

Thus, in noncommutative gauge theories the possible gauge groups and representations are highly restricted~\cite{NoGo}. Later we will see that in the case of electrodynamics charge quantization is a further restriction.

\subsection{Differential calculus}

In order to do gauge theory, we want to define connections. For this, we need a differential calculus. By this we mean the following: Let $\A$ be an algebra. A \emph{differential calculus over} $\A$ is a graded algebra
\begin{equation*}
  \Omega(\A) = \oplus_n \Omega_n(\A),
\end{equation*}
a homomorphism $\rho: \A \to \Omega_0(\A)$ and a differential $\ud$ satisfying
\begin{align*}
  \ud \Omega_n(\A) & \subset \Omega_{n+1}(\A), \\
  \ud^2 & = 0, \\
  \ud( \omega \nu ) & = \ud \omega \nu + (-1)^n \omega \ud \nu.
\end{align*}
In the last equation $n$, is the grade of $\omega$. With $\rho$ and the product in $\Omega(\A)$, $\Omega(\A)$ is an $\A$-bimodule in a natural way. A differential calculus is called \emph{minimal}, if it is generated by $\A$ and $\ud$.

If $\A$ is a unital algebra, there is a differential calculus which has the property that any other minimal differential calculus can be obtained from it by a unique graded algebra homomorphism. It is called the \emph{universal differential calculus}. It can be constructed as follows: Define
\begin{equation*}
  \Lambda(\A) = \oplus_n \Lambda_n(\A) \quad \text{with} \quad \Lambda_n(\A) = \A^{\otimes^{n+1}}
\end{equation*}
and the multiplication
\begin{equation*}
  (a_0 \otimes \dots \otimes a_m, b_0 \otimes \dots \otimes b_n) \mapsto  a_0 \otimes \dots \otimes a_m b_0 \otimes \dots \otimes b_n.
\end{equation*}
Furthermore, define the differential $\ud_u$ by
\begin{equation*}
  \ud_u ( a_0 \otimes \dots \otimes a_m ) = \sum_{k=0}^{m+1} (-1)^k a_0 \otimes \dots \otimes a_{k-1} \otimes \1 \otimes a_k \otimes \dots \otimes a_m.
\end{equation*}
The subalgebra $\Omega_u(\A)$ generated by $\A$ and $\ud_u$ is the universal differential calculus. We note that for practical applications, the universal differential calculus is usually by far too big.

The analog of vector fields on a manifold are the derivations on the algebra $\A$. There is a natural pairing $\Omega_u(\A) \times \Der(\A)^{\otimes n}$ between universal forms and tensor products of derivations:
\begin{equation*}
  \skal{a_0 \ud_u a_1 \dots \ud_u a_m}{\del_1 \otimes \dots \otimes \del_n} = \delta^n_m a_0 \del_1 a_1 \dots \del_m a_m. 
\end{equation*}
This definition is unambiguous, because both $\ud_u$ and $\del$ fulfill the Leibniz rule.

Derivations can be used to reduce the universal differential calculus. Let $\del_\alpha$ be a set of linearly independent derivations. Then the set
\begin{equation*}
  N  = \{ \omega \in \Omega_u^1(\A) | \del_\alpha \omega = 0 \ \forall \alpha \}
\end{equation*}
is an $\A$-bimodule ideal of $\Omega_u^1$, and we can define the reduced differential calculus
\begin{equation*}
 \Omega(\A) = \Omega_u(\A) / N.
\end{equation*}
Note that elements of $\Omega^1(\A)$ can in general only be paired with (linear combinations of) the derivations $\del_\alpha$.

If $\A$ is a $*$-algebra, there is a natural definition of an adjoint on $\Omega_u(\A)$:
\begin{equation*}
  (a_0 \otimes \dots \otimes a_m)^* = (-1)^m a_m^* \otimes \dots \otimes a_0^*.
\end{equation*}
The universal differential $\ud_u$ is real in the sense $(a \ud_u b )^* = \ud_u b^* a^*$. If the derivations $\del_\alpha$ are real in the sense $(\del a)^* = \del a^*$, then the reduced calculus $\Omega(\A)$ also has an adjoint and the differential is real. 

\begin{example}
\label{ex:del_mu}
We consider the unital $*$-algebra $\M$, cf. Section~\ref{sec:E}, describing fields on the noncommutative Minkowski space. Obviously, the derivations $\del_\mu$ are real. Using them to reduce the
universal differential calculus, we get the following relations for one-forms:
\begin{equation*}
  f \ud q^\mu = \ud q^\mu f \qquad \forall f \in \M.
\end{equation*}
Applying $\ud$ once more, we see that the one-forms $\ud q^{\mu}$ anticommute. The pairing of one-forms and the derivations $\del_\mu$ is given by $\skal{f \ud q^\nu}{\del_\mu} = \delta_\mu^\nu f$. This is the differential calculus that is mainly used in the literature.
\end{example}

\subsection{Connections}

Now we may define connections on our analogs of vector bundles. Let $E$ by a right $\A$-module. A \emph{connection} (or \emph{covariant derivative}) $D$ on $E$ is a linear map
\begin{equation*}
  D: E \to E \otimes_\A \Omega^1(\A)
\end{equation*}
fulfilling the Leibniz rule
\begin{equation*}
  D (\phi a) = D \phi a + \phi \otimes \ud a.
\end{equation*}
It can be extended to a map
\begin{equation*}
  D: E \otimes_\A \Omega^n(\A) \to E \otimes_\A \Omega^{n+1}(\A)
\end{equation*}
by setting
\begin{equation*}
  D (\Phi \omega) = D \Phi \omega + (-1)^n \Phi \ud \omega
\end{equation*}
for $\Phi \in E \otimes_\A \Omega^n(\A)$. Note that $E \otimes_\A \Omega^n(\A)$ is also a right $\A$-module. It is straightforward to show that the \emph{field strength}
\begin{equation*}
  F = D^2
\end{equation*}
is a module homomorphism.

If $E$ carries a metric, $D$ is called a \emph{metric connection} if it fulfills
\begin{equation*}
  \ud ( \phi, \psi ) = ( D \phi, \psi ) + ( \phi, D \psi ).
\end{equation*}
The second term on the right hand side is defined by $( \phi, \psi \otimes_\A \omega ) = (\phi, \psi) \omega$, and similarly for the other term.

Every $L \in \End_\A(E)$ can be extended to $L \in \End_{\Omega(\A)}(E \otimes_\A \Omega(\A))$ by
\begin{equation*}
  L( \phi \otimes_\A \omega ) = L(\phi) \otimes_\A \omega.
\end{equation*}
We can thus define the covariant derivative on $\End_\A(E)$ as a map $D: \End_\A(E) \to \Hom_\A(E, E \otimes_\A \Omega^1(\A))$ by
\begin{equation*}
  DL = [D,L].
\end{equation*}
This corresponds to the covariant derivative in the adjoint representation.

In the case of electrodynamics, i.e., $E = \A$, we can choose a basis section $s$. Then the connection defines a vector potential $A \in \Omega^1(\A)$ by $D s = - i e s \otimes_\A A$. Here we introduced a coupling constant $e$. The field strength is then given by $F s = - i e s \otimes_\A (\ud A - i e A^2)$. If the basis section is normalized, i.e., $(s,s) = \1$, the connection is metric iff $A^* = A$. Under a gauge transformation $s' = s \Lambda$, the connection changes as 
\begin{equation}
\label{eq:A_trafo}
  A' = \frac{i}{e} \Lambda^{-1} \ud \Lambda + \Lambda^{-1} A \Lambda.
\end{equation}

In typical situations, we would like to have different matter fields coupled to the same gauge field. This is severely constrained in the present setting. Consider two trivial bundles $E_{1/2} = \A$. Choose normalized sections $s_{1/2}$ such that the connections are given by $D_{1/2} s_{1/2} = - i e_{1/2} s_{1/2} A$ for some $A \in \Omega^1(\A)$.  Now, unless $e_1 = e_2$, for a given gauge transformation $s_1' = s_1 \Lambda_1$ there will be in general no corresponding gauge transformation on $E_2$ such that
\begin{equation*}
  \frac{i}{e_1} \Lambda_1^{-1} \ud \Lambda_1 + \Lambda_1^{-1} A \Lambda_1 = \frac{i}{e_2} \Lambda_2^{-1} \ud \Lambda_2 + \Lambda_2^{-1} A \Lambda_2
\end{equation*}
holds, cf. (\ref{eq:A_trafo}). Thus, there is only one universal electric charge. This has first been noted in~\cite{Hayakawa}. Obviously, this makes the description of quarks quite problematic.

\begin{example}
Considering the noncommutative Minkowski space $\M$ and the reduced differential calculus introduced in example~\ref{ex:del_mu}, we set
\begin{equation*}
  A = \ud q^{\mu} A_\mu.
\end{equation*}
Here the $A_\mu$'s are self-adjoint. The field strength is then given by $F = \frac{1}{2} \ud q^\mu \ud q^\nu F_{\mu \nu}$ with
\begin{equation}
\label{eq:F}
  F_{\mu \nu} = \del_\mu A_\nu - \del_\nu A_\mu - i e [A_\mu, A_\nu].
\end{equation}
\end{example}

\subsection[Covariant Coordinates]{Covariant Coordinates\footnotemark}

\footnotetext{This section is to a large extent based on an unpublished manuscript of S.~Doplicher and K.~Fredenhagen.}

In gauge theories it is always a crucial point to construct local observables, i.e., local gauge invariant quantities. In the present language this means module homomorphisms. We have already seen that the field strength is such a homomorphism. However, it is not localized. Thus, assume we are given a trace $\Tr$ on (a subset of) $\End_\A(E)$. Then $\Tr F_{\mu \nu}$ will in general not exist. Hence, we would like to have a mapping $L:\A \to \End_\A(E)$. The action of $L(a)$ can then be interpreted as a covariant multiplication with~$a$.
If $a$ is appropriately chosen, $\Tr L(a) F_{\mu \nu}$ is well-defined and can be interpreted as the field strength evaluated with a ``test function'' $a$.

We follow an approach in the spirit of Wilson. We define a pairing
\begin{equation*}
 E \otimes_\A \Omega(\A) \times \Der(\A)^{\otimes n} \to E
\end{equation*}
by
\begin{equation*}
  \skal{\phi \otimes_\A \omega}{\del_1 \otimes \dots \otimes \del_n} = \phi \skal{\omega}{\del_1 \otimes \dots \otimes \del_n}.
\end{equation*}
As we already remarked above, in a reduced differential calculus, this is in general only well-defined for the derivations that were used to reduce the calculus. Given a connection $D$, we can interpret the map $\dot U_\del : E \to E$ given by
\begin{equation*}
 \dot U_\del \phi = \skal{D \phi}{\del}
\end{equation*}
as an infinitesimal parallel transport (or covariant derivative) of $\phi$ along the direction given by $\del$. This can formally be integrated to a parallel transport $U_\del$ by defining
\begin{equation*}
  U_\del = \sum_n \frac{1}{n!} \dot{U}_\del^n.
\end{equation*}
Whether the right hand side exists has to be checked in the concrete case. By definition, one has $U_{t \del} U_{s \del} = U_{(t+s) \del}$. Parallel transport is of course no module homomorphism. Instead, one has
\begin{equation*}
  U_\del ( \phi a ) = U_\del \phi V_\del a
\end{equation*}
with $V_\del = e^\del$. It follows that a series $U_{\del_1} \circ \dots \circ U_{\del_n}$ of parallel transports is a module homomorphism iff
\begin{equation*}
  V_{\del_1} \circ \dots \circ V_{\del_n} = \id,
\end{equation*}
i.e., if the path closes. If the derivations $\del$ and $\del'$ commute, one can form a plaquette and recovers the field strength through the formula
\begin{equation*}
  \skal{F}{\del \otimes \del' - \del' \otimes \del} = \frac{\ud^2}{\ud t \ud s} U_{t \del} U_{s \del'} U_{- t \del} U_{-s \del'} |_{t = s = 0}.
\end{equation*}

An important class of derivations are inner derivations, i.e., those which are given by a commutator with an element of the algebra:
\begin{equation*}
  \del_a b = [b,a].
\end{equation*}
Obviously, $\del_a$ is hermitean iff $a^* = - a$. As we have seen in Example~\ref{ex:del_mu}, the derivations $\del_\mu$ on the noncommutative Minkowski space are of this type. For inner derivations, one can also define a module homomorphism by
\begin{equation*}
  W_a \phi = U_{\del_a} \phi e^{-a}.
\end{equation*}
This is a generalization of the open Wilson lines introduced in~\cite{Ishibashi}, see also~\cite{Gross}. Their infinitesimal versions
\begin{equation*}
  L(a) \phi = \frac{\ud}{\ud t} W_{t a} \phi |_{t=0} = \skal{D \phi}{\del_a} - \phi a
\end{equation*}
are the \emph{covariant coordinates}. In a reduced differential calculus, the covariant coordinates are in general only defined for special $a$'s. 

\begin{example}
In the case of electrodynamics on the noncommutative Minkowski space, i.e., $E = \A = \M$, we have $\End_\A(E) = \M$. A trace is given by the integral $\int \ud^4q$ introduced in Section~\ref{sec:Calculus}, which is well-defined on the ideal $\schw_2 \subset \M$.
In the reduced differential calculus
introduced in Example~\ref{ex:del_mu}, the derivations $\del_\mu$ are given as linear combinations of the inner derivations $\del_{q^\nu}$, cf. (\ref{eq:del_mu_commutator}). Thus, the covariant coordinates for the $q^\mu$'s are defined. For a given basis section $s$, they are
\begin{equation}
\label{eq:X}
  L(q^\mu) s = s X^\mu = s ( q^\mu + e \sigma^{\mu \nu} A_\nu ). 
\end{equation}
These are the covariant coordinates introduced in~\cite{Madore}. We note that they fulfill the commutation relations
\begin{equation}
\label{eq:X_comm}
  [X^\mu, X^\nu] = i \sigma^{\mu \nu} - i ( \sigma F \sigma )^{\mu \nu}.
\end{equation}
Furthermore, similarly to (\ref{eq:del_mu_commutator}), one can express the covariant derivative in the adjoint representation by a commutator with the covariant coordinates:
\begin{equation}
\label{eq:X_D}
  D_\mu = \skal{D \ \cdot \ }{\del_\mu} = - i \sigma^{-1}_{\mu \nu} [X^\nu, \ \cdot \ ].
\end{equation}
\end{example}

\section{The Seiberg--Witten map}

We already mentioned that field theory on the noncommutative Minkowski space also arises in the so--called zero--slope limit in the theory of open strings ending on D-branes in the presence of a background $B$-field. But, as was shown by Seiberg and Witten~\cite{SW}, this depends on the regularization. For a stack of $N$ D-branes, one obtains, via a point-splitting regularization, a noncommutative $U(N)$ gauge theory. Using a Pauli--Villars regularator, one finds an ordinary $U(N)$ gauge theory. Since the physics should be regularization independent, there should be a mapping between the two theories. This map, called Seiberg--Witten map, is a solution of an inhomogenous differential equation. It expresses the noncommutative gauge field $A_\mu$ as a function of the commutative gauge field $a_\mu$. The noncommutative gauge parameter, however, is a function of both the commutative gauge parameter $\lambda$ and $a_\mu$. Both functions are formal power series in $\sigma$.

In the case of $U(1)$ gauge theory, i.e., electrodynamics, the Seiberg--Witten map can also be interpreted in a more geometrical fashion. In~\cite{Cornalba} it was shown that it arises in a world-volume reparametrization that eliminates the fluctuating gauge field. Similarly, one can interpret the Seiberg--Witten map as an equivalence relation between different $\star$-products~\cite{Jurco2000}.
Of course an electromagnetic field $f$ could also be put into the background field $B$, which is, in the zero--slope limit, related to $\sigma$ by $\sigma = B^{-1}$. Thus, one would have a new symplectic structure
\begin{equation}
\label{eq:sigma_1}
  \sigma' = (B+f)^{-1} = \sigma (1 + f \sigma)^{-1}.
\end{equation}
However, strictly speaking, this equality only holds for constant $f$.

A similar phenomenon appears in the module approach: Equation~(\ref{eq:X_comm}) can be interpreted as defining a new commutator
\begin{equation}
\label{eq:sigma_2}
  \sigma' = \sigma - \sigma F \sigma
\end{equation}
for the coordinates $X^{\mu}$. But note that the $F$ here and the $f$ in~(\ref{eq:sigma_1}) are not identical, even if they are constant. In fact, equating (\ref{eq:sigma_1}) and (\ref{eq:sigma_2}), we obtain
\begin{equation*}
  f = F (1 - \sigma F)^{-1}.
\end{equation*}
This relation has already been found in~\cite{SW} as the relation between the commutative field strength $f$ and the noncommutative field strength $F$.

Combining Moser's Lemma and the Kontsevich formalism for $\star$-products, one can show~\cite{Jurco2000} that there is a map $L_a$, which is a formal power series in $\sigma$ and intertwines the $\star$-products corresponding to $\sigma$ and $\sigma'$:
\begin{equation*}
  L_a g \star L_a h = L_a ( g \star' h ).
\end{equation*}
In the notation $L_a$ the subscript $a$ refers to the one-form that generates the flow from the symplectic form $B$ to $B + f$, where $f = \ud a$. Interpreting $L_a$ as a covariant coordinate, one can turn equation (\ref{eq:X}) around and obtain $A_\mu$ from $L_a(x^\nu)$. Explicitly, one finds
\begin{equation}
\label{eq:SW_A}
  A_\mu = a_\mu - \frac{1}{2} \sigma^{\nu \lambda} a_\nu ( \del_\lambda a_\mu + f_{\lambda \mu} ) + \order (\sigma^2).
\end{equation}
For the field strength, one obtains
\begin{equation}
\label{eq:SW_F}
  F_{\mu \nu} = f_{\mu \nu} + \sigma^{\lambda \rho} ( f_{\mu \lambda} f_{\nu \rho} - a_\lambda \del_\rho f_{\mu \nu} ) + \order(\sigma^2).
\end{equation}

The strategy is then usually as follows: One expresses the noncommutative action in terms of the commutative fields, up to some (usually first) order in $\sigma$. Concretely, using~(\ref{eq:SW_F}) and the noncommutative Maxwell action~(\ref{eq:NCEDAction}), which we will introduce later, one obtains
\begin{equation}
\label{eq:SW_action}
  S =  \int \ud^4x \ \left( - \frac{1}{4} f_{\mu \nu} f^{\mu \nu} + \frac{1}{8} \sigma^{\alpha \beta} f_{\alpha \beta} f_{\mu \nu} f^{\mu \nu} - \frac{1}{2} \sigma^{\alpha \beta} f_{\mu \alpha} f_{\nu \beta} f^{\mu \nu} \right) + \order(\sigma^2).
\end{equation}
In order to do classical field theory, one can then obtain equations of motion by varying with respect to $a_\mu$. On the quantum level, the terms of higher order in $\sigma$ are interaction terms that can be treated perturbatively. This approach has several advantages: The Seiberg--Witten map can be set up for arbitrary gauge groups~\cite{Jurco2001} and electromagnetic charges~\cite{SW_SM}. It is thus relatively straightforward to construct a noncommutative version of the standard model~\cite{SW_SM}. Furthermore, because of the expansion in $\sigma$, the phase factors that lead to the UV/IR--mixing do not appear. Finally, $f_{\mu \nu}$ is gauge invariant, so one can directly use it as an observable and does not have to bother about covariant coordinates.

However, there are some disadvantages: While the photon two-point function in pure noncommutative electrodynamics is renormalizable to all orders in $\hbar$ and $\sigma$ in this approach~\cite{SW_renormalization}, the theory becomes nonrenormalizable upon introducing fermions, already at first order in $\sigma$ \cite{Wulkenhaar}. From a more conceptual point of view, the validity of the expansion in $\sigma$ is questionable: There is no indication the series converges. But if one expands only to some finite order, the theory is local, in spite of the fact that nonlocality was the main motivation for the introduction of the noncommutative Minkowski space in~\cite{DFR}.

\section{The twist approach}

There is currently no consensus about how to formulate gauge theories in the twisted setting. We limit ourselves to introduce the different points of view.

Regarding a gauge theory as given by a vector bundle, one can consider the module $E$ of sections, as done above. It is then in the spirit of the previous chapter to twist the action of the algebra on the module, i.e., to consider the map
\begin{align*}
  E \otimes_\star \A_\star & \to E \\
  (\phi,a) & \mapsto \phi \star a. 
\end{align*}
In the same spirit, one also has to change the metric. E.g., if $s_i$ is an orthonormal set of basis sections, then the new metric is given by
\begin{equation}
\label{eq:TwistedMetric}
  ( \sum_i s_i \phi_i , \sum_j s_j \psi_j )_\star = \sum_i \bar \phi_i \star \psi_i.
\end{equation}
Also connections can be twisted, they are now maps $E \to E \otimes_\star \Omega^1(\A_\star)$. In this way, one simply recovers the setting of the module approach. This seems to be the point of view taken in~\cite{ChaichianGaugeTwist}. 

A different proposal~\cite{Vassilevich,FrankGaugeTwist} is to retain the gauge group and its action from the commutative case, but to twist its coproduct. Thus, if $\delta_\alpha$ is a gauge transformation, one defines $\Delta_\star (\delta_\alpha) = \F^{-1} \circ \Delta(\delta_\alpha) \circ \F$. It fulfills\footnote{From the point of view of differential geometry, this requirement seems to be a bit strange, since $\delta_\alpha$ should act on bundle sections, which are not multiplied anyway.} $\delta_\alpha \circ \mu_\star = \mu_\star \circ \Delta_\star(\delta_\alpha)$. This approach has the advantage that there are no restrictions on the gauge groups or representations. However, there are other problems. Assuming one has a matrix representation of the gauge groups with generators $T_a$. Then the covariant derivative is given by
\begin{equation*}
  D_\mu \phi^i = \del_\mu \phi^i - i e T_a^{i j} A_\mu^a \star \phi^j.
\end{equation*}
Defining the field strength as the commutator of two such covariant derivatives, one obtains
\begin{equation*}
  F_{\mu \nu} = T_a \left( \del_\mu A^a_\nu - \del_\nu A^a_\mu \right) - i e T_a T_b \left( A^a_\mu \star A^b_\nu - A^a_\nu \star A^b_\mu \right) 
\end{equation*}
This field strength depends on the representation and is in general not in the Lie-algebra generated by the $T_a$'s. Furthermore, considering a $SU(N)$ gauge theory in the fundamental representation, one simply recovers the usual noncommutative $U(N)$ field strength and action.

Yet another point of view was taken in \cite{TwistedNCQED}. There, not only the gauge group and its action are as in the commutative case, but also the covariant derivative. The bundle metric is twisted however, as in (\ref{eq:TwistedMetric}). Thus, one has to use the twisted coproduct on elements of $\bar E \otimes E$, e.g., but the old coproduct on terms like $A_\mu \phi$.

\section{Electrodynamics}

In this section, which is partly based on \cite{NCED}, we study classical electrodynamics on the noncommutative Minkowski space (NCED) in the unexpanded (module) approach, i.e., we have to use covariant coordinates to define local observables. In a couple of publications (e.g. \cite{Berrino,Guralnik,Cai,Abe}), NCED has been treated in the Seiberg--Witten approach to first order in $\sigma$. One of the main results was that the speed of light in a constant background field is modified \cite{Guralnik,Cai,Abe}. In~\cite{Berrino}, corrections to the Coulomb potential were calculated. It is an interesting question whether the module approach leads to the same results. One of the main goals of this section is to answer this question.


An important point we have to clarify is the differential calculus we use. At first sight it seems to be convenient to use the universal calculus, since then the covariant coordinates $L(f)$ are defined for all $f$. However, the following problem appears: The noncommutative analog of the action in the commutative case is
\begin{equation}
\label{eq:NCEDAction}
  S = - \frac{1}{4} \int \ud^4q \  F_{\mu \nu} F^{\mu \nu}
\end{equation}
with
\begin{equation}
\label{eq:F2}
  F_{\mu \nu} = \frac{i}{e} \left( \skal{F}{\del_\mu \otimes \del_\nu} - \skal{F}{\del_\nu \otimes \del_\mu} \right).
\end{equation}
If we choose a basis section and define
\begin{equation}
\label{eq:A2}
  A_\mu = \skal{A}{\del_\mu},
\end{equation}
the definition~(\ref{eq:F2}) coincides with the one given in~(\ref{eq:F}). Variation of the action now yields an equation of motion for $A_\mu$. Suppose we have solved these. In order to compute the covariant coordinate $L(f)$ for an arbitrary $f$, we need to know the universal one-form $A$. This, however, is not determined by $A_\mu$. In order to see this, suppose we have found $A$ such that~(\ref{eq:A2}) holds. We can now simply add terms of the form $f_\mu \ud_u q^\mu -\ud_u q^\mu f_\mu$ to $A$ without changing $A_\mu$. Thus, it seems natural to work in the reduced differential calculus introduced in Example~\ref{ex:del_mu}, because there the knowledge of $A_\mu$ completely determines $A$. But then we can not compute the covariant coordinates for any element of the algebra, only the $X^\mu$'s corresponding to the $q^\mu$'s, cf. (\ref{eq:X}).


Thus, we have to define functions $g(X)$ of the covariant coordinates. This can be done in analogy to the Weyl-Wigner-Moyal calculus: Let $g$ be a Schwartz function on $\R^4$. Then define
\begin{equation}
\label{eq:g_X}
  g(X) := (2\pi)^{-2} \int \ud^4k \ \hat{g}(k) e^{-ik_{\mu} X^{\mu}},
\end{equation}
where $\hat{g}$ is the Fourier transform of $g$. In order to have a well--defined expression, it is crucial that $X^{\mu}$, and thus $A_{\mu}$, is self-adjoint. The field strength can now be evaluated with the observable
\begin{equation}
\label{eq:Observable}
  \int \ud^4q \ h_{\mu \nu}(X) F^{\mu \nu}.
\end{equation}
Here the (tensor valued) test function $h_{\mu \nu}$ encodes the localization properties of the detector.

\begin{remark}
\label{rem:NormalizedObservable}
From a conceptual point of view, it would we nice to have a state on the algebra $\End_\A(E)$. But the observable~(\ref{eq:Observable}) is in general neither normalized nor positive. Note that the the map $g \mapsto g(X)$ does not preserve positivity (this is already the case for $A=0$). Thus, it does not suffice to take a positive test function $g$. In order to have a positive evaluation functional, one could adopt the following procedure: Choose a Schwartz function $f$. Then $g(X) := f(X)^* f(X)$ is positive, and the functional
\begin{equation}
\label{eq:Normalized_Observable}
  F^{\mu \nu} \mapsto  \frac{\int \ud^4q \ g(X) F^{\mu \nu}}{\int \ud^4q \ g(X)}
\end{equation}
is positive and normalized.
The test function $f$ encodes the localization properties of the detector. However, the exact correspondence is not known.
It would of course be desirable to choose $f$ such that the resulting uncertainty is minimal, but this is a difficult problem for general $A_{\mu}$.
Observables of the form~(\ref{eq:Normalized_Observable}) were used in~\cite{NCED}. In the following, we will mostly use the observable~(\ref{eq:Observable}). As long as we are only interested in the frequency content and not the amplitudes, the conclusions we draw will be the same.
\end{remark}

\subsection{The free case}

From the action~(\ref{eq:NCEDAction}) one obtains the equation of motion
\begin{equation}
\label{eq:NCED_eom}
  \del_{\mu} F^{\mu \nu} - i e [A_{\mu}, F^{\mu \nu}] = 0.
\end{equation}
In the following, we construct three solutions of this equation, a constant field, a plane wave, and a superposition of both. We also evaluate the field strength in covariant coordinates. We remark that in~\cite{Bak} a solitonic solution has been treated in the same formalism.


\begin{example}
\label{ex:background}
Setting
\begin{equation}
\label{eq:A}
  A_{\mu} := c_{\mu \nu} q^{\nu}, \qquad c_{\mu \nu} \in \R
\end{equation}
we obtain the field strength
\begin{equation}
\label{eq:Background_F}
  F_{\mu \nu} = c_{\nu \mu} - c_{\mu \nu} + e (c \sigma c^T)_{\mu \nu}.
\end{equation}
The covariant coordinates are
\begin{equation*}
  X_1^{\mu} = q^{\mu} + e (\sigma c)^{\mu}_{\nu} q^{\nu} = (\mathbbm{1} + e \sigma c)^{\mu}_{\nu} q^{\nu}. 
\end{equation*}
For the field strength observable~(\ref{eq:Observable}), we obtain
\begin{equation*}
  \int \ud^4q \ h_{\mu \nu}(X) F^{\mu \nu} = F^{\mu \nu} \int \ud^4q \ h_{\mu \nu}(X) = (2\pi)^2 \betrag{\1+e\sigma c}^{-1} F^{\mu \nu} \hat h_{\mu \nu}(0).
\end{equation*}
Using the normalized observable~(\ref{eq:Normalized_Observable}), we would simply find $F^{\mu \nu}$. Thus, in Example~\ref{ex:superpos}, we will interpret $F^{\mu \nu}$ as the actual (background) field strength.

\end{example}

\begin{example}
\label{ex:PlaneWave}
A plane wave is given by the vector potential
\begin{equation}
\label{eq:plane_wave}
  a_{\mu} = b_{\mu} ( e^{- i k q} + e^{ i k q}), \qquad b_{\mu} \in \R.
\end{equation}
The resulting field strength is (the $f_{\mu \nu}$ here should not be confused with the commutative field strength from the Seiberg-Witten map)
\begin{equation*}
  f_{\mu \nu} = - i ( k_{\mu} b_{\nu} -  k_{\nu} b_{\mu} )  ( e^{- i k q} - e^{i k q} ).
\end{equation*}
In Lorentz gauge $(k_{\mu} b^{\mu}=0)$, the equation of motion is then solved for $k^2=0$. The corresponding covariant coordinates are
\begin{equation*}
  X_2^{\mu} = q^{\mu} + e \sigma^{\mu \nu} b_{\nu} ( e^{-ikq} +  e^{ikq} ).
\end{equation*} 
The field strength observable~(\ref{eq:Observable}) now yields
\begin{equation}
\label{eq:wave_field_strength}
  \int \ud^4q \ h_{\mu \nu}(X_2) f^{\mu \nu} = - 2 i k^\mu b^\nu (2 \pi)^{-2} \int \ud^4p \ \hat{h}_{\mu \nu}(p) \int \ud^4q \  e^{-ipX_2} ( e^{-ikq} - e^{ikq} ).
\end{equation}
Here we assumed $h_{\mu \nu}$ to be antisymmetric.
To evaluate the traces, we express $e^{-ipX_2}$ as
\begin{equation}
\label{eq:BCH_PQ}
   e^{- i pq - i e p \sigma b (e^{- ikq}+e^{ikq}) } = e^{-ipq} e^{- i e p \sigma b  \left( P(p \sigma k) (e^{- i k q}+e^{i k q}) - i Q(p \sigma k) (e^{-i k q}-e^{i k q}) \right) },
\end{equation}
with $P(x)=\frac{\sin x}{x}$ and $Q(x)=\frac{\cos x -1}{x}$. A proof of this equation can be found in Appendix~\ref{app:BCH_PQ}. We expand the second factor in $e$ and obtain
\begin{multline*}
  \int \ud^4q \ e^{-ipX_2} ( e^{-ikq} - e^{ikq} ) = (2\pi)^4 \big\{ \delta(p+k) - \delta(p-k) \\
   + 2 i e k \sigma b ( \delta(p+2k) + \delta(p-2k)) \big\} + \order(e^2).
\end{multline*}
At zeroth order in $e$ we thus find a plane wave with wave vector $k$, as expected. It is easy to see that at $n$th order in $e$ we get a sum of plane waves with wave vectors $(n'+1)k, n' \leq n$, i.e., higher harmonics appear. However, these are suppressed by a factor $(k \sigma b)^n$. In real experiments, $k \sigma b$ is a very small quantity.
For a bunch of the presently must powerful coherent soft X-ray source, FLASH at DESY, it can be estimated to be of the order\footnote{I thank B.~Beutner for the relevant informations.} $10^{18}~\text{m}^{-2}~\lambda_{nc}^2$. This is very small, even if $\lambda_{nc}$ is close to the scale of present-day accelerators.


The appearance of higher harmonics looks like a testable prediction, but there are practical and conceptual difficulties. First of all, in a laboratory one usually produces higher harmonics as well.
Furthermore, we already mentioned in Remark~\ref{rem:NormalizedObservable}, that one should rather use observables of the form~(\ref{eq:Normalized_Observable}). For such observables, the amplitudes depend on the test function used \cite{NCED}. But the exact correspondence between the detector and the test function is not known. Thus the theory does not bear much predictive power concerning the higher harmonics. Nevertheless it is possible, also with observables of the form (\ref{eq:Normalized_Observable}), to determine the wave vector $k$ of the plane wave~(\ref{eq:plane_wave}) by local measurements of the field strength. We will exploit this in the next example.
\end{example}

\begin{example}
\label{ex:superpos}
Since the equation of motion~(\ref{eq:NCED_eom}) is nonlinear, the superposition principle does not hold any more. Nevertheless, it is possible to superpose the constant background field from Example~\ref{ex:background} with a plane wave of the form~(\ref{eq:plane_wave}). But we will see that the wave vector is then in general no longer lightlike. We define the complete vector potential as
\begin{equation*}
  A^c_{\mu} = A_{\mu} + a_{\mu},
\end{equation*}
where $A_{\mu}$ is given by~(\ref{eq:A}) and $a_{\mu}$ is of the form~(\ref{eq:plane_wave}). The complete field strength is then
\begin{equation*}
  F^c_{\mu \nu} = F_{\mu \nu} + \del_{\mu} a_{\nu} - \del_{\nu} a_{\mu} - i e [A_{\mu}, a_{\nu} ] - i e [a_{\mu}, A_{\nu} ] = F_{\mu \nu} + \del'_{\mu} a_{\nu} - \del'_{\nu} a_{\mu}.
\end{equation*}
Here $F_{\mu \nu}$ is given by~(\ref{eq:Background_F}) and we used the notation
\begin{equation*}
  \del'_{\mu} g := \del_{\mu} g - i e [A_{\mu}, g].
\end{equation*}
Inserting this into the equation of motion~(\ref{eq:NCED_eom}), we obtain
\begin{equation*}
  \del_{\mu} \left( \del'^{\mu} a^{\nu} - \del'^{\nu} a^{\mu} \right) - i e \left[ A_{\mu},  \left( \del'^{\mu} a^{\nu} - \del'^{\nu} a^{\mu} \right) \right] = \del'_{\mu} \left( \del'^{\mu} a^{\nu} - \del'^{\nu} a^{\mu} \right) = 0.
\end{equation*}
Thus, in the pseudo Lorentz gauge $\del'_{\mu} a^{\mu} = 0$, we find the equation of motion
\begin{equation}
\label{eq:Background_eom}
  \Box' a^{\nu} = 0.
\end{equation}
In order to solve it, we seek the coordinates $q'$ dual to the derivatives $\del'$, i.e., $\del'_{\mu} q'^{\nu} = \delta_{\mu}^{\nu}$.
Up to an additive constant, these are
\begin{equation*}
  q'^{\mu} = { ( \mathbbm{1} - e \sigma c^T )^{-1} }^{\mu}_{\nu} q^{\nu}.
\end{equation*}
The equation of motion (\ref{eq:Background_eom}) is thus solved by
\begin{equation*}
  a_{\mu} = b_{\mu} ( e^{-ikq'} +  e^{ikq'} ) .
\end{equation*}
with $k \cdot b = 0$ (pseudo Lorentz gauge) and $k^2 = 0$. The complete field strength is now
\begin{equation*}
  F^c_{\mu \nu} = F_{\mu \nu} - i (k_{\mu} b_{\nu} - b_{\mu} k_{\nu}) ( e^{-ikq'} - e^{ikq'} ).
\end{equation*}
Evaluating this in covariant coordinates, we see that the first term gives again the constant field strength $F_{\mu \nu}$. In the second term, we are once more only interested in the frequency content. The computation is completely analogous to the one in Example~\ref{ex:PlaneWave}, we simply have to replace $pq$ by $p(\mathbbm{1}+e\sigma c)q$ and $kq$ by $k (\mathbbm{1}-e\sigma c^T)^{-1} q$. Thus, $p$ is set to nonzero integer multiples of
\begin{equation*}
  k' = k (\mathbbm{1}-e \sigma c^T)^{-1} (\mathbbm{1}+ e \sigma c)^{-1} = k (\mathbbm{1} - e \sigma c^T + e \sigma c - e^2 \sigma c \sigma c^T )^{-1} = k (\mathbbm{1}- e \sigma F)^{-1},
\end{equation*}
which is then the wave vector that is actually measured.
In order to compare this to the results obtained via the Seiberg--Witten map, we absorb $e$ in $A_\mu$: $A'_\mu = e A_\mu$. Thus, $F'=e F$.
Denoting by $\V{e}$ and $\V{m}$ the electric and magnetic components of $\sigma$ and by $\V{E}$ and $\V{B}$ the electric and magnetic field in $F'$, we then have
\begin{align*}
  k_0 & = k'_0 ( 1 - \V{e} \cdot \V{E} ) - \V{k'} \cdot ( \V{m} \times \V{E} ) \\
  \V{k} & = \V{k'} ( 1 - \V{m} \cdot \V{B} ) + k'_0 ( \V{e} \times \V{B} ) - ( \V{k'} \cdot \V{e} ) \V{E} + ( \V{k'} \cdot \V{B} ) \V{m}.
\end{align*}
Assuming $\V{E}=0$ and $k_0' >0$, this leads to
\begin{align*}
  k_0 & = k'_0 \\
  \betrag{\V{k}}^2 & =  \betrag{\V{k'}}^2 \left( 1- 2 \V{{m}_T} \cdot \V{B_T} + 2 \hat{\kappa} \cdot (\V{e} \times \V{B}) \right) + \order((\sigma B)^2),
\end{align*}
where $\hat{\kappa}$ is the unit vector in the direction $\V{k'}$ and the subscript $\V{T}$ stands for the part transversal to $\V{k'}$. Using $k^2=0$ we find the modified dispersion relation
\begin{equation}
\label{eq:disp_rel}
  \omega' = \betrag{\V{k'}} \left( 1 - \V{{m}_T} \cdot \V{B_T} + \hat{\kappa} \cdot (\V{e} \times \V{B}) \right) + \order((\sigma B)^2).
\end{equation}
This is in agreement with the results obtained in~\cite{Guralnik,Cai,Abe} via the Seiberg--Witten map\footnote{Note that~\cite{Guralnik, Cai} were only considering the case of space/space noncommutativity, i.e., $\V{e} = \V{0}$. We also remark that in the original version of~\cite{Abe} a different result was obtained. The calculation was corrected after the calculation presented here had been published in~\cite{NCED}.}. We mention that if one evaluates the field strength not in covariant coordinates but in the usual ones, as in \cite{ChaichianBackground, Mariz}, one obtains a different result. Thus, the covariant coordinates are really necessary to obtain the correct result. We also remark that, as shown in \cite{Mariz}, the group velocity obtained from (\ref{eq:disp_rel}) is not parallel to the wave vector~$\hat \kappa$. Explicitly, one finds
\begin{equation*}
  \V{v_g} = \hat \kappa ( 1 - \V{{m}_T} \cdot \V{B_T} ) + \V{m_T} \hat \kappa \cdot \V B + \V{B_T} \hat \kappa \cdot \V m + \V e \times \V B + \order((\sigma B)^2).
\end{equation*}



In order to discuss possible experimental tests of the modified dispersion relation~(\ref{eq:disp_rel}), we estimate orders of magnitude. We consider a magnetic field of $1~\mathrm{T}$ and assume the noncommutativity scale to be close to the scale of present-day accelerators: $\lambda_{nc} = 10^{-20}~\mathrm{m}$. We find $\sigma B \approx 10^{-24}$. Since this is a very small number, it seems to be necessary to consider astronomical experiments, where small effects have enough time to sum up. One might for example use galactical magnetic fields. In the milky way these are of the order of $10^{-9}~\mathrm{T}$, so the correction would be of order $10^{-33}$. Multiplying by the diameter of the milky way, $10^{5}~\mathrm{ly}$, we find a shift in the arrival time of the order $10^{-20}~\mathrm{s}$. This seems to be far too small to be detectable. There is also the conceptual problem of finding a reference signal. Similar considerations can be found in~\cite{Guralnik}.
\end{example}


\subsection{Coupling to a source}
\label{sec:SourceTerm}

We now discuss how to couple the electromagnetic field to external sources in the unexpanded (module) approach. Of course we want to keep gauge invariance. Thus, we add the term
\begin{equation}
\label{eq:NCED_Coupling}
  \frac{1}{2} \int \ud^4q \ f^{\mu \nu}(X) F_{\mu \nu}
\end{equation}
to the action~(\ref{eq:NCEDAction}). This corresponds to the observable~(\ref{eq:Observable}). The prefactor has been introduced for convenience. Variation of $F_{\mu \nu}$ with respect to $A_\nu$ leads to a contribution $- D_{\mu} f^{\mu \nu}(X)$ to the equation of motion~(\ref{eq:NCED_eom}). However, we also have to vary the $A$ that enters the covariant coordinate $X$. In order to do this, we expand the exponential $e^{-ikX}$ as a power series. Variation with respect to $A_\nu$ then yields
\begin{equation*}
  \sum_{n=1}^\infty \frac{1}{n!} \sum_{m=0}^{n-1} (-ikX)^{n-m-1} (- i e k_\mu \sigma^{\mu \nu} \delta A_\nu ) (-ikX)^m.
\end{equation*}
Using the cyclicity of the integral, we can pull $\delta A_\nu$ to the left, obtaining
\begin{equation*}
  \frac{1}{2} (2 \pi)^{-2} \int \ud^4k \ \hat{f}^{\lambda \rho}(k) (-i e k \sigma)^\nu  \int \ud^4q  \left( \delta A_\nu \sum_{n=1}^\infty \frac{1}{n!} \sum_{m=0}^{n-1} (-ikX)^{n-m-1} F_{\lambda \rho} (-ikX)^m \right).
\end{equation*}
Now we need the following

\begin{lemma}
\label{lemma:NCED}
Let $C, D$ be elements of some algebra. Then we have
\begin{equation*}
  \sum_{n=1}^\infty \frac{1}{n!} \sum_{m=0}^{n-1} C^{n-m-1} D C^m = e^C P_1(- \ad_C ) D = P_1(\ad_C) D e^C.
\end{equation*}
with
\begin{equation}
\label{eq:P_1}
  P_1(x) = \sum_{k=0}^\infty \frac{x^k}{(k+1)!}.
\end{equation}
\end{lemma}

A proof can be found in Appendix~\ref{app:NCED}. Because of~(\ref{eq:X_D}), we have $[ikX,F_{\lambda \rho}] = - k \sigma D F_{\lambda \rho}$. Thus, we finally obtain the equation of motion
\begin{equation}
\label{eq:NCED_source_eom}
  D_\mu F^{\mu \nu} = D_\mu f^{\mu \nu}(X) - \frac{1}{2} (2\pi)^{-2} \int \ud^4k \ \hat{f}^{\lambda \rho}(k) (-i e k \sigma)^\nu e^{-ikX} P_1(-k \sigma D) F_{\lambda \rho}.
\end{equation}
It is highly nonlinear. In order to make it more plausible, consider the commutative limit: $\del_\mu F^{\mu \nu} = \del_\mu f^{\mu \nu}$. Interpreting the right hand side as a current, we have recovered the Maxwell equations. Thus, we propose the following procedure to study the effect of sources. Take a commutative current $j^\nu(x)$ and solve $\del_\mu f^{\mu \nu}(x) = j^\nu(x)$ for the commutative field strength $f^{\mu \nu}$. Now use~(\ref{eq:g_X}) to define $f^{\mu \nu}(X)$ and solve~(\ref{eq:NCED_source_eom}).
This can probably only be done in the sense of formal power series in $e$. We study this approach in the following

\begin{example}
We consider the case of a static source, i.e., $j^\nu(x) = \delta_0^\nu j(\V{x})$. It is convenient to introduce a vector potential $a^\nu$ for $f^{\mu \nu}$. With the ansatz $a^i = 0$, $a^0(x) = a^0(\V{x})$, we get\footnote{Here and in the following the Fourier transform $\hat{f}(\V{k})$ of a time-independent function $f$ is implicitly defined by $\hat{f}(k) = \delta(k_0) \hat{f}(\V{k})$.}
\begin{equation*}
  \hat{a}^0 (\V{k}) = \frac{ \hat{j}(\V{k})}{\betrag{\V{k}}^2}, \quad \hat{f}^{0 i}(\V{k}) = - i \frac{ k^i \hat{j}(\V{k})}{\betrag{\V{k}}^2},
\end{equation*}
with all other components vanishing. Expanding~(\ref{eq:NCED_source_eom}) in the coupling constant $e$, we obtain
\begin{multline*}
  D_\mu F^{\mu \nu} = D_\mu f^{\mu \nu}(q) + \del_\mu (2\pi)^{-2} \int \ud^4k \ \hat{f}^{\mu \nu}(k) (-iek \sigma)_\lambda e^{-ikq} P_1(-k \sigma \del) A^\lambda \\
  - (2 \pi)^{-2} \int \ud^4k \ \hat{f}^{\lambda \rho}(k) (-i e k \sigma)^\nu e^{-ikq} P_1(-k \sigma \del) \del_\lambda A_\rho + \order(e^2).
\end{multline*}
The second term on the left hand side, which is obtained from the expansion of $f^{\mu \nu}(X)$ in $e$, has been calculated again by the use of Lemma~\ref{lemma:NCED}. A simple ansatz is now to set $A^\nu = A^\nu_0 + e A^\nu_1 + \order(e^2)$ with $A^\nu_0 = a^\nu(q)$. This solves the above at zeroth order in $e$. Due to the fact that only the $0$-component of $a$ is nonvanishing, we have $[A_0^\mu, A_0^\nu] = 0$. Then the equation for $A_1$ is 
\begin{multline}
\label{eq:A_1_eom}
  \del_\mu (\del^\mu A_1^\nu - \del^\nu A_1^\mu)  = \del_\mu (2 \pi)^{-2} \int \ud^4k \ \hat{f}^{\mu \nu}(k) (-iek \sigma)_\lambda e^{-ikq} P_1(-k \sigma \del) A_0^\lambda \\
   - (2 \pi)^{-2} \int \ud^4k \ \hat{f}_{\lambda \rho}(k) (-i e k \sigma)^\nu e^{-ikq} P_1(-k \sigma \del)  \del^\lambda A^\rho_0.
\end{multline}
Because of $P_1(x) = e^{x/2} (e^{x/2} - e^{-x/2})/x$, we have
\begin{equation}
\label{eq:P_Equation}
  P_1(ix) e^{- \frac{i}{2} x} = \frac{ \sin x/2 }{ x/2 }.
\end{equation}
With this, one obtains
\begin{equation*}
  \del_\mu (2 \pi)^{-4} \int \ud^4k \ud^4p \ \hat{f}^{\mu \nu}(k) \hat{a}^\lambda(p) (-iek \sigma)_\lambda \frac{\sin \frac{k \sigma p}{2}}{\frac{k \sigma p}{2}} e^{-i(k+p)q}
\end{equation*}
for the first term on the right hand side of~(\ref{eq:A_1_eom}). In a similar fashion we get
\begin{equation*}
 - \frac{1}{4} (\del \sigma)^\nu (2 \pi)^{-4} \int \ud^4k \ud^4p \ \hat{f}_{\lambda \rho}(k) \hat{f}^{\lambda \rho}(p) \frac{\sin \frac{k \sigma p}{2}}{\frac{k \sigma p}{2}} e^{-i(k+p)q}
\end{equation*}
for the second term on the right hand side of~(\ref{eq:A_1_eom}).
Thus, equation~(\ref{eq:A_1_eom}) is of the form
\begin{equation*}
  \del_\mu (\del^\mu A_1^\nu - \del^\nu A_1^\mu)  = \del_\mu C^{\mu \nu} - \frac{1}{4} (\del \sigma)^\nu D
\end{equation*}
with $C$ and $D$ time-independent and $C^{\mu \nu}$ antisymmetric and purely electric. This can be solved by
\begin{equation*}
  \hat{A}_1^\nu(k) = \delta(k_0) \left( -i \delta_0^\nu \frac{ k_i \hat{C}^{i 0}(\V{k})}{\betrag{\V{k}}^2} + \frac{i}{4} \frac{(k \sigma)^\nu \hat{D}(\V{k})}{\betrag{\V{k}}^2} \right).
\end{equation*}
Now we evaluate the field strength in covariant coordinates: $\int \ud^4q \ h_{\mu \nu}(X) F^{\mu \nu}$. In zeroth order in $e$ this is
\begin{equation*}
  \int \ud^4k \ \hat{h}_{\mu \nu}(-k) \hat{F}^{\mu \nu}(k) 
\end{equation*}
which is the same as in the commutative case. At first order in $e$, there is the contribution involving $A_1$:
\begin{multline*}
  2 (2 \pi)^{-2} \int \ud^4k \ud^4p \ \delta(k_0) \delta(p_0) \hat{h}_{\mu \nu}(-k) \frac{k^\mu}{\betrag{\V{k}}^2} \frac{\sin \frac{k \sigma p}{2}}{\frac{k \sigma p}{2}} \\
  \times \Big\{ - \delta_0^\nu k_i \hat{f}^{i 0}(\V{p}) \hat{a}^\lambda(\V{k}-\V{p}) (- i p \sigma)_\lambda + \frac{1}{4} (k \sigma)^\nu \hat{f}_{\lambda \rho}(\V{p}) \hat{f}^{\lambda \rho}(\V{k}-\V{p}) \Big\}.
\end{multline*}
Furthermore, there is an $\order(e)$-contribution from the the covariant coordinate in $h_{\mu \nu}(X)$, which can again be calculated by the use of Lemma~\ref{lemma:NCED}:
\begin{equation*}
  2 (2 \pi)^{-2} \int \ud^4k \ud^4p \ \delta(k_0) \delta(p_0) \hat{h}_{\mu \nu}(-k) \frac{\sin \frac{k \sigma p}{2}}{\frac{k \sigma p}{2}} p^\mu (k \sigma)_\lambda \hat{a}^\nu(\V{p}) \hat{a}^\lambda(\V{k}-\V{p}).
\end{equation*}
Combining these, one obtains
\begin{multline*}
  2 (2 \pi)^{-2} \int \ud^4k \ud^4p \ \delta(k_0) \delta(p_0) \frac{\sin \frac{k \sigma p}{2}}{\frac{k \sigma p}{2}} \hat{a}(\V{p}) \hat{a}(\V{k}-\V{p}) \\
  \times \left\{ \hat{h}_{i 0}(-k) \left( (k \sigma)^0 p^i + k \cdot p (p \sigma)^0 \frac{k^i}{\betrag{\V{k}}^2} \right) - \frac{1}{2} \hat{h}_{i \nu}(-k) p \cdot ( k - p ) (k \sigma)^\nu \frac{k^i}{\betrag{\V{k}}^2} \right\}.
\end{multline*}
Here $a$ stands for $a^0$. We note that the factor containing $k \sigma p$ is $1 + \order((k \sigma p)^2)$, which is $1$ for all practical purposes. Thus, the $\order(e)$ correction to the measured field strength is
\begin{multline}
\label{eq:F_m_CovCoor}
  \hat{F}^{\mu \nu}_{\text m}(k) = (2 \pi)^{-2} \delta(k_0) \int \ud^4p \ \delta(p_0) \hat{a}(\V{p}) \hat{a}(\V{k}-\V{p}) \\
  \times \left\{ \delta^{\nu}_0 \left( (k \sigma)^0 p^\mu + k \cdot p (p \sigma)^0 \frac{k^\mu}{\betrag{\V{k}}^2} \right) - \frac{1}{2} p \cdot ( k - p ) (k \sigma)^\nu \frac{k^\mu}{\betrag{\V{k}}^2} \right\} - \mu \leftrightarrow \nu.
\end{multline}
One could now use this equation as a starting point for phenomenological studies. But  the correction will be very small for $f \sigma \ll 1$, which is the case for realistic field strength. We conclude that it seems highly unlikely that it will be possible to test classical noncommutative electrodynamics experimentally.

Finally, we note that~(\ref{eq:F_m_CovCoor}) is also the term of first order in $\sigma$. We may thus compare it to the result obtained in~\cite{Berrino} via the Seiberg-Witten map. But there the coupling term
\begin{equation*}
  \int \ud^4x \ j^\nu(x) a_\nu(x)
\end{equation*}
was used, with $j^\nu$ a conserved current. This does not coincide with the expansion in $\sigma$ of the coupling term~(\ref{eq:NCED_Coupling}) with~(\ref{eq:SW_A}), (\ref{eq:SW_F}) and $j^\nu = \del_\mu f^{\mu \nu}$ inserted. Hence it is not surprising that the resulting correction to the field strength differs from the one found here.

\end{example}

\begin{remark}
Note that the expansion in $e$ that we use here and the expansion in $\sigma$ that is used in the Seiberg--Witten map are different. The expansion in $e$ of $e^{-ikX}$ brings, due to the form of $X$, a power of $\sigma$ for each power of $e$. However, we do not expand the noncommutative product. E.g., the zeroth order term of the expansion of $e^{-ikX}$ is $e^{-ikq}$ which fulfills the Weyl relation $e^{-ikq} e^{-ipq} = e^{- \frac{i}{2} k \sigma p} e^{-i(k+p)q}$ and is thus still able to yield arbitrary powers of $\sigma$.
\end{remark}

\chapter{NCQFT in the Yang--Feldman formalism}
\label{chapter:NCQFT}

The main goal of this chapter is the development of methods to study noncommutative quantum field theories in the Yang--Feldman formalism. 
We begin by introducing a graphical notation.
Then we consider localized mass terms as interactions and study their adiabatic limit. 
This helps us to understand how to compute modified dispersion relations from perturbative calculations in the truly interacting case.
Finally, we apply these methods to the $\phi^3$ and the Wess--Zumino model. We provide a rigorous definition of the nonplanar loop integral and show that the distortion of the group velocity is moderate for parameters typically expected for the Higgs field.

\section{Graphical rules}
\label{sec:Graphs}

The main idea of the Yang--Feldman formalism is as follows: One starts with some equation of motion, e.g.,
\begin{equation}
\label{eq:eom}
  ( \Box + m^2 ) \phi = \lambda \phi^2
\end{equation}
for the case of the $\phi^3$-model. As in this example, one splits it into a free and an interaction part.
The free part should
admit a well-posed Cauchy problem.
One then defines the interacting field as a formal power series in the coupling constants, i.e., in the case~(\ref{eq:eom}),
\begin{equation*}
  \phi_{int} = \sum_{n=0}^\infty \lambda^n \phi_n.
\end{equation*}
Inserting this in~(\ref{eq:eom}), it is straightforward to see that $\phi_0$ is a solution of the free equation of motion. Thus, it can be represented in Fock space in the usual way. One identifies it with the incoming field. The higher order components of $\phi_{int}$ are then computed recursively. For example, in the case~(\ref{eq:eom}), one has
\begin{equation*}
  \phi_n = \sum_{k=0}^{n-1} \Delta_R \times ( \phi_k \phi_{n-1-k} ).
\end{equation*}
Here $\Delta_R$ is the retarded propagator and $\times$ stands for convolution, cf.~(\ref{eq:Convolution}). The first terms in the noncommutative $\phi^3$ model are then
\begin{subequations}
\begin{align}
\label{subeq:phi_1_q}
  \phi_1(q) = & \int \ud^4 x \ \Delta_R(x) \phi_0(q-x) \phi_0(q-x), \\
\label{subeq:phi_2_q}
  \phi_2(q) = & \int \ud^4 x \ \Delta_R(x) \left( \phi_1(q-x) \phi_0(q-x) + \phi_0(q-x) \phi_1(q-x) \right).
\end{align}
\end{subequations}
In the notation of~\cite{DorosDiss}, which is very similar to Dyson's doubled graph notation \cite{Dyson}, the two terms of $\phi_2(q)$ are graphically represented as
\begin{equation}
\label{graphs:phi_2}
\begin{picture}(100,65)
\Line(49.5,15)(49.5,30)
\Line(50.5,15)(50.5,30)
\Line(50,30)(70,60)
\Line(49.4,30)(39.4,45)
\Line(50.6,30)(40.6,45)
\Line(40,45)(30,60)
\Line(40,45)(50,60)
\Vertex(50,30){1}
\Vertex(40,45){1}
\BCirc(30,60){1}
\BCirc(50,60){1}
\BCirc(70,60){1}
\Text(50,13)[t]{$q$}
\Text(47,30)[r]{$q-x_1$}
\Text(37,45)[r]{$q-x_1-x_2$}
\end{picture}
\begin{picture}(100,65)
\Line(49.5,15)(49.5,30)
\Line(50.5,15)(50.5,30)
\Line(50,30)(30,60)
\Line(49.4,30)(59.4,45)
\Line(50.6,30)(60.6,45)
\Line(60,45)(50,60)
\Line(60,45)(70,60)
\Vertex(50,30){1}
\Vertex(60,45){1}
\BCirc(30,60){1}
\BCirc(50,60){1}
\BCirc(70,60){1}
\Text(50,13)[t]{$q$}
\Text(53,30)[l]{$q-x_1$}
\Text(63,45)[l]{$q-x_1-x_2$}
\end{picture}
\end{equation}
Here the double line stands for the retarded propagator. The open circles symbolize uncontracted free fields, i.e., $\phi_0$. 
The vertex factors can be obtained by considering~$\phi_1$:
\begin{equation}
\label{eq:phi_1}
  \hat{\phi}_1(k_0) = \hat \Delta_R(k_0) \int \ud^4k_1 \ud^4k_2 \ \delta(k_0+k_1+k_2) e^{- \frac{i}{2} k_1 \sigma k_2} \hat \phi_0(-k_1) \hat \phi_0(-k_2).
\end{equation}
By equating the momentum of the incoming line with $k_0$, and the momenta of the two outgoing lines with $k_1$ and $k_2$, respectively, we obtain, in momentum space, the vertex factor
\vspace{-25pt}
\begin{center}
\begin{picture}(50,55)(0,25)
\Line(24.5,15)(24.5,30)
\Line(25.5,15)(25.5,30)
\Line(25,30)(15,45)
\Line(25,30)(35,45)
\Vertex(25,30){1}
\Text(25,13)[t]{$k_0$}
\Text(12,45)[r]{$k_1$}
\Text(38,45)[l]{$k_2$}
\end{picture}
$ = e^{- \frac{i}{2} k_1 \sigma k_2} \delta(k_0+k_1+k_2).$
\end{center}
\vspace{25pt} 
The graphs contributing to $\phi_n$ are then rooted trees with $n$ inner vertices. This means that each graph begins at the root with a double line and each inner vertex is connected to the root via double lines in exactly one way. There are no loops. For simplicity we draw all uncontracted fields on a horizontal line. Since the order is important, lines are not allowed to cross.

Quantum effects, i.e., loops, enter when contractions are considered. The simplest example occurs in $\phi_1$. The contraction of the two free fields in (\ref{eq:phi_1}) is described by the following graph:
\begin{center}
\begin{picture}(50,40)
\Line(24.5,5)(24.5,20)
\Line(25.5,5)(25.5,20)
\Line(25,20)(15,35)
\Line(25,20)(35,35)
\Line(15,35)(35,35)
\Vertex(25,20){1}
\BCirc(15,35){1}
\BCirc(35,35){1}
\end{picture}
\end{center}
The line connecting two open circles, i.e., free fields, is the free two--point function $\Delta_+$. In momentum space,
\vspace{-5pt}
\begin{center}
\begin{picture}(30,15)(0,5)
\Line(10,7)(25,7)
\BCirc(10,7){1}
\BCirc(25,7){1}
\Text(8,7)[r]{$k$}
\end{picture}
$ = (2\pi)^2 \hat \Delta_+(k).$
\end{center}
\vspace{5pt}
The resulting loop integral $\int \ud^4l \ \hat \Delta_+(l)$ is divergent. The easiest way to get rid of it, is to normal order $\phi_1$, i.e., one defines
\begin{equation*}
  \phi_1(q) = \int \ud^4 x \ \Delta_R(x) \WDp{\phi_0(q-x) \phi_0(q-x)}
\end{equation*}
instead of~(\ref{subeq:phi_1_q}). One can then split $\phi_2$ into a normal ordered and a contracted part. Graphically, we have
\begin{center}
\begin{picture}(400,70)
\Line(49.5,15)(49.5,30)
\Line(50.5,15)(50.5,30)
\Line(50,30)(70,60)
\Line(49.4,30)(39.4,45)
\Line(50.6,30)(40.6,45)
\Line(40,45)(30,60)
\Line(40,45)(50,60)
\Vertex(50,30){1}
\Vertex(40,45){1}
\BCirc(30,60){1}
\BCirc(50,60){1}
\BCirc(70,60){1}
\Text(27,60)[r]{$:$}
\Text(53,60)[l]{$:$}
\Text(100,30)[]{$=$}
\Line(149.5,15)(149.5,30)
\Line(150.5,15)(150.5,30)
\Line(150,30)(170,60)
\Line(149.4,30)(139.4,45)
\Line(150.6,30)(140.6,45)
\Line(140,45)(130,60)
\Line(140,45)(150,60)
\Vertex(150,30){1}
\Vertex(140,45){1}
\BCirc(130,60){1}
\BCirc(150,60){1}
\BCirc(170,60){1}
\Text(127,60)[r]{$:$}
\Text(173,60)[l]{$:$}
\Text(200,30)[]{+}
\Line(249.5,15)(249.5,30)
\Line(250.5,15)(250.5,30)
\Line(250,30)(270,60)
\Line(249.4,30)(239.4,45)
\Line(250.6,30)(240.6,45)
\Line(240,45)(230,60)
\Line(240,45)(250,60)
\Line(250,60)(270,60)
\Vertex(250,30){1}
\Vertex(240,45){1}
\BCirc(230,60){1}
\BCirc(250,60){1}
\BCirc(270,60){1}
\Text(300,30)[]{$+$}
\Line(349.5,15)(349.5,30)
\Line(350.5,15)(350.5,30)
\Line(350,30)(370,60)
\Line(349.4,30)(339.4,45)
\Line(350.6,30)(340.6,45)
\Line(340,45)(330,60)
\Line(340,45)(350,60)
\Curve{(330,60)(350,65)(370,60)}
\Vertex(350,30){1}
\Vertex(340,45){1}
\BCirc(330,60){1}
\BCirc(350,60){1}
\BCirc(370,60){1}
\end{picture}
\end{center}
and analogously for the second graph in~(\ref{graphs:phi_2}). As an example, we compute the two contracted graphs. We find
\begin{equation*}
  \hat \phi_2(k) = (2\pi)^2 \hat \Delta_R(k) \hat \phi_0(k) \int \ud^4l \ \hat \Delta_R(k-l) \hat \Delta_+(-l) e^{- \frac{i}{2} (k-l) \sigma l} e^{\mp \frac{i}{2} k \sigma (-l)}.
\end{equation*}
Here the $-$ sign in the second twisting factor refers to the first contraction while the $+$ sign applies to the second. In the first case, the phase factors in the twistings cancel each other, while in the second case they add up. One speaks of a planar and a nonplanar part. Adding the terms coming form the second graph in (\ref{graphs:phi_2}), one obtains for the contracted part of $\phi_2$, as in~\cite{BDFP02}:
\begin{equation}
\label{eq:Sigma_phi_2}
  \hat \phi_2(k) = (2\pi)^2 \hat \Delta_R(k) \hat \phi_0(k) \int \ud^4l \ \hat \Delta_R(k-l) \left\{ \hat \Delta_+(-l) \left( 1 + e^{ - i k \sigma l} \right) + \hat \Delta_+(l) \left( 1 + e^{ + i k \sigma l} \right) \right\}.
\end{equation}
The integral in this expression is not well--defined. Its treatment is discussed later. Assuming that we can give some meaning to the integral, it will be a function $\Sigma(k)$. This is the analog of the self--energy in the conventional approach to QFT. The above equation then looks very similar to what a mass term would contribute at first order. To see this, consider, instead of (\ref{eq:eom}), the equation of motion
\begin{equation}
\label{eq:eom_mu}
  (\Box + m^2) \phi = - \mu \phi
\end{equation}
and take $m$ to be the free mass. Then the term of order $\mu$ of the interacting field is given by
\begin{equation}
\label{eq:mu_phi_1}
  \hat \phi_1(k) = - \hat \Delta_R(k) \hat \phi_0(k).
\end{equation}
Apart from the loop integral, this has the same form as (\ref{eq:Sigma_phi_2}).
We can thus anticipate that $\Sigma(k)$ gives rise a to mass (and also a field strength) renormalization. But before we discuss this, we want to treat another problem: 
The product on the right hand side of (\ref{eq:mu_phi_1}) is not well--defined. The distributions $\hat \Delta_R$ and $\hat \phi_0$ are both singular on the mass shell and their wave front sets are such that their product can not be defined in the sense of H\"ormander~\cite{Hoermander}. In contrast to the problem of defining the loop integral in~(\ref{eq:Sigma_phi_2}), this is an infrared problem. We treat it in the next section.

\section{The adiabatic limit}

We have seen above that the higher order terms in the Yang--Feldman series are not well--defined. One should modify the equation of motion by introducing an infrared cutoff. In a commutative field theory, this can simply be done by multiplying the interaction with a test function $g$, e.g., to consider
\begin{equation*}
  (\Box + m^2) \phi = - \mu g \phi
\end{equation*}
instead of (\ref{eq:eom_mu}). One can then proceed to calculate $n$-point functions of the interacting field. At the end, one removes the cutoff by sending $g \to 1$, i.e., one performs the adiabatic limit. Then the following questions naturally appear:
\begin{enumerate}
\item Does the adiabatic limit exist?
\item One can declare parts of the free equation of motion as belonging to the interaction part. Is the final result independent of this ambiguity, and in which sense?
\item What is an appropriate infrared cutoff in the noncommutative case? 
\end{enumerate}
For the commutative case, the first question has been answered affirmatively by Epstein and Glaser~\cite{EpsteinGlaser}. The second question is related to the principle of perturbative agreement that has been introduced as a renormalization condition by Hollands and Wald~\cite{HollandsWald}. In the following, we will discuss it for the case of a mass term in a commutative theory. The answer to the third question will be given afterwards in Section~\ref{sec:MassNoncommutative}.

\subsection{The commutative case}
\label{sec:MassCommutative}

Our starting point is the equation of motion
\begin{equation}
\label{eq:eom_g_a}
  (\Box + m^2) \phi = - \mu g_a \phi.
\end{equation}
Here $\{g_a\}$ is a sequence of test functions that converges in $\mathcal{O}_M$ (the space of slowly increasing smooth functions, see~\cite{Schwartz}), to the constant function 1.
In this limit, the \emph{adiabatic limit}, the above would be the equation of motion for a free field of mass $\sqrt{m^2+\mu}$. In the Yang--Feldman approach, the interacting field is given by the formal power series
\begin{equation*}
  \phi_{int} = \sum_{n=0}^\infty \mu^n \phi_n,
\end{equation*}
where $\phi_n$ is recursively defined by
\begin{equation*}
  \phi_n = - \Delta_R \times (g_a \phi_{n-1}),
\end{equation*}
$\phi_0$ being the free field. The adiabatic limit of the interacting field does not exist. This can be seen as a consequence of Haag's theorem \cite{HaagsThm}. However, the so-called weak adiabatic limit, i.e., the adiabatic limit of the $n$-point functions of the interacting field, will exist. In the following, we discuss the two--point function
\begin{equation}
\label{eq:2pt}
  \bra{\Omega} \phi_{int}(f) \phi_{int}(h) \ket{\Omega}.
\end{equation}
Here $f$ and $h$ are Schwartz functions and
\begin{equation*}
  \phi(f) = \int \ud^4 x \ \phi(x) f(x).
\end{equation*}
Since the equation of motion (\ref{eq:eom_g_a}) describes a free field of mass $\sqrt{m^2+\mu}$ in the adiabatic limit, we expect to find, at $n$th order in $\mu$,
\begin{equation}
\label{eq:muExpansion}
  \int \ud^4x \ud^4y \ f(x) h(y) \frac{1}{n!} \left( \frac{\del}{\del m^2} \right)^n \Delta_+(x-y, m^2)
\end{equation}
for $a \to \infty$. At zeroth order in $\mu$, this is obviously the case. We take a look at the first order term in~(\ref{eq:2pt}):
\begin{multline}
\label{eq:1stOrder}
  \bra{\Omega} \left\{ \phi_1(f) \phi_0(h) + \phi_0(f) \phi_1(h) \right\} \ket{\Omega} \\
\shoveleft  = - \int \prod_{i=0}^2 \ud^4x_i \ f(x_0) h(x_2) g_a(x_1) \\
  \times \left\{ \Delta_R(x_0-x_1) \Delta_+(x_1-x_2) + \Delta_+(x_0-x_1) \Delta_A(x_1-x_2) \right\}.
\end{multline}
In our graphical notation, the two terms in (\ref{eq:1stOrder}) are expressed as
\begin{center}
\begin{picture}(150,50)
\Line(9.5,15)(9.5,30)
\Line(10.5,15)(10.5,30)
\Line(10,30)(10,45)
\Line(40,15)(40,45)
\Line(10,45)(40,45)
\Vertex(10,30){1}
\BCirc(10,45){1}
\BCirc(40,45){1}
\Text(7,30)[r]{$x_1$}
\Text(10,13)[t]{$x_0$}
\Text(40,13)[t]{$x_2$}
\Text(75,25)[]{$+$}
\Line(139.5,15)(139.5,30)
\Line(140.5,15)(140.5,30)
\Line(110,15)(110,45)
\Line(140,30)(140,45)
\Line(110,45)(140,45)
\Vertex(140,30){1}
\BCirc(110,45){1}
\BCirc(140,45){1}
\Text(143,30)[l]{$x_1$}
\Text(110,13)[t]{$x_0$}
\Text(140,13)[t]{$x_2$}
\end{picture}
\end{center}
In momentum space, the right hand side of (\ref{eq:1stOrder}) is
\begin{equation*}
  - (2 \pi)^2 \int \ud^4k_0 \ud^4k_1 \ \hat f(-k_0) \hat h(k_1) \check g_a(k_1-k_0) \left\{ \hat \Delta_R(k_0) \hat \Delta_+(k_1) + \hat \Delta_+(k_0) \hat \Delta_A(k_1) \right\}.
\end{equation*}
Using~(\ref{eq:Delta_R}) and setting $\pm x = k^0_{0/1} - \omega_{0/1}$, this can be written as
\begin{multline}
\label{eq:1st_order}
  \frac 1{2\pi}\int \frac{\ud^3 \V{k_0}}{2 \omega_0} \frac{\ud^3 \V{k_1}}{2 \omega_1} \ud x \ \check{g}_a(\omega_1 - \omega_0 -x, \V{k_1} - \V{k_0}) \\
 \times \left\{ \hat{f}(-\omega_0-x, -\V{k_0}) \hat{h}(\omega_1, \V{k_1}) \left( \iep{x} - \iep{x + 2 \omega_0} \right) \right. \\
  \left. - \hat{f}(-\omega_0, - \V{k_0}) \hat{h}(\omega_1-x, \V{k_1}) \left( \iep{x} - \iep{x - 2 \omega_1} \right) \right\}.
\end{multline}
We first deal with the two terms involving $\iep{x}$. We Taylor expand $\hat f$ and $\hat h$, i.e., we write
\begin{align*}
  \hat{f}(-\omega_0-x, - \V{k_0}) & = \hat{f}(-\omega_0, - \V{k_0}) - x \tilde{f}(-\omega_0-x, - \V{k_0}), \\
  \hat{h}(\omega_1-x, \V{k_1}) & = \hat{h}(\omega_1, \V{k_1}) - x \tilde{h}(\omega_1-x, \V{k_1}),
\end{align*}
where $\tilde{f}$ and $\tilde{h}$ are smooth functions satisfying $\tilde{f}(-\omega_0,  -\V{k_0}) = \del^0 \hat{f}(-\omega_0, - \V{k_0})$ and $\tilde{h}(\omega_1, \V{k_1}) = \del^0 \hat{h}(\omega_1, \V{k_1})$. Then the terms of zeroth order in $x$ cancel each other in (\ref{eq:1st_order}).  The terms of first order in $x$ now yield
\begin{equation*}
  2\pi\int \ud^3 \V k \ \frac{1}{(2\omega_{\V k})^2}
           \left\{ - \del_0 \hat{f}(-\omega_\V{k}, -\V{k}) \hat{h}(\omega_{\V k}, \V{k})
           + \hat{f}(-\omega_\V{k}, -\V{k}) \del_0 \hat{h}(\omega_{\V k}, \V{k}) \right\}
\end{equation*}
in the adiabatic limit.

\begin{remark}
Note that it was crucial here to consider the sum of the two terms on the left hand side of~(\ref{eq:1stOrder}). The individual terms are divergent. This is a nice illustration of Remark 4 in \cite{EpsteinGlaser}.
\end{remark}

We still have to treat the remaining two terms in~(\ref{eq:1st_order}).
If we assume, in accordance with~\cite{EpsteinGlaser}, that $\check{g}_a$ is supported in a closed subset of $V_1 = \left\{ p \in \R^4 | \betrag{p^0} < 2 m \right\}$, then the singularity $x = \mp 2 \omega_{0/1}$ lies outside the support of $\check{g}_a$. Thus, we may carry out the adiabatic limit and obtain
\begin{equation*}
  -{2\pi}\int \ud^3 \V k \ \frac{2}{(2\omega_{\V k})^3}
           \hat{f}(-\omega_\V{k}, - \V{k}) \hat{h}(\omega_{\V k}, \V{k}).
\end{equation*}
Combining all this, we get
\begin{equation}
\label{eq:delDelta}
 2\pi\int \ud^3 \V k \ \left( - \frac{1}{4 \omega_\V{k}^3} \hat{f}(-k_+) \hat{h}(k_+) - \frac{1}{4 \omega_\V{k}^2} \left\{ \del_0 \hat{f}(-k_+) \hat{h}(k_+)
    - \hat{f}(-k_+) \del_0 \hat{h}(k_+) \right\} \right).
\end{equation}
Here we used the notation $k_+ = (\omega_\V{k}, \V{k})$.
Thus, the adiabatic limit of (\ref{eq:2pt}) exists for $n=1$.
Using
\begin{equation}
\label{eq:delta'}
  \int \ud^4k \ \hat{f}(k) \delta'(k^2-m^2) = \int \ud^3 \V k \ \left( \frac{1}{4 \omega_{\V k}^3} \hat{f}(k_+,\V k) - \frac{1}{4 \omega_{\V k}^2} \del_0 \hat{f}(k_+,\V k) \right), 
\end{equation}
it is easy to check that this coincides with~(\ref{eq:muExpansion}) for $n=1$. That one obtains~(\ref{eq:muExpansion}) also for higher orders has been shown in a joint work with C.~D\"oscher~\cite{Adlim}. The proof involves rather heavy combinatorics and lies somewhat off the main line of this thesis, so we skip it. Further details can be found in~\cite{Claus}. We conclude that Question~2 from above can be answered affirmatively in the commutative case for a mass term.

\begin{remark}
\label{rem:massless}
For the discussion above it was crucial that we were using a massive field. In the massless case, one has to deal with infrared problems. In order to circumvent these, one can restrict to test functions $\hat f$ and $\hat h$ 
that vanish in a neighborhood of the origin.
Furthermore, we assume that the sequence of test functions $\check g_a$ has decreasing support, i.e., for each neighborhood $U$ of the origin, there is an $A \in \mathbbm{N}$ such that $\supp \check g_a \subset U$ for all $a > A$. Now we take a look at~(\ref{eq:1st_order}) again. The terms involving $\iep{x}$ can be treated as before. For the other two terms we notice that, because of the support property of $\hat f$, there is an $\epsilon'$ such that $\betrag{\V{k_0}} > \epsilon'$. Furthermore, we can choose $A$ such that $\check g_a(\betrag{\V{k_1}} - \betrag{\V{k_0}} -x, \V{k_1}-\V{k_0})$ vanishes for $\betrag{x} > \epsilon'$ for all $a > A$. Then the singularity in $\iep{x+2\betrag{\V{k_0}}}$ is not met and one can carry out the adiabatic limit. Obviously, the term involving $\iep{x-2 \betrag{\V{k_1}}}$ can be treated analogously. Thus, we obtain $- (2\pi)^{-1} \theta(k_0) \delta'(k^2)$ at first order in the adiabatic limit. It is then possible to extend this distribution in a Lorentz invariant way to the origin, but this is unambiguous only up to a $\delta$-distribution at the origin, 
see, e.g., \cite{Guettinger}.
\end{remark}


\subsection{The noncommutative case}
\label{sec:MassNoncommutative}

In~\cite{Diplom}, it has been proposed to consider localized interaction terms of the form
\begin{equation*}
 - \frac{\mu}{2}  \int \ud^4q \ g^1_a \phi g^2_a \phi,
\end{equation*}
where $\{g^1_a\}$ and $\{g^2_a\}$ are sequences of elements of $\schw_2$.
With this term, we would get the equation of motion
\begin{equation*}
  ( \Box + m^2 ) \phi = - \frac{\mu}{2} \left( g^1_a \phi g^2_a + g^2_a \phi g^1_a \right).
\end{equation*}
The first order term of the Yang--Feldman series is then, in momentum space,
\begin{multline*}
  \hat{\phi}_1(k) = - (2\pi)^{-2} \hat{\Delta}_R(k) \\ \times \int \prod_{i =1}^3 \ud^4l_i \ \delta(k-\sum_i l_i) \hat{g}^1_a(l_1) \hat{g}^2_a(l_2) \hat{\phi}_0(l_3) \cos (l_3 \sigma (l_2-l_1) + l_1 \sigma l_2).
\end{multline*}
Note that this would also be well--defined for $g_2 = \1$, i.e., when only one cut--off function is used. Computing the two--point function (\ref{eq:2pt}) at first order in $\mu$, where now
\begin{equation*}
  \phi(f) = \int \ud^4q \ \phi(q) f(q),
\end{equation*}
one obtains
\begin{multline*}
  - (2\pi)^{-2} \int \ud^4k_0 \ud^4k_1 \ \hat{f}(-k_0) \hat{h}(k_1) \left\{ \hat{\Delta}_R(k_0) \hat{\Delta}_+(k_1) + \hat{\Delta}_+(k_0) \hat{\Delta}_A(k_1) \right\} \\
  \times \int \ud^4l_1 \ud^4l_2 \ \hat{g}^1_a(l_1) \hat{g}^2_a(l_2) \delta(k_0-k_1-l_1-l_2) \cos ( k_0 \sigma (l_1 - l_2) + l_1 \sigma l_2).
\end{multline*}
The factor in the second line is a test function in $k_0$ and $k_1$ (not in $k_1-k_0$, as in the commutative case). However, it is easy to see that this poses no problems for the adiabatic limit\footnote{In the expression analogous to (\ref{eq:1st_order}), one then has to Taylor expand also the cosine in the expression above. However, the supplementary term vanishes in the adiabatic limit, because of $\sin 0 = 0$. Furthermore, the above restriction on the support of the cut--off functions $\check g_a$ can be implemented by restricting the support of $\hat{g}^1_a$ and $\hat{g}^2_a$ appropriately.}.  Thus, at first order, the above cutoff works and reproduces the expected result in the adiabatic limit. One can also show that this works to all orders~\cite{Adlim}.

We have thus found an answer to Question~3 from above, for the case of a mass term as interaction term. But for truly interacting models problems appear. We are going to discuss these in the following subsection.

\subsection{Interactions}
\label{sec:Interactions}

Again, we begin by considering the situation in the commutative case. As an example, we study the $\phi^3$ model. Thus, we take the equation of motion~(\ref{eq:eom}) and multiply the right hand side with a test function $g$. We also subtract the tadpole from the start, i.e., we normal order $\phi_1$. The first two terms of the interacting field are then
\begin{subequations}
\begin{align}
\label{subeq:phi_1}
  \phi_1 = & \Delta_R \times ( g \WDp{\phi_0 \phi_0} ) \\
\label{subeq:phi_2}
  \phi_2 = & \Delta_R \times ( g \phi_1 \phi_0 + g \phi_0 \phi_1 ).
\end{align}
\end{subequations}
Computing the two--point function at second order in $\lambda$, one finds the three terms
\begin{equation}
\label{eq:ThreeTerms}
  \bra{\Omega} \phi_2(f) \phi_0(h) \ket{\Omega} + \bra{\Omega} \phi_0(f) \phi_2(h) \ket{\Omega} + \bra{\Omega} \phi_1(f) \phi_1(h) \ket{\Omega}.
\end{equation}
For the moment, we focus on the first two terms (the third one will be discussed later). In our graphical notation, they are represented by
\begin{center}
\begin{picture}(400,60)
\Line(39.5,5)(39.5,20)
\Line(40.5,5)(40.5,20)
\Line(40,20)(60,50)
\Line(39.4,20)(29.4,35)
\Line(40.6,20)(30.6,35)
\Line(30,35)(20,50)
\Line(30,35)(40,50)
\Line(80,5)(80,50)
\Line(40,50)(60,50)
\Curve{(20,50)(50,55)(80,50)}
\Vertex(40,20){1}
\Vertex(30,35){1}
\BCirc(20,50){1}
\BCirc(40,50){1}
\BCirc(60,50){1}
\BCirc(80,50){1}
\Text(100,30)[]{$+$}
\Line(139.5,5)(139.5,20)
\Line(140.5,5)(140.5,20)
\Line(139.4,20)(149.4,35)
\Line(140.6,20)(150.6,35)
\Line(140,20)(120,50)
\Line(150,35)(140,50)
\Line(150,35)(160,50)
\Line(180,5)(180,50)
\Line(120,50)(140,50)
\Line(160,50)(180,50)
\Vertex(140,20){1}
\Vertex(150,35){1}
\BCirc(120,50){1}
\BCirc(140,50){1}
\BCirc(160,50){1}
\BCirc(180,50){1}
\Text(200,30)[]{$+$}
\Line(259.5,5)(259.5,20)
\Line(260.5,5)(260.5,20)
\Line(260,20)(280,50)
\Line(259.4,20)(249.4,35)
\Line(260.6,20)(250.6,35)
\Line(250,35)(240,50)
\Line(250,35)(260,50)
\Line(220,5)(220,50)
\Line(260,50)(280,50)
\Line(220,50)(240,50)
\Vertex(260,20){1}
\Vertex(250,35){1}
\BCirc(220,50){1}
\BCirc(240,50){1}
\BCirc(260,50){1}
\BCirc(280,50){1}
\Text(300,30)[]{$+$}
\Line(359.5,5)(359.5,20)
\Line(360.5,5)(360.5,20)
\Line(359.4,20)(369.4,35)
\Line(360.6,20)(370.6,35)
\Line(360,20)(340,50)
\Line(370,35)(360,50)
\Line(370,35)(380,50)
\Line(320,5)(320,50)
\Line(340,50)(360,50)
\Curve{(320,50)(350,55)(380,50)}
\Vertex(360,20){1}
\Vertex(370,35){1}
\BCirc(320,50){1}
\BCirc(340,50){1}
\BCirc(360,50){1}
\BCirc(380,50){1}
\end{picture}
\end{center}
Since we are in the commutative case, the vertex factors do not depend on the order of the momenta, so the other possible contractions give the same result. We thus get a factor $2$ and obtain
\begin{multline}
\label{eq:phi_3_comm_0}
 \int \ud^4x \ud^4y \ f(x) h(y) \int \ud^4z_1 \ud^4z_2 \ g_a(z_1) g_a(z_2) \\
  \times \left\{ \Delta_R(x-z_1) \Delta_+(z_2-y) 2 \Delta_R(z_1-z_2) \Delta^{(1)}(z_1-z_2) \right. \\ \left. + \Delta_+(x-z_1) \Delta_A(z_2-y) 2 \Delta_A(z_1-z_2) \Delta^{(1)}(z_1-z_2) \right\}.
\end{multline}
Here we used the notation $\Delta^{(1)}(x) = \Delta_+(x) + \Delta_+(-x)$. We remark that the first two graphs give rise to the first term in~(\ref{eq:phi_3_comm_0}), while the second term comes from the last two graphs. The two single lines with different orientation in the first (last) two graphs combine to $\Delta^{(1)}$. 
The product of this distribution with $\Delta_{R/A}$ is ill-defined. However, as has been noted in~\cite{BDFP02},
\begin{equation}
\label{eq:DeltaProduct}
  \Delta^{(1)} \Delta_{R/A} 
  = - i \Delta_F \Delta_F - i \Delta_{\mp} \Delta_{\mp}.
\end{equation}
The square of $\Delta_{\mp}$ is a well--defined distribution. The square of the Feynman propagator is a well--defined distribution on test functions vanishing at the origin. But it can be extended to the origin at the expense of a renormalization ambiguity in the form of a $\delta$-distribution at the origin~\cite{BrunettiFredenhagen}. Comparison with~(\ref{eq:1stOrder}) shows that this corresponds to a mass renormalization. We also see that the UV and the IR problem are completely decoupled, since we could renormalize before carrying out the adiabatic limit. The reason why this is possible is that the theory is local~\cite{BrunettiFredenhagen}.

We denote the Fourier transform of the renormalized product of $2 \Delta^{(1)}$ and $\Delta_R$ by $(2\pi)^{-2} \Sigma$. Now, after a change of variables, the above can be written in momentum space as
\begin{multline}
\label{eq:phi_3_comm}
 - (2\pi)^{-3} \int \frac{\ud^3 \V{k_0}}{2 \omega_0} \frac{\ud^3 \V{k_1}}{2 \omega_1} \ud^4l \ud x \ \check{g}_a(l_0 - \omega_0 -x, \V{l} - \V{k_0}) \check{g}_a(\omega_1 - l_0, \V{k_1} - \V{l}) \\
 \times \left\{ \hat{f}(-\omega_0-x, -\V{k_0}) \hat{h}(\omega_1, \V{k_1}) \Sigma(l) \left( \iep{x} - \iep{x + 2 \omega_0} \right) \right. \\
  \left. - \hat{f}(-\omega_0, - \V{k_0}) \hat{h}(\omega_1-x, \V{k_1}) \tilde \Sigma(l_0-x,\V{l}) \left( \iep{x} - \iep{x - 2 \omega_1} \right) \right\}.
\end{multline}
Here we defined $\tilde \Sigma(k) = \Sigma(-k)$. In order to treat the terms involving $\iep{x}$, we expand $\hat{f}, \hat{h}$ and $\tilde \Sigma$ in $x$, as in Section~\ref{sec:MassCommutative}. In order for the adiabatic limit to exist, it is then crucial that the zeroth order terms in this expansion cancel each other, i.e., $\Sigma$ and $\tilde \Sigma$ have to coincide in a neighborhood of the mass shell. This, however, is the case, since $\hat \Delta_F$ is symmetric and the Fourier transform of $\Delta_{\mp}^2$ has support above the $2 m$ mass shell, cf.~(\ref{eq:DeltaProduct}). Hence, we may carry out the adiabatic limit and obtain
\begin{multline}
\label{eq:2pt_Sigma_0}
 - 2\pi \int \ud^3 \V k \ \left( \frac{1}{4 \omega_\V{k}^2} \hat{f}(-k_+) \hat{h}(k_+) \del_0 \Sigma(k_+) - \frac{1}{4 \omega_\V{k}^3} \Sigma(k_+) \hat{f}(-k_+) \hat{h}(k_+) \right. \\
  \left. - \frac{1}{4 \omega_\V{k}^2} \Sigma(k_+) \left\{ \del_0 \hat{f}(-k_+) \hat{h}(k_+)
    - \hat{f}(-k_+) \del_0 \hat{h}(k_+) \right\} \right).
\end{multline}
This can also be written as
\begin{equation}
\label{eq:2pt_Sigma}
  - (2\pi)^2 \int \ud^4 k \ \hat f(-k) \hat h(k) \Sigma(k) \frac{\del}{\del m^2} \hat \Delta_+(k).
\end{equation}
Since $\Sigma$ is Lorentz invariant, we have $\Sigma(k)= \Sigma(k^2)$.
The last three terms in (\ref{eq:2pt_Sigma_0}) are as in~(\ref{eq:delDelta}), and thus correspond to a finite mass renormalization
\begin{equation}
\label{eq:delta_m_comm}
  \delta m^2 = - \lambda^2 \Sigma(m^2).
\end{equation}
The first term is a field strength renormalization
\begin{equation}
\label{eq:delta_Z_comm}
  \delta Z = - \lambda^2 \frac{\del}{\del k^2} \Sigma(m^2).
\end{equation}

It remains to treat the third term in (\ref{eq:ThreeTerms}). In our graphical notation it can be expressed as
\begin{center}
\begin{picture}(80,50)
\Line(19.5,5)(19.5,20)
\Line(20.5,5)(20.5,20)
\Line(20,20)(10,35)
\Line(20,20)(30,35)
\Line(59.5,5)(59.5,20)
\Line(60.5,5)(60.5,20)
\Line(60,20)(50,35)
\Line(60,20)(70,35)
\Line(30,35)(50,35)
\Curve{(10,35)(40,40)(70,35)}
\Vertex(20,20){1}
\Vertex(60,20){1}
\BCirc(10,35){1}
\BCirc(30,35){1}
\BCirc(50,35){1}
\BCirc(70,35){1}
\end{picture}
\end{center}
Again, we suppressed one other contraction. Taking this factor $2$ into account, one obtains
\begin{equation*}
 \int \ud^4x \ud^4y \ f(x) h(y) \int \ud^4z_1 \ud^4z_2 \ g_a(z_1) g_a(z_2)
  \Delta_R(x-z_1) 2 \Delta_+(z_1-z_2)^2 \Delta_A(z_2-y).
\end{equation*}
Here both propagators in the loop are $\Delta_+$, i.e., both loop momenta are on the mass shell. This is only possible if the incoming momentum is above the $2 m$ mass shell. Furthermore, the loop integral is well--defined. Hence, this graph does not require any renormalization and is a contribution to the two--particle spectrum.


Now we come to the noncommutative case, which turns out to be more problematic. In the example of the $\phi^3$ model, a natural infrared cut--off would be to consider the interaction term
\begin{equation}
\label{eq:phi3_NC_cutoff}
  \frac{\lambda}{3} \int \ud^4q \ g_a \phi g_a \phi g_a \phi.
\end{equation}
However, as has been shown in~\cite{Diplom}, the $n$-point functions of the corresponding interacting field are all finite for $g_a \in \schw_2$. Hence, there is no need to subtract anything. But obviously the planar parts will diverge in the adiabatic limit. This is another manifestation of the UV/IR--mixing. In such a situation, one can not proceed as above. Note that it does not help to use only a single test function as infrared cutoff, because then some of the planar parts will be divergent before the adiabatic limit, while others are still regularized by the infrared cutoff. Also the use of the point-wise product with a single test function does not work.
One can try to circumvent this problem by introducing counterterms of the form 
\begin{equation*}
 m_a \int \ud^4q \ g_a \phi g_a \phi,
\end{equation*}
where $m_a$ (and possibly other renormalization constants) is chosen such that the limit $a \to \infty$ exists and fulfills some renormalization conditions. But this seems to work only for logarithmic divergences: Assume the sequence $\{\check g_a\}$ is obtained by scaling from a test function with compact support. Then, if the mass renormalization of the commutative theory is linearly divergent, $m_a$ has to scale as~$\sim a$. Looking at the first order term (\ref{eq:1stOrder}),\footnote{Of course this is the expression from the commutative case. But the reasoning in the noncommutative case is very similar.} it is rather straightforward to see that in the limit $a \to \infty$ also terms involving $\del_0^2 \hat f$ will survive. It is not obvious that such terms cancel each other. This will make the whole procedure very complicated. The $\phi^3$ and Wess--Zumino model, which we will study later in this chapter, are only logarithmically divergent, so one could use the procedure proposed here. However, QED is quadratically divergent by power counting. Only by invoking the Ward identity does it become logarithmically divergent. But since it is not clear how to formulate Ward identities in the present setting, it does not seem to be possible to treat NCQED in this way.

A different idea would be to split the formal expression for $\phi_2$ into a normal ordered and a renormalized contracted part, as in the previous section. The renormalized contracted part could then be cut off in the infrared like the $\phi_1$ from the example of a mass term as interaction, i.e., by multiplication with a single test function $g_a$:
\begin{equation*}
  \hat \phi_2(k) = \hat \Delta_R(k) \left( \hat g_a \times (\Sigma \hat \phi_0) \right)(k),
\end{equation*}
cf. (\ref{eq:Sigma_phi_2}) and (\ref{eq:mu_phi_1}). However, already in the commutative case this procedure would not work. Instead of (\ref{eq:1st_order}), we would find
\begin{multline*}
  \frac 1{2\pi}\int \frac{\ud^3 \V{k_0}}{2 \omega_0} \frac{\ud^3 \V{k_1}}{2 \omega_1} \ud x \ \check{g}_a(\omega_1 - \omega_0 -x, \V{k_1} - \V{k_0}) \\
 \times \left\{ \hat{f}(-\omega_0-x, -\V{k_0}) \hat{h}(\omega_1, \V{k_1}) \Sigma(\omega_0+x, \V{k_0})\left( \iep{x} - \iep{x + 2 \omega_0} \right) \right. \\
  \left. - \hat{f}(-\omega_0, - \V{k_0}) \hat{h}(\omega_1-x, \V{k_1}) \tilde \Sigma(\omega_1-x, \V{k_1}) \left( \iep{x} - \iep{x - 2 \omega_1} \right) \right\}.
\end{multline*}
Contrary to the situation in~(\ref{eq:phi_3_comm}), the difference of the arguments of the two $\Sigma$'s is not just $(x, \V 0)$. After a Taylor expansion in $x$ the arguments do not coincide and the terms involving $\iep{x}$ do not cancel, as in Section~\ref{sec:MassCommutative}. The expression does not have a well--defined adiabatic limit\footnote{This is very similar to the situation when one computes the two--point function in a different state, a KMS-state for example. Then, instead of the retarded propagator in the present case, the $\Delta_+$ is multiplied in momentum space with some function $F(k)$ (and a corresponding negative frequency part is added). Also in this case one does not have a well--defined adiabatic limit, cf.~\cite{Adlim}.}.



In \cite{DorosDiss}, an infrared cutoff was tentatively defined by multiplying the retarded propagator with a test function. But it was shown by C.~D\"oscher \cite{Claus} that with this cutoff the adiabatic limit is not well--defined, even in the commutative case.

In the remainder of this thesis, we adopt the following point of view: In order to compute the two--point function $\bra{\Omega} \phi(f) \phi(h) \ket{\Omega}$ of the interacting field at second order, we calculate the self--energy $\Sigma(k)$ which is implicitly defined by
\begin{equation}
\label{eq:Sigma_def}
  \hat \phi_2(k) = (2\pi)^2 \hat \Delta_R(k) \Sigma(k) \hat \phi_0(k) + \text{ n.o.},
\end{equation}
where ``n.o.'' stands for the normal ordered part. Inserting this in (\ref{eq:2pt_Sigma}) yields the sum of the first two terms in (\ref{eq:ThreeTerms}). The remaining third term in (\ref{eq:ThreeTerms}) can be computed directly without any (implicit) infrared cutoff. 

\begin{remark}
To the best of our knowledge, an appropriate infrared cutoff and its adiabatic limit have not been discussed in the literature on NCQFT in the setting of the modified Feynman rules. In fact one would find the same problems that we discussed above. Our procedure to calculate the two--point function corresponds to the one that is implicitly adopted in the literature.
\end{remark}

\begin{remark}
\label{rem:k4}
The expression (\ref{eq:2pt_Sigma}) will be crucial for the computation of the two--point function, also in the case of electrodynamics. Thus, we want to comment on its validity in the massless case. Assuming that $\Sigma(k) = \Sigma(-k)$ still holds in a neighborhood of the mass shell, the step from (\ref{eq:phi_3_comm}) to (\ref{eq:2pt_Sigma}) can be done as discussed in Remark~\ref{rem:massless} if one restricts to test functions $\hat f$, $\hat h$ that vanish in a neighborhood of the origin. Using a scaling sequence of test functions $\check g_a$ with compact support, one can even show that the adiabatic limit of (\ref{eq:phi_3_comm}) is well--defined and given by (\ref{eq:2pt_Sigma_0}) if $\Sigma(k) - \Sigma(-k) \sim k^4$. A proof can be found in Appendix~\ref{app:k4}.
\end{remark}

\section{Dispersion Relations}
\label{sec:DispRel}

The goal of this section is to discuss, on an abstract level, some consequences of the modified dispersion relations at the one-loop level in noncommutative field theories. The self--energy $\Sigma(k)$ will now be a function of $(k \sigma)^2$ and $k^2$ (and possibly the sign of $k_0$). Analogously to~(\ref{eq:delta_m_comm}) and~(\ref{eq:delta_Z_comm}), one can interpret this as a momentum--dependent mass and field strength renormalization\footnote{In the formula for $\delta Z$ one really only has a partial derivative with respect to $k^2$. Inserting the expression for $\delta m^2$ (instead of $\Sigma$) into (\ref{eq:2pt_Sigma_0}), the first term generates the derivative $\frac{\ud (k \sigma)^2}{\ud k^2} \frac{\del}{\del (k \sigma)^2}$.}:
\begin{align}
\label{eq:M}
  \delta m^2((k \sigma)^2) & = - \lambda^2 \Sigma((k \sigma)^2, m^2), \\
\label{eq:Z}
  \delta Z((k \sigma)^2) & = - \lambda^2 \left( \frac{\del}{\del k^2} \Sigma \right) ((k \sigma)^2, m^2).
\end{align}

\begin{remark}
Although the naming might suggest this, we do not subtract these terms, since they are neither local, nor, in general, divergent. We remark, however, that such a subtraction has been proposed in~\cite{LiaoSibold}. See also the brief discussion in Chapter~\ref{chapter:Summary}.
\end{remark}

One can also interpret the sum of the zeroth order contribution 
\begin{equation}
\label{eq:2pt0order}
  2\pi \int \ud^4k \ \hat{f}(-k) \hat{h}(k) \theta(k^0) \delta(k^2-m^2)
\end{equation}
and the second order term (\ref{eq:2pt_Sigma}) as the expansion (in $\lambda$) of
\begin{equation}
\label{eq:2ptNC}
 2\pi \int \ud^4k \ \hat{f}(-k) \hat{h}(k) \theta(k^0) \delta(k^2-m^2+ \lambda^2 \Sigma((k \sigma)^2), k^2)) + \order (\lambda^4),
\end{equation}
which indicates a modified dispersion relation.

\begin{remark}
\label{rem:LSZ}
This modification of the dispersion relation is a manifestation of the breaking of particle Lorentz invariance, cf. the discussion in Chapter~\ref{chapter:Introduction}. However, particle Lorentz invariance of the asymptotic fields is a crucial ingredient of scattering theory and the LSZ relations, which are part of the foundations of quantum field theory. In this sense, the conceptual basis of the present approach is rather shaky. In the following, we will take a phenomenological standpoint and compute the distortion of the dispersion relation for different models in order to check if they are realistic. 
\end{remark}

\subsection{The group velocity}
\label{sec:GroupVelocity}

We now discuss how to extract the group velocity in the above setting. From (\ref{eq:2ptNC}), and allowing for a
finite mass and field strength renormalization, we get the dispersion relation
\begin{equation}
\label{eq:ImplicitDispRel}
 F(k) = k^2 - m^2 + \lambda^2 \left( \Sigma((k \sigma)^2, k^2) - \alpha + \beta k^2 \right) + \order(\lambda^4) = 0.
\end{equation}
For a given spatial momentum $\V{k}$ we want to compute the corresponding $k^0$ that solves
(\ref{eq:ImplicitDispRel}) as a formal power series in $\lambda$. We find
\begin{equation}
\label{eq:k0}
  k^0 = \omega_{\V k} - \lambda^2 \frac{1}{2 \omega_{\V k}} \left( \Sigma((k_+ \sigma)^2, m^2) - \alpha + \beta m^2 \right) + \order(\lambda^4).
\end{equation}
Note that in $\omega_{\V k} = \sqrt{\betrag{\V{k}}^2 + m^2}$ and $k_+=(\omega_{\V k}, \V{k})$ the bare mass $m$ enters.
The group velocity is then given by
\begin{multline*}
  \nabla k^0 = \frac{\V{k}}{\omega_{\V k}} + \lambda^2 \frac{\V{k}}{2 \omega_{\V k}^3} \left( \Sigma((k_+ \sigma)^2, m^2) - \alpha + \beta m^2 \right) \\ - \lambda^2 \frac{1}{2 \omega_k} \nabla (k_+ \sigma)^2 \frac{\del}{\del (k \sigma)^2} \Sigma((k_+ \sigma)^2, m^2) + \order(\lambda^4).
\end{multline*}
By comparison with (\ref{eq:k0}), we get
\begin{equation*}
  \nabla k^0 = \frac{\V{k}}{k^0} - \lambda^2 \frac{\nabla (k_+ \sigma)^2}{2 k^0}  \frac{\del}{\del (k \sigma)^2} \Sigma((k_+ \sigma)^2, m^2) + \order(\lambda^4).
\end{equation*}
In order to make things more concrete, we choose a particular $\sigma$, namely $\sigma_0$, cf.~(\ref{eq:sigma_0}).
Then we have
\begin{equation}
\label{eq:k_sigma_2}
 (k \sigma_0)^2 = - \lambda_{nc}^{4} \left( k^2 + 2 \betrag{\V{k_{\bot}}}^2 \right)
\end{equation}
with $\V{k_{\bot}} = (k_1, 0, k_3)$. We also define $\V{k_{||}}=(0,k_2,0)$. Thus, in the case $\sigma =
\sigma_0$, we find
\begin{equation}
\label{eq:GroupVelocity}
  \V{\nabla} k^0 = \frac{\V{k_{||}}}{k^0} + \frac{\V{k_{\bot}}}{k^0} \left( 1 +  2 \lambda^2 \lambda_{nc}^4 \frac{\del}{\del (k \sigma)^2} \Sigma((k_+ \sigma_0)^2, m^2) \right) + \order(\lambda^4).
\end{equation}

\begin{remark}
This treatment differs slightly from the one given in~\cite{Quasiplanar}. There, $\Sigma$ is not Taylor expanded in $\lambda$. Then the argument of $\Sigma$ in (\ref{eq:GroupVelocity}) is not restricted to the mass $m$ shell. It follows that by tuning $\alpha$ and $\beta$ one can make the deviation arbitrarily small, which is not possible here.
\end{remark}

\subsection{Acausality}
\label{sec:Acausality}

We recall that the main motivation for the introduction of the noncommutative Minkowski space in \cite{DFR} was the desire to implement the space-time uncertainty relations (\ref{eq:STUR1},b), i.e., some form of nonlocality. It is not surprising that this nonlocality leads to acausal effects, see, e.g., \cite{Causality, WulkenhaarCausality}. However, these are relevant only at the noncommutativity scale and are kinematical in the sense that the nonlocality was put in by hand in the very definition of the noncommutative Minkowski space. Here we want to discuss acausal effects that are created dynamically and are not necessarily limited to the noncommutativity scale.

\begin{remark}
In the literature, one often finds the statement that acausal effects can only occur in the case of space/time noncommutativity. For space/space noncommutativity, however, there is the possibility of an action at a distance. In a different reference frame, this is again an acausal effect.
\end{remark}

The distortion of the group velocity was an effect of the momentum--dependent mass renormalization~(\ref{eq:M}). In the following, we want to discuss the effect of a momentum--dependent field strength renormalization. It multiplies, in momentum space, the free two--point function $\hat{\Delta}_+$. But not only this propagator is modified. Consider a source term $\int \ud^4q \ \phi g$ for the interacting field. We define a new free field $\phi_0' = \phi_0 + \Delta_R \times g$. In the example of the $\phi^3$ model, we then have\footnote{Here we subtracted the tadpole, i.e., we used a normal ordering in the definition of $\phi'_1$.}
\begin{equation*}
  \phi_1' = \Delta_R \times (\WDp{\phi_0' \phi_0'}); \qquad \phi_2' = \Delta_R \times ( \phi_1' \phi_0' + \phi_0' \phi_1' ).
\end{equation*}
At first order in $g$, the vacuum expectation value of $\hat \phi_2'(k)$ is then, cf.~(\ref{eq:Sigma_def}),
\begin{equation*}
 (2\pi)^4 \hat \Delta_R(k) \Sigma(k) \hat \Delta_R(k) \hat g(k).
\end{equation*}
Assume that only a field strength renormalization is present, i.e., $\Sigma(k) = (k^2-m^2) \tilde \Sigma(k)$. Then the above reduces to $- (2\pi)^2 \hat \Delta_R(k) \tilde \Sigma(k) \hat g(k)$. Hence, the effective retarded propagator is
\begin{equation*}
  \hat \Delta_R(k) \left( 1 - \tilde \Sigma(k) \right).
\end{equation*}
Note that here, contrary to~(\ref{eq:Z}), $\tilde \Sigma(k)$ is not restricted to the mass shell. In position space, the retarded propagator is convoluted with a nonlocal kernel. Obviously, this may lead to acausal effects. In Section~\ref{sec:NonlocalEffects} we will discuss these in the case of supersymmetric NCQED. It turns out that the effect is independent of the noncommutativity scale.

\section{The $\phi^3$ model}
\label{sec:phi3}

It was shown in \cite{DorosDiss} that the distortion of the dispersion relation in the $\phi^4$ model is very strong and mainly affects the infrared. The strength of this effect is approximately of the order $m^{-2} \lambda_{nc}^{-2}$. Reasonable dispersion relations are only obtained for $m \sim \lambda_{nc}^{-1}$, which is not acceptable if we want to identify $\lambda_{nc}$ with the Planck length. The reason why the effect is so strong in the $\phi^4$ model, is that it is quadratically divergent. The UV/IR--mixing transforms this into a strong distortion of the dispersion relation in the infrared. It is thus natural to consider models that are only logarithmically divergent. The simplest such model is the $\phi^3$ model. Although it is nonperturbatively not stable, its perturbative treatment is well--defined. Furthermore, the loop integral that we will compute here, will also be important for our study of NCQED.

The noncommutative $\phi^3$-model was studied in~\cite{Minwalla, Raamsdonk} in the context of the modified Feynman rules, in~\cite{DoroUVFinite} in the Hamiltonian formalism, and in~\cite{GrosseSteinacker} in the Euclidean self-dual setting. In \cite{BDFP02} the formal expression for one-loop self--energy in the Yang--Feldman formalism was computed. The aim of this section is to give some meaning to the formal expression and to evaluate it in order to compute the distortion of the group velocity. We also discuss the two--particle spectrum.


\subsection{The self--energy}
\label{sec:phi3SelfEnergy}

A formal expression for the self--energy was already computed in Section~\ref{sec:Graphs}: From~(\ref{eq:Sigma_phi_2}), we read off
\begin{equation*}
  \Sigma(k) = \int \ud^4l \ \hat \Delta_R(k-l) \left\{ \hat \Delta_+(-l) \left( 1 + e^{ - i k \sigma l} \right) + \hat \Delta_+(l) \left( 1 + e^{ + i k \sigma l} \right) \right\}.
\end{equation*}
This can be split into a planar and a nonplanar part. The planar part is exactly as in the commutative case. We recall from Section~\ref{sec:Interactions} that, because of~(\ref{eq:DeltaProduct}), one recovers the usual Feynman loop integral. In particular, one has to perform a mass renormalization, since the loop integral diverges logarithmically.

For later comparison with the nonplanar part, it is nevertheless convenient to compute the planar part of $\Sigma(k)$ formally without recourse to~(\ref{eq:DeltaProduct}). We restrict ourselves to the case of timelike $k$ with positive energy. Because of Lorentz invariance, we may choose $k=(k_0, \V{0})$ with $k_0>0$. Then, with~(\ref{eq:Delta_R}), we obtain
\begin{equation*}
  \Sigma_{pl}(k) = - (2 \pi)^{-3} \int \frac{\ud^3 \V l}{2 \omega_{\V l}} \ \left( \frac{1}{k^2 - 2 \sqrt{k^2} \omega_{\V l} + i \epsilon(k_0-\omega_{\V l})} + \frac{1}{k^2 + 2 \sqrt{k^2} \omega_{\V l} + i \epsilon(k_0+\omega_{\V l})} \right).
\end{equation*}
Since $\omega_{\V l}$ is positive, the singularity in the second term is avoided, so one can ignore the $i \epsilon$-prescription. In order to determine the $i \epsilon$-prescription for the first term, we consider its singularity at $\omega_{\V l} = \sqrt{k^2}/2$. There, the prefactor $k_0-\omega_{\V l}$ is positive. Thus, we may write $i \epsilon$ instead of $i \epsilon (k_0 - \omega_{\V l})$. Changing to spherical coordinates and carrying out the integration over the angles, one obtains
\begin{equation}
\label{eq:F_pl}
  \Sigma_{pl}(k) = - 2 (2 \pi)^{-2} \int_0^{\infty} \ud l \ \frac{l^2}{\omega_l} \frac{1}{k^2 - 4 \omega_l^2+i\epsilon}.
\end{equation}
This diverges logarithmically. For a timelike $k$ with negative energy, one finds the same expression but with the opposite sign for $i \epsilon$. Even though $\Sigma_{pl}$ is not well--defined, we can (formally) compute the field strength renormalization: Differentiating (\ref{eq:F_pl}) with respect to $k^2$, cf.~(\ref{eq:delta_Z_comm}) , one gets a convergent integral. For $k$ in a neighborhood of the mass shell, the singularity at $\omega_l = \sqrt{k^2}/2$ is not met and one obtains
\begin{equation}
\label{eq:Z_pl}
  \delta Z = (2 \pi)^{-2} \frac{3 - \frac{2 \pi}{\sqrt{3}}}{12 m^2}.
\end{equation}

We now want to discuss the nonplanar part of $\Sigma(k)$, i.e.,
\begin{equation}
\label{eq:F_NP}
  \Sigma_{np}(k) = \int \ud^4l \ \hat{\Delta}_+(l) e^{i k \sigma l} \left( \hat{\Delta}_R(k-l) + \hat{\Delta}_R(k+l)
  \right),
\end{equation}
for timelike $k$.
In particular, we want to show that it is finite and that $\Sigma_{np}(k) = \Sigma_{np}(-k)$ in a neighborhood of the mass shell, as this was necessary for the derivation of~(\ref{eq:2pt_Sigma}) in Section~\ref{sec:Interactions}.
Note that the integral~(\ref{eq:F_NP}) is neither absolutely convergent nor a Fourier transformation (since $k$ does not only appear in the phase factor).
In~\cite{NCDispRel} it was shown that~(\ref{eq:F_NP}), for $k$ in a neighborhood of the mass shell, can be defined as an oscillatory integral in the sense of \cite{ReedSimon2}. 
See also~\cite{Claus}. Here, we follow a slightly different approach. We interpret $\Sigma_{np}$ not as a function of one, but of two variables:
\begin{equation}
\label{eq:F_y_k}
  F(y,k) = \int \ud^4l \ \hat{\Delta}_+(l) e^{ i y l} \left( \hat{\Delta}_R(k-l) + \hat{\Delta}_R(k+l) \right).
\end{equation}
This is an element of $\schw'(\R^8)$: The integration of the expression in parentheses with a test function in $k$ yields an element of $\mathcal{O}_M$, the space of smooth functions  that are, together with their derivatives, polynomially bounded. The product of this with $\hat{\Delta}_+(l)$ is then a well--defined tempered distribution, so that after integration over $l$ one obtains a tempered distribution in $y$. The statement that $F$ is an element of $\schw'(\R^8)$ then follows from the nuclear theorem. It is also easy to see that $F$ is invariant under the orthochronous Lorentz transformation $y \mapsto y {\Lambda^{T}}^{-1}, k \mapsto k \Lambda$.

In order to make contact with the original expression~(\ref{eq:F_NP}), we have to set $y =  k \sigma$. It remains to investigate whether this is possible. A first step in this direction is the following
\begin{lemma}
\label{lemma:smooth}
For $k^2 \notin \{ 0, 4m^2 \}$, the product $\hat \Delta_+(\cdot) \hat \Delta_R(k \pm \cdot)$ is well--defined as a tempered distribution. Furthermore, the map
\begin{equation*}
  \{ k \in \R^4 | k^2 \notin \{ 0, 4 m^2 \} \} \ni k \mapsto \hat \Delta_+(\cdot) \hat \Delta_R(k \pm \cdot) \in \schw'(\R^4)
\end{equation*}
is smooth. 
\end{lemma}
The proof can be found in Appendix~\ref{app:smooth}. As an immediate consequence, the map
\begin{equation*}
  \{ k \in \R^4 | k^2 \notin \{ 0, 4 m^2 \} \} \ni k \mapsto F( \cdot , k) \in \schw'(\R^4)
\end{equation*}
is smooth. It remains to show that for fixed $k$, the distribution $y \mapsto F(y, k)$ is smooth in a neighborhood of $k \sigma$. In the following we do this for the case of timelike $k$ with $k^2 \neq 4 m^2$. Later we also comment on the case of spacelike $k$. Because of Lorentz invariance, we may choose $k = (k_0, \V 0)$. Then $(k \sigma)_0 = 0$, because of the antisymmetry of~$\sigma$. Moreover, $(k \sigma)^2 \leq - \lambda_{nc}^4 k^2$ for $\sigma \in \Sigma$, due to (\ref{eq:k_sigma_2}) and Lorentz invariance. Thus, it suffices to show that $y \mapsto F(y, (k_0, \V 0))$ is smooth for $y^2 < 0$.

Integrating $F(y,k)$ with a test function $f(y)$, we obtain
\begin{equation*}
  \int \ud^4l \ \hat{\Delta}_+(l) \left( \hat{\Delta}_R(k-l) + \hat{\Delta}_R(k+l) \right) \int \ud^4y \ e^{i l y} f(y).
\end{equation*}
For $k = (k_0, \V 0)$, $k_0 > 0$, we can proceed as in the formal calculation of the planar part, and obtain
\begin{equation}
\label{eq:f_k}
 (2 \pi)^{-3} \int_0^\infty \ud l \ \frac{l^2}{2 \omega_l} \frac{-2}{k^2 - 4 \omega_l^2 + i \epsilon} \int \ud^2 \Omega \int \ud^4y \ e^{i (\omega_l, \V l) y} f(y).
\end{equation}
Here $\int \ud^2 \Omega$ stands for the integration over the ball of radius $l$. Because of Fubini's theorem, the last two integrations may be interchanged, and one obtains
\begin{equation*}
 2 (2 \pi)^{-2} \int_0^\infty \ud l \ \frac{l^2}{2 \omega_l} \frac{-2}{k^2 - 4 \omega_l^2 + i \epsilon} \int \ud^4y \ e^{i \omega_l y_0} \frac{\sin \betrag{\V y} l}{\betrag{\V y} l} f(y).
\end{equation*}
Using $\iep{x} = \pv{x} - i \pi \delta(x)$, this can be split into the two terms
\begin{equation}
\label{eq:pvTerm}
 (2 \pi)^{-2} \lim_{\epsilon \to 0} \int_{D_\epsilon} \ud l \ \frac{l^2}{2 \omega_l} \frac{1}{l^2 - \frac{k^2}{4} + m^2} \int \ud^4y \ e^{i \omega_l y_0} \frac{\sin \betrag{\V y} l}{\betrag{\V y} l} f(y),
\end{equation}
where $D_\epsilon = \{ l \in \R | 0 \leq l \leq \sqrt{k^2/4-m^2} - \epsilon, \sqrt{k^2/4-m^2} +\epsilon \leq l < \infty \}$,
and
\begin{equation}
\label{eq:deltaTerm}
 (2 \pi)^{-1} i  \int_0^\infty \ud l \ \frac{l}{4 \omega_l} \delta(l-\sqrt{k^2/4 - m^2}) \int \ud^4y \ e^{i \omega_l y_0} \frac{\sin \betrag{\V y} l}{\betrag{\V y} l} f(y).
\end{equation}
From now on we assume that $f$ has compact support $K$ such that $y^2 < 0 \ \forall y \in K$. This is legitimate, since we are only interested in the case of spacelike $y$. Interpreting the $\delta$-distribution as a measure, one can use Fubini's theorem once more. Then the term (\ref{eq:deltaTerm}) reduces to
\begin{equation*}
  (2 \pi)^{-1} i  \theta(k^2 - 4m^2) \frac{\sqrt{k^2-4m^2}}{4\sqrt{k^2}} \int \ud^4 y \  e^{i \frac{1}{2} \sqrt{k^2} y_0} \frac{\sin \frac{1}{2} \betrag{\V y} \sqrt{k^2-4m^2}}{\frac{1}{2} \betrag{ \V y} \sqrt{k^2-4m^2}} f(y).
\end{equation*}
Here $f$ is integrated with a function that is smooth for $\betrag{\V y} \neq 0$, in particular $y^2 < 0$. It remains to treat the term (\ref{eq:pvTerm}). For finite $\epsilon$ one can interchange the integrations. But one still has to interchange the limit $\epsilon \to 0$ with the integration over $y$. This can be done if the sequence of continuous maps
\begin{equation*}
 K \ni y \mapsto \int_{D_\epsilon} \ud l \ \frac{l^2}{2 \omega_l} \frac{1}{l^2 - \frac{k^2}{4} + m^2} e^{i \omega_l y_0} \frac{\sin \betrag{\V y} l}{\betrag{\V y} l}
\end{equation*}
converges uniformly on $K$. That this is the case can be seen by adapting the following calculation:
\begin{equation*}
  \betrag{ \int_\epsilon^{\epsilon'} \ud l \ \frac{g_y(l) - g_y(-l)}{l} } = \betrag{ \int_\epsilon^{\epsilon'} \ud l \ \frac{1}{l} \int_{-l}^l \ud t \ g_y'(t) } \leq 2 ( \epsilon' - \epsilon) \sup_l \betrag{g'_y(l)}.
\end{equation*}
Thus, for the term (\ref{eq:pvTerm}), we obtain
\begin{equation*}
  (2 \pi)^{-2} \int \ud^4 y \ f(y) \int_0^\infty \ud l \ \frac{l^2}{2 \omega_l} \pv{l^2 - \frac{k^2}{4} + m^2} e^{i \omega_l y_0} \frac{\sin \betrag{\V y} l}{\betrag{\V y} l}.
\end{equation*}
It remains to show that the integral over $l$ yields a smooth function of $y$ for spacelike $y$. Unfortunately, we were not able to do this directly by calculating it analytically. However, it can also be interpreted as an oscillatory integral in the sense of \cite{ReedSimon2}. Using \cite[Thm.~IX.47]{ReedSimon2} it is then straightforward to prove that it is smooth for $\betrag{\V y} \neq 0$, $y^2 \neq 0$. Thus, the map $y \mapsto F(y, (k_0, \V 0))$ is smooth for $y^2 < 0$ and the diagonal $F(k \sigma, k)$ is well--defined and smooth for $k$ timelike and $k^2 \neq 4 m^2$.
Using Lorentz invariance, we obtain, for $k$ in the interior of the forward light cone,
\begin{subequations}
\begin{align}
\label{eq:F_np_2}
  \Sigma_{np}(k) = & (2 \pi)^{-1} i  \theta(k^2 - 4m^2) \frac{\sqrt{k^2-4m^2}}{4 \sqrt{k^2}} \frac{\sin \left( \frac{1}{2} \sqrt{-(k \sigma)^2} \sqrt{k^2-4m^2} \right)}{\frac{1}{2} \sqrt{-(k \sigma)^2} \sqrt{k^2-4m^2}} \\
\label{eq:F_np}
  & + (2 \pi)^{-2} \int_0^\infty \ud l \ \frac{l^2}{2 \omega_l} \pv{l^2 - \frac{k^2}{4} + m^2} \frac{\sin \sqrt{-(k \sigma)^2} l}{\sqrt{-(k \sigma)^2} l}.
\end{align}
\end{subequations}
We remark that for $k$ in the interior of the backward light cone one gets a $-$ sign for the term~(\ref{eq:F_np_2}). We also mention that this coincides with the result obtained in \cite{NCDispRel} by interpreting $\Sigma_{np}$ directly as an oscillatory integral.

For the computation of the two--point function, we only need to know $\Sigma_{np}(k)$ in a neighborhood of the mass shell. There, the term (\ref{eq:F_np_2}) does not contribute. The remaining term (\ref{eq:F_np}) is symmetric in $k$, thus we have $\Sigma_{np}(k) = \Sigma_{np}(-k)$ in a neighborhood of the mass shell, as required for the adiabatic limit. Furthermore, $\Sigma_{np}$ is only a function of $(k \sigma)^2$ and $k^2$ (and the sign of $k_0$) as postulated in Section~\ref{sec:DispRel}.

\begin{remark}
The interpretation of $\Sigma_{np}$ as a function of two variables has the following advantage over the interpretation as an oscillatory integral: For spacelike $k$, (\ref{eq:F_NP}) is not an oscillatory integral in the sense of~\cite{ReedSimon2}, see~\cite{NCDispRel, Claus}. But the knowledge of $\Sigma_{np}(k)$ for spacelike $k$ is necessary if one wants to consider higher loop orders. In the present formulation the loop integration is well--defined and yields the tempered distribution $F$. If it can be restricted to $y = k \sigma$ is then a question that can in principle be answered by computing $F$, although this may be quite involved in practice. We also remark that the interpretation of $\Sigma_{np}$ as a function of two variables seems to be necessary if one wants to use nonlocal counterterms to restore the usual dispersion relations, cf. \cite{LiaoSibold} and the brief discussion in Chapter~\ref{chapter:Summary}.
\end{remark}

\begin{figure}
\epsfig{file=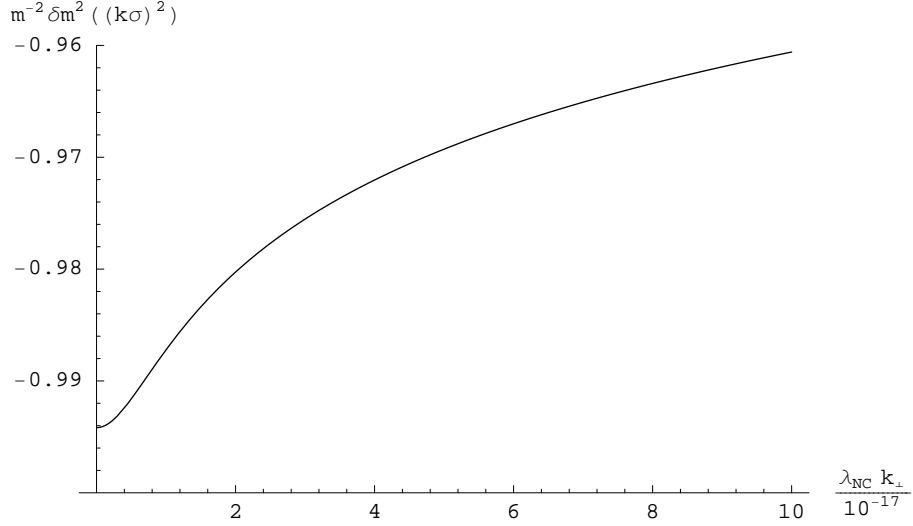,scale=0.85} 
\caption{\label{fig:M}The relative mass correction $m^{-2} \delta m^2((k \sigma)^2)$ as a function of the perpendicular
momentum $k_{\bot}$.}
\end{figure}

In order to estimate the strength of the distortion of the dispersion relation, we calculate $\delta m^2((k \sigma)^2)$ and $\delta Z((k \sigma)^2)$ numerically. We use the parameters $\sigma=\sigma_0$ (cf. (\ref{eq:sigma_0})), $m=10^{-17} \lambda_{nc}^{-1}$ and $\lambda = m$. If $\lambda_{nc}$ is identified with the Planck length, this corresponds to a mass of about $100 \text{GeV}$, i.e., the estimated order of magnitude of the Higgs mass. The chosen value of $\lambda$ is slightly above the expectation for the cubic term in the Higgs potential ($\sim 0.6m$). Figure~\ref{fig:M} shows the relative mass correction $m^{-2} \delta m^2((k \sigma)^2)$ as a function of the perpendicular momentum $k_{\bot}$, obtained with the numerical integration method of {\sc mathematica} (for the definition of $k_\bot$, see Section~\ref{sec:GroupVelocity}). We see that the relative mass shift is of order $1$ for small perpendicular momenta. This might look like a strong effect. However, we have the freedom to apply a finite mass renormalization in order to restore the rest mass. The important question is rather how strong the momentum dependence of the mass renormalization is. As can be seen in Figure~\ref{fig:M}, it is at the \%-level for perpendicular momenta of the order of the mass. As a consequence, also the distortion of the group velocity is of this order, as we will show below.


The plot for $\delta Z((k \sigma)^2)$ for the same parameters is not very interesting, since $\delta Z$ is constant, $-1.32477 \cdot 10^{-3}$, within machine precision. This coincides with the planar contribution (\ref{eq:Z_pl}). The reason
for this is easily understood. Differentiate the integrand in (\ref{eq:F_np}) with respect to $k^2$. One obtains a
function that, even without the factor
\begin{equation*}
  \frac{\sin l \sqrt{-(k \sigma)^2}}{l \sqrt{-(k \sigma)^2}},
\end{equation*}
is integrable. Without this factor, it would coincide with the corresponding planar expression obtained by
differentiating~(\ref{eq:F_pl}). But the above factor deviates from 1 appreciably only for $l \sim (-(k
\sigma)^2)^{-\frac{1}{2}}$, i.e., for very high energies, where the rest of the integrand is negligible.



\begin{figure}
\epsfig{file=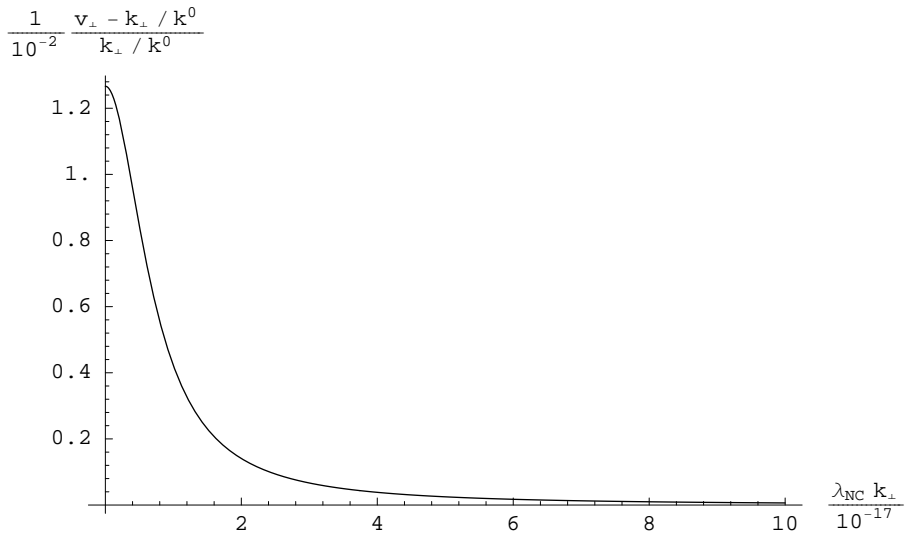,scale=0.85} 
\caption{\label{fig:GroupVel}The distortion of the group velocity in perpendicular direction as a function of
the perpendicular momentum $k_{\bot}$.}
\end{figure}

According to equation (\ref{eq:GroupVelocity}), the deviation of the group velocity from the phase velocity in
the perpendicular direction is, to lowest order in $\lambda$, given by $2 \lambda^2 \lambda_{nc}^4
\frac{\del}{\del (k \sigma)^2} \Sigma_{np}$. Figure~\ref{fig:GroupVel}  shows this quantity for the same parameters
as above.
The deviation is biggest for small perpendicular momenta and at the \%-level.

We see that in the $\phi^3$ model the distortion of the dispersion relation is moderate for realistic masses and couplings. This is in sharp contrast to the situation in the $\phi^4$ model, where realistic dispersion relations could only be obtained for masses close to the noncommutativity scale~\cite{DorosDiss}.

\subsection{The two--particle spectrum}
\label{sec:2_particle_spectrum}

We now discuss the third term in (\ref{eq:ThreeTerms}). There are two graphs that contribute to it:
\begin{center}
\begin{picture}(80,50)
\Line(19.5,5)(19.5,20)
\Line(20.5,5)(20.5,20)
\Line(20,20)(10,35)
\Line(20,20)(30,35)
\Line(59.5,5)(59.5,20)
\Line(60.5,5)(60.5,20)
\Line(60,20)(50,35)
\Line(60,20)(70,35)
\Line(30,35)(50,35)
\Curve{(10,35)(40,40)(70,35)}
\Vertex(20,20){1}
\Vertex(60,20){1}
\BCirc(10,35){1}
\BCirc(30,35){1}
\BCirc(50,35){1}
\BCirc(70,35){1}
\end{picture}
\begin{picture}(80,50)
\Line(19.5,5)(19.5,20)
\Line(20.5,5)(20.5,20)
\Line(20,20)(10,35)
\Line(20,20)(30,35)
\Line(59.5,5)(59.5,20)
\Line(60.5,5)(60.5,20)
\Line(60,20)(50,35)
\Line(60,20)(70,35)
\Curve{(10,35)(30,40)(50,35)}
\Curve{(30,35)(50,40)(70,35)}
\Vertex(20,20){1}
\Vertex(60,20){1}
\BCirc(10,35){1}
\BCirc(30,35){1}
\BCirc(50,35){1}
\BCirc(70,35){1}
\end{picture}
\end{center}
These are a planar and a nonplanar contraction. Their sum is
\begin{equation}
\label{eq:2particle}
  (2 \pi)^4 \int \ud^4k \ \hat f(-k) \hat h(k) \hat \Delta_R(k) \hat \Delta_A(k) \int \ud^4l \ \hat \Delta_+(l) \hat \Delta_+(k-l) \left( 1 + e^{i k \sigma l} \right).
\end{equation}
Because both $l$ and $k-l$ are restricted to the upper mass shell, the loop integral is only over a compact set in momentum space. We can already anticipate that the nonplanar part is very close to the planar part for realistic momenta, because for such momenta the twisting factor is essentially $1$. 
Nevertheless, we want to compute it explicitly. The loop integral is not only well--defined, but also invariant under a proper orthochronous Lorentz transformation $k \to k \Lambda$, $\sigma \to \Lambda^{-1} \sigma {\Lambda^{-1}}^T$. Thus, we can simplify the calculation by choosing $k = (k_0, \V 0)$. Then we compute
\begin{align*}
  & (2 \pi)^{-2} \theta(k_0) \int \frac{\ud^3 \V l}{2 \omega_{\V l}} \ \delta(k^2 - 2 \sqrt{k^2} \omega_{\V l}) e^{i \Vmat{k \sigma} \V l} \\
  = & (2 \pi)^{-1} \theta(k_0) \int_0^\infty \ud l \ \frac{l^2}{2 \omega_l} \delta(k^2 - 2 \sqrt{k^2} \omega_l) \int_{-1}^1 \ud x \ e^{i \betrag{\Vmat{k \sigma}} \betrag{\V l} x } \\
  = & (2 \pi)^{-1} \theta(k_0) \int_0^\infty \ud l \ \frac{l^2}{2 \omega_l} \frac{\omega_l}{2 \sqrt{k^2} l} \delta(l-\sqrt{k^2/4 -m^2}) 2 \frac{\sin \betrag{\Vmat{k \sigma}} l}{ \betrag{\Vmat{k \sigma}} l} \\
  = & (2 \pi)^{-1} \theta(k^2-4m^2) \theta(k_0) \frac{\sin \left( \sqrt{-(k \sigma)^2} \sqrt{\frac{1}{4} k^2 - m^2} \right)}{2 \sqrt{k^2} \sqrt{-(k \sigma)^2}}.
\end{align*}
Here $\Vmat{k \sigma}$ stands for the spatial part of $k \sigma$. In the limit $(k \sigma)^2 \to 0$, this yields the commutative result. Note that deviations from the commutative case become appreciable for $- (k \sigma)^2 \sim k^{-2}$. It follows that if $\lambda_{nc}$ is the Planck scale, the distortion of the two--particle spectrum is negligible for realistic momenta.

Finally, we remark that the loop integral in (\ref{eq:2particle}) has its support, in momentum space, above the $2 m$ mass shell. Thus, the ill-definedness of the product $\hat{\Delta}_R(k) \hat{\Delta}_A(k)$ at $k^2=m^2$ does not matter and (\ref{eq:2particle}) is well--defined.



\section{The Wess--Zumino model}

As we have seen in the previous section, the distortion of the dispersion relation in the noncommutative $\phi^3$ model is moderate for parameters typically expected for the Higgs field. However, the $\phi^3$ model is not realistic: Although it is well--defined on the perturbative level, it is not stable. Furthermore, the comparison with the Higgs field is somewhat misleading, since the Higgs potential also contains a quartic term, which would change the dispersion relation drastically, as shown in \cite{DorosDiss}. Thus, it is natural to study the Wess--Zumino model, which is stable and contains both a cubic and a quartic term. Since it is also only logarithmically divergent in the commutative case, one may hope that the distortion of the dispersion relation is comparable to the one found in the $\phi^3$ model.

The noncommutative Wess--Zumino model was first discussed in~\cite{Rivelles} for space/space noncommutativity in the setting of the modified Feynman rules.
In~\cite{WulkenhaarSuperfield} this was done in the superfield formalism.
It was shown that the UV/IR-mixing is much weaker than in the
$\phi^4$-theory, so that the the theory is renormalizable to all orders. Here we want to treat it in the Yang--Feldman formalism. We use the supersymmetric version of the noncommutative Minkowski space introduced in Section~\ref{sec:SUSYMink}.
In order to arrive at the equations of motion
for the component fields, we start from the Lagrangean in superfield form, taking particular care for the order
of the fields in the different terms\footnote{This is important, since for example the tadpole corresponding to
the interaction term $\phi^* \phi \phi^* \phi$ does not have a twisting factor, in contrast to the interaction
term $\phi^* \phi^* \phi \phi$, as noted in~\cite{UVIRNote}.}.

In superfield form, the Wess--Zumino model is given by the following action\footnote{Our conventions on spinors and supersymmetry are summarized in Appendix~\ref{app:SUSY}.}:
\begin{equation*}
  S = \int \ud^8q \ \bar{\Phi} \Phi + \left\{ \int \ud^6q \ \left( \frac{m}{2} \Phi \Phi + \frac{\lambda}{3} \Phi \Phi \Phi \right) + \text{ h.c.} \right\}.
\end{equation*}
Here $\Phi$ is the chiral superfield
\begin{equation*}
  \Phi = \phi + \sqrt{2} \theta \chi + \theta^2 F - i \theta \sigma^{\mu} \bar{\theta} \del_{\mu} \phi + \frac{i}{\sqrt{2}} \theta^2 \del_{\mu} \chi \sigma^{\mu} \bar{\theta} - \frac{1}{4} \theta^2 \bar{\theta}^2 \Box \phi,
\end{equation*}
while $\phi$ and $F$ are complex scalar fields and $\chi$ is a Weyl spinor.
In component fields, the action is then
\begin{multline*}
  S = \int \ud^4q \ \Bigg( i \bar{\chi} \bar{\sigma}^{\mu} \del_{\mu} \chi - \phi^* \Box \phi + F^* F  \\  + \left\{ \left( m \left( \phi F - \frac{1}{2} \chi \chi \right) + \lambda \left( \phi \phi F - \chi \chi \phi \right) \right) + \text{ h.c.} \right\} \Bigg).
\end{multline*}
This leads to the equations of motion
\begin{gather*}
  F + m \phi^* + \lambda \phi^* \phi^* = 0, \\
  - \Box \phi + m F^* + \lambda (\phi^* F^* + F^* \phi^*) - \lambda \bar{\chi} \bar{\chi} = 0, \\
   i \bar{\sigma}^{\mu} \del_{\mu} \chi - m \bar{\chi} - \lambda (\phi^* \bar{\chi} + \bar{\chi} \phi^* ) = 0.
\end{gather*}
We eliminate the auxiliary field $F$ using its equation of motion. Furthermore, we introduce the Majorana spinors
\begin{equation*}
  \psi = \frac{1}{\sqrt{2}} \begin{pmatrix} \chi_{\alpha} \\ \bar{\chi}^{\dot{\alpha}} \end{pmatrix}, \quad \bar{\psi} = \psi^\dagger \gamma^0 = \frac{1}{\sqrt{2}}  ( \chi^{\alpha},  \bar{\chi}_{\dot{\alpha}} ).
\end{equation*}
Because of $2 \bar{\psi} P_+ \psi = \chi \chi$, where $P_+$ is the chiral projector defined in~(\ref{eq:P_pm}), we can write the equation of motion as
\begin{align*}
  ( \Box + m^2) \phi & =  - 2 \lambda \bar{\psi} P_- \psi - m \lambda ( \phi \phi + \phi^* \phi + \phi \phi^* ) - \lambda^2 (\phi^* \phi \phi + \phi \phi \phi^*), \\
  (  i \dslash - m) \psi & = \lambda P_+ (\phi \psi + \psi \phi) + \lambda P_- (\phi^* \psi + \psi \phi^*).
\end{align*}

\subsection{The SUSY current}

We first want to discuss the changes that the noncommutativity brings in at the classical level. The equations of motion are the same, we only have to replace the usual product by the noncommutative one. But there are some changes for the currents.
As we have seen in Section~\ref{sec:ClassicalFields}, it is an interesting feature of noncommutative interacting theories that the local currents associated to symmetries are in general not conserved.
In the following, we show that the local current associated to the supersymmetry transformation is not conserved in the interacting case, i.e., for $\lambda \neq 0$. We discuss this in terms of the superfield~$\Phi$. Its equation of motion is
\begin{equation*}
 - \frac{1}{4} \bar{D}^2 \bar{\Phi} + m \Phi + \lambda \Phi \Phi = 0.
\end{equation*}
The supercurrent is given by
\begin{equation*}
  V_{\alpha \dot{\alpha}} = \frac{1}{2} [ D_{\alpha} \Phi, \bar{D}_{\dot{\alpha}} \bar{\Phi} ] + i \{ \dslash_{\alpha \dot{\alpha}} \Phi, \bar{\Phi} \} - i \{ \Phi, \dslash_{\alpha \dot{\alpha}} \bar{\Phi} \}.
\end{equation*}
Here we used a symmetrized version of the usual current, since this is usually advantageous in the
noncommutative case.
By standard methods (see, e.g., \cite[Section~15]{Sohnius}) one can show that
\begin{equation*}
  \bar{D}^{\dot{\alpha}} V_{\alpha \dot{\alpha}} = \frac{1}{2} \{ D_{\alpha} \Phi, \bar{D}^2 \bar{\Phi} \} - \frac{1}{4} \{ \Phi, D_{\alpha} \bar{D}^2 \bar{\Phi} \}
\end{equation*}
holds. Using the equation of motion, one obtains
\begin{align*}
  \bar{D}^{\dot{\alpha}} V_{\alpha \dot{\alpha}} & = 2 \left\{ D_{\alpha} \Phi, \left( m \Phi + \lambda \Phi \Phi \right) \right\} - \left\{ \Phi, D_{\alpha}  \left( m \Phi + \lambda \Phi \Phi \right) \right\} \\
  & = m D_{\alpha} \Phi^2 + \lambda [[ D_{\alpha} \Phi, \Phi], \Phi].
\end{align*}
The first term in the second line is already present in the commutative case. It does not affect the charge corresponding to the
supersymmetry transformation, but simply expresses the fact that the theory is not conformal. The second term, however,
is a genuinely noncommutative one. It also affects the SUSY charge.
Like the nonconservation of the local energy--momentum tensor, this effect does not show up in a perturbative treatment of the corresponding quantum theory, at least not at second order.


\subsection{The self--energy}

We now want to compute the self--energy $\Sigma(k)$ 
of the scalar component of the interacting field at second order.
Using the equations of motion, the first terms in the Yang--Feldman series are
\begin{align*}
  \phi_1 & = - \Delta_R \times \left( 2 \bar{\psi}_0 P_- \psi_0 + m ( \phi_0^* \phi_0 + \phi_0 \phi_0^* + \phi_0 \phi_0 )  \right), \\
  \psi_1 & = S_R \times \left( P_+ ( \phi_0 \psi_0 + \psi_0 \phi_0) + P_- ( \phi_0^* \psi_0 + \psi_0 \phi_0^* )
  \right),
\end{align*}
and the analogous formulae for the conjugate fields. Note that we did not employ normal ordering for the definition of $\phi_1$. As we will see below, this is not necessary here, because of the supersymmetry. The second order component of $\phi$ is
\begin{subequations}
\begin{align}
\label{eq:term1}
  \phi_2 = - \Delta_R \times & \Big\{ 2 \bar{\psi}_1 P_- \psi_0 + 2 \bar{\psi}_0 P_- \psi_1 \\
\label{eq:term2}
  & + m ( \phi_1^* \phi_0 + \phi_0^* \phi_1 + \phi_1 \phi_0^* + \phi_0 \phi_1^* + \phi_1 \phi_0 + \phi_0 \phi_1 ) \\
\label{eq:term3}
  & + ( \phi_0 \phi_0 \phi_0^* + \phi_0^* \phi_0 \phi_0 ) \Big\}.
\end{align}
\end{subequations}
For the computation of the graphs involving fermions, we need the formulae\footnote{The factor $1/2$ in the last
line is due to the Majorana nature of the fermions.
}
\begin{align*}
  \hat{S}_R(k) & = ( - \kslash - m) \hat{\Delta}_R(k), \\
  \hat{\bar{S}}_R(k) & = ( \kslash - m) \hat{\Delta}_R(k), \\
  \bra{\Omega} \hat{\bar{\psi}}_\alpha(k) \hat{\psi}_\beta(p) \ket{\Omega} = & \frac{1}{2} (2 \pi)^{2} \delta(k+p) ( \kslash - m )_{\beta \alpha} \hat{\Delta}_+(k).
\end{align*}
We now have to compute the following graphs:
\begin{center}
\begin{picture}(50,50)
\CArc(25,30)(10,0,360)
\Line(10,20.5)(25,20.5)
\Line(10,19.5)(25,19.5)
\Line(25,20)(40,20)
\Vertex(25,20){1}
\end{picture}
\begin{picture}(50,50)
\Line(10,19.5)(25,19.5)
\Line(10,20.5)(25,20.5)
\Vertex(25,20){1}
\CArc(25,35)(5,0,360)
\Line(24.5,20)(24.5,30)
\Line(25.5,20)(25.5,30)
\Vertex(25,30){1}
\Line(25,20)(40,20)
\end{picture}
\begin{picture}(80,50)
\Line(10,19.5)(25,19.5)
\Line(10,20.5)(25,20.5)
\Vertex(25,20){1}
\CArc(40,20)(15.5,180,0)
\CArc(40,20)(14.5,180,0)
\CArc(40,20)(15,0,180)
\Vertex(55,20){1}
\Line(55,20)(70,20)
\end{picture}
\begin{picture}(50,50)
\Line(10,19.5)(25,19.5)
\Line(10,20.5)(25,20.5)
\Vertex(25,20){1}
\DashCArc(25,35)(5,0,360){2}
\Line(24.5,20)(24.5,30)
\Line(25.5,20)(25.5,30)
\Vertex(25,30){1}
\Line(25,20)(40,20)
\end{picture}
\begin{picture}(80,50)
\Line(10,19.5)(25,19.5)
\Line(10,20.5)(25,20.5)
\Vertex(25,20){1}
\DashCArc(40,20)(15.5,180,0){2}
\DashCArc(40,20)(14.5,180,0){1.8}
\DashCArc(40,20)(15,0,180){2}
\Vertex(55,20){1}
\Line(55,20)(70,20)
\end{picture}
\end{center}
Here we did not show all the different contractions that can appear, but rather subsumed them under ``typical'' graphs. Furthermore, we drew the graphs from left to right, not upwards. In fact we will not use the graphical rules for our computation, but compute the contractions in~(\ref{eq:term1},b,c) by hand.

\begin{description}

 \item[The $\phi^4$ tadpole] is obtained from the term (\ref{eq:term3}). There are two planar and two nonplanar contractions. We find the quadratically divergent contribution
\begin{equation*}
  \Sigma_{\phi^4-\text{tp}}(k) = - 2 (2\pi)^{-2} \lambda^2 \int \ud^4 l \ \hat{\Delta}_+(l) \left( 1 + e^{i k \sigma l} \right).
\end{equation*}


\item[The $\phi^3$ tadpole] is obtained from the term (\ref{eq:term2}) by contracting the $\phi_0$ and $\phi_0^*$ in $\phi_1$ or $\phi_1^*$ among themselves. In each $\phi_1$ or $\phi_1^*$ there are two such contractions. And there are four terms in (\ref{eq:term2}) where a $\phi_0$ remains. Due to the retarded propagator with zero momentum connecting the loop with the line, the mass appearing in the interaction term cancels, and we obtain
\begin{equation*}
  \Sigma_{\phi^3-\text{tp}}(k) = 8 (2\pi)^{-2} \lambda^2 \int \ud^4 l \ \hat{\Delta}_+(l).
\end{equation*}
Note that no twisting factor appears.


\item[The $\phi^3$ fish graph] is obtained from the term (\ref{eq:term2}) by contracting the ``outer'' $\phi_0$ (or $\phi_0^*$) with a $\phi_0^*$ (or $\phi_0$) in $\phi_1$ or $\phi_1^*$. One then collects all terms where an uncontracted $\phi_0$ remains.
We find
\begin{equation*}
 \Sigma_{\phi^3-\text{fish}}(k) = 3 m^2 \lambda^2 \int \ud^4 l \ \hat{\Delta}_+(l) \left( 1 + e^{i k \sigma l} \right) \left( \hat{\Delta}_R(k-l) + \hat{\Delta}_R(k+l) \right).
\end{equation*}


\item[The Yukawa tadpole] is obtained from (\ref{eq:term2}) by contracting the fermions in $\phi_1$ or
$\phi_1^*$. With the usual $\gamma$ matrix algebra (see Appendix~\ref{app:SUSY}), we find
\begin{equation*}
  \Sigma_{\text{Yuk}}(k) = - 8 (2\pi)^{-2} \lambda^2 \int \ud^4 l \ \hat{\Delta}_+(l).
\end{equation*}


\item[The fermion fish graph] is obtained from the term (\ref{eq:term1}). The relevant part of $\phi_2$, i.e.,
the part involving $\phi_0$, is
\begin{multline*}
  \hat{\phi}_2(k) = - 4 \hat{\Delta}_R(k) \int \ud^4l \ud^4l' \ \cos \frac{l \sigma l'}{2} \\
 \times \left\{ \hat{\bar{\psi}}_0(k-l) P_- \hat{S}_R(l) P_+ \hat{\psi}_0(l-l') \hat{\phi}_0(l') e^{- \frac{i}{2} k \sigma l} \right. \\
  \left. + \hat{\bar{\psi}}_0(l-l') P_+ \hat{\bar{S}}_R(l) P_- \hat{\psi}_0(k-l) \hat{\phi}_0(l') e^{- \frac{i}{2} l \sigma k} \right\}.
\end{multline*}
Contraction of the fermion fields now yields
\begin{align*}
  & - 2 (2 \pi)^{2} \hat{\Delta}_R(k) \hat \phi_0(k) \int \ud^4l \ \cos \frac{l \sigma k}{2} \\
 & \quad \times \left\{ \tr \left( P_- (-\lslash -m) P_+ (\kslash-\lslash-m) \right) \hat{\Delta}_R(l) \hat{\Delta}_+(k-l) e^{-\frac{i}{2} k \sigma l} \right. \\
 & \qquad \left. + \tr \left( P_+ (\lslash -m) P_- (-\kslash+\lslash-m) \right) \hat{\Delta}_R(l) \hat{\Delta}_+(-k+l) e^{-\frac{i}{2} l \sigma k} \right\} \\
= & - 2 (2 \pi)^{2} \hat{\Delta}_R(k) \hat \phi_0(k) \int \ud^4l \ \cos \frac{l \sigma k}{2} \\
 & \quad \times \left\{ \tr \left( P_- (\lslash - \kslash -m) P_+ (\lslash-m) \right) \hat{\Delta}_R(k-l) \hat{\Delta}_+(l) e^{-\frac{i}{2} l \sigma k} \right. \\
 & \qquad \left. + \tr \left( P_+ (\kslash + \lslash -m) P_- (\lslash-m) \right) \hat{\Delta}_R(k+l) \hat{\Delta}_+(l) e^{-\frac{i}{2} l \sigma k}
 \right\}.
\end{align*}
With a little $\gamma$ matrix algebra, one finds
\begin{equation*}
  \tr \left( P_- (\kslash - m) P_+ (\pslash-m) \right) = 2 k \cdot p.
\end{equation*}
Thus, we get
\begin{multline*}
 \Sigma_{\psi-\text{fish}}(k) = 2 \lambda^2 \int \ud^4 l \ \hat{\Delta}_+(l) \left( 1 + e^{i k \sigma l} \right) \\ \times \left( (k-l) \cdot l  \hat{\Delta}_R(k-l) - (k+l) \cdot l \hat{\Delta}_R(k+l) \right).
\end{multline*}
\end{description}

Now we collect all our terms. The Yukawa tadpole and the $\phi^3$ tadpole cancel (this has to be so in order to
have a vanishing vacuum expectation value of $\phi_1$). Using
\begin{equation*}
  (l^2 - m^2) \hat{\Delta}_+(l) = 0, \quad (l^2 - m^2) \hat{\Delta}_R(l) = - (2 \pi)^{-2},
\end{equation*}
the combination of the other terms yields
\begin{equation*}
  \Sigma(k) = \lambda^2 \left( k^2 + m^2 \right) \int \ud^4l \ \hat{\Delta}_+(l) \left( 1 + e^{i k \sigma l} \right) \left( \hat{\Delta}_R(k-l) + \hat{\Delta}_R(k+l) \right).
\end{equation*}
Apart from the prefactor $(k^2+m^2)$, this is exactly the expression we already found for the $\phi^3$-model. We remark that for the self--energy of the fermion, one obtains the same result.

The prefactor $(k^2+m^2)$ is to be expected: Assuming that the nonrenormalization theorem still holds, we know that only
the $\bar{\Phi} \Phi |_{\theta^2 \bar{\theta}^2}$-term is renormalized. From the free equations of motion
\begin{equation*}
  (1+ \delta Z) F - m \phi^* = 0 , \qquad (1+ \delta Z) \Box \phi + m F^* = 0,
\end{equation*}
one obtains, at first order in $\delta Z$,
\begin{equation*}
  (\Box + m^2) \phi = - \delta Z (\Box - m^2) \phi.
\end{equation*}
Note that in our terminology, this corresponds to both a field strength and a mass renormalization. Explicitly,
we have, after subtracting the planar part,
\begin{subequations}
\begin{align}
\label{eq:WZ_M}
  \delta m^2((k\sigma)^2) = & - 2 m^2 \Sigma_{np}((k \sigma)^2, m^2), \\
\label{eq:WZ_Z}
  \delta Z((k\sigma)^2) = & \Sigma_{np}((k\sigma)^2, m^2) + 2 m^2 \frac{\del}{\del k^2} \Sigma_{np}((k\sigma)^2, m^2).
\end{align}
\end{subequations}
Here we used the $\Sigma_{np}$ from the previous section, cf. equation (\ref{eq:F_NP}). From (\ref{eq:WZ_M}) we
conclude that for $\sigma=\sigma_0, m=10^{-17} \lambda_{nc}^{-1}, \lambda=1$ the distortion of the group
velocity is twice as strong as the one calculated in the previous section for the $\phi^3$-model. Identifying $\phi$ with the Higgs field, an effect of the magnitude might be measurable at the next generation of particle colliders. It could be worthwhile to study this effect in a full-blown phenomenological model like the MSSM, in particular including supersymmetry breaking.




\chapter{Noncommutative quantum electrodynamics}
\label{chapter:NCQED}

Now we bring together the results of the two previous chapters to study noncommutative quantum electrodynamics (NCQED) in the unexpanded (module) approach. In particular, we want to compute the correlation (two--point) function of the interacting field strength in covariant coordinates to second order in $e$. As in the previous section, we use the Yang--Feldman formalism for the definition of the interacting field. To the best of our knowledge, such a computation has not been done before: The calculations in the literature all use the modified Feynman rules\footnote{In~\cite{Ohl} the Hamiltonian approach was used and it was shown that the Ward identity is violated already at tree level.}, and are therefore not valid for the case of space/time noncommutativity. Furthermore, the covariant coordinates were not fully taken into account. Mostly, they were not used at all. In \cite{Gross, Rozali}, only a subclass of the contributions of the covariant coordinates were considered, namely those in which the fields, that the covariant coordinates bring in, all contract among themselves. As we will see later, the contributions from mixed contractions are important.


In~\cite{Hayakawa}, the two--point function $\bra{\Omega} \hat A_\mu(k) \hat A_\nu(p) \ket{\Omega}$, i.e., without covariant coordinates, was calculated at second order in the setting of the modified Feynman rules, and it was shown that the nonplanar part of the photon self--energy is of the form
\begin{equation}
\label{eq:SelfEnergy}
 \Pi^{\mu \nu}_{np}(k) = e^2 \left( (g^{\mu \nu} k^2 - k^{\mu} k^{\nu}) \Sigma_1(k) + \tfrac{(k \sigma)^{\mu} (k \sigma)^{\nu}}{(k \sigma)^4} \Sigma_2(k) \right)
\end{equation}
with
\begin{subequations}
\begin{align}
\label{eq:Hayakawa_Sigma_1}
  \Sigma_1(k) & = - (2\pi)^{-2} \frac{5}{3} \ln (\sqrt{k^2} \sqrt{- (k \sigma)^2} ) + \order(1), \\
\label{eq:Hayakawa_Sigma_2}
  \Sigma_2(k) & = - (2\pi)^{-2} \left( 8 - \frac{1}{3} k^2 (k\sigma)^2 \right) + \order(k^4 (k \sigma)^4).
\end{align}
\end{subequations}
Apart from a different prefactor for the second term in (\ref{eq:Hayakawa_Sigma_2}), the same results were found in~\cite{KhozeTravaglini} with the background field method. From~(\ref{eq:Hayakawa_Sigma_2}) it is obvious that the second term in~(\ref{eq:SelfEnergy}) is quadratically IR--divergent.
This had not been expected, since the commutative theory is only logarithmically UV--divergent. In~\cite{Matusis}, this was explained as follows: The underlying UV--divergence is quadratic by power-counting, only by invoking the Ward identity does it become logarithmic. In the nonplanar diagrams, however, the phase factor with the incoming momentum serves as an UV--regulator. Hence the quadratic IR--divergence in the incoming momentum. It was shown in~\cite{Ruiz} that this term is independent of the gauge chosen.

The second term in~(\ref{eq:SelfEnergy}) is usually interpreted as a severe distortion of the dispersion relation~\cite{Matusis}, leading to tachyonic modes~\cite{Ruiz}. In~\cite{GraciaBondia}, it was argued that in the case of space/space noncommutativity, it leads to ill--defined terms in the effective action. Here we adopt the follwing point of view\footnote{This argument was already used in \cite{DoroOWR} in the context of quasiplanar Wick products for massless fields.}:
In the full expression for the two--point function, $\frac{(k\sigma)^\mu (k\sigma)^\mu}{(k \sigma)^4}$ is multiplied with $\theta(k_0) \delta'(k^2)$. The two distributions have overlapping singular support, thus their product is not well--defined. In order to make it well--defined, one would have to add a nonlocal counterterm, i.e., there are nonlocal renormalization ambiguities (see section~\ref{sec:infrared} for details). Then the theory loses its predictive power.

Here, we compute the two--point function of the interacting field strength at second order in $e$. We consider pure electrodynamics, i.e., we do not include fermionic matter fields\footnote{This can be justified by the fact that at $\order(e^2)$, the fermion contributions are exactly as in the commutative case~\cite{Hayakawa} and thus not of interest for our study.}. As was already anticipated in Chapter~\ref{chapter:NCGaugeTheory}, the effect of the covariant coordinates can be accounted for perturbatively. One may hope that the use of covariant coordinates helps to tame the bad infrared behavior indicated in~(\ref{eq:SelfEnergy}). However, we find that:
\begin{enumerate}
\item In order to set up the free theory, we have to use a modified product of quantum fields, which could also be interpreted as the subtraction of nonlocal counterterms.

\item The contraction of two fields coming from the covariant coordinate yields a nonlocal divergence. Its subtraction can be interpreted as a normal ordering of $e^{ikX}$.

\item From the terms in which one power of $e$ stems from the interaction and one from the covariant coordinates, we get a contribution that is nonlocal and divergent (a nonlocal expression multiplied with a divergent quantity).

\item For the photon self--energy we recover, at leading order in $(k \sigma)^2 \to 0$, the results obtained in the setting of the modified Feynman rules, i.e., the nonplanar contributions are of the form~(\ref{eq:SelfEnergy}) with $\Sigma_{1/2}$ given by~(\ref{eq:Hayakawa_Sigma_1}, \ref{eq:Hayakawa_Sigma_2}).

\end{enumerate}
The subtractions that are necessary in order to get rid of the divergences mentioned in 3. and 4. have to be interpreted as nonlocal counterterms. It follows that the theory can at best be considered as effective.


\section{Setup}

In order to quantize noncommutative electrodynamics via the Yang--Feldman formalism, one needs a well-posed Cauchy problem. Thus, we have to break gauge invariance. We use the BRST formalism and introduce ghosts and antighosts $c$ and $\bar{c}$ and the Nakanishi-Lautrup field $B$. Our Lagrangean is then
\begin{equation*}
  L = - \frac{1}{4} F_{\mu \nu} F^{\mu \nu} + \del_{\mu} B A^{\mu} + \frac{\alpha}{2} B^2 -  \del_{\mu} \bar{c} D^{\mu} c .
\end{equation*}
Here the covariant derivative of $c$ is given by $D_{\mu} c = \del_{\mu} c - i e [A_{\mu}, c]$. The ghosts $c$ and $\bar{c}$ are hermitean, respectively antihermitean, so the Lagrangean is hermitean. The corresponding action is invariant under the BRST transformation
\begin{align*}
  \delta_\xi A_{\mu} & = \xi D_{\mu} c, \\
  \delta_\xi c & = \xi \tfrac{i}{2} e \{c,c\}, \\
  \delta_\xi \bar{c} & = \xi B, \\
  \delta_\xi B & = 0,
\end{align*}
where $\xi$ is an infinitesimal anticommuting parameter. Note that, due to the anticommuting nature of the ghosts, the anticommutator in the second equation can also be expressed by the commutator of the spacetime part:
\begin{align}
  \{c,c\} = & (2\pi)^{-4} \int \ud^4k_1 \ud^4k_2 \ \left( \hat{c}(k_1) \hat{c}(k_2) e^{-ik_1q} e^{-ik_2q} + \hat{c}(k_2) \hat{c}(k_1) e^{-ik_2q} e^{-ik_1q} \right) \nonumber \\
\label{eq:ghost_anticom}  
  = & (2\pi)^{-4} \int \ud^4k_1 \ud^4k_2 \ \hat{c}(k_1) \hat{c}(k_2) [e^{-ik_1q}, e^{-ik_2q}].
\end{align}
The first term in the Lagrangean transforms covariantly,
\begin{equation*}
  F_{\mu \nu} F^{\mu \nu} \to - i e \xi [F_{\mu \nu} F^{\mu \nu},c],
\end{equation*}
so the integral over it is invariant. We now want to check the invariance of the sum of the remaining terms:
\begin{equation*}
 \delta_\xi ( \del_{\mu} B A^{\mu} )  - \delta_\xi ( \del_{\mu} \bar{c} D^{\mu} c ) = - \del_{\mu} \bar{c} \delta_\xi ( D^{\mu} c ) = \xi \del_{\mu} \bar{c} \delta' \delta' A^{\mu}.
\end{equation*}
Here we used the notation $\delta_\xi = \xi \delta'$. That $\delta' \delta' A^{\mu}$ vanishes is something we should check anyway:
\begin{equation*}
 \delta_\xi ( D^{\mu} c ) = \tfrac{i}{2} e \xi D^{\mu} \{c,c\} - i e \xi \{ D^{\mu} c, c \} = 0.
\end{equation*}
Thus we have also shown that $\delta'$ is nilpotent.

From the above action, we obtain the equations of motion
\begin{gather*}
  D_{\mu} F^{\mu \nu} + \del^{\nu} B - i e \{ \del^{\nu} \bar{c}, c\} = 0, \\
  \alpha B - \del_{\mu} A^{\mu} = 0, \\
  \del^{\mu} D_{\mu} c = 0, \\
  D^{\mu} \del_{\mu} \bar{c} = 0.
\end{gather*}

\section{The BRST currents}
\label{sec:BRST}

As discussed in Section~\ref{sec:ClassicalFields}, the issue of current conservation is delicate in noncommutative interacting field theories. Here we investigate whether the local BRST currents are conserved. One could follow the procedure proposed in~\cite{Diplom} to derive the corresponding equation directly from $\delta$ and the Lagrangean. Here, however, we take the simpler approach to postulate a current in analogy to the usual current of ordinary Yang-Mills theory and compute its divergence. 
The result is the same. Thus, we postulate (cf.~\cite{nakanishi-ojima})
\begin{equation*}
  j^{\mu} = \frac{1}{2} \{ B, D^\mu c \} - \frac{1}{2} \{ \del^\mu B, c \} + \frac{i e}{6} \{ c, \del^\mu \bar{c}, c \} - \frac{1}{2} \del_\nu \{ F^{\mu \nu}, c \}.
\end{equation*}
Here $\{ \cdot , \cdot , \cdot \}$ stands for complete symmetrization. Using the equations of motion, the divergence of this is
\begin{equation*}
 \del_\mu  j^{\mu} = - \frac{i e}{2} \{ \del^\mu B, [A_\mu, c] \} - \frac{1}{2} \{ \Box B, c \} + \frac{i e}{3} \{ \del_\mu c, \del^\mu \bar{c}, c \} - \frac{e^2}{6} \{ c, [A_\mu, \del^\mu \bar{c}], c \} .
\end{equation*}
The first term on the right hand side can be treated with the Jacobi identity
\begin{equation*}
  \{ \del^\mu B, [A_\mu, c] \} = [A_\mu, \{ \del^\mu B, c \}] - \{ [A_\mu, \del^\mu B], c\}.
\end{equation*}
Furthermore, due to the equations of motion, we have
\begin{align*}
  i e \{ [A_\mu, \del^\mu B], c\} - \{ \Box B, c \} = & - \{ D_\mu \del^\mu B, c \} \\
  = & \{ D_{\mu} D_\nu F^{\nu \mu}, c \} - i e \{ D_{\mu} \{ \del^\mu \bar{c}, c \} , c\} \\
  = & - \frac{i e}{2} \{ [ F_{\mu \nu}, F^{\nu \mu}], c \} - i e \{ \{ \del^\mu \bar{c}, D_{\mu} c \} , c\}.
\end{align*}
The first term vanishes. Thus, we obtain
\begin{equation*}
 \del_\mu  j^{\mu} = - \frac{i e}{2} [A_\mu, \{ \del^\mu B, c \}] - \frac{i e}{2} \{ \{ \del^\mu \bar{c}, D_{\mu} c \} , c\}  + \frac{i e}{3} \{ \del_\mu c, \del^\mu \bar{c}, c \} - \frac{e^2}{6} \{ c, [A_\mu, \del^\mu \bar{c}], c \} .
\end{equation*}
With a little algebra, one can show that this is
\begin{equation*}
 \del_\mu  j^{\mu} = - \frac{i e}{2} [A_\mu, \{ \del^\mu B, c \}] + \frac{i e}{6} [ \del^\mu \bar{c}, [ c , D_{\mu} c ]] + \frac{i e}{6} [ [ \del^\mu \bar{c}, c] , D_{\mu} c ] - \frac{e^2}{6} [A_\mu, \{ c, \del^\mu \bar{c}, c \} ] .
\end{equation*}
The right hand side does not vanish in general. Thus, the local BRST current is not conserved. In the free case $(e=0)$, however, one finds the usual current conservation.

One can do the same thing for the ghost current
\begin{equation*}
  j^\mu = i \left( \bar{c} D^\mu c - \del^\mu \bar{c} c \right).
\end{equation*}
Using the equations of motion, we find
\begin{equation*}
 \del_\mu  j^\mu = e [A_\mu,  \del^\mu \bar{c} c ].
\end{equation*}
Hence, also the local ghost current is not conserved in the interacting case.

That the local interacting BRST current is not conserved is probably not problematic, as long as $\delta'_{int}$ is still nilpotent. This is the case if the renormalized Lagrangean is still of the appropriate form, i.e., if the usual relations between counterterms hold. In \cite{Martin}, this has been checked at the one-loop level in the setting of the modified Feynman rules. Here, we will not repeat this calculation in the Yang--Feldman formalism. We note however, that in~\cite{Martin} only the local counterterms, i.e., those that arise from planar graphs, were taken into account. These should be the same in the Yang--Feldman formalism. But we will argue in the following that nonlocal counterterms are necessary. Then the situation becomes more involved\footnote{The nonlocal counterterms we will encounter are of the form
\begin{equation*}
  e^2 \int \ud^4k \ \hat{A}^\mu(-k) (k \sigma)_{\mu} (k \sigma)_\nu \hat{A}^\nu(k) \left( \frac{k^2}{(k \sigma)^2} \delta Z + \frac{\Sigma((k \sigma)^2, k^2)}{(k \sigma)^4} \right).
\end{equation*}
Due to the antisymmetry of $\sigma$, this is at least BRST invariant up to $\order(e^3)$.}.

\section{The Yang--Feldman procedure}
\label{sec:YF}

We state once more the equations of motion:
\begin{subequations}
\begin{gather}
\label{eq:eom1}
  \Box A^{\mu} - \del^{\mu} \del_{\nu} A^{\nu} + \del^{\mu} B = i e \del_{\nu} [A^{\nu}, A^{\mu}] + i e [A_{\nu}, F^{\nu \mu}] + i e \{ \del^{\mu} \bar{c}, c\}, \\
\label{eq:eom2}
  \alpha B - \del_{\mu} A^{\mu} = 0, \\
\label{eq:eom3}
  \Box c - i e \del_{\mu} [A^{\mu}, c] = 0, \\
\label{eq:eom4}
  \Box \bar{c} - i e [A^{\mu}, \del_{\mu} \bar{c}] = 0.
\end{gather}
\end{subequations}
Choosing Feynman gauge,
$\alpha =1$, and using~(\ref{eq:eom2}), we can eliminate the Nakanishi-Lautrup field $B$ from~(\ref{eq:eom1}):
\begin{equation}
\label{eq:eom1b}
  \Box A^{\mu} = i e \del_{\nu} [A^{\nu}, A^{\mu}] + i e [A_{\nu}, F^{\nu \mu}] + i e \{ \del^{\mu} \bar{c}, c \}.
\end{equation}
Now we will use equations (\ref{eq:eom1b}), (\ref{eq:eom3}) and (\ref{eq:eom4}) for the Yang--Feldman procedure.
At first order, the interacting fields are
\begin{subequations}
\begin{align}
\label{eq:A1}
  A_1^{\mu} & = i \Delta_R \times \left\{ \del_{\lambda} [A_0^{\lambda}, A_0^{\mu}] + [A_0^{\lambda}, \del_{\lambda} A_0^{\mu}] - [A_0^{\lambda}, \del^{\mu} A_{0 \lambda}] + \{ \del^{\mu} \bar{c}_0, c_0 \} \right\}, \\
\label{eq:c1} 
  c_1 & = i \Delta_R \times \left\{ \del_{\mu} [A_0^{\mu}, c_0] \right\}, \\
\label{eq:c1_bar}
  \bar{c}_1 & = i \Delta_R \times \left\{ [A_0^{\mu}, \del_{\mu} \bar{c}_0] \right\}.
\end{align}
\end{subequations}
The photon field at second order is then
\begin{subequations}
\begin{align}
\label{subeq:A2_1}
  A_2^{\mu} = i \Delta_R \times \big\{ & \del_{\lambda} [A_1^{\lambda}, A_0^{\mu}] + \del_{\lambda} [A_0^{\lambda}, A_1^{\mu}] + [A_1^{\lambda}, \del_{\lambda} A_0^{\mu}] \\
\label{subeq:A2_2}
  & + [A_0^{\lambda}, \del_{\lambda} A_1^{\mu}] - [A_1^{\lambda}, \del^{\mu} A_{0 \lambda}] - [A_0^{\lambda}, \del^{\mu} A_{1 \lambda}] \\
\label{subeq:A2_3} 
  & + \{ \del^{\mu} \bar{c}_1, c_0 \} + \{ \del^{\mu} \bar{c}_0, c_1 \} \\
\label{subeq:A2_4}  
  & - i [A_{0 \lambda}, [A_0^{\lambda}, A_0^{\mu}]] \big\}.
\end{align}
\end{subequations}

Now we have to clarify an important point. 
We recall from Section~\ref{sec:QuantumFields} that quantum fields are formally elements of $\mathfrak{F} \otimes \mathcal{E}_\sigma$, where $\mathfrak{F}$ stands for operators on the Fock space. Their natural product is
\begin{equation}
\label{eq:OldProduct2}
  (\phi \otimes f) \cdot (\psi \otimes g) = \phi \psi \otimes f g.
\end{equation}
Already in Remark~\ref{rem:OldProduct}, we argued that this product is not always appropriate.
Here we show that in the context of NCQED this product is in conflict with some general properties we would like to demand. First of all, in the classical theory, we have
\begin{equation*}
  \int \ud^4q \  [A_{\mu}, E]  = 0,
\end{equation*}
where $E$ stands for any product of a field polynomial with a test function, i.e., an element of $\schw_2$. This is particularly important, since the commutator with $A_{\mu}$ is part of the covariant derivative. However, this equation is no longer true in the quantum theory if one uses the product~(\ref{eq:OldProduct2}). This is a special case of the problem mentioned in Remark~\ref{rem:OldProduct}.
Closely related to this is the fact that the above quantization does not have the right commutative limit, since $A_1^{\mu}$ does not vanish there. Moreover, $A_1^{\mu}$, when quantized with the product~(\ref{eq:OldProduct2}), has a nonvanishing, in fact divergent, vacuum expectation value. As an example, we compute it for the second term in (\ref{eq:A1}):
\begin{multline*}
  \int \ud^4k_1 \ud^4k_2 \   e^{-ik_1q} e^{-ik_2q} 
  \left\{ k_{2 \lambda} \bra{\Omega} \hat{A}^{\lambda}(k_1) \hat{A}^{\mu}(k_2) \ket{\Omega}  - k_{1 \lambda} \bra{\Omega} \hat{A}^{\mu}(k_1) \hat{A}^{\lambda}(k_2) \ket{\Omega}  \right\}  \\
  = 2 \int \ud^4k \ k^{\mu} \hat{\Delta}_+(k).
\end{multline*}
This diverges badly for $\mu=0$ and is independent of $\sigma$, so it does not vanish in the commutative limit. The third and fourth term in (\ref{eq:A1}) contribute similar expressions (the prefactors differ), but these do not cancel each other.


Finally, we mention the following conceptual problem of the product~(\ref{eq:OldProduct2}): Consider classical electrodynamics on the noncommutative Minkowski space. There the commutators appearing in the definition of the field strength and in the field equations refer only to the spacetime noncommutativity and are in this sense ``kinematical''. If one uses the product~(\ref{eq:OldProduct2}), one obtains for the commutator
\begin{equation*}
  [\phi \otimes f, \psi \otimes g] =  \tfrac{1}{2} \left( \{ \phi, \psi \} \otimes [f, g] + [\phi, \psi] \otimes \{ f, g \} \right). 
\end{equation*}
The second term on the right hand side involves a quantum commutator, which is determined by the classical dynamics. It is strange to have this ``dynamical'' object involved in an originally ``kinematical'' one. We remark that precisely this second term is responsible for the problems mentioned above. Thus, it seems natural to replace the product~(\ref{eq:OldProduct2}) by the symmetrized version
\begin{equation}
\label{eq:SymmProd}
  ( \phi \tensor f ) \cdot ( \psi \tensor g ) = \tfrac{1}{2} (\phi \psi + \psi \phi) \tensor f g.
\end{equation}
Note that it is not associative\footnote{Although the product looks similar to that of a Jordan algebra, it does not give rise to a Jordan algebra. It is not even power-associative, i.e., the subalgebra generated by a single element is in general not associative, consider for example $a_1 \otimes b_1 + a_2 \otimes b_2$.
.}.
At least on a practical level, this is not problematic, since the only trilinear term in the equations of motion is a double commutator, which has a natural order of multiplication, i.e.,
\begin{equation}
\label{eq:A3}
  [A_\lambda, [A^\lambda, A^\nu]] = (2\pi)^{-6} \int \prod_{i=1}^3 \ud^4k_i \ \tfrac{1}{2} \{ \hat A_\lambda(k_1) , \hat A^\lambda(k_2) \hat A^\nu(k_3) \} [e^{-ik_1q}, [e^{-ik_2q}, e^{-ik_3q}]].
\end{equation}
Note that, because of the presence of the commutator $[e^{-ik_2q}, e^{-ik_3q}]$, it is not necessary to symmetrize $\hat A^\lambda(k_2)$ and $\hat A^\nu(k_3)$.
Furthermore, the product (\ref{eq:SymmProd}) does not fulfill the Jacobi identity. At first sight this could be problematic, since it might lead to ambiguities in the quantization of the trilinear term in the equation of motion. This, however, is not the case, since
$[A_{\lambda}, [A^{\lambda}, A^{\mu}]]$ becomes, by using the Jacobi identity, $- [A^{\lambda}, [A^{\mu}, A_{\lambda}]]$, which has the same quantization as the original expression.

\begin{remark}
From the point of view of the conventional product~(\ref{eq:OldProduct2}), the symmetrized product~(\ref{eq:SymmProd}) arises by the subtraction of a nonlocal counterterm. To see this, consider as an example the difference of the expressions obtained by using the different products for $(A_\mu c) \cdot A_\nu$. One finds
\begin{equation*}
  \frac{i}{2} (2\pi)^{-4} g_{\mu \nu} \int \ud^4k \ud^4l \ \hat c(k) \hat \Delta(l) e^{-ik(q-\sigma l)}.
\end{equation*}
Here the field $c$ is smeared with the Fourier transform of the commutator function. Such a term is also not $q$--local in the sense of~\cite{DorosDiss}.
\end{remark}

\begin{remark}
In a sense the symmetrized product is implicitly assumed in the modified Feynman rules approach. There, the cubic photon vertex
\begin{center}
\begin{picture}(50,60)
\Photon(25,15)(25,30){1}{3}
\Photon(25,30)(15,40){1}{3}
\Photon(25,30)(35,40){1}{3}
\Text(25,13)[t]{$\mu_1 k_1$}
\Text(15,42)[rb]{$\mu_2 k_2$}
\Text(35,42)[lb]{$\mu_3 k_3$}
\Vertex(25,30){1}
\end{picture}
\end{center}
is obtained from the functional derivative
\begin{equation*}
  i \frac{\delta}{\delta \hat A_{\mu_1}(k_1)} \frac{\delta}{\delta \hat A_{\mu_2}(k_2)} \frac{\delta}{\delta \hat A_{\mu_3}(k_3)} \int \ud^4q \ \del_\mu A_\nu [A^\mu, A^\nu].
\end{equation*}
Here the $\hat A$s, and thus also the functional derivatives, are assumed to commute. Obviously, this is very similar to our symmetrization prescription. We also note that from the above functional derivative, one obtains
\begin{equation*}
  i e \delta(k_1+k_2+k_3) \sin \tfrac{k_2 \sigma k_3}{2} \left( g^{\mu_1 \mu_3} (k_3-k_1)^{\mu_2} + g^{\mu_2 \mu_3} (k_2-k_3)^{\mu_1} + g^{\mu_1 \mu_2} (k_1-k_2)^{\mu_3} \right).
\end{equation*}
As we will show in the next section one obtains the same vertex factor in the Yang--Feldman formalism if one uses the symmetrized product.
However, using the old product (\ref{eq:OldProduct2}), we would find a similar expression, but with $i \sin \frac{k_2 \sigma k_3}{2}$ replaced by $- e^{-i \frac{k_2 \sigma k_3}{2}}$.
\end{remark}

\section{Graphical rules}

Now we want to state the graphical rules that follow from the definition of the interacting fields in the previous section. We begin with the computation of the cubic photon vertex
\begin{center}
\begin{picture}(50,60)
\Photon(24.5,15)(24.5,30){1}{3}
\Photon(25.5,15)(25.5,30){1}{3}
\Photon(25,30)(15,40){1}{3}
\Photon(25,30)(35,40){1}{3}
\Text(25,13)[t]{$\mu_1 k_1$}
\Text(15,42)[rb]{$\mu_2 k_2$}
\Text(35,42)[lb]{$\mu_3 k_3$}
\Vertex(25,30){1}
\end{picture}
\end{center}
where the $k_i$ are incoming momenta. It is obtained from the part of $\hat A^{\mu_1}_1(k_1)$ that is quadratic in $A_0$, i.e., the first three terms in (\ref{eq:A1}):
\begin{multline*}
\hat{A}^{\mu_1}_1(k_1) = - (2\pi)^{-4} \frac{1}{2} \hat{\Delta}_R(k_1) \int \ud^4p_1 \ud^4p_2 \ \int \ud^4q \ [e^{ip_1q}, e^{ip_2q}] e^{ik_1q} \\ 
 \times \left\{ (p_1+2p_2)_{\rho} \left( \hat{A}_0^{\rho}(-p_1) \hat{A}_0^{\mu_1}(-p_2) + \hat{A}_0^{\mu_1}(-p_2) \hat{A}_0^{\rho}(-p_1) \right)  \right. \\
  \left. - p_2^{\mu_1} \left( \hat{A}_0^{\rho}(-p_1) \hat{A}_{0 \rho}(-p_2) + \hat{A}_{0 \rho}(-p_2) \hat{A}_0^{\rho}(-p_1) \right) \right\}. 
\end{multline*}
Now one simply has to equate the momentum and Lorentz index of the left/right field operators with those of the lines leaving the above vertex to the left/right. For the first term, this means replacing $p_1$ by $k_2$, $\rho$ by $\mu_2$, $p_2$ by $k_3$ and multiplying by $g^{\mu_1 \mu_3}$. This way, we find the vertex factor
\begin{equation}
\label{eq:CubicVertex}
  i e \left( g^{\mu_1 \mu_2} (k_1-k_2)^{\mu_3} + g^{\mu_3 \mu_1} (k_3-k_1)^{\mu_2} + g^{\mu_2 \mu_3} (k_2-k_3)^{\mu_1} \right) \sin \tfrac{k_2 \sigma k_3}{2} \delta(k_1+k_2+k_3).
\end{equation}
We remark that this is invariant under permutations of the legs.

In order to find the factor for the quartic vertex
\begin{center}
\begin{picture}(100,60)
\Photon(49.5,15)(49.5,30){1}{3}
\Photon(50.5,15)(50.5,30){1}{3}
\Text(50,13)[t]{$\mu_1 k_1$}
\Vertex(50,30){1}

\Photon(50,30)(35,30){1}{3}
\Text(33,30)[r]{$\mu_2 k_2$}

\Photon(50,30)(65,30){1}{3}
\Text(67,30)[l]{$\mu_4 k_4$}

\Photon(50,30)(50,45){1}{3}
\Text(50,47)[b]{$\mu_3 k_3$}
\end{picture}
\end{center}
one has to consider the term (\ref{subeq:A2_4}), cf. (\ref{eq:A3}):
\begin{multline*}
  \hat A_2^{\mu_1}(k_1) = (2 \pi)^{-6} \hat \Delta_R(k_1) \int \prod_{i=1}^3 \ud^4p_i \ \int \ud^4q \  [e^{i p_1 q}, [e^{i p_2 q}, e^{i p_3}]] e^{ik_1q} \\ \times \frac{1}{2} \left\{ \hat A_{0 \lambda}(-p_1), \hat A_0^\lambda(-p_2) \hat A_0^{\mu_1}(-p_3) \right\}.
\end{multline*}
We find the vertex factor
\begin{multline*}
  (2\pi)^{-2} e^2 \delta(\sum k_i) \left\{ \sin \tfrac{k_1 \sigma k_2}{2} \sin \tfrac{k_3 \sigma k_4}{2} \left( g^{\mu_1 \mu_3} g^{\mu_2 \mu_4} - g^{\mu_1 \mu_4} g^{\mu_2 \mu_3} \right) \right. \\
  \left. + \sin \tfrac{k_2 \sigma k_3}{2} \sin \tfrac{k_4 \sigma k_1}{2}
  \left( g^{\mu_1 \mu_3} g^{\mu_2 \mu_4} - g^{\mu_1 \mu_2} g^{\mu_3 \mu_4} \right) \right\}.
\end{multline*}
Note that this is not symmetric under permutations of the legs and does not coincide with the vertex factor found in the setting of the modified Feynman rules. The reason is that we did not use a total symmetrization of the quantum fields in $A_2$.


It remains to treat the ghost vertices. The relevant part of $A_1$, i.e., the fourth term in (\ref{eq:A1}), is
\begin{multline*}
\hat{A}^{\mu}_1(k_1) = - (2\pi)^{-4} \frac{1}{2} \hat{\Delta}_R(k_1) \int \ud^4p_1 \ud^4p_2 \ \int \ud^4q \ [e^{ip_1q}, e^{ip_2q}] e^{ik_1q}  \\ 
  \times p_1^{\mu} \left( \hat{\bar{c}}_0(- p_1) \hat{c}_0(- p_2) - \hat{c}_0(- p_2) \hat{\bar{c}}_0(-p_1) \right). 
\end{multline*}
Note that we antisymmetrized the ghost field operators. It is easy to read off the vertex factors
\vspace{-20pt}
\begin{center}
\begin{picture}(50,60)(0,20)
\Photon(24.5,15)(24.5,30){1}{3}
\Photon(25.5,15)(25.5,30){1}{3}
\DashArrowLine(14,40)(25,30){2}
\DashArrowLine(25,30)(35,40){2}
\Text(25,13)[t]{$\mu_1 k_1$}
\Text(15,42)[rb]{$k_2$}
\Text(35,42)[lb]{$k_3$}
\Vertex(25,30){1}
\end{picture}
$ = i e k_2^{\mu_1} \sin \frac{k_2 \sigma k_3}{2} \delta(k_1+k_2+k_3) = - $
\begin{picture}(50,60)(0,20)
\Photon(24.5,15)(24.5,30){1}{3}
\Photon(25.5,15)(25.5,30){1}{3}
\DashArrowLine(25,30)(14,40){2}
\DashArrowLine(35,40)(25,30){2}
\Text(25,13)[t]{$\mu_1 k_1$}
\Text(15,42)[rb]{$k_3$}
\Text(35,42)[lb]{$k_2$}
\Vertex(25,30){1}
\end{picture}
\end{center}
\vspace{20pt}
For the vertices with incoming ghost or antighost, we find, using (\ref{eq:c1}) and (\ref{eq:c1_bar}), in exactly the same way,
\vspace{-20pt}
\begin{center}
\begin{picture}(50,60)(0,20)
\DashArrowLine(24.5,15)(24.5,30){2}
\DashArrowLine(25.5,15)(25.5,30){2}
\Photon(14,40)(25,30){1}{3}
\DashArrowLine(25,30)(35,40){2}
\Text(25,13)[t]{$k_2$}
\Text(15,42)[rb]{$\mu_1 k_1$}
\Text(35,42)[lb]{$k_3$}
\Vertex(25,30){1}
\end{picture}
$ = i e k_2^{\mu_1} \sin \frac{k_2 \sigma k_3}{2} \delta(k_1+k_2+k_3) = $
\begin{picture}(50,60)(0,20)
\DashArrowLine(24.5,30)(24.5,15){2}
\DashArrowLine(25.5,30)(25.5,15){2}
\Photon(14,40)(25,30){1}{3}
\DashArrowLine(35,40)(25,30){2}
\Text(25,13)[t]{$k_3$}
\Text(15,42)[rb]{$\mu_1 k_1$}
\Text(35,42)[lb]{$k_2$}
\Vertex(25,30){1}
\end{picture}
\end{center}
\vspace{20pt}
The factors for the vertices where the photon leaves to the right are the same, because photon and ghost fields commute.
The ghost propagators are given by
\vspace{-5pt}
\begin{center}
\begin{picture}(40,15)(0,5)
\DashArrowLine(35,7)(10,7){2}
\BCirc(10,7){1}
\BCirc(35,7){1}
\Text(8,7)[r]{$k$}
\end{picture}
$ = (2\pi)^2 \hat \Delta_+(k) = - $
\begin{picture}(40,15)(0,5)
\DashArrowLine(10,7)(35,7){2}
\BCirc(10,7){1}
\BCirc(35,7){1}
\Text(8,7)[r]{$k$}
\end{picture}
\end{center}
\vspace{5pt}
Finally, one has to take the fermionic nature of the ghosts into account. Each graph is multiplied by $(-1)^I$, where $I$ is the number of intersections of the ghost lines. For example, the contraction
\begin{picture}(40,10)(0,2)

\DashCurve{(5,2)(15,9)(25,2)}{2}
\DashCurve{(15,2)(25,9)(35,2)}{2}

\GCirc(5,2){1}{1}
\GCirc(15,2){1}{1}
\GCirc(25,2){1}{1}
\GCirc(35,2){1}{1}

\end{picture}
yields a factor $-1$. This completes the statement of the graphical rules.

\section{Covariant Coordinates}
\label{sec:CovCoor}

We recall from Chapter~\ref{chapter:NCGaugeTheory} that the covariant coordinates are given by
\begin{equation*}
  X^{\mu} = q^{\mu} + e \sigma^{\mu \nu} A_{\nu}.
\end{equation*}
In the present setting, they transform covariantly under BRST transformations:
\begin{equation*}
  \delta_\xi X^\mu = - i e \xi [X^\mu, c].
\end{equation*}
As in Chapter~\ref{chapter:NCGaugeTheory}, we use observables of the form
\begin{equation}
\label{eq:NCQEDObservable}
  \int \ud^4 q \ f^{\mu \nu}(X) F_{\mu \nu}.
\end{equation}
Because of the cyclicity of the integral, these are BRST invariant.
Again, we define functions of $X$ \`a la Weyl, cf. (\ref{eq:g_X}). It is thus desirable to find an expression for the exponentiated form $e^{ikX}$ as a formal power series in the coupling constant $e$. Note that we are not using the symmetrized product~(\ref{eq:SymmProd}) for the definition of $e^{ikX}$, since it is not clear in which order one should do that (we recall that the symmetrized product is not associative). However, we use it for the product of $f^{\mu \nu}(X)$ and~$F_{\mu \nu}$.

The $N$th order part of $e^{ikX}$ can be found by computing
\begin{equation*}
  \frac{1}{N!} \left( \frac{\ud}{\ud \lambda} \right)^N e^{C+\lambda D} |_{\lambda=0} = \sum_{n_0, \dots , n_{N}}  \frac{C^{n_0} D C^{n_1}  \dots D C^{n_N}}{(n_0 + \dots + n_N+N)!}.
\end{equation*}
Here $C$ and $D$ stand for arbitrary elements of some algebra. We are of course interested in the case where $C$ is replaced by $ikq$ and $D$ by $ik_{\mu} \sigma^{\mu \nu} A_{\nu}$. For the $i$th $A$, we write $A_{\nu_i}(q) = (2\pi)^{-2} \int \ud k_i \ \hat{A}_{\nu_i}(k_i) e^{-ik_i q}$. The $\hat{A}_{\nu_i}(k_i)$'s can be pulled out of the expression. If we then commute all the $e^{-ik_iq}$'s to the right, we will have to replace $(ikq)^{n_i}$ by \mbox{$(i k(q - \sigma \sum_{j \leq i} k_j))^{n_i}$}. In order to deal with the resulting expression, we need the following
\begin{lemma}
\label{lemma:NCQED}
Let $x, y_i$ be pairwise commuting elements of an algebra.
Then
\begin{equation*}
  \sum_{n_0, \dots , n_N = 0}^\infty \frac{x^{n_0} (x+y_1)^{n_1} \dots (x+y_N)^{n_N}}{(n_0 + \dots + n_N+N)!} = e^x \sum_{n_1, \dots , n_N=0}^\infty \frac{y_1^{n_1} \dots y_N^{n_N}}{\left( \sum_{i=1}^N (n_i+1) \right)!}.
\end{equation*}
\end{lemma} 
The proof can be found in Appendix~\ref{app:NCQED}. We define
\begin{equation*}
  P_N(x_1, \dots, x_N) = \sum_{n_1, \dots , n_N=0}^\infty \frac{\prod_{m=1}^N \left( \sum_{n=1}^m x_n \right)^{n_m}}{\left( \sum_{i=1}^N (n_i+1) \right)!}.
\end{equation*}
Note that, with this definition, $P_1$ coincides with the $P_1$ defined in~(\ref{eq:P_1}). Thus, one obtains
\begin{multline}
\label{eq:e_ikX}
  e^{ikX} = e^{ikq} \sum_{N=0}^\infty (ie)^N (2\pi)^{-2N} \int \prod_{i=1}^N \ud^4k_i \ e^{-i k_1q} \dots e^{-ik_Nq} k \sigma \hat A(k_1) \dots k \sigma \hat A(k_N) \\
  \times P_N(-ik\sigma k_1, \dots, -ik\sigma k_N).
\end{multline}
In Section~\ref{sec:FreePart} we will show that the products of fields in this expression are not well--defined and require normal ordering.

\begin{remark}
\label{rem:Equivalence}
Equation~(\ref{eq:e_ikX}) is equivalent to the formula, cf.~\cite{Dhar, BakLeePark},
\begin{equation}
\label{eq:Equivalence}
  e_\star^{ikX} = e^{ikx} \star \bar{\mathrm{P}}_\star e^{i e \int_0^1 \ud t \ k \sigma A(x+t k \sigma)}.
\end{equation}
Here $\bar{\mathrm{P}}_\star$ is the anti-path ordered $\star$-product. This is proven in Appendix~\ref{app:Equivalence}.
\end{remark}

%

It is straightforward to argue that for any operator $E$ that transforms covariantly under the action of the translation group,
e.g., $E = \mathbbm{1}, A_{\mu}$ or $F_{\mu \nu} F_{\lambda \rho}$, we have
\begin{equation}
\label{eq:EIdentity}
  \bra{\Omega} \left( \int \ud^4q \ f(X) E \right) \ket{\Omega} = \bra{\Omega} \left( \int \ud^4q \ f(q) E \right) \ket{\Omega}.
\end{equation}
At $N$th order in $e$, the left hand side is
\begin{multline*}
  (ie)^N (2\pi)^{-2(N+2)} \int \ud^4k \ud^4l \prod_{i=1}^N \ud^4k_i  \ \hat{f}(-k) \tfrac{1}{2} \bra{\Omega} \Big\{ \prod_{j=1}^N k \sigma \hat{A}(k_j), \hat{E}(l) \Big\} \ket{\Omega} \\
 \times P_N(-ik\sigma k_1, \dots, -ik\sigma k_N) \int \ud^4q \ e^{ikq} e^{-ik_1q} \dots e^{-ik_Nq} e^{-ilq}.
\end{multline*}
Because of the translation covariance of $E$, the vacuum expectation value is a multiple of $\delta(l+\sum_{j=1}^N k_j)$. Then the integral over $q$ yields a multiple of $\delta(k)$ and, for $N>0$, the whole expression vanishes due to the presence of the $N$ powers of~$k$.

\section{The two--point function}
\label{sec:2pt}

The goal of the following sections is to compute the two--point correlation function
\begin{equation}
\label{eq:NCQED2pt}
  \bra{\Omega} \left( \int \ud^4q \ f^{\mu \nu}(X) F_{\mu \nu} \right) \left( \int \ud^4q \ h^{\lambda \rho}(X) F_{\lambda \rho} \right) \ket{\Omega}
\end{equation}
of the interacting field to second order in $e$. Here we may assume $f^{\mu \nu}$ and $h^{\lambda \rho}$ to be anti-symmetric. Because of the presence of the commutator term of the field strength and the covariant coordinates, a single observable (\ref{eq:NCQEDObservable})
contains, at order $e^n$, $n+1$ photon fields. Thus, the two--point function (\ref{eq:NCQED2pt}) contains, at order $e^2$, also three- and four-point functions of the photon field. We split the computation of (\ref{eq:NCQED2pt}) into three parts:
\begin{enumerate}

\item We expand the single observable (\ref{eq:NCQEDObservable}) in powers of $e$, which is equivalent to an expansion in the number of photon fields. The result is written in the form
\begin{equation*}
  \int \ud^4q \ f^{\mu \nu}(X) F_{\mu \nu} = \int \ud^4k \ \hat f^{\mu \nu}(-k) K_{\mu \nu}(k)
\end{equation*}
where
\begin{equation*}
  K_{\mu \nu}(k) = \sum_{n=1}^\infty e^{n-1} \int \prod_{i=1}^n \ud^4k_i \ K_{\mu \nu}^{\mu_1 \dots \mu_n}(k ; k_1, \dots k_n)  \hat A_{\mu_1}(k_1) \dots \hat A_{\mu_n}(k_n).
\end{equation*}
For our computation, we need the kernels $K_{\mu \nu}^{\mu_1 \dots \mu_n}(k ; k_1, \dots k_n)$ up to $n=3$.

\item We compute the $n$-point functions
\begin{equation*}
  W_{\mu_1 \dots \mu_n}(k_1, \dots k_n) = \bra{\Omega} \hat A_{\mu_1}(k_1) \dots \hat A_{\mu_n}(k_n) \ket{\Omega}
\end{equation*}
of the photon field. In the following, we call these the \emph{elementary} $n$-point functions. For our computation, we need the elementary two--, three--, and four--point functions to second, first, and zeroth order, respectively.

\item The two--point function (\ref{eq:NCQED2pt}) is now given by
\begin{equation}
\label{eq:2ptCombinatorics}
\sum_{\substack{m=1 \\ n=1}}^\infty e^{m+n-2} \int \ud^4k \ud^4p \ \hat f^{\mu \nu}(-k) \hat h^{\lambda \rho}(-p) \int \ud^{4m} \underline{k} \ud^{4n} \underline{p} \ K_{\mu \nu}^{\underline{\mu}}(k;\underline{k}) K_{\lambda \rho}^{\underline{\nu}}(p;\underline{p}) W_{\underline{\mu} \underline{\nu}}(\underline{k}, \underline{p}).
\end{equation}
Here we used the abbreviations $\underline k = (k_1, \dots k_m)$, $\underline p = (p_1, \dots p_n)$ and analogously for $\underline \mu$ and $\underline \nu$. This will be called the \emph{full} two--point function in the following.
\end{enumerate}

This is the program for the next three sections. We remark that (\ref{eq:2ptCombinatorics}) can be straightforwardly generalized for the computation of higher $n$--point functions.

\section{The computation of $K_{\mu \nu}(k)$}

The zeroth order component of $K_{\mu \nu}(k)$ can be directly read off:
\begin{equation}
\label{eq:K0}
  K_{\mu \nu}^{\mu_1}(k;k_1) = - 2 i \delta(k-k_1) k_\mu \delta_\nu^{\mu_1}.
\end{equation}
Here we used the antisymmetry of $f^{\mu \nu}$. At first order, there are two contributions, one from the commutator term in the field strength and one from the covariant coordinate. We have
\begin{multline*}
 - i e \int \ud^4q \ f^{\mu \nu}(q) [A_\mu, A_\nu] = - e (2 \pi)^{-2} \int \ud^4k \ \hat f^{\mu \nu}(-k) \\ \times \int \ud^4k_1 \ud^4k_2 \ \{ \hat A_\mu(k_1), \hat A_\nu(k_2) \} \sin \tfrac{k_1 \sigma k_2}{2} \delta(k-k_1-k_2).
\end{multline*}
Using once more the antisymmetry of $f^{\mu \nu}$, this is expressed by the kernel
\begin{equation}
\label{eq:K1_1}
  K_{\mu \nu}^{\mu_1 \mu_2}(k; k_1, k_2) = - 2 (2\pi)^{-2} \delta(k-k_1-k_2) \delta^{\mu_1}_\mu \delta^{\mu_2}_\nu \sin \tfrac{k_1 \sigma k_2}{2}.
\end{equation}
In order to find the first order contribution from the covariant coordinate, we have to compute, cf.~(\ref{eq:e_ikX}),
\begin{multline*}
 2 \int \ud^4 q \ f^{\mu \nu}(X)|_e \del_\mu A_\nu = e (2 \pi)^{-6} \int \ud^4k \ud^4k_1 \ud^4k_2 \ \hat f^{\mu \nu}(-k) (i k \sigma)^\alpha (-i k_2)_\mu \\ \times \{ \hat A_\alpha(k_1), \hat A_\nu(k_2) \} P_1(-ik \sigma k_1) \int \ud^4q \ e^{ikq} e^{-ik_1q} e^{-ik_2q}.
\end{multline*}
The integral over $q$ yields $(2\pi)^4 e^{\frac{i}{2} k \sigma k_1} \delta(k-k_1-k_2)$. 
Using (\ref{eq:P_Equation}), we thus find the kernel
\begin{equation}
\label{eq:K1_2}
  K_{\mu \nu}^{\mu_1 \mu_2}(k; k_1, k_2) = (2\pi)^{-2} \delta(k-k_1-k_2) \left\{ (k \sigma)^{\mu_1} k_{2 \mu} \delta^{\mu_2}_\nu + (k \sigma)^{\mu_2} k_{1 \mu} \delta^{\mu_1}_\nu \right\} \frac{\sin \frac{k_1 \sigma k_2}{2}}{\frac{k_1 \sigma k_2}{2}}.
\end{equation}

At second order, there are again two terms. We start with the computation of the one involving the commutator term of the field strength and one supplementary field from the covariant coordinate. In the same manner as above, we find
\begin{multline*}
  - i e \int \ud^4q \ f^{\mu \nu}(X)|_e [A_\mu, A_\nu]  = - i e^2 (2 \pi)^{-4} \int \ud^4k \ \hat f^{\mu \nu}(-k) \int \prod_{i=1}^3 \ud^4k_i \ (k \sigma)^\alpha \\ \times \frac{1}{2} \{ \hat A_\alpha(k_1), \{ \hat A_\mu(k_2), \hat A_\nu(k_3) \} \} \frac{ \sin \frac{k \sigma k_1}{2} }{ \frac{k \sigma k_1}{2} } \sin \tfrac{k_2 \sigma k_3}{2} \delta(k - \sum k_i).
\end{multline*}
This is expressed by the kernel
\begin{multline}
\label{eq:K2_1}
  K_{\mu \nu}^{\underline \mu}(k; \underline k) = - i (2\pi)^{-4} \delta(k- \sum k_i) \\ \times \left\{ (k \sigma)^{\mu_1} \delta^{\mu_2}_\mu \delta^{\mu_3}_\nu \frac{\sin \frac{k \sigma k_1}{2}}{\frac{k \sigma k_1}{2}} \sin \tfrac{k_2 \sigma k_3}{2} + (k \sigma)^{\mu_3} \delta^{\mu_1}_\mu \delta^{\mu_2}_\nu \frac{\sin \frac{k \sigma k_3}{2}}{\frac{k \sigma k_3}{2}} \sin \tfrac{k_1 \sigma k_2}{2} \right\}.
\end{multline}
Here we used again the notation $\underline k = (k_1, k_2, k_3)$.
It remains to compute the term where the covariant coordinate contributes two powers of $e$. We have
\begin{multline*}
 2 \int \ud^4q \ f^{\mu \nu}(X)|_{e^2} \del_\mu A_\nu = e^2 (2 \pi)^{-8} \int \ud^4k \prod_{i=1}^3 \ud^4k_i \ \hat f^{\mu \nu}(-k) (i e k \sigma)^\alpha (i e k \sigma)^\beta (-i k_3)_\mu \\ \times \{ \hat A_\alpha(k_1) \hat A_\beta(k_2), \hat A_\nu(k_3) \} P_2(- i k \sigma k_1, - i k \sigma k_2) \int \ud^4q \ e^{ikq} e^{-ik_1q} e^{-ik_2q} e^{-ik_3q}.
\end{multline*}
Thus, we find the kernel
\begin{align}
\label{eq:K2_2}
  K_{\mu \nu}^{\underline \mu}(k; \underline k) = & i (2\pi)^{-4} \delta(k- \sum k_i) \\ 
  & \times \left\{ (k \sigma)^{\mu_1} (k \sigma)^{\mu_2} k_{3 \mu} \delta^{\mu_3}_\nu P_2(- i k \sigma k_1, - i k \sigma k_2) e^{-\frac{i}{2} k_1 \sigma k_2} e^{- \frac{i}{2} (k_1+k_2) \sigma k_3} \right. \nonumber \\
 & \quad \left. + (k \sigma)^{\mu_2} (k \sigma)^{\mu_3} k_{1 \mu} \delta^{\mu_1}_\nu P_2(- i k \sigma k_2, - i k \sigma k_3) e^{-\frac{i}{2} k_2 \sigma k_3} e^{- \frac{i}{2} (k_2+k_3) \sigma k_1} \right\}. \nonumber
\end{align}

\section{The elementary $n$--point functions}

The next step of the program outlined in Section~\ref{sec:2pt} is the computation of the relevant elementary $n$--point functions. We start with the computation of the elementary two--point function.
At zeroth order in $e$, we have the usual contribution to the photon two--point function:
\begin{equation*}
  W_{\mu \nu}(k,p) = - (2 \pi)^2 g_{\mu \nu} \hat \Delta_+(k) \delta(k+p).
\end{equation*}
There is no first order contribution. At second order, there are three terms, as in the case of the $\phi^3$ model, cf.~(\ref{eq:ThreeTerms}):
\begin{equation}
\label{eq:NCQEDThreeTerms}
  \bra{\Omega} A_{2 \mu}(k) A_{0 \nu}(p) \ket{\Omega} + \bra{\Omega} A_{0 \mu}(k) A_{2 \nu}(p) \ket{\Omega} + \bra{\Omega} A_{1 \mu}(k) A_{1 \nu}(p) \ket{\Omega}.
\end{equation}
As in the previous chapter, we treat the sum of the first two terms by computing the self--energy $\Pi_{\mu \nu}(k)$ and then setting
\begin{equation}
\label{eq:W_SelfEnergy}
  W_{\mu \nu}(k,p) = (2\pi)^2 \delta(k+p) \Pi_{\mu \nu}(k) \frac{\del}{\del m^2} \Delta_+(k).
\end{equation}
We recall from Remark~(\ref{rem:massless}) that this is only well--defined on test functions vanishing in a neighborhood of the origin.

However, we begin with the discussion of the third term in (\ref{eq:NCQEDThreeTerms}). As in the case of the $\phi^3$ model, it gives rise to the two--particle spectrum.

\begin{remark}
\label{rem:k_mu_k_nu}
It turns out that at second order the elementary two--point function is not well--defined because of infrared problems. Most of these can be avoided if one interprets it as a distribution on transversal test functions $\hat f^\mu$, $\hat h^\nu$, i.e., \mbox{$k_\mu \hat f^\mu(k) = k_\nu \hat h^\nu = 0$}. The test functions that are used in the full two--point function at second order are of this form, because of the antisymmetry of $f^{\mu \nu}$ and $h^{\lambda \rho}$, cf. the factor $k_\mu$ in (\ref{eq:K0}).
\end{remark}

\subsection{The two--particle spectrum}

We have to compute these two graphs:
\begin{center}
\begin{picture}(80,53)
\Photon(19.5,5)(19.5,20){1}{3}
\Photon(20.5,5)(20.5,20){1}{3}
\Photon(20,20)(10,35){1}{3}
\Photon(20,20)(30,35){1}{3}
\Photon(59.5,5)(59.5,20){1}{3}
\Photon(60.5,5)(60.5,20){1}{3}
\Photon(60,20)(50,35){1}{3}
\Photon(60,20)(70,35){1}{3}
\Photon(30,35)(50,35){1}{4}
\PhotonArc(40,5)(42.3,45,135){1}{11.5}
\Vertex(20,20){1}
\Vertex(60,20){1}
\BCirc(10,35){1}
\BCirc(30,35){1}
\BCirc(50,35){1}
\BCirc(70,35){1}
\end{picture}
\begin{picture}(80,53)
\Photon(19.5,5)(19.5,20){1}{3}
\Photon(20.5,5)(20.5,20){1}{3}
\DashArrowLine(10,35)(20,20){2}
\DashArrowLine(20,20)(30,35){2}
\Photon(59.5,5)(59.5,20){1}{3}
\Photon(60.5,5)(60.5,20){1}{3}
\DashArrowLine(50,35)(60,20){2}
\DashArrowLine(60,20)(70,35){2}
\DashArrowLine(30,35)(50,35){2}
\DashArrowArc(40,5)(42.3,45,135){2}
\Vertex(20,20){1}
\Vertex(60,20){1}
\BCirc(10,35){1}
\BCirc(30,35){1}
\BCirc(50,35){1}
\BCirc(70,35){1}
\end{picture}
\end{center}
According to the number of possible contractions and orientations, one has to double the first graph and to quadruple the second (it is not necessary to compute the other contractions, since the vertex factors are symmetric under permutations of the legs). Taking this factor $2$ into account, we obtain for the first graph
\begin{multline*}
  - 2 (ie)^2 (2\pi)^4 \delta(k+p) \hat \Delta_R(k) \hat \Delta_A(k) \int \ud^4 l \ \hat \Delta_+(l) \hat \Delta_+(k-l) \sin^2 \tfrac{k \sigma l}{2} \\
  \times g^{\lambda_2 \rho_3} g^{\lambda_3 \rho_2} \Big\{ g_{\mu \lambda_2} (2 k - l)_{\lambda_3} + g_{\mu \lambda_3} (- k - l)_{\lambda_2} + g_{\lambda_2 \lambda_3} (2 l- k)_{\mu} \Big\} \\
  \times \Big\{ g_{\nu \rho_2} (- k - l)_{\rho_3} + g_{\nu \rho_3} (2 k - l)_{\rho_2} + g_{\rho_2 \rho_3} (2 l- k)_{\nu} \Big\}. 
\end{multline*}
The contraction of the last two lines gives
\begin{equation*}
  g_{\mu \nu} (5 k^2 - 2 k \cdot l + 2 l^2) \\ + (d-6) k_{\mu} k_{\nu} + (3-2d) k_{\mu} l_{\nu} + (3-2d) l_{\mu} k_{\nu} + (4d-6) l_{\mu} l_{\nu}.
\end{equation*}
Here $d$ stands for the spacetime dimension. Although it will always be $4$ in the following, write $d$ in order to facilitate comparison with other results. For the second graph, we find, including the factor $4$,
\begin{equation*}
  - 4 (ie)^2 (2\pi)^4 \delta(k+p) \hat \Delta_R(k) \hat \Delta_A(k) \int \ud^4 l \ \hat \Delta_+(l) \hat \Delta_+(k-l) \sin^2 \tfrac{k \sigma l}{2} (k-l)_{\mu} l_{\nu}. 
\end{equation*}
Using $l^2 \hat \Delta_+(l) = 0$ and the change of variables $l \to k -l$, we can combine the two contributions to
\begin{multline}
\label{eq:2particleIntegral}
  2 e^2 (2\pi)^4 \delta(k+p) \hat \Delta_R(k) \hat \Delta_A(k) \int \ud^4 l \ \hat \Delta_+(l) \hat \Delta_+(k-l) \sin^2 \tfrac{k \sigma l}{2} \\
  \times \Big\{ 4 \left( g_{\mu \nu} k^2 - k_{\mu} k_{\nu} \right) + (d-2) (k-2l)_{\mu} (k-2l)_{\nu} \Big\}. 
\end{multline}
Some remarks are in order here. The loop integral is only over a compact set in momentum space and thus well--defined. Using $l^2 \hat \Delta_+(l) = 0$ once more, one can check that it is transversal, i.e., it vanishes after contraction with $k^{\mu}$. Furthermore, using
\begin{equation}
\label{eq:SinSplit}
  \sin^2 \tfrac{k \sigma l}{2} = \frac{1- \cos k \sigma l}{2},
\end{equation}
one can split it into a planar and a nonplanar part. We remark that the $-$ sign in (\ref{eq:SinSplit}) is due to the fact that the interaction term is given by a commutator. It will lead to a cancellation of some infrared divergences. One of these is already present here: The product $\hat \Delta_R(k) \hat \Delta_A(k)$ is not well defined for $k^2 = 0$. We recall that in the massive case this problem does not show up, since there the loop integral has its support above the $2 m$ mass shell, cf. Section~\ref{sec:2_particle_spectrum}. In order to discuss the situation in more detail, we compute the loop integral explicitly. We can proceed as in Section~\ref{sec:2_particle_spectrum}, i.e., we choose $k = (k_0, \V 0)$. One obtains
\begin{equation*}
  \int \ud^4 l \ \hat \Delta_+(l) \hat \Delta_+(k-l) \sin^2 \tfrac{k \sigma l}{2} = \theta(k_0) \theta(k^2) \frac{1}{16\pi} \left( 1 - \frac{\sin \sqrt{-(k \sigma)^2} \sqrt{k^2/4}}{\sqrt{-(k \sigma)^2} \sqrt{k^2/4}} \right).
\end{equation*}
This suffices to treat the first term in (\ref{eq:2particleIntegral}). We define
\begin{equation*}
  \tilde \Sigma_0(k) = \frac{1}{4\pi} \left( 1 - \frac{\sin \sqrt{-(k \sigma)^2} \sqrt{k^2/4}}{\sqrt{-(k \sigma)^2} \sqrt{k^2/4}} \right).
\end{equation*}
In order to compute the second term in (\ref{eq:2particleIntegral}), we note that because of transversality, symmetry, and Lorentz covariance, it will be of the form
\begin{equation*}
 \tilde \Pi_{\mu \nu} (k) = \theta(k_0) \theta(k^2) \left\{ \left( g_{\mu \nu} k^2 -k_{\mu} k_{\nu} \right) \tilde \Sigma_1(k) + \tfrac{(k \sigma)_{\mu} (k \sigma)_{\nu}}{(k \sigma)^4} \tilde \Sigma_2(k) \right\}.
\end{equation*}
Here $\tilde \Sigma_{1/2}$ are only functions of $(k \sigma)^2$ and $k^2$. Then, for $k = (k_0, \V{0})$, $k_0 > 0$, and in an orthogonal coordinate system such that $k \sigma$ is parallel to the spatial unit vector $e_\imath$, we would find
\begin{align*}
  \tilde \Pi_{\mu \nu} (k) & = 0 \quad \text{ for } \mu \neq \nu \text{ or } \mu=\nu=0 \\
  \tilde \Pi_{i i} (k) & = - k^2 \tilde \Sigma_1(k) \quad \forall i \neq \imath \\
  \tilde \Pi_{\imath \imath} (k) & = - k^2 \tilde \Sigma_1(k) - \frac{1}{(k \sigma)^2} \tilde \Sigma_2(k).
\end{align*}
That the first equation is fulfilled can easily be checked. In order to determine $\tilde \Sigma_{1/2}$, it remains to compute $\tilde \Pi^{i i}(k)$ for $i = \imath$ and $i \neq \imath$. In the first case, one finds
\begin{multline*}
  (d-2) \int \ud^4l \ \hat \Delta_+(l) \hat \Delta_+(k-l) \sin^2 \tfrac{k \sigma l}{2} 4 l_\imath^2 \\
  = \frac{8}{2 \pi} \int_0^\infty \ud l \ \frac{l^3}{2} \delta(k^2-2 \sqrt{k^2} l) \int_{-1}^1 \ud x \ x^2 \sin^2 \tfrac{ \sqrt{-(k \sigma)^2} l x}{2}.
\end{multline*}
Because of
\begin{equation*}
  \int_{-1}^1 \ud x \ x^2 \sin^2 a x = \frac{1}{3} - \frac{ \sin a}{a} - 2 \frac{\cos a}{a^2} + 2 \frac{\sin a }{a^3},
\end{equation*}
we obtain
\begin{multline*}
  \tilde \Pi_{\imath \imath} (k) = \theta(k_0) \theta(k^2) \frac{k^2}{8 \pi} \left\{ \frac{1}{3} - \frac{ \sin \sqrt{-(k \sigma)^2} \sqrt{k^2/4}}{\sqrt{-(k \sigma)^2} \sqrt{k^2/4}} \right. \\
  \left. - 2 \frac{\cos \sqrt{-(k \sigma)^2} \sqrt{k^2/4}}{( \sqrt{-(k \sigma)^2} \sqrt{k^2/4} )^2} + 2 \frac{\sin \sqrt{-(k \sigma)^2} \sqrt{k^2/4} }{( \sqrt{-(k \sigma)^2} \sqrt{k^2/4} )^3} \right\}.
\end{multline*}
Similarly, one finds, for $i \neq \imath$,
\begin{equation*}
  \tilde \Pi_{i i} (k) = \theta(k_0) \theta(k^2) \frac{k^2}{8 \pi} \left\{  \frac{1}{3}  + \frac{\cos \sqrt{-(k \sigma)^2} \sqrt{k^2/4}}{( \sqrt{-(k \sigma)^2} \sqrt{k^2/4} )^2} - \frac{\sin \sqrt{-(k \sigma)^2} \sqrt{k^2/4} }{( \sqrt{-(k \sigma)^2} \sqrt{k^2/4} )^3} \right\}.
\end{equation*}
Thus, we obtain
\begin{multline*}
  \tilde \Sigma_1(k) =  - \frac{1}{8 \pi} \left\{ \frac{1}{3} + \frac{\cos \sqrt{-(k \sigma)^2} \sqrt{k^2/4}}{(\sqrt{-(k \sigma)^2} \sqrt{k^2/4})^2} - \frac{\sin \sqrt{-(k \sigma)^2} \sqrt{k^2/4}}{(\sqrt{-(k \sigma)^2} \sqrt{k^2/4})^3} \right\} \\
 \shoveleft{ \tilde \Sigma_2(k) =  - \frac{(k \sigma)^2 k^2}{8 \pi} \left\{ - 3 \frac{\cos \sqrt{-(k \sigma)^2} \sqrt{k^2/4}}{(\sqrt{-(k \sigma)^2} \sqrt{k^2/4})^2} + 3 \frac{\sin \sqrt{-(k \sigma)^2} \sqrt{k^2/4}}{(\sqrt{-(k \sigma)^2} \sqrt{k^2/4})^3}
   \right. } \\ 
    \left. - \frac{\sin \sqrt{-(k \sigma)^2} \sqrt{k^2/4}}{\sqrt{-(k \sigma)^2} \sqrt{k^2/4}} \right\}.
\end{multline*}
We conclude that the contribution to the two--point function is given by
\begin{multline}
\label{eq:W0}
  W_{\mu \nu}(k,p) = 2 e^2 (2\pi)^4 \delta(k+p) \hat \Delta_R(k) \hat \Delta_A(k)  \\ \times \theta(k_0) \theta(k^2) \left\{ \left( g_{\mu \nu} k^2 -k_{\mu} k_{\nu} \right) \left( \tilde \Sigma_0(k) + \tilde \Sigma_1(k) \right) + \tfrac{(k \sigma)_{\mu} (k \sigma)_{\nu}}{(k \sigma)^4} \tilde \Sigma_2(k) \right\}.
\end{multline}
We want to discuss the well--definedness of this expression. As already mentioned, the product $\hat \Delta_R(k) \hat \Delta_A(k)$ is not well defined on the light cone and singular as $k^{-4}$. Thus, we integrate $W_{\mu \nu}(k,p)$ with test functions $\tilde f^\mu(k)$, $\tilde h^\nu(p)$ that vanish in a neighborhood of the light cone. In order to check whether it is possible to extend the distribution to the light cone, we need to know the asymptotic behavior of the $\tilde \Sigma$'s. It is straightforward to check that for small $k^2$, $\tilde \Sigma_{0/1} \sim k^2 (k \sigma)^2$ and $\tilde \Sigma_2 \sim k^4 (k \sigma)^4$. It follows that for the term involving $\tilde \Sigma_2$ the singularity is cancelled and the distribution can be extended to the light cone. Using the theory of scaling degrees at submanifolds, cf. \cite{BrunettiFredenhagen} and Section~\ref{sec:infrared}, one can show that this extension is unique.
The same argument can be used for the term involving $g_{\mu \nu} k^2$. Thus, only the term proportional to $k_{\mu} k_{\nu}$ is problematic.
However, if we restrict to transversal test functions, as proposed in Remark~\ref{rem:k_mu_k_nu}, this term vanishes right from the start, i.e., before extending the distribution. The extension of a vanishing distribution is unambiguous and vanishes, too.
We may thus conclude that (\ref{eq:W0}) can be defined unambiguously on transversal test functions.
Once more, we remark that the cancellation of infrared divergences is due to the fact that the interaction is given by a commutator.


\subsection{The self--energy}

It remains to treat the first two terms in (\ref{eq:NCQEDThreeTerms}), i.e., to compute the self--energy. We start with the computation of the tadpole, i.e., the following graphs:
\begin{center}
\begin{picture}(50, 45)
\Photon(24.5,5)(24.5,20){1}{3}
\Photon(25.5,5)(25.5,20){1}{3}
\Photon(25,20)(10,40){1}{6}
\Photon(25,20)(25,40){1}{4}
\Photon(25,20)(40,40){1}{6}
\Photon(25,40)(40,40){1}{3}
\Vertex(25,20){1}
\BCirc(10,40){1}
\BCirc(25,40){1}
\BCirc(40,40){1}
\Text(27,5)[l]{$\mu$}
\end{picture}
\begin{picture}(50, 50)
\Photon(24.5,5)(24.5,20){1}{3}
\Photon(25.5,5)(25.5,20){1}{3}
\Photon(25,20)(10,40){1}{6}
\Photon(25,20)(25,40){1}{4}
\Photon(25,20)(40,40){1}{6}
\PhotonArc(25,25)(21,45,135){1}{8.5}
\Vertex(25,20){1}
\BCirc(10,40){1}
\BCirc(25,40){1}
\BCirc(40,40){1}
\Text(27,5)[l]{$\mu$}
\end{picture}
\begin{picture}(50, 50)
\Photon(24.5,5)(24.5,20){1}{3}
\Photon(25.5,5)(25.5,20){1}{3}
\Photon(25,20)(10,40){1}{6}
\Photon(25,20)(25,40){1}{4}
\Photon(25,20)(40,40){1}{6}
\Photon(10,40)(25,40){1}{3}
\Vertex(25,20){1}
\BCirc(10,40){1}
\BCirc(25,40){1}
\BCirc(40,40){1}
\Text(27,5)[l]{$\mu$}
\end{picture}
\end{center}
\vspace{2pt}
For the first and the third graph, we find
\begin{equation*}
 - e^2 (d-1) \hat A^\mu_0 (k) \hat \Delta_R(k) \int \ud^4 l \ \hat \Delta_+(l) \sin^2 \tfrac{k \sigma l}{2},
\end{equation*}
respectively. For the second graph, we find this twice. Thus, the tadpole contribution to the self--energy is
\begin{equation*}
  \Pi_{\mu \nu}(k) = - 4 (2\pi)^{-2} e^2 (d-1) g_{\mu \nu} \int \ud^4 l \ \hat \Delta_+(l) \sin^2 \tfrac{k \sigma l}{2}.
\end{equation*}
In the following, we write all the contributions to the self--energy in the form
\begin{equation}
\label{eq:Pi}
   \Pi_{\mu \nu}(k) = 2 e^2 \int \ud^4l \ \hat{\Delta}^{(1)}(l) \hat{\Delta}_R(k-l) \sin^2 \tfrac{k \sigma l}{2}  \pi_{\mu \nu}(k,l).
\end{equation}
For the tadpole this can be done using $k^2 \hat{\Delta}_R(k) = - (2\pi)^{-2}$ and the symmetry of $\hat \Delta^{(1)}$:
\begin{equation*}
  \pi^{tp}_{\mu \nu}(k,l) = (d-1) g_{\mu \nu} (k-l)^2.
\end{equation*}

Now we come to the photon fish graph. We have to compute the following graphs:
\begin{center}
\begin{picture}(60,55)
\Photon(29.5,5)(29.5,20){1}{3}
\Photon(30.5,5)(30.5,20){1}{3}
\Photon(30,20)(10,50){1}{8}
\Photon(29.5,20.3)(39.5,35.3){1}{4}
\Photon(30.5,19.7)(40.5,34.7){1}{4}
\Photon(40,35)(30,50){1}{4}
\Photon(40,35)(50,50){1}{4}
\Photon(10,50)(30,50){1}{4}
\Vertex(30,20){1}
\Vertex(40,35){1}
\BCirc(10,50){1}
\BCirc(30,50){1}
\BCirc(50,50){1}
\Text(32,5)[l]{$\mu$}
\end{picture}
\begin{picture}(60,55)
\Photon(29.5,5)(29.5,20){1}{3}
\Photon(30.5,5)(30.5,20){1}{3}
\Photon(30,20)(50,50){1}{8}
\Photon(29.5,19.7)(19.5,34.7){1}{4}
\Photon(30.5,20.3)(20.5,35.3){1}{4}
\Photon(20,35)(10,50){1}{4}
\Photon(20,35)(30,50){1}{4}
\Photon(30,50)(50,50){1}{4}
\Vertex(30,20){1}
\Vertex(20,35){1}
\BCirc(10,50){1}
\BCirc(30,50){1}
\BCirc(50,50){1}
\Text(32,5)[l]{$\mu$}
\end{picture}
\end{center}
\vspace{2pt}
These have to be counted twice in order to account for the other possible contraction. Similarly to the computation in the previous subsection, one obtains for the first graph
\begin{multline*}
  - e^2 (2 \pi)^2 \hat A^\nu_0 (k) \hat \Delta_R(k) \int \ud^4 l \ \hat \Delta_+(l) \hat \Delta_R(k-l) \sin^2 \tfrac{k \sigma l}{2} \\
  \times \left\{ g_{\mu \nu} (5 k^2 - 2 k \cdot l) + (d-6) k_\mu  k_\nu + (2d-3) \left( 2 l_\mu l_\nu - k_\mu l_\nu - l_\mu k_\nu \right) \right\}.
\end{multline*}
For the second graph, one finds nearly the same expression, but with $\hat \Delta_+(l)$ replaced by $\hat \Delta_+(-l)$. Taking the factor $2$ into account, we thus obtain for the photon fish graph
\begin{equation*}
  \pi^{pf}_{\mu \nu}(k,l) = - g_{\mu \nu} (5 k^2 - 2  k \cdot l ) - (d-6) k_{\mu} k_{\nu} + (2d-3) \left\{ (k-l)_{\mu} l_{\nu} + l_{\mu} (k-l)_{\nu} \right\}.
\end{equation*}

It remains to treat the ghost fish graphs:
\begin{center}
\begin{picture}(60,55)
\Photon(29.5,5)(29.5,20){1}{3}
\Photon(30.5,5)(30.5,20){1}{3}
\DashArrowLine(10,50)(30,20){2}
\DashArrowLine(29.5,20.3)(39.5,35.3){2}
\DashArrowLine(30.5,19.7)(40.5,34.7){2}
\DashArrowLine(40,35)(30,50){2}
\Photon(40,35)(50,50){1}{4}
\DashArrowLine(30,50)(10,50){2}
\Vertex(30,20){1}
\Vertex(40,35){1}
\BCirc(10,50){1}
\BCirc(30,50){1}
\BCirc(50,50){1}
\Text(32,5)[l]{$\mu$}
\end{picture}
\begin{picture}(60,55)
\Photon(29.5,5)(29.5,20){1}{3}
\Photon(30.5,5)(30.5,20){1}{3}
\DashArrowLine(30,20)(10,50){2}
\DashArrowLine(39.5,35.3)(29.5,20.3){2}
\DashArrowLine(40.5,34.7)(30.5,19.7){2}
\DashArrowLine(30,50)(40,35){2}
\Photon(40,35)(50,50){1}{4}
\DashArrowLine(10,50)(30,50){2}
\Vertex(30,20){1}
\Vertex(40,35){1}
\BCirc(10,50){1}
\BCirc(30,50){1}
\BCirc(50,50){1}
\Text(32,5)[l]{$\mu$}
\end{picture}
\begin{picture}(60,55)
\Photon(29.5,5)(29.5,20){1}{3}
\Photon(30.5,5)(30.5,20){1}{3}
\DashArrowLine(50,50)(30,20){2}
\DashArrowLine(29.5,19.7)(19.5,34.7){2}
\DashArrowLine(30.5,20.3)(20.5,35.3){2}
\Photon(20,35)(10,50){1}{4}
\DashArrowLine(20,35)(30,50){2}
\DashArrowLine(30,50)(50,50){2}
\Vertex(30,20){1}
\Vertex(20,35){1}
\BCirc(10,50){1}
\BCirc(30,50){1}
\BCirc(50,50){1}
\Text(32,5)[l]{$\mu$}
\end{picture}
\begin{picture}(60,55)
\Photon(29.5,5)(29.5,20){1}{3}
\Photon(30.5,5)(30.5,20){1}{3}
\DashArrowLine(30,20)(50,50){2}
\DashArrowLine(19.5,34.7)(29.5,19.7){2}
\DashArrowLine(20.5,35.3)(30.5,20.3){2}
\Photon(20,35)(10,50){1}{4}
\DashArrowLine(30,50)(20,35){2}
\DashArrowLine(50,50)(30,50){2}
\Vertex(30,20){1}
\Vertex(20,35){1}
\BCirc(10,50){1}
\BCirc(30,50){1}
\BCirc(50,50){1}
\Text(32,5)[l]{$\mu$}
\end{picture}
\end{center}
\vspace{2pt}
We have to count these graphs twice, in order to account for the case where the photon leaves the second vertex to the other side.
For the first graph, one obtains
\begin{equation*}
 - e^2 (2 \pi)^2 \hat A^\nu_0 (k) \hat \Delta_R(k) \int \ud^4 l \ \hat \Delta_+(l) \hat \Delta_R(k-l) \sin^2 \tfrac{k \sigma l}{2} l_\mu (k-l)_\nu.
\end{equation*}
The second graph yields a similar expression, but with $\mu$ and $\nu$ interchanged in the loop integral. For the last two graphs, one finds the same results, but with $\hat \Delta_+(l)$ replaced by $\hat \Delta_+(-l)$. Thus, the ghost loop contribution is
\begin{equation*}
  \pi^{gh}_{\mu \nu}(k,l) = - (k-l)_{\mu} l_{\nu} - l_{\mu} (k-l)_{\nu}.
\end{equation*}
Adding all this up, we find, using $l^2 \hat{\Delta}^{(1)}(l) = 0$,
\begin{subequations}
\begin{align}
\label{subeq:f_tot_1}
  \pi^{tot}_{\mu \nu}(k,l) = & - (6-d) \left( g_{\mu \nu} k^2 - k_{\mu} k_{\nu} \right) \\ 
\label{subeq:f_tot_2}
   & - 2 (d-2) \left( g_{\mu \nu} k \cdot l - (k-l)_{\mu} l_{\nu} - l_{\mu} (k-l)_{\nu} \right).
\end{align}
\end{subequations}
In the remainder of this subsection, we want to compute $\Pi_{\mu \nu}(k)$ explicitly. We do this separately for the planar and the nonplanar part, where the split is again given by~(\ref{eq:SinSplit}).


\subsubsection{The planar part}

We first focus on the term~(\ref{subeq:f_tot_1}). It already has the usual tensor structure. The loop integral is the same as that of the fish graph in the massless $\phi^3$ model and corresponds, in position space, to computing the point-wise product $\Delta^{(1)} \Delta_R$.
Using (\ref{eq:DeltaProduct}) and the well--known expression for the square of the Feynman propagator (see, e.g., \cite{Itzy}), we obtain, after renormalization,
\begin{equation}
\label{eq:planarIR1}
  \Pi_{\mu \nu}(k) = e^2 (2\pi)^{-2} ( g_{\mu \nu} k^2 - k_\mu k_\nu ) \left( \ln \frac{\sqrt{k^2}}{\mu} - \theta(k^2) \varepsilon(k_0) \frac{i \pi}{2} \right).
\end{equation}
Here $\mu$ is some mass scale that depends on the renormalization condition, and $\varepsilon$ is the sign function.


In order to treat the term~(\ref{subeq:f_tot_2}) in a similar fashion, we need the identity
\begin{equation*}
  \del_\mu \Delta_R(x) = \del_\mu \theta(x^0) \Delta(x) = \theta(x^0) \del_{\mu} \Delta(x).
\end{equation*}
Here we used $\delta(x^0) \Delta(x) = 0$. Similarly, we obtain
\begin{equation*}
  \del_\mu \Delta_F(x) = \theta(x^0) \del_\mu \Delta_+(x) + \theta(-x^0) \del_\mu \Delta_-(x)
\end{equation*}
with $\Delta_-(x) = \Delta_+(-x)$.
One can now show that
\begin{multline*}
  g^{\mu \nu} \del_\lambda \Delta_R \del^\lambda \Delta^{(1)} - \del^\mu \Delta_R \del^\nu \Delta^{(1)} - \del^\nu \Delta_R \del^\mu \Delta^{(1)} \\ = - i g^{\mu \nu} \del_\lambda \Delta_F \del^\lambda \Delta_F + 2 i \del^\mu \Delta_F \del^\nu \Delta_F - i g^{\mu \nu} \del_\lambda \Delta_- \del^\lambda \Delta_- + 2 i \del^\mu \Delta_- \del^\nu \Delta_-
\end{multline*}
holds. 
The first two terms on the right hand side are terms that one also finds in the two--point function of a nonabelian gauge theory. As there, their sum can be made well--defined (renormalized) in such a way that the Ward identity (transversality) is fulfilled. One then obtains the contribution
\begin{equation}
\label{eq:planarIR2}
  \Pi_{\mu \nu}(k) = \frac{2}{3} (2\pi)^{-2} e^2 ( g_{\mu \nu} k^2 - k_\mu k_\nu ) \left( \ln \frac{\sqrt{k^2}}{\mu} - \theta(k^2) \varepsilon(k_0) \frac{i \pi}{2} \right).
\end{equation}
We can already anticipate some potential infrared problems arising from (\ref{eq:planarIR1}) and (\ref{eq:planarIR2}): The expression is not well--defined for $k^2 \to 0$. Furthermore, $\Pi_{\mu \nu}(k)$ and $\Pi_{\mu \nu}(-k)$ do not coincide in a neighborhood of the forward light cone, because of the imaginary part. Such difficulties are typical for nonabelian gauge theories. We come back to this problem later.

\subsubsection{The nonplanar part}
\label{sec:nonplanar}

Now we take a look at the nonplanar part. The loop integral corresponding to the term~(\ref{subeq:f_tot_1}) is the same as for the massless $\phi^3$ fish graph\footnote{Apart from the fact that the twisting factor already comes in as a cosine, but this does not change anything.}. We already treated this integral in the massive case in Section~\ref{sec:phi3SelfEnergy} by interpreting it as a distribution $F(y,k)$ in two variables, cf.~(\ref{eq:F_y_k}). For the present case, we define
\begin{equation*}
  F(y,k) = 2 \int \ud^4l \ \hat \Delta_+(l) \cos yl \left( \hat \Delta_R(k-l) + \Delta_R(k+l) \right).
\end{equation*}
The proof that $F(k\sigma, k)$ is well--defined for $k^2>0$ goes through also in the massless case. Thus, the term~(\ref{subeq:f_tot_1}) yields a contribution to the self--energy of the form\footnote{For our purposes, we need to know the self--energy in a neighborhood of the light cone, thus in principle also for $k^2 < 0$. However, it was not yet possible to do a rigorous calculation in this range (but see Remark~\ref{rem:SigmaSpacelike}). Therefore, we compute the asymptotics $k^2 \to 0$ only from inside the light cone.}
\begin{equation*}
  \Pi_{\mu \nu} (k) = \left( g_{\mu \nu} k^2 -k_{\mu} k_{\nu} \right) \Sigma_0(k).
\end{equation*}
But in contrast to the massive case, we can even compute $\Sigma_0$ analytically: Due to Lorentz invariance, we can simplify the calculation by choosing $k = (k_0, \V 0)$.  Instead of~(\ref{eq:f_k}), one now has
\begin{equation*}
  F(y,k) = 4 (2\pi)^{-2} \int_0^\infty \ud l \ l \frac{-1}{k^2 - 4 l^2 + i \epsilon \varepsilon(k_0)}  \int \ud^4 y \ f(y) \cos y_0 l \frac{\sin \betrag{\V y} l}{\betrag{\V y} l}.
\end{equation*}
We recall that $\varepsilon$ is the sign function.
As in Section~\ref{sec:phi3SelfEnergy}, we can interchange the integrations. The integral over $l$ can be written as
\begin{equation}
\label{eq:F_NCQED}
 F(y,k) = - \frac{1}{8 \pi^2} \frac{1}{\betrag{\V y}} \int_0^\infty \ud l \ \frac{1}{k^2/4 - l^2 + i \epsilon \varepsilon(k_0)} \left\{ \sin [ ( \betrag{\V y} + y_0 ) l ] + \sin [ ( \betrag{\V y} - y_0 ) l ] \right\}.
\end{equation}
We define
\begin{align*}
  G_0^\pm(a,b,c) := & \frac{1}{a} \int_0^\infty \ud l \ \frac{\sin [(a+b)l]}{c^2 \pm i \epsilon - l^2} \\
  = & \mp \frac{i \pi}{2} \frac{\sin (a+b)c}{ac} + \frac{1}{a} \int_0^\infty \ud l \ \sin (a+b)l \pv{c^2 - l^2}.
\end{align*}
The second term is a standard integral. With \cite[Eq.~(3.723.8)]{Gradshteyn}, we obtain
\begin{equation*}
  G_0^\pm(a,b,c) = \mp \frac{i \pi}{2} \frac{\sin (a+b)c}{ac} + \frac{1}{ac} \left[ \sin (a+b)c \ci (a+b)c - \cos (a+b)c \si (a+b)c  \right].
\end{equation*}
Here $\si$ and $\ci$ are the sine and cosine integral, see~(\ref{eq:si}). We have
\begin{equation*}
  F(y,k)= - \frac{1}{2} (2\pi)^{-2} \left\{ G_0^{\varepsilon(k_0)}(\betrag{\V y}, y_0, \sqrt{k^2/4}) + G_0^{\varepsilon(k_0)}(\betrag{\V y}, - y_0, \sqrt{k^2/4}) \right\}.
\end{equation*}
Applying the expression for $G_0^\pm$ to (\ref{eq:F_NCQED}), we obtain, using Lorentz invariance, for $k$ timelike,
\begin{equation}
\label{eq:Sigma0}
  \Sigma_0(k) = - (2\pi)^{-2} G_0^{\varepsilon(k_0)}(\sqrt{-(k \sigma)^2}, 0, \sqrt{k^2/4}).
\end{equation}

It remains to treat the terms~(\ref{subeq:f_tot_2}). We define
\begin{equation*}
 F_{\mu \nu}(y,k) = 4 \int \ud^4l \ \hat \Delta^{(1)}(l) \hat \Delta_R(k-l) \cos yl \left\{ g_{\mu \nu} k \cdot l - (k-l)_\mu l_\nu - l_\mu (k-l)_\nu \right\}.
\end{equation*}
This is invariant under proper orthochronous Lorentz transformations, i.e., 
\begin{equation}
\label{eq:F_mu_nu_Lorentz}
  F_{\mu' \nu'}(y {\Lambda^T}^{-1}, k \Lambda) {\Lambda^{-1}}^{\mu'}_{\ \mu} {\Lambda^{-1}}^{\nu'}_{\ \nu} = F_{\mu \nu}(y,k).
\end{equation}
Using $\hat \Delta^{(1)}(l) = \hat \Delta_+(l) + \hat \Delta_+(-l)$, we obtain for the integration of $F_{\mu \nu}(y,k)$ with a test function $f(y)$, for $k = (k_0, \V 0)$,
\begin{multline*}
 - 4 (2\pi)^{-2} \int \ud^4l \ \hat \Delta_+(l) \left\{ \frac{g_{\mu \nu} k \cdot l - (k-l)_\mu l_\nu - l_\mu (k-l)_\nu}{k^2 - 2 \sqrt{k^2} l_0 + i \epsilon \varepsilon(k_0)} \right. \\ + \left. \frac{-g_{\mu \nu} k \cdot l + (k+l)_\mu l_\nu + l_\mu (k-l)_\nu}{k^2 + 2 \sqrt{k^2} l_0 + i \epsilon \varepsilon(k_0)} \right\} \int \ud^4y \ f(y) \cos y l.
\end{multline*}
The expression in curly brackets can be rewritten as
\begin{equation*}
  \frac{4}{k^2-4l_0^2 + i \epsilon \varepsilon(k_0)} \left\{ g_{\mu \nu} l_0^2 + l_\mu l_\nu - \delta_\mu^0 l_\nu l_0 - \delta_\nu^0 l_\mu l_0 \right\}.
\end{equation*}
One may replace the powers of $l_\rho$ by derivatives acting on $f$. Therefore, we can compute exactly the same integral as above and then pull the derivatives back onto 
$F_{\mu \nu}$ (this is possible if we assume $f$ to have compact support). Thus, we have
\begin{equation*}
  F_{\mu \nu}(y,k) = - 4 \left\{ g_{\mu \nu} \del_{y_0}^2 + \del_{y_\mu} \del_{y_\nu} - \delta_\mu^0 \del_{y_0} \del_{y_\nu} - \delta_\nu^0 \del_{y_0} \del_{y_\mu} \right\} F(y,k).
\end{equation*}
As $F(y,k)$, this is smooth for $k^2>0$, $y^2<0$.
That this vanishes for $\mu = \nu = 0$ is obvious. We define
\begin{equation*}
  G_n^\pm(a,b,c)=\frac{\del^n}{\del b^n} G_0^\pm(a,b,c). 
\end{equation*}
Because of 
\begin{equation*}
  \del_{y_0} F(y, k) = - \frac{1}{2} (2\pi)^{-2} \left\{ G_1^{\varepsilon(k_0)}(\betrag{\V y}, y_0, \sqrt{k^2/4}) - G_1^{\varepsilon(k_0)}(\betrag{\V y}, - y_0, \sqrt{k^2/4}) \right\},
\end{equation*}
the terms with $\mu=0$, $\nu=i$ and vice versa vanish for $y = (0, \V y)$, which is the case of interest for us.
It remains to treat the case $\mu=i$, $\nu=j$. 
The following identities hold:
\begin{subequations}
\begin{align}
\label{eq:del_F_1}
  \del_{y_0} \del_{y_0} F((0,\V y),k) = & - (2\pi)^{-2} G_2^{\varepsilon(k_0)}(\betrag{\V y}, 0, \sqrt{k^2/4}), \\
\label{eq:del_F_2}
  \del_{y_i} \del_{y_j} F((0,\V y),k) = & \left( \frac{\delta_{i j}}{\betrag{\V y}^2} - 3 \frac{y_i y_j}{\betrag{\V y}^4} \right) (2\pi)^{-2} G_0^{\varepsilon(k_0)}(\betrag{\V y},0,\sqrt{k^2/4}) \\
  & - \left( \frac{\delta_{i j}}{\betrag{\V y}} - 3 \frac{y_i y_j}{\betrag{\V y}^3} \right) (2\pi)^{-2} G_1^{\varepsilon(k_0)}(\betrag{\V y}, 0, \sqrt{k^2/4}) \nonumber \\
  & - \frac{y_i y_j}{\betrag{\V y}^2} (2\pi)^{-2} G_2^{\varepsilon(k_0)}(\betrag{\V y}, 0, \sqrt{k^2/4}). \nonumber
\end{align}
\end{subequations}
The contribution to the self--energy is given by $\Pi_{\mu \nu}(k) = e^2 F_{\mu \nu}(k\sigma, k)$. We have shown that, for $k= (k_0, \V 0)$, it is of the form
\begin{equation*}
  \Pi_{i j}(k) = - e^2 \delta_{i j} k^2 \Sigma_1(k) + e^2 \tfrac{ (k \sigma)_i (k \sigma_j) }{(k\sigma)^4} \Sigma_2(k),
\end{equation*}
with all other components vanishing. The functions $\Sigma_{1/2}$ can be read off straightforwardly from (\ref{eq:del_F_1}) and (\ref{eq:del_F_2}).
Because of Lorentz covariance, cf. (\ref{eq:F_mu_nu_Lorentz}), we then obtain, for timelike $k$,
\begin{equation*}
  \Pi_{\mu \nu} (k) = e^2 \left( g_{\mu \nu} k^2 -k_{\mu} k_{\nu} \right) \Sigma_1(k) + e^2 \tfrac{(k \sigma)_{\mu} (k \sigma)_{\nu}}{(k \sigma)^4} \Sigma_2(k)
\end{equation*}
as the contribution of the term (\ref{subeq:f_tot_2}) to the self--energy.
For $\Sigma_{1/2}$, we find
\begin{multline*}
  \Sigma_1(k) = - \frac{4}{(2\pi)^{2} k^2} \Big\{ - G_2^{\varepsilon(k_0)}(\sqrt{-(k\sigma)^2}, 0,\sqrt{ k^2/4})  \\
   \shoveright{  + \tfrac{1}{\sqrt{-(k\sigma)^2}} G_1^{\varepsilon(k_0)}(\sqrt{-(k\sigma)^2}, 0, \sqrt{ k^2/4}) + \tfrac{1}{(k\sigma)^2} G_0^{\varepsilon(k_0)}(\sqrt{-(k\sigma)^2}, 0, \sqrt{ k^2/4}) \Big\} , } \\
  \shoveleft{ \Sigma_2(k) = \frac{4 (k \sigma)^2}{(2\pi)^2} \Big\{ - G_2^{\varepsilon(k_0)}(\sqrt{-(k\sigma)^2}, 0, \sqrt{ k^2/4})  } \\
   + 3 \tfrac{1}{\sqrt{-(k\sigma)^2}} G_1^{\varepsilon(k_0)}(\sqrt{-(k\sigma)^2}, 0, \sqrt{ k^2/4})  + 3 \tfrac{1}{(k\sigma)^2} G_0^{\varepsilon(k_0)}(\sqrt{-(k\sigma)^2}, 0, \sqrt{ k^2/4}) \Big\}.
\end{multline*}
We emphasize that the nonplanar loop integrals are completely solved. To the best of our knowledge, this has not been achieved before. In the literature, only the leading behaviour for $(k\sigma)^2 \to 0$ was computed. We want to compare with these results. Using the series expansions (\ref{eq:siExpansion}) of $\si$ and $\ci$, one finds
\begin{align*}
  G_0^\pm(a,0,c) & = \left( \gamma -1 + \ln ac \mp \frac{i\pi}{2} \right) - \frac{a^2 c^2}{6} \left( \gamma - \frac{11}{6} + \ln ac \mp \frac{i\pi}{2} \right) + \order(a^4 c^4), \\
  G_1^\pm(a,0,c) & = \frac{1}{a} \left( \gamma + \ln ac \mp \frac{i\pi}{2} \right) - \frac{a c^2}{2} \left( \gamma - \frac{3}{2} + \ln ac \mp \frac{i\pi}{2} \right) + \order(a^3 c^4), \\
  G_2^\pm(a,0,c) & = \frac{1}{a^2} - c^2 \left( \gamma - 1 + \ln ac \mp \frac{i\pi}{2} \right) + \order(a^2 c^4).
\end{align*}
Here $\gamma$ is the Euler-Mascheroni constant.
Thus, we obtain
\begin{align}
\label{eq:nonplanarIR1}
  \Sigma_0(k) = & - (2\pi)^{-2} \left( \ln \sqrt{-(k\sigma)^2} \sqrt{ k^2/4} +\gamma-1 - \varepsilon(k_0) \frac{i\pi}{2} \right) + \order((k\sigma)^2 k^2), \\
\label{eq:nonplanarIR2}
  \Sigma_1(k) = & - (2\pi)^{-2} \left( \frac{2}{3} \ln \sqrt{-(k\sigma)^2} \sqrt{ k^2/4} + \frac{2}{3} \gamma - \frac{5}{9} - \varepsilon(k_0) \frac{2}{3} \frac{i\pi}{2} \right) + \order((k\sigma)^2 k^2), \\
\label{eq:nonplanarIR3}
  \Sigma_2(k) = & - (2\pi)^{-2} \left( 8 - \frac{1}{3} (k\sigma)^2 k^2 \right) + \order((k\sigma)^4 k^4).
\end{align}
This is in agreement with the results obtained in \cite{Hayakawa} in the setting of the modified Feynman rules, cf.~(\ref{eq:Hayakawa_Sigma_1},b). Adding~(\ref{eq:nonplanarIR1}) and~(\ref{eq:nonplanarIR2}) to the planar terms (\ref{eq:planarIR1}) and (\ref{eq:planarIR2}), we obtain the following contribution to the self--energy
\begin{equation}
\label{eq:Pi_0}
  \Pi_{\mu \nu}(k) = - \frac{5}{3} (2\pi)^{-2} e^2 (g_{\mu \nu} k^2 - k_\mu k_\nu) \left( \ln \mu \sqrt{-(k \sigma)^2} + \order( k^2 (k \sigma)^2) \right).
\end{equation}
We see that the problematic term proportional to $\ln k^2$ drops out. Also the imaginary parts cancel between the planar and the nonplanar part at zeroth order in $k^2$. Once more, this cancellation is due to the fact that we compute the difference between the planar and the nonplanar part, cf. (\ref{eq:SinSplit}), which is a consequence of the interaction term being a commutator\footnote{To the best of our knowledge, this cancellation of infrared divergences between the planar and nonplanar parts has not been noticed before.}. As discussed in Remark~\ref{rem:k4}, we need $\Pi_{\mu \nu}(k) - \Pi_{\mu \nu}(-k) \sim k^4$ in the neighborhood of the light cone in order to have a well--defined adiabatic limit. Since this difference is given by the imaginary part, which is of order $k^2$, the condition is fulfilled for the part proportional to $g_{\mu \nu} k^2$. Formally, the term proportional to $k_\mu k_\nu$ drops out when we restrict to transversal test functions, as proposed in Remark~\ref{rem:k_mu_k_nu}.
Rigorously, one should check whether for such test functions a self--energy contribution proportional to $k_\mu k_\nu$ vanishes in the adiabatic limit in a formula analogous to (\ref{eq:phi_3_comm}). This is done in Appendix~\ref{app:k_mu_k_nu} for scaling sequences of infrared cut--off functions $\check g_a$ with compact support.
Thus, for transversal test functions, we obtain the following contribution to the elementary two--point function, cf. (\ref{eq:W_SelfEnergy}),
\begin{equation}
\label{eq:W1}
  W_{\mu \nu}(k,p) = - \frac{5}{3} e^2 g_{\mu \nu}  \ln \mu \sqrt{-(k \sigma)^2}  \hat \Delta_+(k) \delta(k+p).
\end{equation}
Here we used $k^2 \frac{\del}{\del m^2} \hat \Delta_+(k) = \hat \Delta_+(k)$.
In Section~\ref{sec:infrared}, we will discuss the well--definedness of the product of distributions in this expression.

As is obvious from (\ref{eq:nonplanarIR3}), the imaginary part of $\Sigma_2(k)$ is proportional to $k^4$. Thus, the condition $\Pi_{\mu \nu}(k) - \Pi_{\mu \nu}(-k) \sim k^4$ is fulfilled for this term and we obtain the following contribution to the elementary two--point function:
\begin{equation}
\label{eq:W2}
  W_{\mu \nu}(k,p) = - e^2 \delta(k+p) \frac{(k\sigma)_\mu (k\sigma)_\nu}{(k\sigma)^4} \left( 8 \frac{\del}{\del m^2} \hat \Delta_+(k) - \frac{(k\sigma)^2}{3} \hat \Delta_+(k) \right).
\end{equation}
The well--definedness of the products of distributions in this expression will be discussed later in Section~\ref{sec:infrared}.

\begin{remark}
\label{rem:SigmaSpacelike}
Above, we computed the self--energy $\Pi_{\mu \nu}(k)$ on the light cone by computing it in the interior and then taking the limit $k^2 \to 0$. In principle one should check whether one gets the same result by continuation from $k^2<0$. The knowledge of $\Pi_{\mu \nu}(k)$ for spacelike $k$ is also important if one wants to consider higher loop orders. However, it was not yet possible to rigorously compute the nonplanar part of $\Pi_{\mu \nu}(k)$ in this range. In Appendix~\ref{app:SigmaSpacelike}, we show that a formal calculation of $\Sigma_0(k)$ for $k^2 < 0$, $(k \sigma)^2 >0$ leads to
\begin{equation}
\label{eq:SigmaSpacelike}
  \Sigma_0(k) = - \frac{1}{2} (2\pi)^{-2} \left\{ G_0^+(\sqrt{-(k \sigma)^2},0,\sqrt{-k^2/4}) + G_0^-(\sqrt{-(k \sigma)^2},0,\sqrt{-k^2/4}) \right\}.
\end{equation}
This is the real part of the expression for timelike $k$, cf. (\ref{eq:Sigma0}), which may be seen as an indication that the approach taken here is consistent.
\end{remark}

\subsection{The elementary three-- and four--point functions}

We need the elementary three--point function at first order. Using the cubic photon vertex (\ref{eq:CubicVertex}), it is straightforward to obtain
\begin{multline*}
  W_{\underline \mu}(\underline k) =  2 i e (2\pi)^4 \left( g_{\mu_1 \mu_2} (k_1-k_2)_{\mu_3} + g_{\mu_3 \mu_1} (k_3-k_1)_{\mu_2} + g_{\mu_2 \mu_3} (k_2-k_3)_{\mu_1} \right) \\ \times \sin \tfrac{k_2 \sigma k_3}{2} \delta(\sum k_i) \left( \hat \Delta_R(k_1) \hat \Delta_+(-k_2) \hat \Delta_+(-k_3) \right. \\
  \left. + \hat \Delta_+(k_1) \hat \Delta_R(k_2) \hat \Delta_+(-k_3) +\hat \Delta_+(k_1) \hat \Delta_+(k_2) \hat \Delta_R(k_3) \right).
\end{multline*}
The elementary four--point function is only needed at zeroth order: 
\begin{align}
\label{eq:4pt}
  W_{\underline \mu}(\underline k) = (2 \pi)^4 & \left( g_{\mu_1 \mu_2} g_{\mu_3 \mu_4} \delta(k_1+k_2) \delta(k_3+k_4) \hat \Delta_+(k_1) \hat \Delta_+(k_3) \right. \\
  & + g_{\mu_1 \mu_3} g_{\mu_2 \mu_4} \delta(k_1+k_3) \delta(k_2+k_4) \hat \Delta_+(k_1) \hat \Delta_+(k_2) \nonumber \\
  & \left. + g_{\mu_1 \mu_4} g_{\mu_2 \mu_3} \delta(k_1+k_4) \delta(k_2+k_3) \hat \Delta_+(k_1) \hat \Delta_+(k_2) \right). \nonumber
\end{align}

\section{The full two--point function}

Having calculated $K_{\mu \nu}(k)$ and the relevant elementary $n$-point functions, we now come to the third point of our list in Section~\ref{sec:2pt}, the computation of the full two--point function (\ref{eq:NCQED2pt}). At zeroth order, we find
\begin{equation*}
  - 4 (2\pi)^2 \int \ud^4 k \ \hat f^{\mu \nu}(-k) \hat h^\lambda_{\ \nu}(k) k_\mu k_\lambda \hat \Delta_+(k). 
\end{equation*}
There is no first order contribution. In the following, we compute all second order terms. Since they are not the main point of interest, the loop integrals leading to finite contributions to the continuous spectrum will not be computed explicitly.
We order the presentation of the various contributions by the powers of $e$ that the elementary two--point functions $W_{\underline \mu}(\underline k)$ contribute. 

\subsection{Zeroth order}
\label{sec:FreePart}

If the elementary $n$-point function does not contribute any power of $e$, both powers of $e$ must stem from the two kernels $K_{\mu \nu}^{\underline \mu}(k;\underline k)$.
We start by considering the case where both kernels are of first order. Using the kernel (\ref{eq:K1_1}) for both observables, one finds
\begin{multline*}
  4 e^2 \int \ud^4k \ud^4p \ \hat f^{\mu \nu}(-k) \hat h^{\lambda \rho}(-p) \int \ud^8 \underline k \ud^8 \underline p \\
  \times \delta(k - \sum k_i) \delta(p - \sum p_j) \sin \tfrac{k_1 \sigma k_2}{2} \sin \tfrac{p_1 \sigma p_2}{2} \hat \Delta_+(k_1) \hat \Delta_+(k_2) \\
  \times \big\{ g_{\mu \lambda} g_{\nu \rho} \delta(k_1+p_1) \delta(k_2+p_2) + g_{\mu \rho} g_{\nu \lambda} \delta(k_1+p_2) \delta(k_2+p_1) \big\}.
\end{multline*}
Here we suppressed the first term in the elementary four--point function (\ref{eq:4pt}), since its contribution vanishes because of the antisymmetry of $\sigma$. Carrying out the integrations and using the antisymmetry of $f^{\mu \nu}$ and $h^{\lambda \rho}$, one obtains
\begin{equation}
\label{eq:Finite_1}
  8 e^2 \int \ud^4k \ \hat{f}^{\mu \nu}(-k) \hat{h}_{\mu \nu}(k) \int \ud^4l \ \hat{\Delta}_+(l) \hat{\Delta}_+(k-l) \sin^2 \tfrac{k \sigma l}{2}.
\end{equation}
This is a contribution to the continuous part of the spectrum.

Now we consider the term where the kernel (\ref{eq:K1_2}) is used for the observable involving $f^{\mu \nu}$ and the kernel (\ref{eq:K1_1}) for the observable involving $h^{\lambda \rho}$. One obtains
\begin{multline*}
 -2 e^2 \int \ud^4k \ud^4p \ \hat f^{\mu \nu}(-k) \hat h^{\lambda \rho}(-p) \int \ud^8 \underline k \ud^8 \underline p \ \delta(k - \sum k_i) \delta(p - \sum p_j) \\
 \times \big\{ (k \sigma)^{\mu_1} k_{2 \mu} \delta_\nu^{\mu_2} + (k \sigma)^{\mu_2} k_{1 \mu} \delta_\nu^{\mu_1} \big\} \frac{ \sin \frac{k_1 \sigma k_2}{2} }{\frac{k_1 \sigma k_2}{2}} \sin \tfrac{p_1 \sigma p_2}{2} \hat \Delta_+(k_1) \hat \Delta_+(k_2) \\
  \times \big\{ g_{\mu_1 \lambda} g_{\mu_2 \rho} \delta(k_1+p_1) \delta(k_2+p_2) + g_{\mu_1 \rho} g_{\mu_2 \lambda} \delta(k_1+p_2) \delta(k_2+p_1) \big\}.
\end{multline*}
Again, we suppressed the first term in (\ref{eq:4pt}). Carrying out the integrations and the sum over the indices, one finds
\begin{equation}
\label{eq:Finite_2}
  - 8 e^2 \int \ud^4k \ \hat f^{\mu \nu}(-k) \hat{h}^{\lambda}_{\ \nu}(k) \int \ud^4l \ \hat \Delta_+(l) \hat \Delta_+(k-l) \frac{\sin^2 \frac{k \sigma l}{2}}{\frac{k \sigma l}{2}}
  (k \sigma)_\lambda l_\mu.
\end{equation}
For the term where the kernel (\ref{eq:K1_1}) is used for the observable involving $f^{\mu \nu}$ and the kernel (\ref{eq:K1_2}) for the observable involving $h^{\lambda \rho}$ one obtains, in the same way,
\begin{equation}
\label{eq:Finite_3}
  - 8 e^2 \int \ud^4k \ \hat f^{\mu \nu}(-k) \hat{h}^{\lambda}_{\ \nu}(k) \int \ud^4l \ \hat \Delta_+(l) \hat \Delta_+(k-l) \frac{\sin^2 \frac{k \sigma l}{2}}{\frac{k \sigma l}{2}}
  (k \sigma)_\mu l_\lambda.
\end{equation}
Both are contributions to the continuous part of the spectrum.



It remains to calculate the contribution where in both observables the kernel (\ref{eq:K1_2}) is used. One finds
\begin{multline*}
  e^2 \int \ud^4k \ud^4p \ \hat f^{\mu \nu}(-k) \hat h^{\lambda \rho}(-p) \int \ud^8 \underline k \ud^8 \underline p \ \delta(k - \sum k_i) \delta(p - \sum p_j) \hat \Delta_+(k_1) \hat \Delta_+(k_2) \\
 \times \big\{ (k \sigma)^{\mu_1} k_{2 \mu} \delta_\nu^{\mu_2} + (k \sigma)^{\mu_2} k_{1 \mu} \delta_\nu^{\mu_1} \big\} \big\{ (p \sigma)^{\nu_1} p_{2 \lambda} \delta_\rho^{\nu_2} + (p \sigma)^{\nu_2} p_{1 \lambda} \delta_\rho^{\nu_1} \big\} \frac{ \sin \frac{k_1 \sigma k_2}{2} }{\frac{k_1 \sigma k_2}{2}} \frac{\sin \frac{p_1 \sigma p_2}{2}}{\frac{p_1 \sigma p_2}{2}} \\
  \times \big\{ g_{\mu_1 \nu_1} g_{\mu_2 \nu_2} \delta(k_1+p_1) \delta(k_2+p_2) + g_{\mu_1 \nu_2} g_{\mu_2 \nu_1} \delta(k_1+p_2) \delta(k_2+p_1) \big\}.
\end{multline*}
Again, we suppressed the first term in (\ref{eq:4pt}). That this is possible is a consequence of (\ref{eq:EIdentity}). One obtains
\begin{multline}
\label{eq:Finite_4}
  4 e^2 \int \ud^4k \ \hat f^{\mu \nu}(-k) \hat h^{\lambda \rho}(k) \int \ud^4l \ \hat \Delta_+(l) \hat \Delta_+(k-l) \frac{\sin^2 \frac{k \sigma l}{2}}{(\frac{k \sigma l}{2})^2} \\
\times \left\{ g_{\nu \rho} l_{\mu} l_\lambda (k \sigma)^2 + l_\mu (k-l)_\lambda (k \sigma)_\rho (k \sigma)_\nu \right\}.
\end{multline}
The integral over $l$ is well--defined and gives a contribution to the continuous part of the spectrum.

Now we consider the terms involving the second order kernel (\ref{eq:K2_1}) and the zeroth order kernel (\ref{eq:K0}). Using the kernel (\ref{eq:K2_1}) in the first observable, we obtain
\begin{multline*}
 - 2 e^2 \int \ud^4k \ud^4p \ \hat f^{\mu \nu}(-k) \hat h^{\lambda \rho}(-p) \int \ud^{12} \underline k \ \delta(k - \sum k_i) p_\lambda \\
 \shoveleft{ \quad \times \big\{ (k \sigma)^{\mu_1} \delta_\mu^{\mu_2} \delta_\nu^{\mu_3} \frac{ \sin \frac{k \sigma k_1}{2} }{\frac{k \sigma k_1}{2}} \sin \tfrac{k_2 \sigma k_3}{2} + (k \sigma)^{\mu_3} \delta_\mu^{\mu_1} \delta_\nu^{\mu_2} \frac{ \sin \frac{k \sigma k_3}{2} }{\frac{k \sigma k_3}{2}} \sin \tfrac{k_1 \sigma k_2}{2} \big\} } \\
  \times \big\{ \hat \Delta_+(k_1) \hat \Delta_+(k_2) \left[ g_{\mu_1 \mu_3} g_{\mu_2 \rho} \delta(k_1+k_3) \delta(k_2+p) + g_{\mu_1 \rho} g_{\mu_2 \mu_3} \delta(k_1+p) \delta(k_2+k_3) \right] \\
   + \hat \Delta_+(k_1) \hat \Delta_+(k_3) g_{\mu_1 \mu_2} g_{\mu_3 \rho} \delta(k_1+k_2) \delta(k_3+p) \big\}.
\end{multline*}
After carrying out the integrations and the sum over the indices, four terms remain. These cancel each other. The same is true for the term where the kernel (\ref{eq:K2_1}) is used for the observable involving $h^{\lambda \rho}$.

Finally, we consider the terms involving the kernel (\ref{eq:K2_2}). Using it in the first observable, we find 
\begin{multline*}
  2 e^2 \int \ud^4k \ud^4p \ \hat f^{\mu \nu}(-k) \hat h^{\lambda \rho}(-p) \int \ud^{12} \underline k \ \delta(k - \sum k_i) p_\lambda \\
\shoveleft{ \quad  \times \left\{ (k \sigma)^{\mu_1} (k \sigma)^{\mu_2} k_{3 \mu} \delta^{\mu_3}_\nu P_2(- i k \sigma k_1, - i k \sigma k_2) e^{-\frac{i}{2} k_1 \sigma k_2} e^{- \frac{i}{2} (k_1+k_2) \sigma k_3} \right. } \\
\shoveright{ \quad  \left. + (k \sigma)^{\mu_2} (k \sigma)^{\mu_3} k_{1 \mu} \delta^{\mu_1}_\nu P_2(- i k \sigma k_2, - i k \sigma k_3) e^{-\frac{i}{2} k_2 \sigma k_3} e^{- \frac{i}{2} (k_2+k_3) \sigma k_1} \right\} } \\
  \times \big\{ \hat \Delta_+(k_1) \hat \Delta_+(k_2) \left[ g_{\mu_1 \mu_3} g_{\mu_2 \rho} \delta(k_1+k_3) \delta(k_2+p) + g_{\mu_1 \rho} g_{\mu_2 \mu_3} \delta(k_1+p) \delta(k_2+k_3) \right] \\
   + \hat \Delta_+(k_1) \hat \Delta_+(k_3) g_{\mu_1 \mu_2} g_{\mu_3 \rho} \delta(k_1+k_2) \delta(k_3+p) \big\}.
\end{multline*}
This can be reduced to
\begin{multline*}
  - 2 e^2 \int \ud^4k \ \hat f^{\mu \nu}(-k) \hat h^{\lambda \rho}(k) k_\lambda \hat \Delta_+(k) \int \ud^4l \ \hat \Delta_+(l) \Big( 2 g_{\nu \rho} k_\mu (k \sigma)^2 P_2(-ik \sigma l, ik\sigma l) \\
  +  (k\sigma)_\nu (k\sigma)_\rho l_\mu \left\{ P_2(ik\sigma l, 0) e^{-ik\sigma l} + P_2(0, ik\sigma l) - P_2(-ik\sigma l, 0) e^{ik\sigma l} - P_2(0, -ik\sigma l) \right\} \Big).
\end{multline*}
Using
\begin{equation*}
  P_2(x,0) = \frac{1-e^x+x e^x}{x^2}, \quad P_2(0,x) = \frac{-1+e^x-x}{x^2},
\end{equation*}
one can show that the expression in curly brackets vanishes. Thus, we are left with
\begin{equation}
\label{eq:CovCoorO2_1_result}
- 4 e^2 \int \ud^4k \ \hat{f}^{\mu \nu}(-k) \hat{h}^{\lambda}_{\ \nu}(k) (k \sigma)^2  k_{\mu} k_{\lambda}  \hat{\Delta}_+(k) \int \ud^4 l \ \hat{\Delta}_+(l) P_2(0, -i k \sigma l).
\end{equation}
Here we used $P_2(x,-x) = P_2(0,x)$.
From the term where the kernel (\ref{eq:K2_2}) is used in the observable involving $h^{\lambda \rho}$, one obtains the same contribution, but with $P_2(0,-ik\sigma l)$ replaced by $P_2(0,ik\sigma l)$. We have
\begin{equation*}
  P_2(0,i x) + P_2(0,-ix) = 2 \frac{1 - \cos x}{x^2} = \left( \frac{\sin \frac{x}{2}}{\frac x 2} \right)^2.
\end{equation*}
The combination of both terms thus gives
\begin{equation}
\label{eq:divergence1}
- 4 e^2 \int \ud^4k \ \hat{f}^{\mu \nu}(-k) \hat{h}^{\lambda}_{\ \nu}(k) (k \sigma)^2  k_{\mu} k_{\lambda}  \hat{\Delta}_+(k) \int \ud^4 l \ \hat{\Delta}_+(l) \frac{\sin^2 \frac{k \sigma l}{2}}{(\frac{k \sigma l}{2})^2}.
\end{equation}
The integral over $l$ can not be split into a planar (local) and a finite nonplanar part. There is no obvious way to define it rigorously. We want to compute it at least formally. We choose $k=(k_0,\V 0)$ and obtain
\begin{equation*}
  \int_0^\infty \ud l \ \frac{l}{2} \int_{-1}^1 \ud x \ \frac{\sin^2 \frac{\betrag{ \underline{k \sigma} } l x}{2}}{\frac{\betrag{ \underline{k \sigma} } l x}{2}} = 4 \int_0^\infty \ud l \ \frac{l}{2} \frac{-1 + \cos (\betrag{ \underline{k \sigma} } l) + \betrag{ \underline{k \sigma} } l \si (\betrag{ \underline{k \sigma} } l)}{(\betrag{ \underline{k \sigma} } l)^2}.
\end{equation*}
Here $\underline{k \sigma}$ is the spatial part of $k \sigma$. This diverges linearly. We are faced with a nonlocal divergence.
We remark that this term is obtained from the contraction of the two fields stemming from the covariant coordinates. Thus, subtracting it can be interpreted as a normal ordering of $e^{i k X}$. More precisely, we would redefine $f^{\mu \nu} \to f^{\mu \nu} + f_c^{\mu \nu}$ with
\begin{equation*}
  \hat{f}^{\mu \nu}_c(-k) = (2 \pi)^{-2} e^2 \hat{f}^{\mu \nu}(-k)  (k \sigma)^2 \int \ud^4l \ \hat{\Delta}_+(l) P_2(0,-ik \sigma l),
\end{equation*}
and analogously for $h^{\lambda \rho}$. Alternatively, one could modify the action with a nonlocal field strength counterterm.
That the contraction of two fields stemming from the covariant coordinate leads to a linear divergence has already been noticed in~\cite{Gross}.

\subsection{First order}
\label{sec:FirstOrder}

If the elementary $n$-point function contributes one power of $e$, then another one must come from the kernels.
We first consider the kernel (\ref{eq:K1_1}). Using it for the observable involving $f^{\mu \nu}$ and the zeroth order kernel (\ref{eq:K0}) for the second observable, one obtains
\begin{multline*}
  8 (2 \pi)^2 e^2 \int \ud^4k \ \hat f^{\mu \nu}(-k) \hat h^{\lambda \rho}(k) k_\lambda \int \ud^{8} \underline k \ \delta(k - \sum k_i) \sin \tfrac{k_1 \sigma k_2}{2} \sin \tfrac{k \sigma k_2}{2} \\
  \times \left( g_{\nu \rho} (k_2 + k)_\mu - g_{\mu \rho} (k_1 + k)_\nu \right) \left\{ \hat \Delta_R(k_1) \hat \Delta^{(1)}(k_2) \hat \Delta_+(k) + \hat \Delta_+(k_1) \hat \Delta_+(k_2) \hat \Delta_A(k) \right\}.
\end{multline*}
Here we already used the antisymmetry of $f^{\mu \nu}$
to reduce the number of terms. For the two terms in curly brackets, one finds, respectively,
\begin{subequations}
\label{subeq:1}
\begin{gather}
\label{subeq:A1_1_1_res}
  24 (2\pi)^2 e^2 \int \ud^4 k \ \hat{f}^{\mu \nu}(-k) \hat{h}^{\lambda}_{\ \nu}(k) k_\mu k_\lambda \hat{\Delta}_+(k) \int \ud^4 l \ \hat{\Delta}^{(1)}(l) \hat{\Delta}_R(k-l) \sin^2 \tfrac{k \sigma l}{2}, \\
\label{subeq:A1_1_2_res}
  24 (2\pi)^2 e^2 \int \ud^4 k \ \hat{f}^{\mu \nu}(-k) \hat{h}^{\lambda}_{\ \nu}(k) k_\mu k_\lambda \hat{\Delta}_A(k) \int \ud^4 l \ \hat{\Delta}_+(l) \hat{\Delta}_+(k-l) \sin^2 \tfrac{k \sigma l}{2}.
\end{gather}
For the term where the kernel (\ref{eq:K1_1}) is used in the second observable, one obtains, in the same way, 
\begin{gather}
\label{subeq:A1_1_4_res}
  24 (2\pi)^2 e^2 \int \ud^4 k \ \hat{f}^{\mu \nu}(-k) \hat{h}^{\lambda}_{\ \nu}(k) k_\mu k_\lambda \hat{\Delta}_+(k) \int \ud^4 l \ \hat{\Delta}^{(1)}(l) \hat{\Delta}_A(k-l) \sin^2 \tfrac{k \sigma l}{2}, \\
\label{subeq:A1_1_3_res}
  24 (2\pi)^2 e^2 \int \ud^4 k \ \hat{f}^{\mu \nu}(-k) \hat{h}^{\lambda}_{\ \nu}(k) k_\mu k_\lambda \hat{\Delta}_R(k) \int \ud^4 l \ \hat{\Delta}_+(l) \hat{\Delta}_+(k-l) \sin^2 \tfrac{k \sigma l}{2}.
\end{gather}
\end{subequations}
The terms (\ref{subeq:A1_1_2_res}) and (\ref{subeq:A1_1_3_res}) are finite contributions to the continuous spectrum. 
The loop integrals in~(\ref{subeq:A1_1_1_res}) and~(\ref{subeq:A1_1_4_res}) were already computed in the previous section. Thus, after a field strength renormalization for the planar part, one obtains
\begin{equation}
\label{eq:2ptIR0}
 12 e^2 \int \ud^4k \ \hat{f}^{\mu \nu}(-k) \hat{h}^{\lambda}_{\ \nu}(k) k_\mu k_\lambda \hat{\Delta}_+(k) \ln \mu \sqrt{-(k \sigma)^2}
\end{equation}
for the sum of (\ref{subeq:A1_1_1_res}) and (\ref{subeq:A1_1_4_res}).
Here $\mu$ is a mass scale that depends on the renormalization condition.


It remains to discuss the terms involving the kernel (\ref{eq:K1_2}). Using it in the observable with $f^{\mu \nu}$, we obtain
\begin{multline*}
 - 4 (2 \pi)^2 e^2 \int \ud^4k \ \hat f^{\mu \nu}(-k) \hat h^{\lambda \rho}(k) k_\lambda \int \ud^{8} \underline k \ \delta(k - \sum k_i) \frac{\sin \frac{k_1 \sigma k_2}{2} }{\frac{k_1 \sigma k_2}{2} } \sin \tfrac{k \sigma k_2}{2} \\
 \times \left\{ (k\sigma)^{\mu_1} k_{2 \mu} \delta_\nu^{\mu_2} + (k \sigma)^{\mu_2} k_{1 \mu} \delta_\nu^{\mu_1} \right\} \left[ g_{\mu_1 \mu_2} (k_1-k_2)_\rho - g_{\mu_1 \rho} (k + k_1)_{\mu_2} + g_{\mu_2 \rho} (k_2+k)_{\mu_1} \right] \\
  \times \left\{ \left( \hat \Delta_R(k_1) \hat \Delta_+(-k_2) + \hat \Delta_+(k_1) \hat \Delta_R(k_2) \right) \hat \Delta_+(k) + \hat \Delta_+(k_1) \hat \Delta_+(k_2) \hat \Delta_A(k) \right\}.
\end{multline*}
For the two terms in curly brackets, we find, respectively,
\begin{subequations}
\label{subeq:2}
\begin{align}
\label{eq:1stOrderNonlocal}
 - 8 (2\pi)^2 e^2 \int \ud^4 k \ \hat{f}^{\mu \nu}(-k) \hat{h}^{\lambda \rho}(k) k_\mu k_\lambda \hat{\Delta}_+(k) \int \ud^4 l  & \ \hat{\Delta}^{(1)}(l) \hat{\Delta}_R(k-l) \frac{\sin^2 \frac{k \sigma l}{2}}{\frac{k \sigma l}{2}} \\
 & \times \left\{ 2 (k \sigma)_\rho l_\nu - (k \sigma)_\nu l_\rho + \tfrac{k \sigma l}{2} g_{\nu \rho} \right\}, \nonumber \\
\label{subeq:A1_3_3_res}
 - 8 (2\pi)^2 e^2 \int \ud^4 k \ \hat{f}^{\mu \nu}(-k) \hat{h}^{\lambda \rho}(k) k_\mu k_\lambda \hat{\Delta}_A(k) \int \ud^4 l & \ \hat{\Delta}_+(l) \hat{\Delta}_+(k-l) \frac{\sin^2 \frac{k \sigma l}{2}}{\frac{k \sigma l}{2}} \\
 & \times \left\{ 2 (k \sigma)_\rho l_\nu - (k \sigma)_\nu l_\rho + \tfrac{k \sigma l}{2} g_{\nu \rho} \right\}. \nonumber
\end{align}
For the contribution where the kernel (\ref{eq:K1_2}) is used in the second observable, we obtain
\begin{align}
\label{eq:1stOrderNonlocal_A}
 - 8 (2\pi)^2 e^2 \int \ud^4 k \ \hat{f}^{\mu \nu}(-k) \hat{h}^{\lambda \rho}(k) k_\mu k_\lambda \hat{\Delta}_+(k) \int \ud^4 l & \ \hat{\Delta}^{(1)}(l) \hat{\Delta}_A(k-l) \frac{\sin^2 \frac{k \sigma l}{2}}{\frac{k \sigma l}{2}} \\
 & \times \left\{ 2 (k \sigma)_\nu l_\rho - (k \sigma)_\rho l_\nu + \tfrac{k \sigma l}{2} g_{\nu \rho} \right\}, \nonumber \\
\label{subeq:A1_3_2_res}
 - 8 (2\pi)^2 e^2 \int \ud^4 k \ \hat{f}^{\mu \nu}(-k) \hat{h}^{\lambda \rho}(k) k_\mu k_\lambda \hat{\Delta}_R(k) \int \ud^4 l & \ \hat{\Delta}_+(l) \hat{\Delta}_+(k-l) \frac{\sin^2 \frac{k \sigma l}{2}}{\frac{k \sigma l}{2}} \\
 & \times \left\{ 2 (k \sigma)_\nu l_\rho - (k \sigma)_\rho l_\nu + \tfrac{k \sigma l}{2} g_{\nu \rho} \right\}. \nonumber
\end{align}
\end{subequations}
The loop integrals in (\ref{subeq:A1_3_3_res}) and (\ref{subeq:A1_3_2_res}) are well--defined and contribute to the continuous spectrum.
The third term in curly brackets in (\ref{eq:1stOrderNonlocal}), respectively (\ref{eq:1stOrderNonlocal_A}), gives rise to a term proportional to~(\ref{subeq:A1_1_1_res}), respectively (\ref{subeq:A1_1_4_res}). For the sum of these, one obtains,
cf. (\ref{eq:2ptIR0}),
\begin{equation}
\label{eq:2ptIR0_2}
 - 4 e^2 \int \ud^4k \ \hat{f}^{\mu \nu}(-k) \hat{h}^{\lambda}_{\ \nu}(k) k_\mu k_\lambda \hat{\Delta}_+(k) \ln \mu \sqrt{-(k \sigma)^2}.
\end{equation}
The first two terms in curly brackets in (\ref{eq:1stOrderNonlocal}) and (\ref{eq:1stOrderNonlocal_A}), are quite unusual, however. Because of the twisting factor in the denominator, it is not possible to split this contribution into a planar and a nonplanar part. Even worse, there is no obvious way to define this integral rigorously.
But we want to compute it at least formally.
The integral over $l$ formally yields an expression of the form $\Pi_{\nu \rho}(k, \sigma)$. If it is well--defined it should transform properly under Lorentz transformations, i.e.,
\begin{equation*}
  \Pi_{\nu \rho}(k,\sigma) = {\Lambda^{-1}}^{\nu'}_{ \ \nu} {\Lambda^{-1}}^{\rho'}_{ \ \rho} \Pi_{\nu' \rho'}(k \Lambda , \Lambda^{-1} \sigma {\Lambda^{-1}}^T).
\end{equation*}
We consider a timelike $k$ (and later discuss the limit $k^0 \to \betrag{\V{k}}$). With the above formula it suffices to compute $\Pi_{\nu \rho}$ for $k = (k_0, \V{0})$, $k_0>0$ and arbitrary $\sigma$.
For this $k$
\begin{equation}
\label{eq:chiExpression}
  \int \ud^4 l \ \hat{\Delta}^{(1)}(l) \hat{\Delta}_R(k-l) \frac{\sin^2 \frac{k \sigma l}{2}}{\frac{k \sigma l}{2}} l_\nu
\end{equation}
vanishes for $\nu = 0$, because the integrand is antisymmetric under space reflection. The same is true for $\nu = i$ if $e_i$ is perpendicular to $k \sigma$ (now the integrand is antisymmetric under reflection in the direction $k \sigma$). It follows that~(\ref{eq:chiExpression}) is of the form $(k \sigma)_\nu \chi(k)$. We want to compute the function $\chi$ formally. If $e_\nu$ is parallel to and in the same direction as $k \sigma$, we obtain
\begin{align*}
  & - (2\pi)^{-3} \int \ud^3 \V l \ \frac{l_\nu}{\betrag{\V l} (k_0^2 - 4 \betrag{\V l}^2 + i \epsilon)} \frac{\sin^2 \frac{ \Vmat{ k \sigma} \cdot \V{l}}{2}}{\frac{ \Vmat{ k \sigma} \cdot \V{l}}{2}} \\
  = & - (2 \pi)^{-2} \int_0^\infty \ud l \ \frac{l^2}{l (k_0^2 - 4 l^2 + i \epsilon)} \int_{-1}^1 \ud x \ l x \frac{\sin^2 \frac{ \betrag{\Vmat{ k \sigma}} l x}{2}}{\frac{ \betrag{\Vmat{ k \sigma}} l x}{2}} \\
  = & \frac{- 2 (2 \pi)^{-2}}{\betrag{\Vmat{ k \sigma}}} \int_0^\infty \ud l \ \frac{l}{k_0^2 - 4 l^2 + i \epsilon} \left( 1 - \frac{\sin l \betrag{\Vmat{ k \sigma}}}{l \betrag{\Vmat{ k \sigma}}} \right).
\end{align*}
Here $\Vmat{ k \sigma}$ is the spatial part of $k \sigma$. Thus, we have
\begin{equation}
\label{eq:chi}
  \chi(k) = \frac{2 (2\pi)^{-2}}{(k \sigma)^2} \int_0^\infty \ud l \ \frac{l}{k^2 - 4 l^2 + i \epsilon} \left( 1 - \frac{\sin l \sqrt{-(k \sigma)^2}}{l \sqrt{-(k \sigma)^2}} \right).
\end{equation}
The integral is again the formal expression for the difference of the planar and the nonplanar part of the fish graph in the massless $\phi^3$ model, cf. Sections~\ref{sec:phi3} and~\ref{sec:nonplanar}.
Thus, the sum of the first two terms in curly brackets in~(\ref{eq:1stOrderNonlocal}) is formally given by
\begin{equation}
\label{eq:CovCoorDivergence}
 8 (2\pi)^2 \int \ud^4 k \ \hat{f}^{\mu \nu}(-k) \hat{h}^{\lambda \rho}(k) k_\mu k_\lambda \hat{\Delta}_+(k) \frac{(k \sigma)_{\nu} (k \sigma)_{\rho}}{(k \sigma)^2} \left( \Sigma_{pl}(k) - \Sigma_{np}(k) \right), 
\end{equation}
where $\Sigma_{pl}$ and $\Sigma_{np}$ are the planar and nonplanar part of the self--energy of the massless $\phi^3$ model at second order, i.e., they are given by (\ref{eq:F_pl}) and (\ref{eq:F_np_2},b), respectively.
We recall that $\Sigma_{pl}(k)$ is logarithmically divergent. Thus, we have found a divergent quantity that is multiplied with the nonlocal expression $(k \sigma)^{-2}$. It seems as if a nonlocal counterterm is unavoidable.

We recall from Section~\ref{sec:nonplanar} that in the difference between the planar and the nonplanar part in (\ref{eq:CovCoorDivergence}) the imaginary part cancels for $k^2 \to 0$. Thus, the sign of the $i \epsilon$-prescription in (\ref{eq:chi}) is not relevant. It follows that for the sum of the first two terms in curly brackets in (\ref{eq:1stOrderNonlocal_A}), one also obtains~(\ref{eq:CovCoorDivergence}).


\subsection{Second order}

From the contribution (\ref{eq:W0}) to the elementary two--point function and the zeroth order kernel (\ref{eq:K0}), we obtain
\begin{equation*}
  8 e^2 \int \ud^4 k \ \hat f^{\mu \nu}(-k) \hat h^{\lambda \rho}(k) k_\mu k_\lambda \theta(k_0) \theta(k^2) \left\{ \tfrac{g_{\nu \rho}}{k^2} \left( \tilde \Sigma_0(k) + \tilde \Sigma_1(k) \right) + \tfrac{(k \sigma)_\nu (k \sigma)_\rho}{k^4 (k\sigma)^4} \tilde \Sigma_2(k) \right\}.
\end{equation*}
This is a well--defined contribution to the continuous spectrum. From the term (\ref{eq:W1}) we obtain
\begin{equation}
\label{eq:2ptIR1}
 - \frac{20}{3} e^2 \int \ud^4k \ \hat{f}^{\mu \nu}(-k) \hat{h}^{\lambda}_{\ \nu}(k) k_\mu k_\lambda \hat{\Delta}_+(k) \ln \mu \sqrt{-(k \sigma)^2}. 
\end{equation}
Finally, the term (\ref{eq:nonplanarIR3}) yields
\begin{equation}
\label{eq:2ptIR2}
 - 4 e^2 \int \ud^4k \ \hat{f}^{\mu \nu}(-k) \hat{h}^{\lambda \rho}(k) \left( 8 \frac{\del}{\del m^2} \hat{\Delta}_+(k) - \frac{(k \sigma)^2}{3} \hat \Delta_+(k) \right) k_{\mu} k_{\lambda} \frac{(k \sigma)_{\nu} (k \sigma)_{\rho}}{(k \sigma)^4}.
\end{equation}
Once more, we remark that the distribution $\frac{\del}{\del m^2} \hat{\Delta}_+(k)$ is only well--defined on test functions vanishing in a neighborhood of the origin. 

Whether the products of distribution in (\ref{eq:2ptIR1}) and (\ref{eq:2ptIR2}) are well--defined will be discussed in the next section.

\subsection{Summary}

Let us summarize the results of this section. Apart from the contributions (\ref{eq:Finite_1}), (\ref{eq:Finite_2}), (\ref{eq:Finite_3}), (\ref{eq:Finite_4}), (\ref{subeq:A1_1_2_res}), (\ref{subeq:A1_1_3_res}), (\ref{subeq:A1_3_3_res}) and (\ref{subeq:A1_3_2_res}) to the continuous part of the spectrum, we found the following terms:
\begin{itemize}
\item The term (\ref{eq:divergence1}) came from the contraction of the two photons coming in through the covariant coordinates. It is a nonlocal divergence whose subtraction can be interpreted as a normal ordering of the covariant coordinates.

\item The terms (\ref{eq:2ptIR0}), (\ref{eq:2ptIR0_2}) and (\ref{eq:2ptIR1}) are a momentum-dependent field strength renormalization. They sum up to
\begin{equation}
\label{eq:2ptIR1_All}
 - \frac{4}{3} e^2 \int \ud^4k \ \hat{f}^{\mu \nu}(-k) \hat{h}^{\lambda}_{\ \nu}(k) k_\mu k_\lambda \hat{\Delta}_+(k) \ln \mu \sqrt{-(k \sigma)^2}. 
\end{equation}
As we will show in the next section, this expression is well--defined.

\item The term (\ref {eq:CovCoorDivergence}) was obtained formally from (\ref{eq:1stOrderNonlocal}) and (\ref{eq:1stOrderNonlocal_A}). These were the contributions where one power of $e$ came from the covariant coordinate and one from the interacting field. It is a nonlocal expression multiplied with a divergent quantity.

\item The first term in (\ref{eq:2ptIR2}), which arose from the contribution $\Sigma_2$ to the self--energy, is formally a momentum-dependent mass renormalization. As such it was treated in the literature, cf.~\cite{Matusis}. In the next section we will show that this expression is not well--defined even for test functions $\hat{f}$ and $\hat{h}$ vanishing in a neighborhood of the origin. Giving some meaning to this expression introduces nonlocal renormalization ambiguities.
\end{itemize}

\section{Products of distributions and nonlocal renormalization ambiguities}
\label{sec:infrared}

In the preceding section, in (\ref{eq:2ptIR1_All}) and (\ref{eq:2ptIR2}), we encountered products of distributions like $\theta(k_0) \delta(k^2) \ln (- (k \sigma)^2)$ or $\theta(k_0) \delta'(k^2) \frac{1}{(k \sigma)^4}$. Here, we want to discuss the well--definedness of such expressions. We focus on products of the form
\begin{equation}
\label{eq:K}
  \theta(k_0) \delta(k^2) K(-(k\sigma)^2),
\end{equation}
where $K(x)$ might be, e.g., $\ln x$ or $x^{-1}$, i.e., a smooth function apart from a singularity at $x=0$. Products where $\delta(k^2)$ is replaced by $\delta'(k^2)$ can be discussed analogously. Since the tip of the light cone, i.e., the origin $k=0$ might pose additional problems, we (momentarily) exclude it from the following considerations by restricting to test functions that vanish in a neighborhood of the origin.

We start by noticing that, as such, the product (\ref{eq:K}) is not well--defined: The wave front set \cite{Hoermander} of $\theta(k_0) \delta(k^2)$, excluding the origin, is
\begin{equation*}
  \left\{ (k, p_k)| k^2 = 0, k_0 > 0, p_k = \lambda k, \lambda \in \R \setminus \{0\} \right\}.
\end{equation*}
Note that the sign of $\lambda$ is not restricted. The wave front set of $K(-(k\sigma)^2)$ is contained in (once more we excluded the origin)
\begin{equation*}
  \left\{ (k, p_k)| (k\sigma)^2 = 0, k \neq 0, p_k = \lambda k, \lambda \in \R \setminus \{0\} \right\}.
\end{equation*}
This might be further restricted by
$\lambda \gtrless 0$, depending on some $i \epsilon$-prescription. For convenience, we restrict our considerations to the case $\sigma = \sigma_0$, cf.~(\ref{eq:sigma_0}). Using Lorentz invariance, the result then applies to all $\sigma \in \Sigma$. Now for $k=(\kappa,0,\pm \kappa,0), \kappa >0$, we have $k^2=0, k_0>0$ and $(k \sigma_0)^2=0$.
Thus, there is an overlap
\begin{equation*}
  N = \{ k \in \R^4 | k_1 = k_3 = 0 \}
\end{equation*}
of the singular supports. Furthermore, the cotangent components of the wave front sets at some fixed $k \in N$ can always add up to zero, even if there is a restriction on the sign of $\lambda$ in the wave front set of $K(-(k \sigma)^2)$. The reason is that there is no such restriction in the wave front set of $\theta(k_0) \delta(k^2)$. Hence the product is not well--defined in the sense of H\"ormander~\cite{Hoermander}.

However, we may take the following point of view. The product~(\ref{eq:K}) is well--defined on test functions $f$ that vanish in a neighborhood of $N$. Explicitly, we find
\begin{equation*}
  \int \frac{\ud^3 \V k}{2 \betrag{\V{k}}} \ K( 2 \lambda_{nc}^4 \betrag{ \V{k_\bot} }^2) f(\betrag{\V{k}}, \V{k}),
\end{equation*}
where $\V{k_\bot} = (k_1, 0, k_3)$.
Since $f$ vanishes in a neighborhood of the origin, we may define the test function $\tilde f(\V k) = \frac{1}{2 \betrag{\V k}} f( \betrag{\V k}, \V k)$ and write the above as
\begin{equation*}
  \int \ud^3 \V k \ K( 2 \lambda_{nc}^4 \betrag{ \V{k_\bot} }^2) \tilde{f}(\V{k}).
\end{equation*}
We may now ask for the possibility to extend the distribution $K(2 \lambda_{nc}^4 \betrag{ \V{k_\bot} }^2)$ to
\begin{equation*}
\dot N = \{ \V{k} \in \R^3 | \betrag{\V{k_\bot}}=0, \betrag{k_2} > 0 \}.
\end{equation*}
The possibility to do this is governed by its scaling degree
at $\dot N$ \cite{BrunettiFredenhagen}. Since $K$ is only singular at the origin, and the wave front set of $K(2 \lambda_{nc}^4 \betrag{ \V{k_\bot} }^2)$ is orthogonal to the normal bundle of $\dot N$, the scaling degree can be computed simply by scaling $\V{k_\bot}$. In the case $K(x) = \ln x$, we obtain $0$, while in the case $K(x) = 1/x$, we get $2$. Since the codimension of $\dot N$ in $\R^3$ is $2$, the extension is unique in the first case, but nonunique in the second \cite[Thm.~6.9]{BrunettiFredenhagen}. Thus, in the second case, a nonlocal counterterm is needed. Finally, after extending the distribution to $\dot N$, we can further extend it to the origin. This extension is still unique in the case $K(x) = \ln x$, corresponding to (\ref{eq:2ptIR1_All}) and nonunique in the case $K(x) = 1/x$, which corresponds to the second term in (\ref{eq:2ptIR2}). It is easy to see that the problems become even worse when the product $\theta(k_0) \delta'(k^2) \frac{1}{(k \sigma)^4}$ that occurs in the first term in (\ref{eq:2ptIR2}) is considered.

We have thus shown that~(\ref{eq:2ptIR1_All}) is well--defined, while the extension of the product of distributions in~(\ref{eq:2ptIR2}) has nonlocal ambiguities, i.e., a continuum of renormalization conditions. Note that it does not help to assume only space/space noncommutativity. Then $(k \sigma)^2 = - \lambda_{nc}^4 \betrag{ \V{k_\bot} }^2$, where $\V{k_\bot}$ is the projection of $k$ on the plane spanned by the noncommuting directions. It follows that $\frac{(k \sigma)_\mu (k \sigma)_\nu}{(k \sigma)^4}$ is proportional to $\frac{1}{\betrag{\V{k_\bot}}^2}$ for $\mu$ and $\nu$ in the noncommuting directions. This is still too singular.


\chapter{Supersymmetric NCQED}
\label{chapter:SNCQED}

The upshot of the preceding chapter was that NCQED is plagued by numerous nonlocal divergences and can thus at best be considered as an effective theory. It is a natural question whether a supersymmetric version of the theory behaves better. This is the subject of the present chapter.

It has been noticed in \cite{Matusis} that the problematic term~(\ref{eq:Hayakawa_Sigma_2}) disappears in a supersymmetric version of the theory, at least in the setting of the modified Feynman rules\footnote{\label{footnote:BrokenSUSY}However, this seems to break down if supersymmetry is broken in such a way that $M^2$, the supertrace over the squared masses, does not vanish~\cite{Carlson, Alvarez-GaumeGeneral}. One then obtains $\Sigma_2((k \sigma)^2, k^2) \propto (k \sigma)^2 M^2$. This gives rise to a product $\theta(k_0) \delta'(k^2) \frac{1}{(k \sigma)^2}$, which also has nonlocal ambiguities.
}. Because of our experience with the Wess--Zumino model, it is reasonable to hope that this is also true in the Yang--Feldman formalism.

A simple way to introduce supersymmetry is to add a Weyl fermion~$\lambda$, the photino, and an auxiliary field $D$. Both fields transform in the adjoint representation. Then one can postulate the action
\begin{equation}
\label{eq:SNCQED_Action}
  \int \ud^4q \ \left( - \frac{1}{4} F_{\mu \nu} F^{\mu \nu} + i \bar \lambda \bar \sigma^\mu D_\mu \lambda + 2 D^2 \right)
\end{equation}
and work out the consequences. For the elementary two--point function at second order in $e$ there would be one extra term, the photino loop. We will show in Section~\ref{sec:SNCQED_SelfEnergy} that this loop cancels the $\Sigma_2$-term (\ref{eq:Hayakawa_Sigma_2}) also in the Yang--Feldman formalism. But obviously, the problems with the terms stemming from the covariant coordinates remain. However, we notice that while the observable $\int \ud^4q \ f^{\mu \nu}(X) F_{\mu \nu}$ is gauge invariant, it is not invariant under the supersymmetry transformation\footnote{Note that this transformation is nonlinear. This is due to the fact that we implicitly adopted the Wess--Zumino gauge.}
\begin{align*}
  \delta_\xi A^\mu & = i \xi \sigma^\mu \bar \lambda + i \bar \xi \bar \sigma^\mu \lambda, \\
  \delta_\xi \lambda_\alpha & = - (\sigma^{\mu \nu})_\alpha^{\ \beta} F_{\mu \nu} \xi_\beta - 2 i \xi_\alpha D, \\
  \delta_\xi D & = \frac{1}{2} ( \bar \xi \bar \sigma^\mu D_\mu \lambda - \xi \sigma^\mu D_\mu \bar \lambda ).
\end{align*}
Here $\xi$ is an infinitesimal anticommuting parameter. For our conventions on Weyl spinors and supersymmetry, see Appendix~\ref{app:SUSY}. In order to construct observables that are invariant under supersymmetry transformations, it is advantageous to work in the superfield formalism. Thus, we embed our fields in the vector multiplet $V$ by defining
\begin{equation*}
  V = - \theta \sigma^\mu \bar \theta A_\mu + i \theta^2 \bar \theta \bar{\lambda} - i \bar{\theta}^2 \theta \lambda + \theta^2 \bar{\theta}^2 D.
\end{equation*}
Because of the anticommutativity of the $\theta$'s, we have $V^3=0$. An infinitesimal gauge transformation can now be written as
\begin{equation}
\label{eq:delta_V}
  \delta_\Lambda V = \frac{i}{2e} (\bar{\Lambda} - \Lambda) - \frac{i}{2} [V, \bar{\Lambda}+ \Lambda],
\end{equation}
where $\Lambda$ is a chiral field given by
\begin{equation*}
  \Lambda(q, \theta, \bar \theta) = e^{-i \theta \sigma^\mu \bar \theta \del_\mu} \chi(q).
\end{equation*}
Here $\chi$ is the usual infinitesimal gauge parameter. Because of the rather complicated form of the gauge transformation~(\ref{eq:delta_V}), it is advantageous to introduce yet another superfield, namely
\begin{equation*}
  W_\alpha = - \frac{1}{4e} \bar{D}^2 (e^{-2eV} D_\alpha e^{2eV}) = - \frac{1}{2} \bar{D}^2 ( D_\alpha V - e [V, D_\alpha V] ).
\end{equation*}
It transforms in the adjoint representation, i.e., as
\begin{equation}
\label{eq:delta_W}
  \delta_\Lambda W_\alpha = i [\Lambda, W_\alpha].
\end{equation}
Because of the anticommutativity of the $\bar D^\dotalpha$'s, $W_\alpha$ is chiral. In component form, it is given by
\begin{equation}
\label{eq:W_alpha_Components}
  W_\alpha = - 2 i \lambda_\alpha + 2 i {{\sigma^{\mu \nu}}_\alpha}^\beta \theta_\beta F_{\mu \nu} + 4 \theta_\alpha D - 2 \theta^2 \sigma^\mu_{\alpha \dotalpha} D_\mu \bar \lambda^\dotalpha + \order(\bar \theta).
\end{equation}
The action~(\ref{eq:SNCQED_Action}) can now be expressed as
\begin{equation*}
  S = \frac{1}{16} \int \ud^6q \ W^\alpha W_\alpha + \text{ h.c.}
\end{equation*}
Properly, one should also embed the ghosts into (chiral) supermultiplets. However, their superpartners will not contribute to the two--point function at second order, in which we are mainly interested.

\section{The Yang--Feldman procedure}

From the action~(\ref{eq:SNCQED_Action}), we obtain the following equations of motion for the photino:
\begin{equation*}
  \bar \sigma^\mu \del_\mu \lambda - i e \bar \sigma^\mu [A_\mu, \lambda] = 0, \quad
  \sigma^\mu \del_\mu \bar \lambda - i e \sigma^\mu [A_\mu, \bar \lambda] = 0.
\end{equation*}
Thus, the first order contribution to the interacting field is
\begin{equation*}
  \lambda_1 = S_R \times \bar \sigma^\mu [A_{\mu 0}, \lambda_0], \quad
  \bar \lambda_1 = \bar S_R \times \sigma^\mu [A_{\mu 0}, \bar \lambda_0].
\end{equation*}
Here we used the notation
\begin{equation*}
  S_R = i \sigma^\nu \del_\nu \Delta_R, \quad
  \bar S_R = i \bar \sigma^\nu \del_\nu \Delta_R.
\end{equation*}
In the same manner as in the preceding chapter, we obtain the vertex factors
\vspace{-20pt}
\begin{center}
\begin{picture}(50,60)(0,20)
\ArrowLine(24.5,15)(24.5,30)
\ArrowLine(25.5,15)(25.5,30)
\Photon(14,40)(25,30){1}{3}
\ArrowLine(25,30)(35,40)
\Text(25,13)[t]{$\dotalpha k_1$}
\Text(15,42)[rb]{$\mu k_2$}
\Text(35,42)[lb]{$\alpha k_3$}
\Vertex(25,30){1}
\end{picture}
$ = - i e \sigma^\mu_{\alpha \dot \alpha} \sin \frac{k_2 \sigma k_3}{2} \delta(k_1+k_2+k_3) = $
\begin{picture}(50,60)(0,20)
\ArrowLine(24.5,30)(24.5,15)
\ArrowLine(25.5,30)(25.5,15)
\Photon(14,40)(25,30){1}{3}
\ArrowLine(35,40)(25,30)
\Text(25,13)[t]{$\alpha k_1$}
\Text(15,42)[rb]{$\mu k_2$}
\Text(35,42)[lb]{$\dotalpha k_3$}
\Vertex(25,30){1}
\end{picture}
\end{center}
\vspace{20pt}
The photino propagators are given by
\vspace{-5pt}
\begin{center}
\begin{picture}(50,15)(0,5)
\ArrowLine(40,7)(15,7)
\BCirc(15,7){1}
\BCirc(40,7){1}
\Text(13,7)[r]{$\dot \alpha k$}
\Text(42,7)[l]{$\alpha$}
\end{picture}
$ = (2\pi)^2 \sigma^\mu_{\alpha \dotalpha} k_\mu \hat \Delta_+(k) = $
\begin{picture}(50,15)(0,5)
\ArrowLine(15,7)(40,7)
\BCirc(15,7){1}
\BCirc(40,7){1}
\Text(13,7)[r]{$\alpha k$}
\Text(42,7)[l]{$\dot \alpha$}
\end{picture}
\end{center}
\begin{center}
\begin{picture}(50,15)(0,5)
\ArrowLine(40,6.5)(15,6.5)
\ArrowLine(40,7.5)(15,7.5)
\Text(13,7)[r]{$\dot \alpha k$}
\Text(42,7)[l]{$\alpha$}
\end{picture}
$ = (2\pi)^2 \sigma^\mu_{\alpha \dotalpha} k_\mu \hat \Delta_R(k) = $
\begin{picture}(50,15)(0,5)
\ArrowLine(15,6.5)(40,6.5)
\ArrowLine(15,7.5)(40,7.5)
\Text(13,7)[r]{$\alpha k$}
\Text(42,7)[l]{$\dot \alpha$}
\end{picture}
\end{center}
\vspace{5pt}

The introduction of the photino changes the equation of motion~(\ref{eq:eom1b}) for $A^\mu$: To the right hand side we have to add
\begin{equation}
\label{eq:ExtraTerm}
  e \sigma^\mu_{\alpha \dot \alpha} \{ \lambda^\alpha, \bar \lambda^{\dot \alpha} \}.
\end{equation}
This leads to the vertex factors
\vspace{-20pt}
\begin{center}
\begin{picture}(50,60)(0,20)
\Photon(24.5,15)(24.5,30){1}{3}
\Photon(25.5,15)(25.5,30){1}{3}
\ArrowLine(14,40)(25,30)
\ArrowLine(25,30)(35,40)
\Text(25,13)[t]{$\mu_1 k_1$}
\Text(15,42)[rb]{$\dot \alpha k_2$}
\Text(35,42)[lb]{$\alpha k_3$}
\Vertex(25,30){1}
\end{picture}
$ = - i e \sigma^\mu_{\alpha \dot \alpha} \sin \frac{k_2 \sigma k_3}{2} \delta(k_1+k_2+k_3) = - $
\begin{picture}(50,60)(0,20)
\Photon(24.5,15)(24.5,30){1}{3}
\Photon(25.5,15)(25.5,30){1}{3}
\ArrowLine(25,30)(14,40)
\ArrowLine(35,40)(25,30)
\Text(25,13)[t]{$\mu_1 k_1$}
\Text(15,42)[rb]{$\alpha k_3$}
\Text(35,42)[lb]{$\dot \alpha k_2$}
\Vertex(25,30){1}
\end{picture}
\end{center}
\vspace{20pt}

\begin{remark}
We recall from the previous chapter that the old product (\ref{eq:OldProduct2}) of quantum fields led to a divergent vacuum expectation value of $A_1$, which was one of the reasons for the introduction of the symmetrized product. We note that the extra term (\ref{eq:ExtraTerm}) would cancel this divergence. But since the product (\ref{eq:OldProduct2}) also had other (conceptual) problems, we will still use the symmetrized product in the following.
\end{remark}

The above vertices and propagators lead to the elementary three-point functions
\begin{subequations}
\begin{align}
\label{eq:W3_SUSY_1}
  W_{\dotalpha \alpha \mu}(\underline k) = & 2 i e (2\pi)^4 (\sigma_\rho \bar \sigma_\mu \sigma_\lambda)_{\alpha \dotalpha} k_1^\lambda k_2^\rho \sin \tfrac{k_2 \sigma k_3}{2} \delta(\sum k_i) \\
 \times & \left( \left[ \hat \Delta_R(k_1) \hat \Delta_+(-k_2) + \hat \Delta_+(k_1) \hat \Delta_R(k_2) \right] \hat \Delta_+(-k_3) +\hat \Delta_+(k_1) \hat \Delta_+(k_2) \hat \Delta_R(k_3) \right), \nonumber \\
\label{eq:W3_SUSY_2}
  W_{\alpha \dotalpha \mu}(\underline k) = & 2 i e (2\pi)^4 (\sigma_\lambda \bar \sigma_\mu \sigma_\rho)_{\alpha \dotalpha} k_1^\lambda k_2^\rho \sin \tfrac{k_2 \sigma k_3}{2} \delta(\sum k_i) \\
 \times & \left( \left[ \hat \Delta_R(k_1) \hat \Delta_+(-k_2) + \hat \Delta_+(k_1) \hat \Delta_R(k_2) \right] \hat \Delta_+(-k_3) +\hat \Delta_+(k_1) \hat \Delta_+(k_2) \hat \Delta_R(k_3) \right), \nonumber \\
\label{eq:W3_SUSY_3}
 W_{\mu \dotalpha \alpha}(\underline k) = & 2 i e (2\pi)^4 (\sigma_\rho \bar \sigma_\mu \sigma_\lambda)_{\alpha \dotalpha} k_2^\lambda k_3^\rho \sin \tfrac{k_2 \sigma k_3}{2} \delta(\sum k_i) \\
  \times & \left( \left[ \hat \Delta_R(k_1) \hat \Delta_+(-k_2) + \hat \Delta_+(k_1) \hat \Delta_R(k_2) \right] \hat \Delta_+(-k_3) +\hat \Delta_+(k_1) \hat \Delta_+(k_2) \hat \Delta_R(k_3) \right), \nonumber \\
\label{eq:W3_SUSY_4}
 W_{\mu \alpha \dotalpha}(\underline k) = & 2 i e (2\pi)^4 (\sigma_\lambda \bar \sigma_\mu \sigma_\rho)_{\alpha \dotalpha} k_2^\lambda k_3^\rho \sin \tfrac{k_2 \sigma k_3}{2} \delta(\sum k_i) \\
  \times & \left( \left[ \hat \Delta_R(k_1) \hat \Delta_+(-k_2) + \hat \Delta_+(k_1) \hat \Delta_R(k_2) \right] \hat \Delta_+(-k_3) +\hat \Delta_+(k_1) \hat \Delta_+(k_2) \hat \Delta_R(k_3) \right). \nonumber
\end{align}
\end{subequations}
We also have the elementary four-point functions
\begin{align}
\label{eq:W4_SUSY_1}
  W_{\dotalpha \alpha \mu \nu}(\underline k) = & (2 \pi)^4 g_{\mu \nu} \sigma^\lambda_{\alpha \dotalpha} k_{1 \lambda} \hat \Delta_+(k_1) \hat \Delta_+(k_3) \delta(k_1+k_2) \delta(k_3+k_4), \\
\label{eq:W4_SUSY_2}
  W_{\dotalpha \alpha \dotbeta \beta}(\underline k) = & (2 \pi)^4 \left( \sigma^\lambda_{\alpha \dotalpha} k_{1 \lambda} \sigma^\rho_{\beta \dotbeta} k_{3 \rho} \hat \Delta_+(k_1) \hat \Delta_+(k_3) \delta(k_1+k_2) \delta(k_3+k_4) \right. \\
  & \left. + \sigma^\lambda_{\alpha \dotbeta} k_{1 \lambda} \sigma^\rho_{\beta \dotalpha} k_{2 \rho} \hat \Delta_+(k_1) \hat \Delta_+(k_2) \delta(k_1+k_4) \delta(k_2+k_3) \right). \nonumber
\end{align}

\section{Covariant Coordinates}

We want to construct observables for the field strength that are not only invariant under gauge, but also under supersymmetry transformations. The easiest way to achieve the latter is to express
\begin{equation*}
  \int \ud^4q \ f^{\mu \nu}(q) F_{\mu \nu}
\end{equation*}
in superfield form. Using the component form (\ref{eq:W_alpha_Components}) of $W_\alpha$, one can show that with
\begin{equation}
\label{eq:f_alpha}
  f_\alpha = - \frac{i}{2} {{\sigma^{\mu \nu}}_\alpha}^\beta \theta_\beta f_{\mu \nu},
\end{equation}
one obtains
\begin{equation*}
  \int \ud^4q \ f^{\mu \nu}(q) F_{\mu \nu} = \int \ud^6q \ f^\alpha(q) W_\alpha + \text{ h.c.}
\end{equation*}
It remains to find the appropriate covariant coordinates. For this, we do not proceed analogously to Chapter~\ref{chapter:NCGaugeTheory} and start with connections on projective modules over noncommutative superspace.
The reason is that the path from general connections in superspace to gauge theories as we know them is rather involved, already in the commutative case (see, e.g., \cite[Chapter~10]{Sohnius}). Thus, we will take the simpler approach to directly look for $X^\mu$'s that transform covariantly, as~(\ref{eq:delta_W}). On the basis of our experience from Chapter~\ref{chapter:NCGaugeTheory}, it will not be hard to guess the right object. We make the ansatz
\begin{equation*}
  X^\mu = q^\mu + e \sigma^{\mu \nu} Y_\nu.
\end{equation*}
In order for $X^\mu$ to transform covariantly, $Y_\nu$ must transform as
\begin{equation}
\label{eq:delta_Y}
  \delta_\Lambda Y_\nu = \frac{1}{e} \del_\nu \Lambda + i [\Lambda, Y_\nu].
\end{equation}
We also know that $Y_\nu$ should involve the vector potential $A_\nu$. It is then not so difficult to guess
\begin{equation*}
  Y_\nu = \frac{1}{4e} \bar{\sigma}_\nu^{\dot \alpha \alpha} \bar{D}_{\dot \alpha} \left( e^{-2eV} D_\alpha e^{2eV} \right) = \frac{1}{2} \bar{\sigma}_\nu^{\dot \alpha \alpha} \bar{D}_{\dot \alpha} ( D_\alpha V - e [V, D_\alpha V] ).
\end{equation*}
As can be shown straightforwardly, this transforms under $\delta_\Lambda$ to
\begin{equation*}
 - \frac{i}{4e} \bar{\sigma}_\nu^{\dot \alpha \alpha} \bar{D}_{\dot \alpha} D_\alpha \Lambda + i [\Lambda, Y_\nu].
\end{equation*}
Using~(\ref{eq:D_anticomm}) and~(\ref{eq:sigma_trace}), we exactly recover~(\ref{eq:delta_Y}). In components, we have
\begin{equation}
\label{eq:Y}
  Y_\nu = A_\nu - i \theta \sigma_\nu \bar \lambda + i \lambda \sigma_\nu \bar \theta + \text{ higher orders in } \theta, \bar \theta.
\end{equation}
Thus, the body of $Y_\nu$ is just $A_\nu$, as one would expect. Note that $Y_\nu$ is, as $W_\alpha$, not real. But this does not make the definition of $f^\alpha(X)$ problematic, since one simply Taylor expands in $\theta, \bar \theta$, similarly to the use of chiral coordinates. Since $A_\nu$ is real, the coefficient at each order of $\theta, \bar \theta$ can be defined \`a la Weyl.

Therefore, instead of $\int \ud^4q \ f^{\mu \nu}(X) F_{\mu \nu}$, we consider the observable
\begin{equation}
\label{eq:SUSY_F_mu_nu_X}
 \int \ud^6q \ f^\alpha(X) W_\alpha + \text{ h.c.},
\end{equation}
with $f^\alpha$ given by~(\ref{eq:f_alpha}), i.e., containing only a $\theta$-component. But, as we saw in~(\ref{eq:Y}), $Y_\nu$ also has a $\theta$-component, $-i\theta \sigma_\nu \bar \lambda$. It follows that $f^\alpha(X)$ has a $\theta^2$-component that involves $\bar \lambda$. This, together with the body of $W_\alpha$, also contributes to~(\ref{eq:SUSY_F_mu_nu_X}). As can be seen from~(\ref{eq:W_alpha_Components}), the body of $W_\alpha$ is $- 2 i \lambda_\alpha$. Thus, the observable~(\ref{eq:SUSY_F_mu_nu_X}) also contains a component with a product of $\bar \lambda$ and $\lambda$. We compute it explicitly for $f^\alpha$ given by~(\ref{eq:f_alpha}), to first order in $e$:
\begin{multline*}
 - (2\pi)^{-2} \int \ud^4k \ ( - \tfrac{i}{2} )( \sigma^{\mu \nu})_\alpha^{\ \beta}  \hat{f}_{\mu \nu}(-k) (i e k\sigma)_\lambda \sigma^\lambda_{\gamma \dot \alpha} \\
  \times \int \ud^6q \ \theta_\beta e^{ikq} \theta^\gamma (-i) P_1(k \sigma \del) \bar \lambda^{\dot \alpha} (-2i) \lambda^\alpha + \text{ h.c.}
\end{multline*}
This can be brought to the form
\begin{equation}
\label{eq:SUSY_CovCoor}
 - \frac{e}{2} (2\pi)^{-2} \int \ud^4k \ \hat{f}^{\mu \nu}(-k) (k\sigma)^\lambda  ( \bar \sigma_{\mu \nu} \bar \sigma_\lambda)^{\dotalpha \alpha} \int \ud^4q \ e^{ikq} P_1(k \sigma \del) \bar{\lambda}_{\dot \alpha} \lambda_\alpha + \text{ h.c.}
\end{equation}
Because of (\ref{eq:3sigmas_2}), we have
\begin{equation}
\label{eq:2sigmas}
  \bar \sigma_{\mu \nu} \bar \sigma_\lambda = \tfrac{1}{2} \left( - g_{\mu \lambda} \bar \sigma_\nu + g_{\nu \lambda} \bar \sigma_\mu - i \epsilon_{\mu \nu \lambda \kappa} \bar \sigma^\kappa \right).
\end{equation}
When we add the hermitean conjugate in (\ref{eq:SUSY_CovCoor}), the first two terms in (\ref{eq:2sigmas}) drop out (this is due to the presence of $(k \sigma)^\lambda$, which changes sign under conjugation). Employing the symmetrized product of $f^\alpha(X)$ and $W_\alpha$, we obtain the supplementary first order kernels
\begin{subequations}
\begin{align}
\label{eq:K1_3}
  K_{\mu \nu}^{\dotalpha \alpha}(k;k_1,k_2) = & \frac{i}{4} (2\pi)^{-2} \delta(k-k_1-k_2) (k\sigma)^\lambda \epsilon_{\mu \nu \lambda \kappa} ( \bar \sigma^\kappa )^{\dotalpha \alpha} \frac{\sin \frac{k \sigma k_1}{2}}{\frac{k \sigma k_1}{2}}, \\
\label{eq:K1_4}
  K_{\mu \nu}^{\alpha \dotalpha}(k;k_1,k_2) = & - \frac{i}{4} (2\pi)^{-2} \delta(k-k_1-k_2) (k\sigma)^\lambda \epsilon_{\mu \nu \lambda \kappa} ( \bar \sigma^\kappa )^{\dotalpha \alpha} \frac{\sin \frac{k \sigma k_1}{2}}{\frac{k \sigma k_1}{2}}.
\end{align}
\end{subequations}

Each further $\bar \lambda$ coming in through the covariant coordinate would bring in another $\theta$, so there are no terms with more than two photinos. However, the covariant coordinate can provide for arbitrary powers of $A_\mu$. We find, in a calculation similar to the one that led to (\ref{eq:K2_2}), the second order kernels
\begin{subequations}
\begin{align}
\label{eq:K2_3}
  K_{\mu \nu}^{\rho \dotalpha \alpha}(k; \underline k) = & \frac{i}{4} (2\pi)^{-4} \delta(k-\sum k_i) (k \sigma)^\rho (k \sigma)^\lambda \epsilon_{\mu \nu \lambda \kappa} ( \bar \sigma^\kappa )^{\dotalpha \alpha} e^{-\frac{i}{2} (k_1+k_2) \sigma k_3} \\
  & \times \left( P_2(-ik\sigma k_1, - i k\sigma k_2) e^{- \frac{i}{2} k_1 \sigma k_2} + P_2(-ik\sigma k_2, - i k\sigma k_1) e^{ \frac{i}{2} k_1 \sigma k_2} \right), \nonumber \\
\label{eq:K2_4}
  K_{\mu \nu}^{\rho \alpha \dotalpha}(k; \underline k) = & - \frac{i}{4} (2\pi)^{-4} \delta(k-\sum k_i) (k \sigma)^\rho (k \sigma)^\lambda \epsilon_{\mu \nu \lambda \kappa} ( \bar \sigma^\kappa )^{\dotalpha \alpha} e^{-\frac{i}{2} (k_1+k_3) \sigma k_2} \\
  & \times \left( P_2(-ik\sigma k_1, - i k\sigma k_3) e^{- \frac{i}{2} k_1 \sigma k_3} + P_2(-ik\sigma k_3, - i k\sigma k_1) e^{ \frac{i}{2} k_1 \sigma k_3} \right). \nonumber
\end{align}
\end{subequations}




\section{The two--point function}

Now we want to compute the new contributions to the full two--point function (\ref{eq:NCQED2pt}) at second order. Our hope is that these cancel the nonlocal divergences found in the previous chapter. 
That the problematic term $\Sigma_2$ in the the self--energy is cancelled by the photino loop was shown in \cite{Matusis} in the setting of the modified Feynman rules.
We prove that this is the case in the Yang--Feldman formalism, too. We also show that the contributions coming from the first order kernels (\ref{eq:K1_3},b) cancel the nonlocal divergence (\ref{eq:CovCoorDivergence}). However, the term (\ref{eq:divergence1}), which arose from the contraction of two photons coming in through the covariant coordinate, remains.

\subsection{The self--energy}
\label{sec:SNCQED_SelfEnergy}

We have to compute the graphs
\begin{center}
\begin{picture}(60,55)
\Photon(29.5,5)(29.5,20){1}{3}
\Photon(30.5,5)(30.5,20){1}{3}
\ArrowLine(10,50)(30,20)
\ArrowLine(29.5,20.3)(39.5,35.3)
\ArrowLine(30.5,19.7)(40.5,34.7)
\ArrowLine(40,35)(30,50)
\Photon(40,35)(50,50){1}{4}
\ArrowLine(30,50)(10,50)
\Vertex(30,20){1}
\Vertex(40,35){1}
\BCirc(10,50){1}
\BCirc(30,50){1}
\BCirc(50,50){1}
\Text(32,5)[l]{$\mu$}
\end{picture}
\begin{picture}(60,55)
\Photon(29.5,5)(29.5,20){1}{3}
\Photon(30.5,5)(30.5,20){1}{3}
\ArrowLine(30,20)(10,50)
\ArrowLine(39.5,35.3)(29.5,20.3)
\ArrowLine(40.5,34.7)(30.5,19.7)
\ArrowLine(30,50)(40,35)
\Photon(40,35)(50,50){1}{4}
\ArrowLine(10,50)(30,50)
\Vertex(30,20){1}
\Vertex(40,35){1}
\BCirc(10,50){1}
\BCirc(30,50){1}
\BCirc(50,50){1}
\Text(32,5)[l]{$\mu$}
\end{picture}
\begin{picture}(60,55)
\Photon(29.5,5)(29.5,20){1}{3}
\Photon(30.5,5)(30.5,20){1}{3}
\ArrowLine(50,50)(30,20)
\ArrowLine(29.5,19.7)(19.5,34.7)
\ArrowLine(30.5,20.3)(20.5,35.3)
\Photon(20,35)(10,50){1}{4}
\ArrowLine(20,35)(30,50)
\ArrowLine(30,50)(50,50)
\Vertex(30,20){1}
\Vertex(20,35){1}
\BCirc(10,50){1}
\BCirc(30,50){1}
\BCirc(50,50){1}
\Text(32,5)[l]{$\mu$}
\end{picture}
\begin{picture}(60,55)
\Photon(29.5,5)(29.5,20){1}{3}
\Photon(30.5,5)(30.5,20){1}{3}
\ArrowLine(30,20)(50,50)
\ArrowLine(19.5,34.7)(29.5,19.7)
\ArrowLine(20.5,35.3)(30.5,20.3)
\Photon(20,35)(10,50){1}{4}
\ArrowLine(30,50)(20,35)
\ArrowLine(50,50)(30,50)
\Vertex(30,20){1}
\Vertex(20,35){1}
\BCirc(10,50){1}
\BCirc(30,50){1}
\BCirc(50,50){1}
\Text(32,5)[l]{$\mu$}
\end{picture}
\end{center}
\vspace{2pt}
These have to be counted twice in order to account for the graphs where the photon leaves the second vertex to the other side. With this factor, we obtain, respectively,
\begin{align*}
 & - 2 (2\pi)^2 e^2 \hat \Delta_R(k) \hat A_0^\nu(k) \tr ( \bar \sigma_\lambda \sigma_\mu \bar \sigma_\rho \sigma_\nu ) \int \ud^4 l \ \hat \Delta_R(k-l) \hat \Delta_+(l) (k-l)^\lambda l^\rho \sin^2 \tfrac{k \sigma l}{2}, \\
 & - 2 (2\pi)^2 e^2 \hat \Delta_R(k) \hat A_0^\nu(k) \tr ( \bar \sigma_\rho \sigma_\mu \bar \sigma_\lambda \sigma_\nu ) \int \ud^4 l \ \hat \Delta_R(k-l) \hat \Delta_+(l) (k-l)^\lambda l^\rho \sin^2 \tfrac{k \sigma l}{2}, \\
 & - 2 (2\pi)^2 e^2 \hat \Delta_R(k) \hat A_0^\nu(k) \tr ( \bar \sigma_\lambda \sigma_\mu \bar \sigma_\rho \sigma_\nu ) \int \ud^4 l \ \hat \Delta_R(k-l) \hat \Delta_+(-l) (k-l)^\lambda l^\rho \sin^2 \tfrac{k \sigma l}{2}, \\
 & - 2 (2\pi)^2 e^2 \hat \Delta_R(k) \hat A_0^\nu(k) \tr ( \bar \sigma_\rho \sigma_\mu \bar \sigma_\lambda \sigma_\nu ) \int \ud^4 l \ \hat \Delta_R(k-l) \hat \Delta_+(-l) (k-l)^\lambda l^\rho \sin^2 \tfrac{k \sigma l}{2}.
\end{align*}
Thus, the photino contribution to the self--energy is
\begin{equation*}
  \pi_{\mu \nu}(k,l) = - (k-l)^\lambda l^\rho \left\{ \tr ( \bar \sigma_\lambda \sigma_\mu \bar \sigma_\rho \sigma_\nu ) + \tr ( \bar \sigma_\rho \sigma_\mu \bar \sigma_\lambda \sigma_\nu ) \right\}.
\end{equation*}
Because of~(\ref{eq:sigma_trace}) and~(\ref{eq:3sigmas_1}), we have
\begin{equation}
\label{eq:sigmaTrace}
  \tr ( \sigma^\kappa \bar \sigma^\lambda \sigma^\mu \bar \sigma^\nu ) = 2 \left( - g^{\kappa \mu} g^{\lambda \nu} +  g^{\lambda \mu} g^{\kappa \nu} + g^{\kappa \lambda} g^{\mu \nu} + i \epsilon^{\kappa \lambda \mu \nu} \right).
\end{equation}
In the sum above the imaginary parts cancel each other, and we obtain
\begin{equation*}
  \pi_{\mu \nu}(k,l) = 4 g_{\mu \nu} k \cdot l - 4 (k-l)_\mu l_\nu - 4 (k-l)_\mu l_\nu.
\end{equation*}
This cancels the term (\ref{subeq:f_tot_2}).
Therefore we have shown that the term $\Sigma_2$, which requires nonlocal counterterms, vanishes upon introducing supersymmetry, also in the Yang--Feldman formalism. The final expression for the self--energy is thus
\begin{equation}
\label{eq:Pi_final}
  \Pi_{\mu \nu}(k) = - (2\pi)^{-2} e^2 (g_{\mu \nu} k^2 - k_\mu k_\nu) \left( \ln \mu \sqrt{-(k \sigma)^2} + \order( k^2 (k \sigma)^2) \right).
\end{equation}
Its contribution to the full two--point function (\ref{eq:NCQED2pt}) is
\begin{equation}
\label{eq:2ptSUSY}
  - 4 e^2 \int \ud^4k \ \hat{f}^{\mu \nu}(-k) \hat{h}^\lambda_{\ \nu}(k) k_\mu k_\lambda \hat{\Delta}_+(k) \ln \mu \sqrt{-(k \sigma)^2}.
\end{equation}

\subsection{The two--particle spectrum}

We have to compute the graph
\begin{center}
\begin{picture}(80,53)
\Photon(19.5,5)(19.5,20){1}{3}
\Photon(20.5,5)(20.5,20){1}{3}
\ArrowLine(10,35)(20,20)
\ArrowLine(20,20)(30,35)
\Photon(59.5,5)(59.5,20){1}{3}
\Photon(60.5,5)(60.5,20){1}{3}
\ArrowLine(50,35)(60,20)
\ArrowLine(60,20)(70,35)
\ArrowLine(30,35)(50,35)
\ArrowArc(40,5)(42.3,45,135)
\Vertex(20,20){1}
\Vertex(60,20){1}
\BCirc(10,35){1}
\BCirc(30,35){1}
\BCirc(50,35){1}
\BCirc(70,35){1}
\end{picture}
\end{center}
and quadruple the result, in order to account for the different possible directions of the arrows at the vertices. We obtain
\begin{equation*}
  4 e^2 (2\pi)^4 \delta(k+p) \hat \Delta_R(k) \hat \Delta_A(k) \int \ud^4 l \ \hat \Delta_+(l) \hat \Delta_+(k-l) \sin^2 \tfrac{k \sigma l}{2} \tr( \sigma_\lambda \bar \sigma_\mu \sigma_\rho \bar \sigma_\nu) (k-l)_\lambda l_\rho.
\end{equation*}
Using (\ref{eq:sigmaTrace}), this reduces to
\begin{multline*}
  4 e^2 (2\pi)^4 \delta(k+p) \hat \Delta_R(k) \hat \Delta_A(k) \int \ud^4 l \ \hat \Delta_+(l) \hat \Delta_+(k-l) \sin^2 \tfrac{k \sigma l}{2} \\
  \times \left\{ - g_{\mu \nu} k^2 + 2 (k-l)_\mu l_\nu + 2 l_\mu (k-l)_\nu \right\}.
\end{multline*}
Adding this to the contribution (\ref{eq:2particleIntegral}), we obtain
\begin{equation*}
  4 e^2 (2\pi)^4 \delta(k+p) \hat \Delta_R(k) \hat \Delta_A(k) ( g_{\mu \nu} k^2 - k_\mu k_\nu ) \int \ud^4 l \ \hat \Delta_+(l) \hat \Delta_+(k-l) \sin^2 \tfrac{k \sigma l}{2}.
\end{equation*}
As the self--energy (\ref{eq:Pi_final}), this has now the usual tensor structure. In the full two--point function, it leads to the following contribution to the continuous spectrum:
\begin{equation*}
  16 e^2 \int \ud^4k \ \hat{f}^{\mu \nu}(-k) \hat{h}^\lambda_{\ \nu}(k) k_\mu k_\lambda \frac{1}{k^2} \int \ud^4 l \ \hat \Delta_+(l) \hat \Delta_+(k-l) \sin^2 \tfrac{k \sigma l}{2}.
\end{equation*}

\subsection{First order}

Using the first order kernel (\ref{eq:K1_3}) in the observable involving $f^{\mu \nu}$, the free kernel (\ref{eq:K0}) in the second observable, and the elementary three-point function (\ref{eq:W3_SUSY_1}),
we obtain
\begin{multline*}
  - i (2 \pi)^2 e^2 \int \ud^4k \ \hat f^{\mu \nu}(-k) \hat h^{\lambda \rho}(k) k_\lambda \int \ud^{8} \underline k \ \delta(k - \sum k_i)  \\
  \times (k\sigma)^\chi \epsilon_{\mu \nu \chi \kappa} \tr \left( \sigma_\xi \bar \sigma_\rho \sigma_\tau \bar \sigma^\kappa \right) k_1^\tau k_2^\xi \frac{\sin \frac{k \sigma k_1}{2}}{\frac{k \sigma k_1}{2}} \sin \tfrac{k_1 \sigma k_2}{2} \\ 
  \times \left\{ \left( \hat \Delta_R(k_1) \hat \Delta_+(-k_2) + \hat \Delta_+(k_1) \hat \Delta_R(k_2) \right) \hat \Delta_+(k) +\hat \Delta_+(k_1) \hat \Delta_+(k_2) \hat \Delta_A(k) \right\}.
\end{multline*}
For the term where the kernel (\ref{eq:K1_4}) (in the first observable) and the elementary three-point function (\ref{eq:W3_SUSY_2}) are used, one obtains nearly the same result, but with the opposite sign and $\xi$ and $\tau$ interchanged in the traces. Using (\ref{eq:sigmaTrace}),
\begin{equation}
\label{eq:epsilon_sigma}
   \epsilon^{\mu \nu \lambda \xi} g_{\xi \xi'} \epsilon^{\kappa \rho \tau \xi'} = \left( - g^{\mu \kappa} g^{\nu \rho} g^{\lambda \tau} + g^{\mu \kappa} g^{\lambda \rho} g^{\nu \tau} - g^{\lambda \kappa} g^{\mu \rho} g^{\nu \tau}  \right) - \mu \leftrightarrow \nu,
\end{equation}
and the antisymmetry of $f^{\mu \nu}$, we get, for the two terms in curly brackets
\begin{subequations}
\label{subeq:3}
\begin{align}
\label{subeq:1stSUSY_1}
 8 (2\pi)^2 e^2 \int \ud^4 k \ \hat{f}^{\mu \nu}(-k) \hat{h}^{\lambda \rho}(k) k_\mu k_\lambda \hat{\Delta}_+(k) \int \ud^4 & l \ \hat{\Delta}^{(1)}(l) \hat{\Delta}_R(k-l) \frac{\sin^2 \frac{k \sigma l}{2}}{\frac{k \sigma l}{2}} \\
 & \times \left\{ (k \sigma)_\rho l_\nu + k \sigma l g_{\nu \rho} \right\}, \nonumber \\
\label{subeq:1stSUSY_2}
 8 (2\pi)^2 e^2 \int \ud^4 k \ \hat{f}^{\mu \nu}(-k) \hat{h}^{\lambda \rho}(k) k_\mu k_\lambda \hat{\Delta}_A(k) \int \ud^4 l & \ \hat{\Delta}_+(l) \hat{\Delta}_+(k-l) \frac{\sin^2 \frac{k \sigma l}{2}}{\frac{k \sigma l}{2}} \\
 & \times \left\{ (k \sigma)_\rho l_\nu + k \sigma l g_{\nu \rho} \right\}, \nonumber
\end{align}
respectively. For the terms where (\ref{eq:K1_3},b) are used in the second observable and combined with the elementary two--point functions (\ref{eq:W3_SUSY_3},d), we find, in the same way,
\begin{align}
\label{subeq:1stSUSY_3}
 8 (2\pi)^2 e^2 \int \ud^4 k \ \hat{f}^{\mu \nu}(-k) \hat{h}^{\lambda \rho}(k) k_\mu k_\lambda \hat{\Delta}_+(k) \int \ud^4 l & \ \hat{\Delta}^{(1)}(l) \hat{\Delta}_A(k-l) \frac{\sin^2 \frac{k \sigma l}{2}}{\frac{k \sigma l}{2}} \\
 & \times \left\{ (k \sigma)_\nu l_\rho + k \sigma l g_{\nu \rho} \right\}, \nonumber \\
\label{subeq:1stSUSY_4}
 8 (2\pi)^2 e^2 \int \ud^4 k \ \hat{f}^{\mu \nu}(-k) \hat{h}^{\lambda \rho}(k) k_\mu k_\lambda \hat{\Delta}_R(k) \int \ud^4 l & \ \hat{\Delta}_+(l) \hat{\Delta}_+(k-l) \frac{\sin^2 \frac{k \sigma l}{2}}{\frac{k \sigma l}{2}} \\
 & \times \left\{ (k \sigma)_\nu l_\rho + k \sigma l g_{\nu \rho} \right\}, \nonumber
\end{align}
\end{subequations}
Adding up all the first order contributions, i.e., (\ref{subeq:1}), (\ref{subeq:2}), and (\ref{subeq:3}), we get, ignoring the prefactor $8 (2\pi)^2 e^2$ for a moment,
\begin{subequations}
\begin{align}
\label{subeq:1stAll_1}
  \int \ud^4 k \ \hat{f}^{\mu \nu}(-k) \hat{h}^{\lambda \rho}(k) k_\mu k_\lambda \hat{\Delta}_+(k) \int \ud^4 & l \ \hat{\Delta}^{(1)}(l) \hat{\Delta}_R(k-l) \frac{\sin^2 \frac{k \sigma l}{2}}{\frac{k \sigma l}{2}} \{ (k \sigma)_\nu l_\rho - (k\sigma)_\rho l_\nu \} \\
\label{subeq:1stAll_2}
 \int \ud^4 k \ \hat{f}^{\mu \nu}(-k) \hat{h}^{\lambda \rho}(k) k_\mu k_\lambda \hat{\Delta}_A(k) \int \ud^4 l & \ \hat{\Delta}_+(l) \hat{\Delta}_+(k-l) \frac{\sin^2 \frac{k \sigma l}{2}}{\frac{k \sigma l}{2}} \{ (k \sigma)_\nu l_\rho - (k\sigma)_\rho l_\nu \} \\
\label{subeq:1stAll_3}
 \int \ud^4 k \ \hat{f}^{\mu \nu}(-k) \hat{h}^{\lambda \rho}(k) k_\mu k_\lambda \hat{\Delta}_+(k) \int \ud^4 l & \ \hat{\Delta}^{(1)}(l) \hat{\Delta}_A(k-l) \frac{\sin^2 \frac{k \sigma l}{2}}{\frac{k \sigma l}{2}} \{ (k \sigma)_\rho l_\nu - (k\sigma)_\nu l_\rho \} \\
\label{subeq:1stAll_4}
 \int \ud^4 k \ \hat{f}^{\mu \nu}(-k) \hat{h}^{\lambda \rho}(k) k_\mu k_\lambda \hat{\Delta}_R(k) \int \ud^4 l & \ \hat{\Delta}_+(l) \hat{\Delta}_+(k-l) \frac{\sin^2 \frac{k \sigma l}{2}}{\frac{k \sigma l}{2}} \{ (k \sigma)_\rho l_\nu - (k\sigma)_\nu l_\rho \},
\end{align}
\end{subequations}
respectively. The sum of (\ref{subeq:1stAll_1}) and (\ref{subeq:1stAll_3}) is
\begin{equation*}
  \int \ud^4 k \ \hat{f}^{\mu \nu}(-k) \hat{h}^{\lambda \rho}(k) k_\mu k_\lambda \hat{\Delta}_+(k) \int \ud^4 l \ \hat{\Delta}^{(1)}(l) \hat{\Delta}(k-l) \frac{\sin^2 \frac{k \sigma l}{2}}{\frac{k \sigma l}{2}} \{ (k \sigma)_\nu l_\rho - (k\sigma)_\rho l_\nu \}.
\end{equation*}
The integrand of the loop integral vanishes: $k,l$ and $k-l$ are forced to lie on the light cone, which is only possible if they are parallel, but then $k \sigma l = 0$. The terms (\ref{subeq:1stAll_2}) and (\ref{subeq:1stAll_4}) cancel for similar reasons. We have thus shown that the contributions to the full two--point function, that involve one kernel of first order, cancel each other. In particular,  the nonlocal divergence (\ref{eq:CovCoorDivergence}) does not appear.

\subsection{Zeroth order}

We start with the case where both observables involve a first order kernel. The combinations of a kernel $K_{\mu \nu}^{\mu_1 \mu_2}(k;k_1,k_2)$ with the kernel $K_{\mu \nu}^{\dotalpha \alpha}(k;k_1,k_2)$ all vanish. The reason is that the only possible contraction is the one among the fields belonging to the same observable. Such contractions vanish because of (\ref{eq:EIdentity}). The only new contributions of this type are thus the four combinations of the two first order kernels (\ref{eq:K1_3},b). For them, we find
\begin{multline*}
  \frac{e^2}{4} \int \ud^4 k \ \hat{f}^{\mu \nu}(-k) \hat{h}^{\lambda \rho}(k) (k\sigma)^\tau (k\sigma)^{\tau'} \epsilon_{\mu \nu \tau \kappa} \epsilon_{\lambda \rho \tau' \kappa'} \tr (\bar \sigma^\kappa \sigma^\xi \bar \sigma^{\kappa'} \sigma^{\xi'}) \\
  \times \int \ud^4 l \ l_\xi (k-l)_{\xi'} \hat \Delta_+(l) \hat \Delta_+(k-l) \frac{\sin^2 \frac{k\sigma l}{2}}{(\frac{k\sigma l}{2})^2}.
\end{multline*}
Using (\ref{eq:sigmaTrace}) and (\ref{eq:epsilon_sigma}), this can be written as
\begin{multline*}
  e^2 \int \ud^4 k \ \hat{f}^{\mu \nu}(-k) \hat{h}^{\lambda \rho}(k) \int \ud^4 l \ \hat \Delta_+(l) \hat \Delta_+(k-l) \frac{\sin^2 \frac{k\sigma l}{2}}{(\frac{k\sigma l}{2})^2} \\
  \times 
 \left\{ g_{\mu \lambda} g_{\mu \rho} k \cdot l (k \sigma)^2 - 2 g_{\mu \lambda} k \cdot l (k\sigma)_\nu (k\sigma)_\rho + \epsilon_{\mu \nu \tau \kappa} \epsilon_{\lambda \rho \tau' \kappa'} (k\sigma)^\tau (k\sigma)^{\tau'} l^\kappa (k-l)^{\kappa'} \right\}.
\end{multline*}
This is a well--defined contribution to the continuous spectrum.

Using the two second order kernels (\ref{eq:K2_3},b) in the observable involving $f^{\mu \nu}$ and the elementary four-point function (\ref{eq:W4_SUSY_1}), we find
\begin{multline*}
 \frac{1}{2} e^2 \int \ud^4 k \ \hat{f}^{\mu \nu}(-k) \hat{h}^{\lambda \rho}(k) k_\lambda (k\sigma)_\rho (k\sigma)^\tau \hat \Delta_+(k) \epsilon_{\mu \nu \tau \kappa} \tr (\bar \sigma^\kappa \sigma^\xi) \int \ud^4 l \ l_\xi \hat{\Delta}_+(l) \\
 \times \left\{ P_2(ik\sigma l, 0) e^{-ik\sigma l} + P_2(0, ik\sigma l) - P_2(-ik\sigma l, 0) e^{ik\sigma l} - P_2(0, -ik\sigma l) \right\}.
\end{multline*}
As in Section~\ref{sec:FreePart}, the expression in curly brackets vanishes. Thus, the divergent contribution (\ref{eq:CovCoorO2_1_result}) is not cancelled. In a sense this had to be expected, since it arose from the contraction of the two photons that came in through the covariant coordinate. Here, we contract the $\lambda_\alpha$ coming from the field strength $W_\alpha$ and the $\bar \lambda_\dotalpha$ from the covariant coordinate.
Thus, the two terms have a different structure and it is not surprising that they do not cancel.

\section{Acausal effects}
\label{sec:NonlocalEffects}

We have seen that in the case of exact supersymmetry there are, apart from the necessary normal ordering of the covariant coordinates, no nonlocal divergences, and 
the modification of the singular part of the two--point function is given by the momentum--dependent field strength renormalization~(\ref{eq:2ptSUSY}).
As discussed in Section~\ref{sec:Acausality}, this should give rise to acausal effects. Here, we want to investigate these in a rather heuristic fashion.

First, we recall that we need to know the self--energy $\Pi_{\mu \nu}(k)$ also for spacelike momenta $k$. As discussed in Remark~\ref{rem:SigmaSpacelike}, these could up to now only be treated formally for $(k \sigma)^2>0$. There, however, we recovered exactly the same expression as for $k$ timelike. We will thus simply assume that the nonplanar part of the self--energy is given, for all $k^2 < 0$, $(k \sigma)^2 \neq 0$, by (\ref{eq:SigmaSpacelike}). As in Section~\ref{sec:Acausality}, we can consider the effect of a source $j^\nu$ for $A^\nu$.\footnote{As discussed in Section~\ref{sec:SourceTerm}, such a source term is not gauge invariant. Properly, one should couple the field strength to a function of the covariant coordinates. But since the corrections are of order $e$ and our discussion is heuristic, we ignore them.} At second order in $e$ and first order in $j$, we find, as effect of this source, the following vacuum expectation value of $\hat A_\mu(k)$:
\begin{equation*}
  (2\pi)^2 \hat \Delta_R(k) \left\{ g_{\mu \nu} - e^2 (g_{\mu \nu} k^2 - k_\mu k_\nu ) \left( \ln \mu \sqrt{-(k \sigma)^2} + \order(k^2 (k \sigma)^2) \right) \hat{\Delta}_R(k) \right\} j^\nu(k).
\end{equation*}
The part proportional to $k_\mu k_\nu$ is cancelled if one calculates the field strength\footnote{Strictly speaking, this is only the case for the part $\del_\mu A_\nu - \del_\nu A_\mu$ of the field strength. But the commutator part is again of order $e$.}, so we discard it.
The above then reduces to
\begin{equation*}
  (2\pi)^2 \hat \Delta_R(k) \left\{ 1 + e^2 (2\pi)^{-2} \ln \mu \sqrt{-(k \sigma)^2} + e^2 \order(k^2 (k \sigma)^2 ) \right\} j_\mu(k).
\end{equation*}
We also discard the terms of $\order(k^2 (k \sigma)^2 )$, since they are not propagated (the factor $k^2$ cancels the retarded propagator). Thus, it remains to compute the Fourier transform of $\ln \mu \sqrt{-(k \sigma)^2}$.


We begin by noticing that $\mu$ and the noncommutativity scale are irrelevant. They can be pulled out of the logarithm and can be absorbed in a local field strength renormalization. There is thus no natural scale connected to this effect and one expects a power-law decay. It has been shown in \cite[Eq.~(7.14)]{Guettinger}, that
\begin{equation*}
  (2\pi)^{-2} \int \ud^4k \ \ln \sqrt{k^2} e^{-ikx} = - 2 \pi \delta'(x^2) + c \delta(x).
\end{equation*}
Here $c$ depends on the extension of $\delta'(x^2)$ to the origin. It corresponds to a local wave function renormalization and is thus irrelevant for our us. Thus, the nonlocal kernel is given by
\begin{equation*}
  \frac{e^2}{2\pi} \delta'((\tilde{\sigma}^{-1} x)^2).
\end{equation*}
Here $\tilde \sigma$ is a descaled version of $\sigma$, such that $\betrag{\tilde \sigma} = 1$. Convoluting this with a source $f$, we obtain
\begin{equation*}
  \frac{e^2}{2\pi} \int \ud^4y \ f(x- \tilde \sigma y) \delta'(y^2).
\end{equation*}
We recall from (\ref{eq:k_sigma_2}) that if $y$ is lightlike, then $\sigma_0 y$ is spacelike or lightlike. Since $\Sigma$ is the orbit of $\sigma_0$ under Lorentz transformations, the same is true for all $\sigma \in \Sigma$. Now assume that $f$ is localized in a region of typical space-time extension $\Delta z$ around the origin. We want to determine its effect at $x$ where $x = \tilde \sigma y$ for $y$ lightlike. We consider $x = (0, \V x)$ (for these $x$ the effect is strongest), i.e., an action at a distance. We then have $\betrag{\V x}^2 = 2 \betrag{ y_0 }^2$. Furthermore, we assume $\betrag{ \V x } \gg \Delta z$.  Thus, using (\ref{eq:delta'}), we can estimate the relative strength at $x$ to be of the order
\begin{equation*}
  e^2 \frac{(\Delta z)^2}{\betrag{\V x}^2}.
\end{equation*}
Here the second term in (\ref{eq:delta'}) was the leading one.

The effect seems to be rather weak, although it might be possible to detect it with the present-day advanced techniques in quantum optics, for example. However, we recall that we had to assume unbroken supersymmetry, which is not realistic. As already mentioned at the beginning of this chapter in Footnote~\ref{footnote:BrokenSUSY}, in the case of broken supersymmetry, one will again need nonlocal counterterms. 

\chapter{Summary and Outlook}
\label{chapter:Summary}

In this thesis we studied several topics in field theory on the noncommutative Minkowski space. The main goal was to compute the distortion of the dispersion relation of the photon in the Yang--Feldman formalism, using the concept of covariant coordinates. As a preparation, we discussed the formulation of gauge theories on noncommutative spaces and studied the example of classical electrodynamics on the noncommutative Minkowski space in some detail. In particular, we computed the modification of the speed of light due to the presence of a background electromagnetic field. It turned out that the effect is very small for realistic values of the background field and the noncommutativity scale. We also studied the quantization of field theories on the noncommutative Minkowski space using the Yang--Feldman formalism. In particular, we discussed the adiabatic limit and showed how to compute the distortion of the dispersion relation. These tools were then applied to the $\phi^3$ and the Wess--Zumino model and it was shown that the distortion is moderate for parameters that are typically expected for the Higgs field. In Chapter~\ref{chapter:NCQED}, we combined the knowledge gained on classical electrodynamics on the noncommutative Minkowski space and the Yang--Feldman formalism to compute the two--point correlation function of the field strength. We not only found the quadratic infrared divergence already known from computations using the modified Feynman rules, but also new divergences stemming from the covariant coordinates. Unfortunately, they did not cancel. We also gave a new interpretation to the quadratic infrared divergences by showing that they require nonlocal counterterms.
This introduces a lot of ambiguities and makes it impossible to compute the dispersion relation. Finally, we showed that a supersymmetric version of the theory behaves much better. For this, we had to adapt the covariant coordinates to noncommutative superspace. Along the way we also discussed the twist approach to NCQFT and pointed out some conceptual difficulties. Furthermore, we showed that the local supersymmetry, respectively BRST, current is not conserved in the Wess--Zumino model, respectively NCQED.

Future research might proceed along the following lines:
\begin{itemize}
\item The computation of the self--energy for spacelike outer momentum (and also the adiabatic limit in interacting models) could only be done formally. It would be very desirable to overcome this restriction.

\item As mentioned in Remark~\ref{rem:LSZ}, the conceptual basis of the present approach is not very solid, since we use the machinery of quantum field theory, even though it is not clear whether the distorted dispersion relations admit the construction of a scattering theory. These issues deserve a thorough investigation.

\item In view of the results obtained here, it seems necessary to consider nonlocal counterterms if one wants to take NCQED serious. This would of course be a major deviation from the standard formalism of quantum field theory. Perhaps this can be justified by considering only counterterms that are, in momentum space, functions of $(k \sigma )^2$, as proposed in~\cite{LiaoSibold}. In the commutative limit, these would be local. One could then impose the standard dispersion relations as a kind of renormalization condition. At the one--loop level the only remaining effect of the noncommutativity would then be the modification of the two--particle spectrum, which is in general very small. However, it remains to be investigated whether this can be done consistently at higher loop orders.

\item For the fulfillment of the space-time uncertainty relations~(\ref{eq:STUR1},b), it is not necessary that the commutators are central. A model where the commutators are no longer central has been proposed in \cite{DoplicherTalk}. This will modify the Weyl relation and might thus weaken the UV/IR--mixing.

\item Another very interesting topic would be the study of field theories on the locally noncommutative spaces introduced in \cite{LocalNC}. It is conceivable that the UV/IR--mixing is much weaker if the noncommutativity is cut off in the infrared.

\item Recently, it has been proposed to discuss the localization of events in operational terms by constructing ``coordinates'' from quantum fields \cite{Events}. Applying this formalism to interacting noncommutative fields, one could investigate whether the coordinate commutators are renormalized. 

\end{itemize}

\appendix

\chapter{Conventions}

\section{Basics}

We use the metric $g^{\mu \nu} = \text{diag} (+,-,-,-)$. Fourier transformation is defined by
\begin{align*}
  \hat{f}(k) = & (2\pi)^{-2} \int \ud^4 x \ f(x) e^{ikx}, \\
  \check{f}(k) = & (2\pi)^{-2} \int \ud^4 x \ f(x) e^{-ikx}.
\end{align*}
The sine and cosine integrals are defined as
\begin{equation}
\label{eq:si}
  \si x = \int_0^x \ud t \ \frac{\sin t}{t}, \quad \ci x = - \int_x^\infty \ud t \ \frac{\cos t}{t}.
\end{equation}
They have the series expansion
\begin{equation}
\label{eq:siExpansion}
  \si x = \sum_{k=1}^{\infty} (-1)^{k+1} \frac{x^{2k-1}}{(2k-1) (2k-1)!}, \quad
  \ci x = \gamma + \ln x + \sum_{k=1}^{\infty} (-1)^k \frac{x^{2k}}{2k (2k)!}.
\end{equation}

\section{Propagators}

The retarded (advanced) propagator $\Delta_{R/A}$ is the unique retarded (advanced) solution of
\begin{equation*}
  (\Box + m^2) \phi = \delta.
\end{equation*}
In momentum space, it is given by
\begin{align}
  \hat{\Delta}_{R/A}(k) = & \frac{1}{(2\pi)^{2}} \frac{-1}{k^2 - m^2 \pm i \epsilon k_0} \nonumber \\
\label{eq:Delta_R}
  = & \frac{1}{(2\pi)^{2}} \frac{1}{2\omega_k} \left( \frac{1}{k_0 + \omega_k \pm i \epsilon} - \frac{1}{k_0 - \omega_k \pm i \epsilon} \right).
\end{align}
The commutator function is given by
\begin{equation*}
  \Delta = \Delta_R - \Delta_A.
\end{equation*}
Its positive frequency part times $-i$ is the free two--point function $\Delta_+$. In momentum space it is
\begin{equation*}
  \hat{\Delta}_+(k) = \frac{1}{2\pi} \theta(k_0) \delta(k^2-m^2).
\end{equation*}
In the Yang-Feldman formalism, one often encounters
\begin{equation*}
  \Delta^{(1)}(x) = \Delta_+(x) + \Delta_+(-x).
\end{equation*}
The Feynman propagator is defined as
\begin{equation*}
  \Delta_F(x) = i \left( \theta(x_0) \Delta_+(x) + \theta(-x_0) \Delta_+(-x) \right).
\end{equation*}

\section{Spinors and supersymmetry}
\label{app:SUSY}

We mainly use the conventions of~\cite{WessBagger}. However, we use another sign for the metric and for $\sigma^0$ (and thus also for $\gamma^0$ and $\gamma^5$). We also changed a sign in the definition of $D_\alpha$ and $\bar D_\dotalpha$.

Weyl spinors are anticommuting. Their indices are raised and lowered with the help of the totally antisymmetric $\epsilon$-tensor:
\begin{gather*}
  \chi^\alpha = \epsilon^{\alpha \beta} \chi_\beta, \quad \chi_\alpha = \epsilon_{\alpha \beta} \chi^\beta, \quad \epsilon^{1 2} = \epsilon_{2 1} = 1, \\
  \bar \chi^\dotalpha = \epsilon^{\dotalpha \dotbeta} \chi_\dotbeta, \quad \bar \chi_\dotalpha = \epsilon_{\dotalpha \dotbeta} \bar \chi^\dotbeta, \quad \epsilon^{\dot 1 \dot 2} = \epsilon_{\dot 2 \dot 1} = 1.
\end{gather*}
Products of Weyl spinors are defined as
\begin{gather*}
  \lambda \chi = \lambda^\alpha \chi_\alpha = - \lambda_\alpha \chi^\alpha = \chi^\alpha \lambda_\alpha = \chi \lambda, \\
  \bar \lambda \bar \chi = \bar \lambda_\dotalpha \bar \chi^\dotalpha = - \bar \lambda^\dotalpha \bar \chi_\dotalpha = \bar \chi_\dotalpha \bar \lambda^\dotalpha = \bar \chi \bar \lambda.
\end{gather*}
The $\sigma$-matrices are defined as
\begin{equation*}
  \sigma^\mu_{\alpha \dotalpha} = ( \1, \sigma^i )_{\alpha \dot \alpha}, \quad \bar \sigma^{\mu \ \dotalpha \alpha} = \epsilon^{\dotalpha \dotbeta} \epsilon^{\alpha \beta} \sigma^\mu_{\beta \dotbeta} = ( \1, - \sigma^i )^{\dotalpha \alpha}.
\end{equation*}  
Here $\sigma^i$ are the usual Pauli matrices. One also defines
\begin{equation*}
  ( \sigma^{\mu \nu} )_\alpha^{\ \beta} = \tfrac{1}{4} \left( \sigma^\mu \bar \sigma^\nu - \sigma^\nu \bar \sigma^\mu \right)_\alpha^{\ \beta}, \quad ( \bar \sigma^{\mu \nu} )^\dotalpha_{\ \dotbeta} = \tfrac{1}{4} \left( \bar \sigma^\mu \sigma^\nu - \bar \sigma^\nu \sigma^\mu \right)^\dotalpha_{\ \dotbeta}.
\end{equation*}
Using the definition above, one finds
\begin{equation*}
  \lambda \sigma^\mu \bar \chi = \lambda^\alpha \sigma^\mu_{\alpha \dotalpha} \bar \chi^\dotalpha = - \epsilon^{\dotalpha \dotbeta} \epsilon^{\alpha \beta} \sigma^\mu_{\alpha \dotalpha} \bar \chi_\dotbeta \lambda_\beta = - \bar \chi \bar \sigma^\mu \lambda.
\end{equation*}
Furthermore, the following identities hold:
\begin{align}
  \sigma^\mu_{\alpha \dotalpha} \bar \sigma^{\nu \ \dotalpha \beta} & = g^{\mu \nu} \delta_\alpha^\beta + 2 ( \sigma^{\mu \nu} )_\alpha^{\ \beta}, \nonumber \\
\label{eq:sigma_trace}
  \tr ( \sigma^\mu \bar \sigma^\nu ) & = 2 g^{\mu \nu}, \\
\label{eq:3sigmas_1}
  \sigma^\mu \bar \sigma^\nu \sigma^\lambda & = - g^{\mu \lambda} \sigma^\nu + g^{\nu \lambda} \sigma^\mu + g^{\mu \nu} \sigma^\lambda + i \epsilon^{\mu \nu \lambda \kappa} \sigma_\kappa, \\
\label{eq:3sigmas_2}
  \bar \sigma^\mu \sigma^\nu \bar \sigma^\lambda & = - g^{\mu \lambda} \bar \sigma^\nu + g^{\nu \lambda} \bar \sigma^\mu + g^{\mu \nu} \bar \sigma^\lambda - i \epsilon^{\mu \nu \lambda \kappa} \bar \sigma_\kappa.
\end{align}
The anticommuting superspace coordinates $\theta$, $\bar \theta$ fulfill
\begin{gather*}
  \theta^\alpha \theta^\beta = - \tfrac{1}{2} \epsilon^{\alpha \beta} \theta^2, \quad \theta_\alpha \theta^\beta = - \tfrac{1}{2} \delta_\alpha^\beta \theta^2, \quad \theta^\alpha \theta_\beta = \tfrac{1}{2} \delta^\alpha_\beta \theta^2, \quad \theta_\alpha \theta_\beta = \tfrac{1}{2} \epsilon_{\alpha \beta} \theta^2, \\
  \bar \theta^\dotalpha \bar \theta^\dotbeta = \tfrac{1}{2} \epsilon^{\dotalpha \dotbeta} \bar \theta^2, \quad \bar \theta_\dotalpha \bar \theta^\dotbeta = \tfrac{1}{2} \delta_\dotalpha^\dotbeta \bar \theta^2, \quad \bar \theta^\dotalpha \bar \theta_\dotbeta = - \tfrac{1}{2} \delta^\dotalpha_\dotbeta \bar \theta^2, \quad \bar \theta_\dotalpha \bar \theta_\dotbeta = - \tfrac{1}{2} \epsilon_{\dotalpha \dotbeta} \bar \theta^2.
\end{gather*}
One defines covariant spinor derivates by
\begin{equation*}
  D_\alpha = \del_\alpha - i \sigma^\mu_{\alpha \dotalpha} \bar \theta^\dotalpha \del_\mu, \quad \bar D_\dotalpha = - \bar \del_\dotalpha + i \theta^\alpha \sigma^\mu_{\alpha \dotalpha} \del_\mu.
\end{equation*}
The partial spinor derivatives are given by
\begin{gather*}
  \del_\alpha \theta^\beta = \delta_\alpha^\beta, \quad   \del_\alpha \theta_\beta = \epsilon_{\beta \alpha}, \quad   \del^\alpha \theta_\beta = \delta_\beta^\alpha, \quad   \del^\alpha \theta^\beta = \epsilon^{\beta \alpha},  \\
  \bar \del_\dotalpha \bar\theta^\dotbeta = \delta_\dotalpha^\dotbeta, \quad   \bar \del_\dotalpha \bar \theta_\dotbeta = \epsilon_{\dotbeta \dotalpha}, \quad   \bar \del^\dotalpha \bar \theta_\dotbeta = \delta_\dotbeta^\dotalpha, \quad  \bar \del^\dotalpha \bar \theta^\dotbeta = \epsilon^{\dotbeta \dotalpha}. 
\end{gather*}
Thus,
\begin{align}
\label{eq:D_anticomm}
  \{ D_\alpha, \bar D_\dotalpha \} & = 2 i \sigma^\mu_{\alpha \dotalpha} \del_\mu, \\
  \del_\alpha \theta^2 & = 2 \theta_\alpha, \nonumber \\
  \bar \del_\dotalpha \bar \theta^2 & = - 2 \bar \theta_\dotalpha. \nonumber
\end{align}
We also define the $\gamma$-matrices
\begin{equation*}
  \gamma^\mu = \begin{pmatrix} 0 & \sigma^\mu \\ \bar \sigma^\mu & 0 \end{pmatrix}, \quad \gamma_5 = \gamma^5 = \gamma^0 \gamma^1 \gamma^2 \gamma^3 = i \begin{pmatrix} \1 & 0 \\ 0 & - \1 \end{pmatrix},
\end{equation*}
and the chiral projectors
\begin{equation}
\label{eq:P_pm}
  P_\pm = \frac{1 \mp i \gamma_5}{2}.
\end{equation}
The $\gamma$-matrices fulfill the usual identities
\begin{align*}
  \{ \gamma^\mu, \gamma^\nu \} & = 2 g^{\mu \nu}, \\
  \gamma_5^2 & = - \1, \\
  \{ \gamma^\mu, \gamma_5 \} & = 0, \\
  \tr \gamma^\mu & = 0, \\
  \tr \gamma_5 & = 0, \\
  \tr (\gamma_5 \gamma^\mu) & = 0, \\
  \tr (\gamma^\mu \gamma^\nu) & = 4 g^{\mu \nu}, \\
  \tr (\gamma^\mu \gamma^\nu \gamma^\kappa \gamma^\lambda) & = 4 (g^{\mu \nu} g^{\kappa \lambda} - g^{\mu \kappa} g^{\nu \lambda} + g^{\mu \lambda} g^{\nu \kappa} ). 
\end{align*}

\chapter{Proofs}
\label{app:Proofs}

\section{Proof of (\ref{eq:BCH_PQ})}
\label{app:BCH_PQ}

We define
\begin{equation*}
  A = pq; \quad B = e p \sigma b (e^{-ikq} + e^{ikq}); \quad C = -i e p \sigma b (e^{-ikq} - e^{ikq}); \quad x = p \sigma k.
\end{equation*}
We have
\begin{equation*}
  [A,B] = i x C; \quad [A,C] = - i x B; \quad [B,C] = 0.
\end{equation*}
Now we make the ansatz
\begin{equation*}
  e^{-i(A+B)t} = e^{-itA} e^{-iP(xt) t B - i Q(xt) t C}.
\end{equation*}
Differentiating with respect to $t$, we obtain
\begin{multline*}
  e^{-i(A+B)t} (A+B) = e^{-itA} A e^{-iP(xt) t B - i Q(xt) t C} \\ + e^{-i(A+B)t} \left\{ x t P'(xt) B + P(xt) B + x t Q'(xt) C + Q(xt) C \right\}.
\end{multline*}
We have
\begin{equation*}
  [ A, e^{-iP(xt) t B - i Q(xt) t C} ] = e^{-iP(xt) t B - i Q(xt) t C} ( xtP(xt) C - xt Q(xt) B ).
\end{equation*}
Thus, we obtain
\begin{equation*}
  A + B = A + C( xtP(xt) + x t Q'(xt) + Q(xt) ) - B( xt Q(xt) - x t P'(xt) - P(xt)) .
\end{equation*}
Since $B$ and $C$ are linearly independent, we find the system of differential equations
\begin{align*}
  x Q(x) - x P'(x) - P(x) +1 & = 0 \\
  x P(x) + x Q'(x) + Q(x) & = 0.
\end{align*}
It is easy to see that this is solved by
\begin{equation*}
  P(x) = \frac{ \sin x }{x} ; \quad Q(x) = \frac{ \cos x -1}{x}.
\end{equation*}
Furthermore, these have the right boundary values
\begin{equation*}
  P(0) = 1; \quad Q(0) = 0; \quad Q'(0) = - \frac{1}{2}
\end{equation*}
known from the Baker--Campbell--Hausdorff series.

\section{Proof of Lemma~\ref{lemma:NCED}}
\label{app:NCED}

We have
\begin{equation*}
  D C^m = \sum_{l=0}^m {m \choose l} C^{m-l} (- \ad_C)^l D.
\end{equation*}
Thus,
\begin{align*}
  \sum_{n=1}^\infty \frac{1}{n!} \sum_{m=0}^{n-1} C^{n-m-1} D C^m = & \sum_{n=1}^\infty \frac{1}{n!} \sum_{m=0}^{n-1} \sum_{l=0}^m {m \choose l} C^{n-l-1} (- \ad_C)^l D \\
  = & \sum_{n=1}^\infty \frac{1}{n!} \sum_{l=0}^{n-1} C^{n-l-1} (- \ad_C)^l D \sum_{m=l}^{n-1} {m \choose l} \\
  = & \sum_{n=1}^\infty \frac{1}{n!} \sum_{l=0}^{n-1} {n \choose l+1} C^{n-l-1} (-\ad_C)^l D \\
  = & \sum_{n=1}^\infty \sum_{l=0}^{n-1} \frac{C^{n-l-1}}{(n-l-1)!} \frac{(- \ad_C)^l}{(l+1)!} D \\
  = & \sum_{n=0}^\infty \frac{C^n}{n!} \sum_{l=0}^{\infty} \frac{(-\ad_C)^l}{(l+1)!} D.
\end{align*}

\section{Proof of Remark~\ref{rem:k4}}
\label{app:k4}


The terms involving $\iep{x+2 \betrag{\V{k_0}}}$ and $\iep{x-2 \betrag{\V{k_1}}}$ can be treated as sketched in Remark~\ref{rem:massless}. 
The problematic terms are those proportional to $\iep{x}$. This singularity is not cancelled after the Taylor expansion in $x$ if $\Sigma(k) - \Sigma(-k)$ does not vanish in a neighborhood of the forward light cone. If this difference behaves as $k^4$ for $k^2 \to 0$, we are left, after Taylor expansion in $x$, with the expression, cf.~(\ref{eq:phi_3_comm}),
\begin{equation}
\label{eq:RestTerm}
  \int \frac{\ud^3 \V{k_0}}{2 \betrag{\V{k_0}}} \frac{\ud^3 \V{k_1}}{2 \betrag{\V{k_1}}} \ud^4 l \ud x \ \check g_a(l_0-\betrag{\V{k_0}}-x,\V l - \V{k_0}) \check g_a(k_1^+-l)
  \hat f(-k_0^+) \hat h(k_1^+) l^4 \iep{x}.
\end{equation}
For simplicity, we replaced $\Sigma(k) - \Sigma(-k)$ by its asymptotic behavior for $k^2 \to 0$. It is straightforward to generalize the following arguments for the case where $\Sigma(k) - \Sigma(-k) = k^4 \Sigma'(k)$ with smooth and bounded $\Sigma'$.
We have to show that (\ref{eq:RestTerm}) vanishes in the adiabatic limit. Now we assume that $\check g_a$ scales, i.e., $\check g_a(k) = a^4 \check g(a k)$, where $\check g$ has compact support $S \subset \R^4$. Then one obtains
\begin{equation*}
 a^{-4}  \int \frac{\ud^3 \V{k_0}}{2 \betrag{\V{k_0}}} \frac{\ud^3 \V{k_1}}{2 \betrag{\V{k_1}}} \ud^4 l \ud x \ \check g(l_0-\betrag{\V{k_0}}-x,\V l - \V{k_0}) \check g(k_1^+ - l)
  \hat f(- a^{-1} k_0^+) \hat h(a^{-1} k_1^+) l^4 \iep{x}.
\end{equation*}
Now we write
\begin{align*}
 & \int \ud^4l \ \check g(l_0-\betrag{\V{k_0}}-x,\V l - \V{k_0}) \check g(k_1^+ - l) l^4 \\
 = & \int \ud^4l \ \check g(l_0 + \betrag{\V{k_1}} -\betrag{\V{k_0}}-x,\V l + \V{k_1} - \V{k_0}) \check g(-l) \left( l^2 + 2 l \cdot k_1^+ \right)^2 \\
  = & \int \ud^4l \ \check g(l_0 + \betrag{\V{k_1}} -\betrag{\V{k_0}}-x,\V l + \V{k_1} - \V{k_0}) \check g(-l) \left( l^4 + 4 l^2 l \cdot k_1^+ + 4 (l \cdot k_1^+)^2 \right) \\
  = & \tilde g(\betrag{\V{k_1}} -\betrag{\V{k_0}}-x, \V{k_1} - \V{k_0}) + k_1^{+ \mu} \tilde g_\mu (\betrag{\V{k_1}} -\betrag{\V{k_0}}-x, \V{k_1} - \V{k_0}) \\
  & + k_1^{+ \mu} k_1^{+ \nu} \tilde g_{\mu \nu} (\betrag{\V{k_1}} -\betrag{\V{k_0}}-x, \V{k_1} - \V{k_0}).
\end{align*}
Here $\tilde g$, $\tilde g_{\mu}$ and $\tilde g_{\mu \nu}$ are test functions with support in the compact set
\begin{equation*}
  \bar S = \{k \in \R^4| \exists l \in S, k-l \in S\}.
\end{equation*}
Carrying out the integration over $x$, we obtain smooth bounded functions $\tilde{\tilde g}$, $\tilde{\tilde{g}}_{\mu}$ and $\tilde{\tilde g}_{\mu \nu}$ that have support in $\R \times S'$, where $S'$ is the spatial projection of $\bar S$. Thus, we get
\begin{multline*}
 a^{-4} \int \frac{\ud^3 \V{k_0}}{2 \betrag{\V{k_0}}} \frac{\ud^3 \V{k_1}}{2 \betrag{\V{k_1}}} \ \hat f(- a^{-1} k_0^+) \hat h(a^{-1} k_1^+) 
 \\ \times \left( \tilde{ \tilde g}(k_1^+ - k_0^+) + k_1^{+ \mu} \tilde{ \tilde g}_\mu (k_1^+ - k_0^+) + k_1^{+ \mu} k_1^{+ \nu} \tilde{ \tilde g}_{\mu \nu} (k_1^+ - k_0^+)  \right).
\end{multline*}
Reversing the change of variables, this is
\begin{multline*}
 \int \frac{\ud^3 \V{k_0}}{2 \betrag{\V{k_0}}} \frac{\ud^3 \V{k_1}}{2 \betrag{\V{k_1}}} \ \hat f(- k_0^+) \hat h(k_1^+)
 \\ \times \left( \tilde{ \tilde g}(a(k_1^+ - k_0^+)) + a k_1^{+ \mu} \tilde{ \tilde g}_\mu (a(k_1^+ - k_0^+)) + a^2 k_1^{+ \mu} k_1^{+ \nu} \tilde{ \tilde g}_{\mu \nu} (a(k_1^+ - k_0^+))  \right).
\end{multline*}
Keeping $\V{k_0}$ fixed, the volume $a^{-1} S'$, over which $\V{k_1}$ is integrated, scales as $a^{-3}$. Thus, the three terms scale as $a^{-3}$, $a^{-2}$ and $a^{-1}$, respectively. It follows that they vanish in the adiabatic limit. It is straightforward to see that $\Sigma(k) - \Sigma(-k) \sim k^2$ would not be enough, since then we would find a term scaling as~$a^0$.

\section{Proof of Lemma~\ref{lemma:smooth}}
\label{app:smooth}

We only treat the case $\hat{\Delta}_+( \cdot ) \hat \Delta_R(k- \cdot )$. The proof for $\hat{\Delta}_+( \cdot ) \hat \Delta_R(k + \cdot )$ is completely analogous.
We begin by showing that $\hat{\Delta}_+( \cdot ) \hat \Delta_R(k- \cdot )$ is well-defined as an element of $\mathcal{D}'(\R^4)$ for $k^2 \notin \{ 0, 4 m^2 \}$. The wave front sets of the distributions are
\begin{align*}
  WF(\hat \Delta_+) & = \{ (l, p_l) | l^2= m^2, l_0>0, p_l=\lambda l, \lambda \neq 0 \}, \\
  WF(\hat \Delta_R(k- \cdot )) & \subset \{ (l, p_l) | (k-l)^2= m^2, p_l=\lambda (k-l), \lambda \neq 0 \}.
\end{align*}
A theorem due to H\"ormander \cite[Thm.~8.2.10]{Hoermander} now states that the product is well-defined as an element of $\mathcal{D}'(\R^4)$, unless there is an $l \in \R^4$ such that $(l, p^1_l) \in WF(\hat \Delta_+)$,  $(l, p^2_l) \in WF(\hat \Delta_R(k- \cdot)$ and $p^1_l + p^2_l = 0$. In the massive case it would follow from $l^2 = (k-l)^2 = m^2$ and $\lambda l = k-l$ that $l = \pm (k-l)$ and thus $k = 0$ or $k = 2l$. But then we would have $k^2 = 0$ or $k^2 = 4 m^2$, respectively, which we excluded. In the massless case the conditions $l^2 = (k-l)^2 = 0$ and $\lambda l = k-l$ would imply $k^2 = 0$.


Now we show that the product is even a tempered distribution for $k^2 \notin \{ 0, 4m^2 \}$. We start by integrating it with a test function $f \in \mathcal{D}(\R^4)$:
\begin{equation*}
  \int \ud^4 l \ \hat \Delta_+(l) \hat \Delta_R(k-l) f(l).
\end{equation*}
This is invariant under a simultaneous orthochronous Lorentz transformation of $k$ and $f$. If $k$ is timelike, we bring it to the form $k = (k_0, \V 0)$. Then the above yields
\begin{equation*}
  (2 \pi)^{-3} \int \frac{\ud^3 \V l}{2 \omega_{\V l}} \ \frac{-1}{k^2 - 2 \sqrt{k^2} \omega_{\V l} + i ( \sqrt{k^2} - \omega_\V{l}) \epsilon} f_k(\omega_{\V l}, \V l).
\end{equation*}
Here $f_k$ is obtained from $f$ by the same Lorentz transformation that brings $k$ to the form $(k_0, \V 0)$. Since the singularity is met for $\omega_{\V l} = \frac{1}{2} \sqrt{ k^2}$, we can drop the prefactor of the $\epsilon$. Carrying out the angular integration, we obtain
\begin{equation*}
  (2 \pi)^{-3} \int_0^\infty \ud l \ \frac{l^2}{2 \omega_l} \frac{-1}{k^2 - 2 \sqrt{k^2} \omega_l + i \epsilon} \tilde f_k(l).
\end{equation*}
where $\tilde f_k(l)$ is the mean of $f_k(\omega_{\V l}, \V l)$ over the sphere of radius $l$. Obviously, this can be extended to Schwartz functions and thus also defines a tempered distribution. Furthermore, the map $k \mapsto \hat{\Delta}_+(\cdot) \hat \Delta_R(k- \cdot)$ is smooth for $k^2 \notin \{ 0, 4 m^2 \}$. In the case where $k$ is spacelike, we choose $k = (0,0,0,k_3)$. We then obtain
\begin{equation*}
  (2 \pi)^{-3} \int \ud l_3 \ \frac{-1}{k^2 + 2 \sqrt{-k^2} l_3 - i \epsilon} \int \ud l_1 \ud l_2 \ \frac{f_k(\omega_{\V l}, \V l)}{2 \omega_{\V l}}. 
\end{equation*}
This also defines a tempered distribution and for $k^2 < 0$ the map $k \mapsto \hat{\Delta}_+(\cdot) \hat \Delta_R(k- \cdot)$ is smooth.

\section{Proof of Lemma~\ref{lemma:NCQED}}
\label{app:NCQED}

We proceed by induction. The case $N=0$ is obvious. The left hand side can be written as
\begin{multline*}
  \sum_{n_0, \dots , n_N=0}^\infty \frac{x^{n_0} (x+y_1)^{n_1} \dots (x+y_{N-1})^{n_{N-1}}}{(n_0+ \dots + n_N + N)!}
   \sum_{l=0}^{n_N} (x+y_{N-1})^{n_N-l} (y_N-y_{N-1})^l {n_N \choose l} \\
= \sum_{n_0, \dots , n_{N-1}=0}^\infty \frac{x^{n_0} (x+y_1)^{n_1} \dots (x+y_{N-2})^{n_{N-2}}}{(n_0+ \dots + n_{N-1} + N)!} \sum_{k=0}^{n_{N-1}} (x+y_{N-1})^{n_{N-1}-k} \\
  \times \sum_{l=0}^{k} (x+y_{N-1})^{k-l} (y_N-y_{N-1})^l {k \choose l}.
\end{multline*}
Here we introduced the new variables $n_{N-1} = n_{N-1} + n_N$, $k = n_N$. Changing the order of summations, we obtain
\begin{multline*}
  \sum_{n_0, \dots , n_{N-1}=0}^\infty \frac{x^{n_0} (x+y_1)^{n_1} \dots (x+y_{N-2})^{n_{N-2}}}{(n_0+ \dots + n_{N-1} + N)!} \\
 \times \sum_{l=0}^{n_{N-1}} (x+y_{N-1})^{n_{N-1}-l} (y_N-y_{N-1})^l \sum_{k=l}^{n_{N-1}}  {k \choose l}.
\end{multline*}
Using $\sum_{k=l}^n {k \choose l} = {n+1 \choose l+1}$ and $\sum_{l=0}^n A^{n-l} B^l { n+1 \choose l+1 } = \frac{1}{B} \left( (A+B)^{n+1} - A^{n+1} \right)$, this can be written as
\begin{align*}
  & \frac{1}{y_N-y_{N-1}} \sum_{n_0, \dots , n_{N-1}=0}^\infty  \frac{x^{n_0} (x+y_1)^{n_1} \dots (x+y_{N-2})^{n_{N-2}}}{(n_0+ \dots + n_{N-1} + N)!} \\
  & \qquad \times \left( (x+y_N)^{n_{N-1}+1} - (x+y_{N-1})^{n_{N-1}+1} \right) \\
=  & \frac{1}{y_N-y_{N-1}} \sum_{n_0, \dots , n_{N-1}=0}^\infty \frac{x^{n_0} (x+y_1)^{n_1} \dots (x+y_{N-2})^{n_{N-2}}}{(n_0+ \dots + n_{N-1} + N-1)!} \\
  & \qquad \times \left( (x+y_N)^{n_{N-1}} - (x+y_{N-1})^{n_{N-1}} \right) \\
=  & \frac{e^x}{y_N-y_{N-1}} \sum_{n_1, \dots , n_{N-1}=0}^\infty \frac{y_1^{n_1} \dots y_{N-2}^{n_{N-2}}}{\left( \sum_{i=1}^{N-1} (n_i+1) \right)!}  \left( y_N^{n_{N-1}} - y_{N-1}^{n_{N-1}} \right) \\
= & e^x \sum_{n_1, \dots , n_N=0}^\infty \frac{1}{\left( \sum_{i=1}^N (n_i+1) \right)!} y_1^{n_1} \dots y_N^{n_N}.
\end{align*}
In the first step, we relabeled $n_{N-1} = n_{N-1}+1$. In the second step we employed the induction hypothesis and in the last step
\begin{equation*}
  \frac{A^n-B^n}{A-B} = \sum_{k=0}^{n-1} A^k B^{n-1-k}. 
\end{equation*}

\section{Proof of Remark~\ref{rem:Equivalence}}
\label{app:Equivalence}

We rewrite (\ref{eq:Equivalence}) explicitly:
\begin{multline*}
 e^{ikX}_\star = e^{ikx} \star \sum_{N=0}^\infty (ie)^N (2\pi)^{2N} \int \prod_{i=1}^N \ud^4k_i \ e^{-ik_1x} \star \dots \star e^{-ik_Nx} k \sigma \hat A(k_1) \dots k \sigma \hat A(k_N) \\
 \times \int_0^1 \ud t_1 \dots \int_0^{t_{N-2}} \ud t_{N-1} \int_0^{t_{N-1}} \ud t_N \ e^{-it_1 k \sigma k_1} \dots e^{-it_N k \sigma k_N}.
\end{multline*}
The claim that this is equivalent to (\ref{eq:e_ikX}) now means that the second line is identical to $P_N(-ik\sigma k_1, \dots , -ik\sigma k_N)$. This can be shown by induction. For $N=1$, we obtain
\begin{equation*}
  \int_0^1 \ud t_1 \ e^{-it_1 k \sigma k_1} = \frac{e^{-ik \sigma k_1} - 1}{-ik\sigma k_1} = \sum_{n=0}^\infty \frac{(-ik \sigma k_1)^n}{(n+1)!} = P_1(-ik\sigma k_1).
\end{equation*}
For arbitrary $N$, we find
\begin{align*}
  & \int_0^1 \ud t_1 \dots \int_0^{t_{N-2}} \ud t_{N-1} \int_0^{t_{N-1}} \ud t_N \ e^{-it_1 k \sigma k_1} \dots e^{-it_N k \sigma k_N} \\
 = &  \int_0^1 \ud t_1 \dots \int_0^{t_{N-2}} \ud t_{N-1} \ e^{-it_1 k \sigma k_1} \dots e^{-it_{N-1} k \sigma k_{N-1}} \frac{1}{-ik\sigma k_N} \left( e^{-i t_{N-1} k \sigma k_N} - 1 \right) \\
 = & \frac{1}{-ik\sigma k_N} \left( P_{N-1}(-ik\sigma k_1, \dots , -i k \sigma k_{N-1} -ik \sigma k_N) - P_{N-1}(-ik\sigma k_1, \dots , -i k \sigma k_{N-1}) \right).
\end{align*}
Thus, we have to show
\begin{equation*}
  P_N(x_1, \dots , x_N) = \frac{1}{x_N} \left( P_{N-1}(x_1, \dots , x_{N-1}+x_N) - P_{N-1}(x_1, \dots , x_{N-1}) \right).
\end{equation*}
The right hand side can be written as
\begin{equation*}
  \sum_{n_1, \dots , n_{N-1}=0}^\infty \frac{x_1^{n_1} \dots (x_1 + \dots x_{N-2})^{n_{N-2}}}{(n_1 + \dots + n_{N-1} + N - 1)!} \frac{(x_1 + \dots + x_N)^{n_{N-1}} - (x_1 + \dots + x_{N-1})^{n_{N-1}}}{x_N}.
\end{equation*}
Using
\begin{equation*}
  \frac{(A+B)^n-A^n}{B} = \sum_{m=0}^{n-1} A^{n-1-m} B^m { n \choose m+1 } = \sum_{m=0}^{n-1} A^{n-1-m} (A+B)^m,
\end{equation*}
this can be written as
\begin{equation*}
  \sum_{n_1, \dots , n_{N}=0}^\infty \frac{x_1^{n_1} \dots (x_1 + \dots + x_{N-1})^{n_{N-1}}  (x_1 + \dots + x_{N})^{n_{N}}}{(n_1 + \dots + n_{N} + N)!},
\end{equation*}
which is precisely the definition of $P_N(x_1, \dots , x_N)$.

\section{Proof of (\ref{eq:W1})}
\label{app:k_mu_k_nu}

For simplicity, we concentrate on the case $\Pi_{\mu \nu}(k) - \Pi_{\mu \nu}(-k) = k_\mu k_\nu$ in a neighborhood of the light cone. If the difference scales as $k_\mu k_\nu k^2$, as in our concrete case, the convergence will be even faster. The proof proceeds similarly to that of Remark~\ref{rem:k4} given in Appendix~\ref{app:k4}. Instead of (\ref{eq:RestTerm}), we here have
\begin{equation*}
 - \int \frac{\ud^3 \V{k_0}}{2 \betrag{\V{k_0}}} \frac{\ud^3 \V{k_1}}{2 \betrag{\V{k_1}}} \ud^4 l \ud x \ \check g_a(l_0-\betrag{\V{k_0}}-x,\V l - \V{k_0}) \check g_a(k_1^+-l) \hat f^{\mu}(-k_0^+) \hat h^{\nu}(k_1^+) l_\mu l_\nu \iep{x}.
\end{equation*}
Because of the transversality of $\hat f^{\mu}$ and $\hat h^{\nu}$, we can replace $-l_\mu l_\nu$ by
\begin{equation*}
 (l-k_0^+)_\mu (k_1^+-l)_\nu.
\end{equation*}
As in Appendix~\ref{app:k4}, we assume that $\check g_a$ scales, $\check g_a(k) = a^4 \check g(ak)$, with $\check g$ having compact support $S$. We obtain
\begin{multline*}
 a^{-2} \int \frac{\ud^3 \V{k_0}}{2 \betrag{\V{k_0}}} \frac{\ud^3 \V{k_1}}{2 \betrag{\V{k_1}}} \ud^4 l \ud x \ \check g(l_0-\betrag{\V{k_0}}-x,\V l - \V{k_0}) \check g(k_1^+-l) \hat f^{\mu}(- a^{-1} k_0^+) \hat h^{\nu}(a^{-1} k_1^+) \\
 \times (l-k_0^+)_\mu (k_1^+-l)_\nu \iep{x}.
\end{multline*}
Integrating over $l$ and $x$ as in Appendix~\ref{app:k4}, we obtain
\begin{equation*}
 a^{-2} \int \frac{\ud^3 \V{k_0}}{2 \betrag{\V{k_0}}} \frac{\ud^3 \V{k_1}}{2 \betrag{\V{k_1}}} \ 
 \left( (k_1^+ - k_0^+)_\mu \tilde{\tilde g}_\nu(k_1^+-k_0^+) + \tilde{\tilde g}_{\mu \nu}(k_1^+-k_0^+) \right) \hat f^{\mu}(- a^{-1} k_0^+) \hat h^{\nu}(a^{-1} k_1^+),
\end{equation*}
where $\tilde{\tilde g}_\nu$ and $\tilde{\tilde g}_{\mu \nu}$ are smooth bounded functions with support in $\R \times S'$, where $S'$ is compact, cf. Appendix~\ref{app:k4}. After a change of variables, this is
\begin{equation*}
 a^2 \int \frac{\ud^3 \V{k_0}}{2 \betrag{\V{k_0}}} \frac{\ud^3 \V{k_1}}{2 \betrag{\V{k_1}}} \ 
 \left( a (k_1^+ - k_0^+)_\mu \tilde{\tilde g}_\nu(a(k_1^+-k_0^+)) + \tilde{\tilde g}_{\mu \nu}(a(k_1^+-k_0^+)) \right) \hat f^{\mu}(- k_0^+) \hat h^{\nu}(k_1^+).
\end{equation*}
Keeping $\V{k_0}$ fixed, the volume over which $\V{k_1}$ is integrated scales as $a^{-3}$. Thus, the above scales as $a^{-1}$ and vanishes in the adiabatic limit.

\section{Proof of (\ref{eq:SigmaSpacelike})}
\label{app:SigmaSpacelike}

We want to formally compute the massless nonplanar fish graph loop for spacelike outer momentum $k$. We have to compute
\begin{equation*}
  (2 \pi)^{-3} \int \frac{\ud^3 \V l}{2 \betrag{\V{l}}} \ \left( \frac{ - 1}{(k-l_+)^2 + i \epsilon (k_0-\betrag{\V{l}})} + \frac{ - 1}{(k+l_+)^2 + i \epsilon (k_0+\betrag{\V{l}})} \right) \cos ( y_0 \betrag{\V{l}} - \V y \cdot \V{l} ).
\end{equation*}
For the moment, we keep $\epsilon > 0$ finite. We are interested in the case $\V y  = \V 0$ and $k = (0, \V{k})$. We get
\begin{equation*}
  (2\pi)^{-2} \int_0^\infty \ud l \int_{-1}^1 \ud x \ \frac{l^2}{2 l} \left( \frac{ -1}{-\betrag{\V{k}}^2 + 2 \betrag{\V{k}} l x - i \epsilon l } + \frac{ -1}{ - \betrag{\V{k}}^2 - 2 \betrag{\V{k}} l x + i \epsilon l } \right) \cos y_0 l.
\end{equation*}
Carrying out the integration over $x$, we obtain
\begin{multline*}
  (2\pi)^{-2} \frac{1}{4 \betrag{\V{k}}} \int_0^\infty \ud l \left( - \ln ( ( 2 \betrag{\V{k}} - i \epsilon) l - \betrag{\V{k}}^2 ) + \ln ( ( 2 \betrag{\V{k}} + i \epsilon) l + \betrag{\V{k}}^2 ) \right. \\ \left. + \ln ( (2 \betrag{\V{k}} - i \epsilon) l + \betrag{\V{k}}^2 ) - \ln ( ( 2 \betrag{\V{k}} + i \epsilon) l - \betrag{\V{k}}^2 ) \right) \cos y_0 l.
\end{multline*}
We compute this by introducing a cutoff $L$. We have
\begin{multline*}
  \int_0^L \ud l \ \ln (a l + b) \cos c l = \frac{1}{c} \left( - \ci \frac{bc}{a} \sin \frac{bc}{a} + \ci \frac{(b+aL)c}{a} \sin \frac{bc}{a} \right. \\ + \left. \ln (b+aL) \sin(cL) + \cos \frac{bc}{a} \si \frac{bc}{a} - \cos \frac{bc}{a} \si \frac{(b+aL)c}{a} \right).
\end{multline*}
Using this for the expression above, it is easy to see that the terms involving $L$ cancel each other in the limit $L \to \infty$. In the remaining terms, one may take the limit $\epsilon \to 0$. We are then left with
\begin{equation*}
  \frac{(2\pi)^{-2}}{(k \sigma)_0 \betrag{\V{k}}} \left( - \ci \frac{ \betrag{\V{k}} (k \sigma)_0}{2} \sin \frac{ \betrag{\V{k}} (k \sigma)_0}{2} + \cos \frac{ \betrag{\V{k}} (k \sigma)_0}{2} \si \frac{ \betrag{\V{k}} y_0}{2} \right)
\end{equation*}
Setting $y = k \sigma$ and extending it in a Lorentz invariant way, we obtain
\begin{equation*}
  \frac{-(2\pi)^{-2}}{\sqrt{\betrag{(k \sigma)^2 k^2}}} \left( \ci \frac{ \sqrt{ \betrag{ k^2 (k \sigma)^2}}}{2} \sin \frac{ \sqrt{ \betrag{ k^2 (k \sigma)^2}}}{2} - \cos \frac{ \sqrt{ \betrag{ k^2 (k \sigma)^2}}}{2} \si \frac{ \sqrt{ \betrag{ k^2 (k \sigma)^2}}}{2} \right).
\end{equation*}

\chapter*{Danksagung}

F\"ur die fachkundige Betreuung dieser Arbeit m\"ochte ich mich ganz herzlich bei Professor Fredenhagen bedanken. Vielen Dank auch an Claus D\"oscher f\"ur die gute Zusammenarbeit. F\"ur hilfreiche Anregungen und Diskussionen bedanke ich mich bei Dorothea Bahns, Romeo Brunetti, J\"org J\"ackel und Christoph L\"udeling.

Grosser Dank geb\"uhrt Lutz Osterbrink und Martin Porrmann f\"ur das sorgf\"altige Korrekturlesen.

Tante grazie anche a Sergio Doplicher per l'ospilit\`a e l'assistenza a Roma.

Vielen Dank f\"ur die gemeinsam verbrachte Zeit an alle Mitglieder der Arbeitsgruppe ``Algebraische Quantenfeldtheorie'' sowie an Jan Hamann.

Diese Arbeit wurde erm\"oglicht durch ein Stipendium des Graduiertenkollegs ``Zuk\"unftige Entwicklungen in der Teilchenphysik''. Auch daf\"ur vielen Dank.

Ein besonderes Dankesch\"on geht an Judith f\"ur ihre Liebe und Unterst\"utzung.

\end{document}